\documentclass[12pt,letterpaper]{article}
\usepackage{bbm}
\usepackage[toc,page]{appendix}
\usepackage{feynmf}
\usepackage{feyn}
\usepackage{amsmath, amsthm, amsfonts, amssymb}
\usepackage{mathrsfs} 
\usepackage{hyperref}
\usepackage{cleveref}
\usepackage{slashed}
\usepackage{dsfont}
\usepackage{wrapfig}
\usepackage{minibox}
\usepackage{graphicx}
\usepackage{graphics}
\usepackage{epsfig}
\usepackage{xcolor}
\usepackage{array}
\usepackage{tikz}
\usepackage[greek,english]{babel}
\usetikzlibrary{decorations.pathmorphing,calc}
\usepackage{authblk}

\def\a{\alpha}
\def\b{\beta}
\def\g{\gamma}

\def\d{\delta}
\def\ve{\varepsilon}
\def\m{\mu}
\def\n{\nu}
\def\l{\lambda}

\def\thintablerule{\hrule height0.4pt}

\newcommand{\be}{\begin{equation}}
\newcommand{\ee}{\end{equation}}
\newcommand{\bea}{\begin{eqnarray}}
\newcommand{\eea}{\end{eqnarray}}
\newcommand{\bal}{\begin{aligned}}
\newcommand{\eal}{\end{aligned}}
\newcommand{\eq}[1]{Eq.~(\ref{#1})}
\newcommand{\fig}[1]{Fig.~\ref{#1}}

\newcommand{\sect}[1]{Section~\ref{#1}}

\numberwithin{equation}{section}

\allowdisplaybreaks

\begin{document}

\centerline
{\LARGE Renormalization of the Abelian-Higgs Model}
{\LARGE\centerline  {in the $R_\xi$ and Unitary gauges and}} 
{\LARGE\centerline  {the physicality of its scalar potential}}

\vskip 2 cm
\centerline{\large Nikos Irges and Fotis Koutroulis}
\vskip1ex
\vskip.5cm
\centerline{\it Department of Physics}
\centerline{\it National Technical University of Athens}
\centerline{\it Zografou Campus, GR-15780 Athens, Greece}
\vskip .4cm
\begin{center}
{\it e-mail: irges@mail.ntua.gr, fkoutroulis@central.ntua.gr}
\end{center}
\vskip 1.5 true cm
\centerline {\bf Abstract}
\vskip 3.0ex
\thintablerule
\vskip 2.0ex

We perform an old school, one-loop renormalization of the Abelian-Higgs model in the Unitary and $R_\xi$ gauges,
focused on the scalar potential and the gauge boson mass. 
Our goal is to demonstrate in this simple context the validity of the Unitary gauge at the quantum level,
which could open the way for an until now (mostly) avoided framework for loop computations.
We indeed find that the Unitary gauge is consistent and equivalent to the $R_\xi$ gauge at the level of $\b$-functions.
Then we compare the renormalized, finite, one-loop Higgs potential in the two gauges and we again find equivalence.
This equivalence needs not only a complete cancellation of the gauge fixing parameter $\xi$ from the $R_\xi$ gauge potential
but also requires its $\xi$-independent part to be equal to the Unitary gauge result.
We follow the quantum behaviour of the system by plotting Renormalization Group trajectories and 
Lines of Constant Physics, with the former the well known curves and with the latter, determined by the finite
parts of the counter-terms, particularly well suited for a comparison with non-perturbative studies.

\vskip 2.0ex
\thintablerule

\vskip-0.2cm
\newpage

\tableofcontents

\tikzset{
photon/.style={decorate, decoration={snake}, draw=black},
electron/.style={draw=black, postaction={decorate},
decoration={markings,mark=at position .55 with {\arrow[draw=black]{>}}}},
gluon/.style={decorate, draw=black,
decoration={coil,amplitude=4pt, segment length=5pt}} 
}

\newpage
\section{Introduction}

Spontaneously broken gauge theories are of great physical interest, notably because of the 
Brout-Englert-Higgs (or simply Higgs) mechanism \cite{BEH}.
Even though classicaly there is a simple qualitative description of the mechanism that one
can find in textbooks, at the quantum level where these theories are more precisely defined, ambiguities arise. 
These ambiguities are related to the question of whether one can consider the quantum scalar potential itself
as a physical quantity. In this work we touch on this issue approaching it from two angles, in the context of the Abelian-Higgs model \cite{AbHiggs}.

One angle of approach has to do with the basis of representation for the Higgs field.
Recall that in most loop calculations the (complex) Higgs field is written in a Cartesian basis: $H=\phi_1+i\phi_2$ \cite{Peskin}.
The real part of the field $\phi_1$ is the physical Higgs particle, while $\phi_2$ represents the unphysical Nambu-Goldstone (or simply Goldstone) boson.
For dimensional reasons the resulting $\b$-functions for the Higgs mass and quartic coupling must have the form 
$\b^{\rm Cart.}_{m_H}=c_m^{\rm Cart.} \l m_H^2+\cdots$ and $\b^{\rm Cart.}_{\l}=c_\l^{\rm Cart.}\l^2+\cdots$ respectively.
The numerical coefficients $c_m^{\rm Cart.}$ and $c_\l^{\rm Cart.}$ are determined after the loop calculation has been performed. 
The former coefficient affects the behaviour of the Higgs mass under Renormalization Group (RG) flow and the latter that of the quartic coupling, that
in turn affect triviality and in the presence of fermions, also instability bounds. Taking these bounds seriously at a quantitative level means that the scalar potential
is considered to be a physical quantity. Here we perform our computations in a Polar basis, $H=\phi e^{i\chi/v}$, where now
the physical Higgs field is $\phi=\sqrt{\phi_1^2+\phi_2^2}$ and $\chi$ is the Goldstone boson ($v$ is some vacuum expectation value).
In this basis one expects to find $\b^{\rm Pol.}_{m_H}=c_m^{\rm Pol.} \l m_H^2+\cdots$ and $\b^{\rm Pol.}_{\l}=c_\l^{\rm Pol.}\l^2+\cdots$
with a pending question if $c_m^{\rm Cart.} $ agrees with $c_m^{\rm Pol.}$ and if $c_\l^{\rm Cart.}$ agrees with $c_\l^{\rm Pol.}$.
We find that these coefficients do not exactly agree, introducing possibly a small but computable 
ambiguity in the RG flows and the above mentioned bounds. The Polar basis may have though also a deeper impact on the quantum potential,
having to do with its gauge (in)dependence. Some of these issues were noticed already in \cite{Tye} in the context of the effective potential.

The other angle of our approach has to do with the quantization scheme. In particular, there is an infinitum of possible gauge fixing functions that one can use during quantization,
each one of them introducing at least one gauge fixing parameter, say $\xi$.  A typical representative of these schemes is $R_\xi$ gauge fixing. 
It is a well known fact that any sensible quantization scheme
should produce a gauge independent set of physical quantities. Such quantities are for sure the masses of physical fields and the independent dimensionless
couplings like $\l$. Gauge independence of physical quantities in a scheme like the $R_\xi$ gauge fixing scheme 
is then ensured if the corresponding $\b$-functions are $\xi$-independent and this is indeed the case 
in every consistent loop calculation. The scalar effective potential computed with the background field method on the other hand is notoriously known to be $\xi$-dependent
already at one-loop, which renders its physicality (and the relevance of the precision triviality and instability bounds derived from it) a delicate matter \cite{EffectivePotentialSM,Schwartz}.
It is therefore an open issue how to define an unambiguous, physical, quantum potential in spontaneously broken gauge theories.
For this reason, we choose the Unitary gauge \cite{Weinberg} as our quantization scheme. The Unitary gauge is one where only physical
fields propagate and it is commonly used in textbooks in order to demonstrate the physical spectrum of spontaneously broken gauge theories at the classical level.
It is rarely used though for loop calculations, in fact we are not aware of a complete renormalization work in this gauge.
A possible obstruction to completing reliably such a program may be the high momentum behaviour of the Unitary gauge boson propagator resulting in
integrals that often diverge worse than quadratically with a cut-off, even in a renormalizable theory. We call these integrals, "$U$-integrals".
For this reason the Unitary gauge is sometimes called a non-renormalizable gauge.
A necessary condition therefore for a Unitary gauge calculation to make sense is that the physical quantities that are $\xi$-independent 
in the $R_\xi$ gauge, to coincide with the corresponding quantities derived in the Unitary gauge.
If this condition is fulfilled then through the renormalization procedure one automatically obtains a version of the scalar potential, 
the Unitary gauge scalar potential, that is by definition gauge independent. 
A question we would like to answer here is if this version of the scalar potential can be used, in principle, to derive competitive with respect to the $R_\xi$ gauge
physical predictions. Of course, an important issue is to understand the connection between the Unitary and $R_\xi$ gauge potentials.
The standard connection between these two gauges is to take the "Unitary gauge limit" $\xi\to\infty$ at the level of the Feynman rules,
before loop integrals have been performed. Clearly this is not what we want to do here. Instead, we would like to compare the $R_\xi$ gauge potential
with the Unitary gauge potential, after loop integration. This is non-trivial, as the $\xi\to\infty$ limit and the loop integration 
may not commute. For recent studies of related issues, in the context of the $H\to \g\g$ decay, see \cite{Marciano,Jegerlehner,Dedes,WuWu}
and for some earlier works on the Unitary gauge and the Abelian-Higgs model, see \cite{Tye,Sonoda}.

In this work, we consider the Abelian-Higgs model and perform the one-loop renormalization of the gauge boson mass and of the
Higgs potential in both the $R_\xi$ and Unitary gauges. In addition to the above mentioned comparison reasons, this double computation allows us 
to monitor and ensure the credibility of the Unitary gauge calculation.
We will be able to show that as far as the $\b$-functions (determined by the divergent parts of the one-loop amplitudes) is concerned, the Unitary
gauge is equally consistent, in fact equivalent to the $R_\xi$ gauge. 
Then we consider the scalar potential (determined by the finite parts of the amplitudes after renormalization) and ask whether it could also be physical. 
Let the finite part of the one-loop value of a quantity $\star$ be defined as $( \star )_f$ with the subscript $f$ denoting finite part. 
Such quantities will be the various loop corrections entering in the renormalized Higgs potential.
Then schematically we have that in general
\be\label{starcomm}
\left[\left (\star \right)_f, \lim_{\xi\to\infty} \right] = \lim_{\xi\to\infty} g(\l,m_H,m_Z,\m,\xi)\, ,
\ee
where the behaviour of the function $g$ at $\xi=\infty$ is one thing we would like to understand.
The background field method in the ${\overline {\rm MS}}$ scheme for example gives an effective potential where 
$\lim_{\xi\to\infty} g (\xi) = \infty$  for a large class of gauge fixing functions \cite{Schwartz}, rendering the Unitary gauge limit after loop integration, singular.
The singularity implies that the Unitary gauge is disconnected from the space of $R_\xi$ gauges and essentially that
the Higgs potential is unphysical away from its extrema. Our calculation instead shows that $\lim_{\xi\to\infty} g (\xi) = 0$, perhaps the most striking result of this work,
as it implies that the Higgs potential can be made gauge invariant, hence physical, even away from its extrema. For formal aspects of the quantization of the Abelian-Higgs model, see \cite{B.Lee, Stora}

In Section 2 we review the classical Abelian-Higgs model and discuss some basic conventions in our calculation.
In Section 3 we perform in detail the $R_\xi$-gauge computation and in Section 4 the Unitary gauge computation.
In Section 5 we renormalize the AH model. 
In Sections 6 and 7 we present a numerical analysis of our results.
In Section 8 we state our conclusions.
We also have a number of Appendices where auxiliary material can be found.

\section{The classical theory and some basics}
\label{basics}

The bare Lagrangean of the AH model is
\be\label{clas.Lag.1}
{\cal L}_0 =  - \frac{1}{4}F_{0,\mu \nu }^2  
+ {\left| {{D_\mu }H_0} \right|^2} + m_0^2{\left| H_0 \right|^2} - \lambda_0{\left| H_0 \right|^4}\, .
\ee
Zero subscripts or superscripts denote bare quantities.
As usual, the covariant derivative is $D_\m = \partial_\m + i g_0 A_\m^0$ and the gauge field strength is $F_{0,\m\n}=\partial_\m A_\n^0 - \partial_\n A_\m^0$.
The Higgs field $H_0$ is a complex scalar field.
The Lagrangean is invariant under the $U(1)$ gauge transformations
\bea
A^0_\m(x) &\to& A^0_\m(x) + \frac{1}{g_0} \partial _\mu \theta(x) \nonumber\\
H_0(x)  &\to&  H_0(x)e^{i \theta(x)}
\eea
with $\theta(x)$ a gauge transformation function and the discrete, global $Z_2$  symmetry
\bea
H_0(x)  &\to&  -H_0(x) 
\eea
We assume that both $m_0$ and $\l_0$ are positive quantities.

As it stands, the part of the Lagrangean that corresponds to a potential
\be
V_0 = - m_0^2{\left| H_0 \right|^2} + \l_0{\left| H_0 \right|^4}
\ee
triggers SSB. Minimization yields the vev
\be
\left\langle H_0 \right\rangle  =  \pm  \frac{m_0}{\sqrt{2 \lambda_0}} = \pm \frac{v_0}{\sqrt{2}}\, .
\ee
The second of the above equations defines the bare vacuum expectation value (vev) parameter $v_0$.
The field $H_0$ can be expanded around its vev as
\be\label{PolarH}
H_0\left( x \right) = \frac{{\left( {v_0+\phi_0 (x)} \right){e^{i \frac{\chi_0(x)}{v_0} }}}}{{\sqrt 2 }}\, ,
\ee
where now $\phi_0$ is the new Higgs field fluctuation and $\chi_0$ is a massless Goldstone boson.
\eq{PolarH} is the so called "Polar basis" for the Higgs field, as opposed to the "Cartesian basis"
$H_0(x) = 1/\sqrt{2}(v_0 + \phi_{1,0}(x) + i \phi_{2,0}(x))$.
The former is typically used for demonstrating the physical spectrum while the latter for quantization.
Here we will stick to the Polar basis for both.
Replacing the vev into Eq.\eqref{clas.Lag.1}, the Lagrangean takes the form 
\bea\label{clas.Lag.2}
{\cal L}_0 &=& - \frac{1}{4}F_{0,\mu \nu }^2 + \frac{1}{2}\left( {{\partial _\mu }\phi_0 } \right)\left( {{\partial ^\mu }\phi_0 } \right) + 
\frac{1}{2}\left( {{\partial _\mu }\chi_0 } \right)\left( {{\partial ^\mu }\chi_0 } \right)  
 + \frac{1}{2}m_{Z_0}^2{A^0_\mu }{A^{0\mu} } \nonumber\\
&+& m_{Z_0}(\partial ^\mu \chi_0) A^0_\m +2 \frac{m_{Z_0}}{m_{H_0}}\sqrt{2 \l_0} A^0_\m (\partial ^\mu \chi_0) \phi_0 + 
\frac{\sqrt{2 \l_0}}{m_{H_0}}(\partial _\mu \chi_0)^2 \phi_0 + \frac{2 \l_0 m_{Z_0}}{m^2_{H_0}} A^0_\m (\partial ^\mu \chi_0)  \phi_0^2  \nonumber\\
&+&  \frac{ \l_0 }{m^2_{H_0}} (\partial _\mu \chi_0)^2 \phi_0^2  + {{{g^{\mu \nu }}}}{}{\frac{ \l_0 m^2_{Z_0}}{m^2_{H_0}}}{A^0_\mu }{A^0_\nu } {{\phi_0 ^2} {}} 
+ {g^{\mu \nu }}{\frac{m^2_{Z_0}}{m_{H_0}}\sqrt{2 \l_0}}\phi_0 {A^0_\mu }{A^0_\nu }   \nonumber\\
&-& \frac{1}{2}{m_{H_0}^2}{\phi_0 ^2} -  \sqrt{\frac{\lambda_0}{2}} {m_{H_0}}{\phi_0 ^3} - \frac{\lambda_0 }{{4}}{\phi_0 ^4} +  const.\, ,
\eea
with $g^{\m\n}$ the Minkowski space metric. The bare gauge boson ($Z$ boson) mass is defined as $m_{Z_0} = g_0 v_0$
and the bare Higgs mass as  ${m_{H_0} = \sqrt {2} m_0} = \sqrt{2\l_0} v_0 $.
As implied by the above expression, we have decided to use as independent parameters the masses and the quartic coupling $\l_0$.
This means that we have eliminated the gauge coupling according to $g_0 = \frac{m_{Z_0}}{m_{H_0}}\sqrt{2 \l_0}$ 
and the vev according to $v_0 = \frac{m_{H_0}}{\sqrt{2 \l_0}}$ wherever they appear.
Setting the Goldstone field $\chi_0$  to zero, gives us the Unitary gauge Lagrangean which will be the topic of a separate section.
For now we will keep the Goldstone in the spectrum. A consequence of having a Goldstone in the spectrum is a mixing term between the $Z$ and $\chi_0$ in \eq{clas.Lag.2}.

We will compute all one-loop Feynman diagrams that contribute to the renormalization of the $Z$ mass and of the scalar potential, using Dimensional Regularization (DR) \cite{DR}.
We assume basic knowledge of the DR technology that we therefore do not review here, except from some necessary basic facts that can be found in the Appendices.
The renormalization scale parameter of DR is denoted by $\m$.
The small expansion parameter $\ve$ of DR is defined via
\be
\ve = 4-d \, .
\ee
Since we are using DR in our renormalization scheme, a consequence is that the trace of the metric is $g_{\m\n} g^{\m\n} = d $. 

Now, each diagram comes with a symmetry factor.
Consider a one-loop diagram containing \textit{n} vertices with $k_n$ lines on each vertex. These $k_n$ lines are 
divided into $k^{in}_n$ and $k^{ext}_n$ for the internal and the external lines respectively. 
The procedure to obtain the correct symmetry factor is given for example in \cite{Rodigast}:  
\begin{itemize}
\item 
For each of the \textit{n} vertices, count all possible ways that the $k_n$ lines can be connected to
the external legs of a given diagram. This is $k^{ext}_n$. 
Doing this for every vertex defines $n_O$ as the product 
of the $k^{ext}_n$'s. The remaining lines belong to $k^{in}_n$.
\item 
At each of the \textit{n} vertices there are $k^{in}_n$ lines. Count all the possible ways
the loop can be constructed using these lines. The product of the $k^{in}_n$'s defines $n_I$.
\item 
For each of the \textit{n} vertices, count the number of $k_i$ lines that are equivalent. This defines $\ell_i $.
\item 
Finally, for a given diagram count all the possible equivalent vertices of type \textit{j}, defining $v_j$.
\end{itemize}
The symmetry factor of the diagram is then
\bea\label{sym.fact.}
{\cal S}^a_{b}= \frac{{{n_O}{n_I}}}{{\prod\limits_i {{\ell _i}}! \prod\limits_j {{v_j}}! }}\, ,
\eea
where $a$, $b$ are indicators specifying the diagram. 

A large part of the one-loop diagram expressions is dominated by the set of basic integrals called Passarino-Veltman (PV) integrals \cite{PV}.
We collect in Appendix \ref{PassarinoVeltman} the basics of the formulation of PV integrals, following mostly \cite{Kunszt}.
In our calculation several non-standard integrals emerge as well. The divergent ones we call collectively $U$-integrals and we compute them in Appendix \ref{Uint}.
Finite integrals and finite parts of divergent integrals are harder to classify systematically so we will be dealing with them as we proceed.

We also introduce some useful notation.
Our convention for the name of a one-loop Feynman diagram $F$ is
\be
F^{G,L}_{E}
\ee
$G=R_\xi, U$ specifies either the $R_\xi$ or the Unitary gauge where the diagram is computed. $L$ is a list containing the field(s) running 
in the loop in the direction of the loop momentum flow, starting from the vertex on the left side of the diagram. In case the diagram is an irreducible Box,
we start the list from the upper left vertex. $E=H, Z$ specifies the external legs. In the case $E=Z$, there may be additional Lorentz indices following $E$.
We know that one-loop Feynman diagrams can be either finite or divergent. In the first case the corresponding integrals contain 
only finite terms and in the second case they include both divergent and finite parts. Furthermore, in each of the above cases, in the $R_\xi$ gauge, the corresponding 
parts could be either gauge (i.e. $\xi$) dependent or gauge independent. In the Unitary gauge there is no such distinction.
Therefore, every set of one-loop diagrams contributing to the same correlator, where the sum over the index $L$ has been performed and the
Lorentz indices (if present) have been appropriately contracted, can be expressed as
\bea\label{general FD}
(4\pi)^{d/2} {F}^{G}_{E} = \m^\ve \left( \left[ F^{G}_{E} \right]_{\ve} + \left\{F^{G}_{E} \right\}_{\ve} + \left[ F^{G}_{E} \right]_{f} + \left\{ F^{G}_{E} \right\}_{f} \right) \, ,
\eea
where the square brackets denote $\xi$-independent part, the curly brackets 
denote $\xi$-dependent part and the subscripts $\ve$ and $f$ denote divergent and finite part respectively. 
The $1/\ve$ factor is absorbed in the definitions of $[ F^{G}_{E} ]_{\ve}$ and $\left\{F^{G}_{E} \right\}_{\ve}$.
A word of caution here is that the above separation of diagrams into $\xi$-independent and $\xi$-dependent parts is clearly not unique. 
Nevertheless it is a very useful notation (once we get used to it) since it allows to perform algebra with diagrams easily and also
facilitates the comparison with the Unitary gauge. In fact, in most cases the $\xi$-independent part of sums of diagrams is just the corresponding Unitary gauge result.
The sum of the gauge independent and gauge dependent finite parts is denoted as
\be\label{entirefinite}
\left( F^{G}_{E} \right)_{f} = \left[ F^{G}_{E} \right]_{f} + \left\{ F^{G}_{E} \right\}_{f}\, .
\ee
Round, square and curly brackets appearing in any other context have their usual meaning.
All entirely finite integrals are computed as described in Appendix \ref{FiniteIntegrals}.
We will be giving the divergent parts and the finite parts of sums of groups of diagrams explicitly in the main text, leaving some of the (increasingly cumbersome) expressions 
for finite (parts of) diagrams to Appendix \ref{FiniteParts}. There are some finite parts sitting inside the $U$-integrals too, which we compute
together with their divergent parts and do not show explicitly, in order to avoid repetitions.

We perform renormalization away from the usual ${\overline {\rm MS}}$ scheme.
This means that our renormalization conditions force us to keep some non-trivial finite terms, that eventually enter
in the renormalized Higgs potential. What finite terms are kept is of course not a unique choice.
We have not checked if the gauge invariance of the potential could also be achieved in a 
pure ${\overline {\rm MS}}$ scheme. If however ${\overline {\rm MS}}$ (or any other scheme) would result in a gauge dependent potential, 
it would mean that in spontaneously broken gauge theories scheme dependence kicks in already at one loop.
Recall that in QCD (a not spontaneously broken theory) scheme dependence appears at three loops.
This would be strange because it would relate gauge dependence to scheme dependence.
Our feeling is that any physical renormalization subtraction scheme applied to our calculation should produce a
gauge invariant Higgs potential.

Furthermore, we perform renormalization off-shell, at zero external momenta. 
External momenta will be generically denoted by $p$.
In 2-point, 3-point and 4-point functions zero external momenta means $p_i=0$ with $i=1,2$, $i=1,2,3$ and $i=1,2,3,4$ respectively.
This choice, beyond being a huge simplifying factor, is justified since we are interested in terms of the Lagrangean
without derivatives. It would be equally strange if after renormalization the Higgs potential would pick up an
external momentum dependence. In other words we believe that even if we had performed renormalization
on-shell for example, we would have obtained the same results, alas in a more complicated way:
not only individual diagrams would become much more involved, but also non-1PI diagrams may have to be added
in order to arrive at gauge invariant $\b$-functions.
A physical argument for zero external momenta could be that if one is interested in the high energy limit, is entitled to set the masses 
of external particles equal to zero, that is $p_i^2=0$.
But this choice, using momentum conservation in 2, 3 and 4-point functions is the same as setting the momenta themselves to zero.
A subtle point of setting external momenta to zero is that in a diagram
a term of the form $p_\m p_\n T^{\m\n} (p,m_i)$, with $p$ an external momentum, $m_i$ mass parameters and $T^{\m\n}$ a tensor (DR) integral,
may appear. If we set $p=0$ before integration, this term is zero. If on the other hand we do the integral first, then contract and then
take the limit $p\to 0$, we may find a non-zero result. 
The latter procedure is the correct one.
Expressions of the form of \eq{general FD} in the main text will be thus given at zero external momenta.
Nevertheless, for completeness and for potential future use, we collect some on-shell expressions in Appendix \ref{Onshell}.\\

\section{$R_\xi$ gauge}\label{AHL}

Quantization of \eq{clas.Lag.2} requires gauge fixing. The gauge fixing term 
\be
{\cal L}_{\rm gf} = \frac{1}{2\xi}( {\partial ^\mu }A^0_\m - \xi g_0 v_0 \chi_0 )^2
\ee
that removes the Goldstone-gauge boson mixing defines the $R_\xi$ gauge and under the standard Faddeev-Popov procedure introduces the extra ghost contribution
\bea
{\cal L}_{ghost} =  \left( {{\partial _\mu }\bar c } \right)\left( {{\partial ^\mu }c } \right)  - \xi m^2_{Z_0} \bar c c \, .
\eea
The sum ${\cal L}_{R_\xi} = {\cal L}_0 + {\cal L}_{\rm gf} + {\cal L}_{ghost}$
\bea\label{LRx}
{\cal L}_{R_\xi}&=& - \frac{1}{4}F_{\mu \nu }^2 + \frac{1}{2}\left( {{\partial _\mu }\phi_0 } \right)\left( {{\partial ^\mu }\phi_0 } \right) + \frac{1}{2}\left( {{\partial _\mu }\chi_0 } \right)
\left( {{\partial ^\mu }\chi_0 } \right)  - \frac{1}{{2 {  \xi } }}{\left( {{\partial ^\mu }A^0_\mu} \right)^2}
 + \frac{1}{2}m_{Z_0}^2{A^0_\mu }{A^{0\mu} } \nonumber\\
&+& 2 \frac{m_{Z_0}}{m_{H_0}}\sqrt{2 \l_0} A^0_\m (\partial ^\mu \chi_0) \phi_0 + \frac{\sqrt{2 \l_0}}{m_{H_0}}(\partial _\mu \chi_0)^2 \phi_0 + 
\frac{2 \l_0 m_{Z_0}}{m^2_{H_0}} A^0_\m (\partial ^\mu \chi_0)  \phi_0^2  + \frac{ \l_0 }{m^2_{H_0}} (\partial _\mu \chi_0)^2 \phi_0^2 \nonumber\\
&+& {{{g^{\mu \nu }}}}{}{\frac{ \l_0 m^2_{Z_0}}{m^2_{H_0}}}{A^0_\mu }{A^0_\nu } {{\phi_0 ^2} {}} + 
{g^{\mu \nu }}{\frac{m^2_{Z_0}}{m_{H_0}}\sqrt{2 \l_0}}\phi_0 {A^0_\mu }{A^0_\nu } -\frac{1}{2}\xi m^2_{Z_0} \chi_0^2  \nonumber\\
&-& \frac{1}{2}{m_{H_0}^2}{\phi_0 ^2} -  \sqrt{\frac{\lambda_0}{2}} {m_{H_0}}{\phi_0 ^3} - \frac{\lambda_0 }{{4}}{\phi_0 ^4} + {\cal L}_{ghost} + const.
\eea
yields the final expression from which the $R_\xi$ gauge Feynman rules can be derived.
Notice that the Goldstone has acquired an unphysical, gauge dependent mass $m_{\chi_0} = \sqrt{\xi} m_{Z_0}$.
With this gauge fixing choice the ghost fields have apart from a kinetic term, a mass term equal to that of the Goldstone boson's but they are not coupled to the Higgs field or to the gauge boson. 

The Feynman rules arising from Eq.\eqref{LRx} are the following:
\begin{itemize}
\item Gauge boson propagator
\begin{center}
\begin{tikzpicture}[scale=0.8]
\draw[photon] (0,-0.19)--(2.5,-0.19) ;
\node at (6,0) {$=\displaystyle
 \frac{i \left(-g^{\m\n} + \frac{(1-\xi)k^\m k^\n}{k^2 - \xi m^2_{Z_0}}  \right)}{k^2 - m^2_{Z_0} + i\varepsilon }$};
\end{tikzpicture}
\end{center}
\item Higgs propagator 
\begin{center}
\begin{tikzpicture}[scale=0.8]
\draw[dashed,thick] (-1,0)--(1.5,0) ;
\node at (5,0) {$=\displaystyle \frac{i}{{{k^2} - m_{H_0}^2 + i\varepsilon }} $};
\end{tikzpicture}
\end{center}
\item Goldstone propagator
\begin{center}
\begin{tikzpicture}[scale=0.8]
\draw[] (-1,0)--(1.5,0) ;
\node at (5,0) {$=\displaystyle \frac{i}{{{k^2} - \xi m_{Z_0}^2 + i\varepsilon }} $};
\end{tikzpicture}
\end{center}
\item Ghost propagator
\begin{center}
\begin{tikzpicture}[scale=0.8]
\draw[->,thick] (-0.1,0)--(0.25,0);
\draw[] (-1,0)--(1.5,0) ;
\node at (5,0) {$=\displaystyle \frac{i}{{{k^2} - \xi m_{Z_0}^2 + i\varepsilon }} $};
\end{tikzpicture}
\end{center}
\item $\phi$-$\chi$-$Z$ vertex
\begin{center}
\begin{tikzpicture}[scale=0.7]
\draw [dashed,thick] (-2.5,1.5)--(-1,0);
\draw [] (-2.5,-1.5)--(-1,0);
\draw [<-] (-1.5,0.8)--(-2.2,1.5);
\node at (-2.5,1.1) {$p_1$};
\draw [->] (-1.5,-0.8)--(-2.2,-1.5);
\node at (-2.5,-1) {$k_a$};
\draw [->] (-0.5,0.2)--(0.3,0.2);
\node at (0,-0.4) {$p_2$};
\draw[photon] (-1,0)--(1,0) ;
\node at (4,0) {$= \displaystyle -\frac{2m_{Z_0}}{m_{H_0}}\sqrt{2 \l_0} k_a^\mu$};
\end{tikzpicture}
\end{center}
\item $\phi$-$Z$-$Z$ vertex
\begin{center}
\begin{tikzpicture}[scale=0.7]
\draw [photon] (-2.5,1.5)--(-1,0);
\draw [photon] (-2.5,-1.5)--(-1,0);
\draw [<-] (-1.5,0.8)--(-2.2,1.5);
\node at (-2.6,1.1) {$p_1$};
\draw [<-] (-1.5,-0.8)--(-2.2,-1.5);
\node at (-2.6,-1.1) {$p_2$};
\draw [->] (-0.5,0.2)--(0.3,0.2);
\node at (0,-0.3) {$p_3$};
\draw[dashed,thick] (-1,0)--(1,0) ;
\node at (4,0) {$=  \displaystyle 2i{g^{\mu \nu }}{\frac{m^2_{Z_0}}{m_{H_0}}\sqrt{2 \l_0}}$};
\end{tikzpicture}
\end{center}
\item $\phi$-$\phi$-$\phi$ vertex
\begin{center}
\begin{tikzpicture}[scale=0.7]
\draw [dashed,thick] (-2.5,1.5)--(-1,0);
\draw [dashed,thick] (-2.5,-1.5)--(-1,0);
\draw [<-] (-1.5,0.8)--(-2.2,1.5);
\node at (-2.6,1.1) {$p_1$};
\draw [<-] (-1.5,-0.8)--(-2.2,-1.5);
\node at (-2.6,-1.1) {$p_2$};
\draw [->] (-0.5,0.2)--(0.3,0.2);
\node at (0,-0.3) {$p_3$};
\draw[dashed,thick] (-1,0)--(1,0) ;
\node at (4,0) {$= \displaystyle  - 6i\sqrt{\frac{\lambda_0}{2}} {m_{H_0}}$};
\end{tikzpicture}
\end{center}
\item $\phi$-$\chi$-$\chi$ vertex
\begin{center}
\begin{tikzpicture}[scale=0.7]
\draw [] (-2.5,1.5)--(-1,0);
\draw [] (-2.5,-1.5)--(-1,0);
\draw [<-] (-1.5,0.8)--(-2.2,1.5);
\node at (-2.7,1.1) {$k^a$};
\draw [->] (-1.5,-0.8)--(-2.2,-1.5);
\node at (-2.5,-1) {$k^b$};
\draw [->] (-0.5,0.2)--(0.3,0.2);
\node at (0,-0.3) {$p_1$};
\draw[dashed,thick] (-1,0)--(1,0) ;
\node at (4,0) {$=  \displaystyle  2 i g^{\m\n}\frac{\sqrt{2 \lambda_0}}{{m_{H_0}}} k_\m^a k_\n^b$};
\end{tikzpicture}
\end{center}
\item $\phi$-$\phi$-$Z$-$Z$ vertex
\begin{center}
\begin{tikzpicture}[scale=0.5]
\draw [dashed,thick] (0,0)--(1.5,1.4);
\draw [dashed,thick] (0,0)--(1.5,-1.4);
\draw [photon] (-1.5,1.4)--(0,0);
\draw [photon] (-1.5,-1.4)--(0,0);
\draw [->]  (0.7,0.3)--(1.3,0.9);
\node at (1.7,1.7) {$p_3$};
\draw [->]  (0.9,-0.4)--(1.5,-1);
\node at (1.7,-1.7) {$p_4$};
\draw [<-] (-0.9,0.3)--(-1.6,1);
\node at (-1.7,1.7) {$p_1$};
\draw [<-] (-0.8,-0.4)--(-1.5,-1);
\node at (-1.7,-1.7) {$p_2$};
\node at (6,0) {$=\displaystyle 4i{{{} \frac{ \l_0 m^2_{Z_0}}{m^2_{H_0}}g^{\mu \nu }}}$};
\end{tikzpicture}
\end{center}
\item $\phi$-$\phi$-$\phi$-$\phi$ vertex
\begin{center}
\begin{tikzpicture}[scale=0.5]
\draw [dashed,thick] (0,0)--(1.5,1.4);
\draw [dashed,thick] (0,0)--(1.5,-1.4);
\draw [dashed,thick] (-1.5,1.4)--(0,0);
\draw [dashed,thick] (-1.5,-1.4)--(0,0);
\draw [->]  (0.7,0.3)--(1.3,0.9);
\node at (1.7,1.7) {$p_3$};
\draw [->]  (0.9,-0.4)--(1.5,-1);
\node at (1.7,-1.7) {$p_4$};
\draw [<-] (-0.9,0.3)--(-1.6,1);
\node at (-1.7,1.7) {$p_1$};
\draw [<-] (-0.8,-0.4)--(-1.5,-1);
\node at (-1.7,-1.7) {$p_2$};
\node at (6,0) {$= \displaystyle -6i \lambda_0$};
\end{tikzpicture}
\end{center}
\item $\phi$-$\phi$-$\chi$-$Z$ vertex
\begin{center}
\begin{tikzpicture}[scale=0.5]
\draw [photon] (0,0)--(1.5,1.4);
\draw [] (0,0)--(1.5,-1.4);
\draw [dashed,thick] (-1.5,1.4)--(0,0);
\draw [dashed,thick] (-1.5,-1.4)--(0,0);
\draw [->]  (0.7,0.3)--(1.3,0.9);
\node at (1.7,1.7) {$p_3$};
\draw [->]  (0.9,-0.4)--(1.5,-1);
\node at (1.7,-1.7) {$k_a$};
\draw [<-] (-0.9,0.3)--(-1.6,1);
\node at (-1.7,1.7) {$p_1$};
\draw [<-] (-0.8,-0.4)--(-1.5,-1);
\node at (-1.7,-1.7) {$p_2$};
\node at (6,0) {$=  \displaystyle -4{{{} \frac{ \l_0 m_{Z_0}}{m^2_{H_0}}k_a^\mu}}$};
\end{tikzpicture}
\end{center}
\item $\phi$-$\phi$-$\chi$-$\chi$ vertex
\begin{center}
\begin{tikzpicture}[scale=0.5]
\draw [] (0,0)--(1.5,1.4);
\draw [] (0,0)--(1.5,-1.4);
\draw [dashed,thick] (-1.5,1.4)--(0,0);
\draw [dashed,thick] (-1.5,-1.4)--(0,0);
\draw [->]  (0.7,0.3)--(1.3,0.9);
\node at (1.7,1.7) {$k_a$};
\draw [<-]  (0.9,-0.4)--(1.5,-1);
\node at (1.7,-1.7) {$k_b$};
\draw [<-] (-0.9,0.3)--(-1.6,1);
\node at (-1.7,1.7) {$p_1$};
\draw [<-] (-0.8,-0.4)--(-1.5,-1);
\node at (-1.7,-1.7) {$p_2$};
\node at (6,0) {$=  \displaystyle 4ig^{\m\n} \frac{\l_0 }{m^2_{H_0}} k^a_\m k^b_\n$};
\end{tikzpicture}
\end{center}
\end{itemize}
In the above rules, $k_{a,b}$ denotes the momenta of the Goldstone bosons in a vertex. 
In the case that we have two Goldstones in a diagram, we choose the convention where one of them gets in and the other gets out of the vertex
(since we will encounter the Goldstone only in loops).

The momentum dependence of the vertices including at least one Goldstone boson
is a direct consequence of our choice to use the Polar instead of the Cartesian basis for the Higgs field,
as in the latter case there are no such vertices. 
As a result, the corresponding loop integrals will have extra loop momentum factors in their numerator and this
triggers the appearance of $U$-integrals (also) in the $R_\xi$ gauge.

In what follows, we present only the final expressions for the divergent and finite parts of the various sectors given at zero external momenta. The explicit calculation steps that we followed are given in the Appendix \ref{Explicitcalculation}.

\subsection{Tadpoles}\label{Tadpoles in Rx}

One-point functions are also called tadpoles. The first such diagram is
\vskip .5cm
\begin{center}
\begin{tikzpicture}[scale=0.7]
\draw [dashed,thick] (0,0) circle [radius=1];
\draw [dashed,thick] (-2.5,0)--(-1,0);
\draw [->]  (-2.2,0.2)--(-1.5,0.2);
\node at (-2.9,0) {$p$};
\draw [->]  (0.45,-0.6) arc [start angle=-45, end angle=-135, radius=0.6cm];
\node at (0,-0.4) {$k$};
\node at (3,0) {$=\, \, i{\cal T}_H^{R_\xi,\phi}$};
\end{tikzpicture}
\end{center}
and analytically evaluates to
\be\label{final Tad1Rx}
{\cal T}_H^{R_\xi, \phi} ={3  \sqrt{\frac{\l_0}{2}}m_{H_0}{\mu ^{\ve}}}A_0(m_{H_0}) \, .
\ee

The next tadpole comes with a gauge boson loop:
\vskip .5cm
\begin{center}
\begin{tikzpicture}[scale=0.7]
\draw [photon] (0,0) circle [radius=1];
\draw [dashed,thick] (-2.5,0)--(-1,0);
\draw [->]  (-2.2,0.2)--(-1.5,0.2);
\node at (-2.9,0) {$p$};
\draw [->]  (0.45,-0.5) arc [start angle=-45, end angle=-135, radius=0.6cm];
\node at (0,-0.3) {$k$};
\node at (3,0) {$=\, \, i{\cal T}_H^{R_\xi, Z}$};
\end{tikzpicture}
\end{center}
and it is equal to
\bea\label{final Tad2Rx}
{\cal T}_H^{R_\xi, Z} &=& \frac{m^2_{Z_0}}{m_{H_0}}\sqrt{2 \l_0} {\mu^{\ve}} \Biggl \{  3A_0(m_{Z_0}) + \xi A_0(\sqrt{\xi} m_{Z_0}) \Biggr \}\, . 
\eea
The last tadpole has a Goldstone loop:
\vskip .5cm
\begin{center}
\begin{tikzpicture}[scale=0.7]
\draw [] (0,0) circle [radius=1];
\draw [dashed,thick] (-2.5,0)--(-1,0);
\draw [->]  (-2.2,0.2)--(-1.5,0.2);
\node at (-2.9,0) {$p$};
\draw [->]  (0.45,-0.6) arc [start angle=-45, end angle=-135, radius=0.6cm];
\node at (0,-0.4) {$k$};
\node at (3,0) {$=\, \, i{\cal T}_H^{R_\xi,\chi}$};
\end{tikzpicture}
\end{center}
Using Eq.\eqref{gmnJmn1} where ${U}_{\cal T}(m_{\chi_0})$ is calculated we obtain that the above diagram is equal to 
\be\label{final Tad3Rx}
{\cal T}_H^{R_\xi, \chi}= -\frac{\sqrt{2 \l_0}}{m_{H_0}} \m^\ve m_{\chi_0}^2 A_0(m_{\chi_0}) \, .
\ee
The total tadpole value is the sum of the above three contributions:
\bea\label{full Tad}
{\cal T}_H^{R_\xi} &=& \mu ^{\ve} \Biggl ( 3 \sqrt{\frac{\l_0}{2}}m_{H_0}  A_0(m_H)  + 3 \frac{\sqrt{2 \l_0}m^2_{Z_0}}{m_{H_0}} A_0(m_Z)  \Biggr ) \, .
\eea
Exploiting our notation it can also be expressed as
\be\label{full Tad2}
(4\pi)^{d/2}{\cal T}_H^{R_\xi} = \mu ^{\ve} \Biggl ( [{\cal T}_H^{R_\xi} ]_{\varepsilon} + \left \{ {\cal T}_H^{R_\xi} \right \}_{\varepsilon} +
[ {\cal T}_H^{R_\xi} ]_{f} + \left \{ {\cal T}_H^{R_\xi} \right \}_{f}   \Biggr )\, , \nonumber\\
\ee
with 
\bea\label{Tade}
\ve [{\cal T}_H^{R_\xi} ]_{\ve}  &=& 6 \sqrt{\frac{\l_0}{2}}m^3_{H_0} + 6 \frac{\sqrt{2 \l_0}m^4_{Z_0}}{m_{H_0}} \nonumber\\
\left\{ {\cal T}_H^{R_\xi} \right\}_{\ve} &=& 0
\eea
and 
\bea\label{Tadf}
[{\cal T}_H^{R_\xi}]_f &=&  3 \sqrt{\frac{\l_0}{2}}m^3_{H_0} + 3 \frac{\sqrt{2 \l_0}m^4_{Z_0}}{m_{H_0}} + 3 \sqrt{\frac{\l_0}{2}}m^3_{H_0} \ln \frac{\m^2}{m_{H_0}^2} + 3\frac{\sqrt{2 \l_0}m^4_{Z_0}}{m_{H_0}} \ln \frac{\m^2}{m_{Z_0}^2} \nonumber\\
\left\{ {\cal T}_H^{R_\xi} \right\}_{f} &=& 0 \, .
\eea
These expressions show that the tadpole sum is $\xi$-independent both in its divergent and in its finite part.
Note that tadpoles are external momentum independent objects.\\A detailed descriptions of the steps that we followed is presented in Appendix \ref{RGD}.

\subsection{Corrections to the gauge boson mass}
\label{Corrections to the Z mass in Rxi}

Starting the 2-point function calculations, before we compute the Higgs 2-point function, we first move out of the way
the $Z$ 2-point function. It will be needed for the renormalization of the $Z$ mass.

A gauge-boson vacuum polarization amplitude can be Lorentz-covariantly split into a transverse and a longitudinal part, as we show in Appendix \ref{RGD}. So, following this procedure we obtain that the quantity that enters in the renormalization of the mass of the $Z$ gauge boson is therefore
\bea\label{calMZRxi}
{\cal M}_{Z} = -\frac{1}{3} \left( g^{\m\n} - \frac{p^\m p^\n}{p^2} \right) {\cal M}_{Z, \m\n}(p) \, .
\eea
We now start computing the one-loop Feynman diagrams contributing to ${\cal M}^{R_\xi}_{Z, \m\n}$.

The first contributing diagram has a Higgs running in the loop:
\vskip .5cm
\begin{center}
\begin{tikzpicture}[scale=0.7]
\draw [photon] (0,0)--(1.8,0);
\draw [dashed,thick] (0,0.9) circle [radius=0.9];
\draw [photon] (-1.8,0)--(-0,0);
\draw [->]  (-1.7,0.2)--(-1,0.2);
\node at (-2.2,0.2) {$p$};
\draw [->]  (0.45,0.3) arc [start angle=-45, end angle=-135, radius=0.6cm];
\node at (0,0.4) {$k$};
\draw [->]  (1.7,0.2)--(1,0.2);
\node at (4.5,0) {$=\, \, i{\cal M}^{R_\xi,\phi}_{Z,\mu\nu}$};
\end{tikzpicture}
\end{center}
and, in DR, it is equal to
\be\label{M1ZRxi}
{\cal M}^{R_\xi,\phi}_{Z,\mu\nu} = -2{{g_{\mu \nu }}}\frac{m^2_{Z_0}}{m^2_{H_0}} \l_0 \mu^{\ve} A_0(m_{H_0}) \, .
\ee
Next we meet a couple of "sunset" diagrams. The first is 
\vskip .5cm
\begin{center}
\begin{tikzpicture}[scale=0.7]
\draw [photon] (-2.3,0)--(2.3,0);
\draw [->]  (-1.9,0.2)--(-1.2,0.2);
\node at (-2.5,0.2) {$p$};
\draw [<-]  (0.45,0.7) arc [start angle=45, end angle=135, radius=0.6cm];
\node at (0,1.3) {$k+p$};
\draw [<-] (-0.4,0.2)--(0.4,0.2);
\node at (0,-0.3) {$k$};
\draw [->]  (1.9,0.2)--(1.2,0.2);
\node at (4.5,0) {$=\,\,  i{\cal M}^{R_\xi,\phi Z}_{Z,\mu\nu}$ };
\draw  [dashed,thick] (-1,0) .. controls (-1,0.555) and (-0.555,1) .. (0,1)
.. controls (0.555,1) and (1,0.555) .. (1,0);
\end{tikzpicture}
\end{center}
In DR and using \eq{explicit Cmn2}, it can be expressed as
\be\label{M2ZRxi2}
{\cal M}^{R_\xi,\phi Z}_{Z,\mu\nu}  = 8{}{{}\frac{m^4_{Z_0}}{m^2_{H_0}} \l_0\mu^{\ve}} \Biggl \{- g_{\mu \nu }B_0(p,m_{Z_0},m_{H_0})
+ ( 1- \xi ) C^{1}_{\m\n}(p,m_{Z_0}, m_{H_0},m_{\chi_0}) \Biggr \}\, .  \nonumber\\
\ee
Notice that the $C$-type integral above is a special PV case, computed in Appendix \ref{PassarinoVeltman} as well.

The next sunset diagram is the last that contributes to the one-loop correction of the gauge boson propagator:
\vskip .5cm
\begin{center}
\begin{tikzpicture}[scale=0.7]
\draw [photon] (-2.3,0)--(-1,0);
\draw [photon] (1,0)--(2.3,0);
\draw [] (-1,0)--(1,0);
\draw [->]  (-1.9,0.2)--(-1.2,0.2);
\node at (-2.5,0.2) {$p$};
\draw [<-]  (0.45,0.7) arc [start angle=45, end angle=135, radius=0.6cm];
\node at (0,1.3) {$k+p$};
\draw [<-] (-0.4,0.2)--(0.4,0.2);
\node at (0,-0.3) {$k$};
\draw [->]  (1.9,0.2)--(1.2,0.2);
\node at (4.5,0) {$=\,\,  i{\cal M}^{R_\xi,\chi\phi}_{Z,\mu\nu}$ };
\draw  [dashed,thick] (-1,0) .. controls (-1,0.555) and (-0.555,1) .. (0,1)
.. controls (0.555,1) and (1,0.555) .. (1,0);
\end{tikzpicture}
\end{center}
and its explicit form reads
\be
{\cal M}^{R_\xi,\chi\phi}_{Z,\mu\nu} = 8 \frac{m^2_{Z_0}}{m^2_{H_0}} \l_0 \mu^{\ve} B_{\m\n}(p,m_{\chi_0},m_{H_0}) \, . 
\ee
Adding up all contributions we obtain
\bea\label{full MZRxi}
{\cal M}^{R_\xi}_{Z,\mu\nu} &=& \frac{m^2_{Z_0}}{m^2_{H_0}} \l_0 \mu^{\ve} \Biggl \{  -2{{g_{\mu \nu }}}A_0(m_{H_0}) - 8 g_{\mu \nu } m_{Z_0}^2B_0(p,m_{Z_0},m_{H_0}) \nonumber\\
&+& 8( 1- \xi ) m_{Z_0}^2 C^{1}_{\m\n}(m_{Z_0}, m_{H_0},m_{\chi_0}) + 8 B_{\m\n}(p,m_{\chi_0},m_{H_0})   \Biggr \} \, .
\eea 
The contraction that we need is \eq{calMZRxi}, which for general $p$ has the form
\bea\label{fullMZRxi}
{\cal M}^{R_\xi}_{Z}(p) &=& -\frac{1}{3}  \frac{m^2_{Z_0}}{m^2_{H_0}} \l_0 \frac{ \mu^{\ve} }{16\pi^{2}} \Biggl(   \Bigl \{  -2{{(d+\varepsilon )}}A_0(m_{H_0}) - 8 (d+\varepsilon ) m_{Z_0}^2 B_0(p,m_{Z_0},m_{H_0}) \nonumber\\
&+& 8( 1- \xi )m_{Z_0}^2 g^{\m\n}C^{1}_{\m\n}(p,m_{Z_0},m_{H_0},m_{\chi_0}) + 8 g^{\m\n} B_{\m\n}(p,m_{\chi_0},m_{H_0})   \Bigr \} \nonumber\\
&-& \Bigl \{ -2 A_0(m_H) - 8 m_{Z_0}^2 B_0(p,m_{Z_0},m_{H_0}) \nonumber\\
&+& 8( 1- \xi )m_{Z_0}^2 \frac{p^\m p^\n}{p^2} C^{1}_{\m\n}(p,m_{Z_0},m_{H_0},m_{\chi_0}) + 8  \frac{p^\m p^\n}{p^2} B_{\m\n}(p,m_{\chi_0},m_{H_0}) 
 \Bigr \}  \Biggr)\, . \nonumber\\
\eea
Specializing now to $p=0$, we can express the result as
\bea\label{fullMZRxi2}
(4\pi)^{d/2} {\cal M}^{R_\xi}_{Z} &=& \mu ^{\ve} \left( [{\cal M}^{R_\xi}_{Z} ]_{\varepsilon} + \left \{ {\cal M}^{R_\xi}_{Z} \right \}_{\varepsilon} +[ {\cal M}^{R_\xi}_{Z}]_{f} + \left \{ {\cal M}^{R_\xi}_{Z} \right \}_{f}   \right)\, ,
\eea
with 
\bea\label{MZRxinotation}
\ve [{\cal M}^{R_\xi}_{Z} ]_{\varepsilon}  &=& 12 \frac{\l_0 m_{Z_0}^4}{m_{H_0}^2} \nonumber\\
\left\{ {\cal M}^{R_\xi}_{Z} \right\}_{\varepsilon} &=& 0 \, .
\eea
The anomalous dimension of $Z$ can be also determined now, through the relation
\bea\label{adMZRxi}
\d A^{R_\xi} & =&- {\left. {\frac{d  {\cal M}^{R_\xi}_{Z}(p) }{d p^2}} \right|_{{p^2} = 0}} = \frac{\m^\ve}{(4\pi)^{d/2}}\left( [\d A^{R_\xi} ]_{\varepsilon} + \left\{ \d A^{R_\xi} \right\}_{\varepsilon} +[\d A^{R_\xi} ]_{f} + \left\{ \d A^{R_\xi} \right\}_{f} \right) \, ,\nonumber\\
\eea
with 
\bea\label{dARxinotation}
\ve [\d A^{R_\xi} ]_{\varepsilon}  &=& - \frac{4}{3} \l_0 \frac{m_{Z_0}^2}{m_{H_0}^2}  \nonumber\\
\left\{ \d A^{R_\xi} \right\}_{\varepsilon} &=& 0 \, .
\eea
Evidently, all divergent parts in this sector are $\xi$-independent.

Regarding the finite parts, we have
\bea\label{MZRxfinite}
 ({\cal M}^{R_\xi}_Z )_{f}  &=& -\frac{\lambda_0  m_{Z_0}^4 (\xi -19)}{3 m_{H_0}^2}+\frac{8 \lambda_0  m_{Z_0}^6 (\xi -1) \xi  \ln \frac{m_{Z_0}^2}{m_{H_0}^2}}{3 \left(m_{H_0}^2-m_{Z_0}^2\right) \left(m_{H_0}^2-m_{Z_0}^2 \xi \right)}\nonumber\\
 &+&\frac{2 \lambda_0  m_{Z_0}^4 \ln \left(m_{H_0}^2\right) \left(m_{H_0}^2 (\xi -9)+4 m_{Z_0}^2 \xi  (\xi +1)\right)}{3 \left(m_{H_0}^2-m_{Z_0}^2\right) \left(m_{H_0}^2-m_{Z_0}^2 \xi \right)} \nonumber\\
 &-& \frac{2 \lambda_0  m_{Z_0}^4 \ln \left(m_{Z_0}^2\right) \left(m_{Z_0}^4 \xi +m_{H_0}^2 m_{Z_0}^2 \left(4 \xi ^2-5 \xi -9\right)+9 m_{Z_0}^4 \xi \right)}{3 m_{H_0}^2 \left(m_{H_0}^2-m_{Z_0}^2\right) \left(m_{H_0}^2-m_{Z_0}^2 \xi \right)} \nonumber\\
 &+& \frac{6 \lambda_0 m_{Z_0}^4 \ln\mu ^2}{m_{H_0}^2}+\frac{8 \lambda_0  m_{Z_0}^4}{3 m_{H_0}^2}-\frac{2 \lambda_0  m_{Z_0}^4 \xi  \ln \xi }{3 m_{H_0}^2-3 m_{Z_0}^2 \xi }-\frac{4 \lambda_0  m_{Z_0}^2}{3}  \, 
\eea
and
\bea\label{dARxfinite}
 (\d A^{R_\xi} )_{f}  &=& -\frac{2}{3}\frac{ m_{Z_0}^2 \l_0}{( m_{H_0}^2 -  m_{Z_0}^2 \xi  )} 
 \left(\left(1+\ln \frac{\m^2}{m_{H_0}^2}\right) - \frac{m_{Z_0}^2}{m_{H_0}^4} \left(1+\xi  \ln \frac{\m^2}{\xi m_{Z_0}^2} \right)\right)\, .
\eea
The finite parts here are gauge dependent and we do not separate them further to $\xi$-independent and $\xi$-dependent parts.
The limit $\xi\to\infty$ is divergent.

\subsection{Corrections to the Higgs mass}\label{Corrections to the Higgs mass in Rx}

We move on to the Higgs propagator. The first diagram we encounter is
\vskip .5cm
\begin{center}
\begin{tikzpicture}[scale=0.7]
\draw [dashed,thick] (0,0)--(1.8,0);
\draw [photon] (0,0.9) circle [radius=0.9];
\draw [dashed,thick] (-1.8,0)--(-0,0);
\draw [->]  (-1.7,0.2)--(-1,0.2);
\node at (-2.2,0.2) {$p$};
\draw [->]  (0.45,0.50) arc [start angle=-45, end angle=-135, radius=0.6cm];
\node at (0,0.8) {$k$};
\draw [->]  (1.7,0.2)--(1,0.2);
\node at (4.5,0) {$=\, \,i{\cal M}_H^{R_\xi, Z}$};
\end{tikzpicture}
\end{center}
and after reductions performed in Appendix \ref{RGD}, finally the above integral is written as
\bea\label{full MH1b}
{\cal M}_H^{R_\xi, Z}&=&  \frac{m^2_{Z_0}}{m^2_{H_0}} \l_0 {\mu^{\ve}} \Biggl \{  6A_0(m_{Z_0}) +2 \xi A_0( m_{\chi_0}) \Biggr \}\, . 
\eea 
The next contribution comes from the diagram 
\vskip .5cm
\begin{center}
\begin{tikzpicture}[scale=0.7]
\draw [dashed,thick] (0,0)--(1.8,0);
\draw [dashed,thick] (0,0.9) circle [radius=0.9];
\draw [dashed,thick] (-1.8,0)--(-0,0);
\draw [->]  (-1.7,0.2)--(-1,0.2);
\node at (-2.2,0.2) {$p$};
\draw [->]  (0.45,0.4) arc [start angle=-45, end angle=-135, radius=0.6cm];
\node at (0,0.5) {$k$};
\draw [->]  (1.7,0.2)--(1,0.2);
\node at (4.5,0) {$=\, \, i{\cal M}^{R_\xi, \phi}_H$};
\end{tikzpicture}
\end{center}
and its explicit form is given by
\be\label{full MH2}
{\cal M}_H^{R_\xi,\phi}={3 \l_0  {\mu^{\ve}}{}}A_0(m_{H_0})
\ee 
in DR.

Next comes the Goldstone loop
\vskip .5cm
\begin{center}
\begin{tikzpicture}[scale=0.7]
\draw [dashed,thick] (0,0)--(1.8,0);
\draw [] (0,0.9) circle [radius=0.9];
\draw [dashed,thick] (-1.8,0)--(-0,0);
\draw [->]  (-1.7,0.2)--(-1,0.2);
\node at (-2.2,0.2) {$p$};
\draw [->]  (0.45,0.4) arc [start angle=-45, end angle=-135, radius=0.6cm];
\node at (0,0.5) {$k$};
\draw [->]  (1.7,0.2)--(1,0.2);
\node at (4.5,0) {$=\, \, i{\cal M}^{R_\xi,\chi}_H$};
\end{tikzpicture}
\end{center}
which, following Appendix \ref{RGD}, is equal to
\be\label{full MH3}
{\cal M}_H^{R_\xi,\chi}=-{ \frac{2 \l_0}{m_{H_0}^2} {\mu^{\ve}} m_{\chi_0}^2 } A_0(m_{\chi_0}) \, .
\ee 
It is easy to check that all of the above three diagrams are reducible Tadpoles corresponding to the three Tadpoles of \sect{Tadpoles in Rx}.
 
A few vacum polarization diagrams are in order. The first is
\vskip .5cm
\begin{center}
\begin{tikzpicture}[scale=0.7]
\draw [dashed,thick] (0.9,0)--(2.2,0);
\draw [dashed,thick] (0,0) circle [radius=0.9];
\draw [dashed,thick] (-2.2,0)--(-0.9,0);
\draw [->]  (-1.9,0.2)--(-1.2,0.2);
\node at (-2.6,0.2) {$p$};
\draw [<-]  (0.45,0.6) arc [start angle=45, end angle=135, radius=0.6cm];
\node at (0,1.2) {$k+p$};
\draw [->]  (0.45,-0.6) arc [start angle=-45, end angle=-135, radius=0.6cm];
\node at (0,-1.2) {$k$};
\draw [->]  (1.9,0.2)--(1.2,0.2);
\node at (4.5,0) {$=\, \, i{\cal M}^{R_\xi,\phi\phi}_H$};
\end{tikzpicture}
\end{center}
and in DR, is equal to 
\bea\label{full MH4}
{\cal M}^{R_\xi,\phi\phi}_H&= &9 \l_0 m^2_{H_0}{\mu ^{\ve}}B_0(p,m_{H_0},m_{H_0})\, .
\eea

The Goldstone loop contribution
\vskip .5cm
\begin{center}
\begin{tikzpicture}[scale=0.7]
\draw [dashed,thick] (0.9,0)--(2.2,0);
\draw [] (0,0) circle [radius=0.9];
\draw [dashed,thick] (-2.2,0)--(-0.9,0);
\draw [->]  (-1.9,0.2)--(-1.2,0.2);
\node at (-2.6,0.2) {$p$};
\draw [<-]  (0.45,0.6) arc [start angle=45, end angle=135, radius=0.6cm];
\node at (0,1.2) {$k+p$};
\draw [->]  (0.45,-0.6) arc [start angle=-45, end angle=-135, radius=0.6cm];
\node at (0,-1.2) {$k$};
\draw [->]  (1.9,0.2)--(1.2,0.2);
\node at (4.5,0) {$=\, \, i{\cal M}^{R_\xi,\chi\chi}_H$};
\end{tikzpicture}
\end{center}
which, following Appendix \ref{RGD}, takes in DR the final form  
\bea\label{full MH5}
{\cal M}^{R_\xi,\chi\chi}_H&= & 4 \frac{\l_0 }{m^2_{H_0}} {\mu ^{\ve}}\Biggl \{ m_{\chi_0}^2 A_0(m_{\chi_0}) + 
(m_{\chi_0}^2 - p^2 ) g_{\m\n} B^{\m\n}(p,m_{\chi_0},m_{\chi_0}) + p_\m p_\n B^{\m\n}(p,m_{\chi_0},m_{\chi_0}) \Biggr \}\, . \nonumber\\
\eea
Slightly more complicated is the gauge boson loop
\vskip .5cm
\begin{center}
\begin{tikzpicture}[scale=0.7]
\draw [dashed,thick] (0.9,0)--(2.2,0);
\draw [photon] (0,0) circle [radius=0.9];
\draw [dashed,thick] (-2.2,0)--(-0.9,0);
\draw [->]  (-1.9,0.2)--(-1.2,0.2);
\node at (-2.6,0.2) {$p$};
\draw [<-]  (0.45,0.6) arc [start angle=45, end angle=135, radius=0.6cm];
\node at (0,1.3) {$k+p$};
\draw [->]  (0.45,-0.40) arc [start angle=-45, end angle=-135, radius=0.6cm];
\node at (0,-1.1) {$k$};
\draw [->]  (1.9,0.2)--(1.2,0.2);
\node at (4.5,0) {$=\, \, i{\cal M}^{R_\xi,ZZ}_H$};
\end{tikzpicture}
\end{center}
and expressing it by standard steps in terms of PV integrals, it becomes 
\bea\label{full MH7}
{\cal M}^{R_\xi,ZZ}_H&= &4  \frac{m^4_{Z_{0}}}{m^2_{H_0}}\l_0  {\mu ^{\ve}} \Biggl \{d B_0(p,m_{Z_0},m_{Z_0}) - (1-\xi)  \Bigl \{g_{\m\n} C^{1\m\n}(p,m_{Z_0},m_{Z_0},m_{\chi_0}) \nonumber\\
&+& g_{\m\n} C^{\m\n}(p,p,m_{Z_0},m_{Z_0},m_{\chi_0})  \Bigr \} \nonumber\\ 
&+& (1-\xi)^2 \Bigl \{ g_{\m\n} C^{1\m\n}(p,m_{Z_0},m_{Z_0},m_{\chi_0})  \nonumber\\
&+& (m_{Z_0}^2 - p^2) g_{\m\n}D^{\m\n}(p,m_{Z_0},m_{Z_0},m_{\chi_0},m_{\chi_0}) + p_\m p_\n D^{\m\n}(p,m_{Z_0},m_{Z_0},m_{\chi_0},m_{\chi_0})\, ,
\Bigr \}   \Biggr \} \nonumber\\
\eea
where the $a = 1,2,3$ superscripts on the $C_0$-integrals correspond to the different combinations of the denominators
according to \eq{C012ab} of Appendix \ref{PassarinoVeltman}. The $D^{\m\n}$ integrals are defined in \eq{DPV}.

The last contribution to the one-loop correction of the Higgs mass comes from the sunset
\vskip .5cm
\begin{center}
\begin{tikzpicture}[scale=0.7]
\draw [dashed,thick] (-2.3,0)--(-1,0);
\draw [dashed,thick] (1,0)--(2.3,0);
\draw [] (-1,0)--(1,0);
\draw [->]  (-1.9,0.2)--(-1.2,0.2);
\node at (-2.5,0.2) {$p$};
\draw [<-]  (0.45,0.7) arc [start angle=45, end angle=135, radius=0.6cm];
\node at (0,1.6) {$k+p$};
\draw [<-] (-0.4,0.2)--(0.4,0.2);
\node at (0,-0.5) {$k$};
\draw [->]  (1.9,0.2)--(1.2,0.2);
\node at (4.5,0) {$=\,\, i{\cal M}^{R_\xi,\chi Z}_H$ };
\draw  [photon] (-1,0) .. controls (-1,0.555) and (-0.555,1) .. (0,1)
.. controls (0.555,1) and (1,0.555) .. (1,0);
\end{tikzpicture}
\end{center}
where using again the standard steps, we are allowed to write this as
\bea\label{full MH8}
{\cal M}^{R_\xi, \chi Z}_H  &=& 8 \l_0 \frac{m^2_{Z_{0}}}{m^2_{H_0}}  {\mu ^{\ve}}  \Biggl \{ - g_{\m\n} B^{\m\n}(p,m_{\chi_0},m_{Z_0}) \nonumber\\
&+& (1-\xi )\Bigl \{ g_{\m\n} B^{\m\n}(p,m_{\chi_0},m_{\chi_0}) + (m_{Z_0}^2 - p^2) g_{\m\n} C^{1\m\n}(p,m_{Z_0},m_{\chi_0},m_{Z_0}) \nonumber\\
&+&p_\m p_\n C^{1\m\n}(p,m_{\chi_0},m_{Z_0},m_{\chi_0}) \Bigr \} \Biggr \}\, , 
\eea
where $C^1_{\m\n}$ is defined in \eq{explicit Cmn1} in Appendix \ref{PassarinoVeltman}.

Finally, summing up all contributions into ${\cal M}^{R_\xi}_H$ we obtain 
\bea\label{full MHRxi}
{\cal M}^{R_\xi }_H (p) &=& {\mu ^{\ve}}  \Biggl \{ 3 \l_0 A_0(m_{H_0}) +6 \frac{m^2_{Z_{0}}}{m^2_{H_0}}\l_0   A_0(m_{Z_0}) + 2 \xi \frac{m^2_{Z_{0}}}{m^2_{H_0}}\l_0 A_0(m_{\chi_0})  \nonumber\\
&-& 2 \frac{\l_0}{m_{H_0}^2} m_{\chi_0}^2 A_0(m_{\chi_0})  + 9 \l_0 m^2_{H_0} B_0(p,m_{H_0},m_{H_0}) \nonumber\\
&+& 4 \frac{\l_0}{m_{H_0}^2} \left(  m_{\chi_0}^2 A_0(m_{\chi_0})  + (m_{\chi_0}^2 - p^2) g_{\m\n} B^{\m\n}(p,m_{\chi_0},m_{\chi_0}) + p_\m p_\n B^{\m\n}(p,m_{\chi_0},m_{\chi_0})  \right)  \nonumber\\
&+& 
 4 d \frac{m^4_{Z_{0}}}{m^2_{H_0}}\l_0  B_0(p,m_{Z_0},m_{Z_0})  
  - 8  \l_0 \frac{m^2_{Z_{0}}}{m^2_{H_0}} g_{\m\n} B^{\m\n}(p,m_{\chi_0},m_{Z_0}) \nonumber\\
&+&  \frac{m^2_{Z_{0}}}{m^2_{H_0}}\l_0
(1-\xi)\Bigl \{ -4 m_{Z_0}^2 g_{\m\n} C^{1\m\n}(p,m_{Z_0},m_{Z_0},m_{\chi_0}) - 4 m_{Z_0}^2 g_{\m\n} C^{\m\n}(p,p,m_{Z_0},m_{Z_0},m_{\chi_0})  \nonumber\\
&+& 8 g_{\m\n} B^{\m\n}(p,m_{\chi_0},m_{\chi_0}) + 8 (m_{Z_0}^2 - p^2)g_{\m\n} C^{1\m\n}(p,m_{\chi_0},m_{Z_0},m_{\chi_0}) \nonumber\\
&+& 8p_\m p_\n C^{1\m\n}(p,m_{\chi_0},m_{Z_0},m_{\chi_0})  \Bigr \}   \nonumber\\
&+&4\frac{m^4_{Z_{0}}}{m^2_{H_0}}\l_0(1-\xi)^2  \Bigl \{ g_{\m\n} C^{1\m\n}(p,m_{Z_0},m_{Z_0},m_{\chi_0}) \nonumber\\
&+& (m_{Z_0}^2 - p^2) g_{\m\n}D^{\m\n}(p,m_{Z_0},m_{Z_0},m_{\chi_0},m_{\chi_0}) + p_\m p_\n D^{\m\n}(p,m_{Z_0},m_{Z_0},m_{\chi_0},m_{\chi_0}) \Bigr \}    \Biggr \}\, . \nonumber\\
\eea
Note that the reduction of the $g_{\m\n} D^{\m\n}$ and $p_\m p_\n D^{\m\n} $ terms give only $C_0$ and $D_0$ contributions which are finite.
We then have that
\bea\label{fullMHRxi2}
(4\pi)^{d/2} {\cal M}^{R_\xi }_H &=& \mu ^{\ve} \Biggl ( [{\cal M}^{R_\xi }_H ]_{\ve} + \left\{ {\cal M}^{R_\xi }_H \right\}_{\ve} +[ {\cal M}^{R_\xi }_H ]_{f} + \left\{ {\cal M}^{R_\xi }_H \right\}_{f}   \Biggr )
\eea
with 
\bea\label{MHRxinotation}
\ve [{\cal M}^{R_\xi }_H ]_{\varepsilon}  &=&24 \l_0 m_{H_0}^2 + 36 \frac{\l_0 m_{Z_0}^4}{m_{H_0}^2}  \nonumber\\
\left\{ {\cal M}^{R_\xi }_H \right\}_{\varepsilon} &=&  0
\eea
and 
\bea\label{MHRxifinite}
[{\cal M}^{R_\xi }_H]_f &=&  3 \l_0 m_{H_0}^2 + 6 \frac{ \l_0 m^4_{Z_0}}{m_{H_0}^2} + 12 \l_0 m_{H_0}^2 \ln \frac{\m^2}{m_{H_0}^2} + 18\frac{ \l_0 m^4_{Z_0}}{m_{H_0}^2} \ln \frac{\m^2}{m_{Z_0}^2} \nonumber\\
\left\{ {\cal M}^{R_\xi }_H \right\}_{f} &=& 0 
\eea
at $p=0$.

We are also able to compute the anomalous dimension of the Higgs, determined by
\bea\label{adMHRxi}
\d \phi^{R_\xi} & =&- {\left. {\frac{d  {\cal M}^{R_\xi }_H(p) }{d p^2}} \right|_{{p^2} = 0}} = 
\frac{\m^\ve}{(4\pi)^{d/2}}\left( [\d \phi^{R_\xi} ]_{\varepsilon} + \left\{ \d \phi^{R_\xi} \right\}_{\varepsilon} +[\d \phi^{R_\xi} ]_{f} + \left\{ \d \phi^{R_\xi} \right\}_{f} \right)\nonumber
\eea
with 
\bea\label{dphiRxinotation}
\ve [\d \phi^{R_\xi} ]_{\varepsilon}  &=& 12 \l_0 \frac{m_{Z_0}^2}{m_{H_0}^2}  \nonumber\\
\left\{ \d \phi^{R_\xi} \right\}_{\varepsilon} &=& 0 \, .
\eea
and
\bea\label{dphiRxifinite}
 [\d \phi^{R_\xi} ]_{f}  &=& 2 \l_0 \frac{m_{Z_0}^2}{m_{H_0}^2} + 6\l_0 \frac{m_{Z_0}^2}{m_{H_0}^2}\ln \frac{\m^2}{m_{Z_0}^2}  \nonumber\\
\left\{ \d \phi^{R_\xi} \right\}_{f} &=& 0 \, .
\eea
This sector turns out to be $\xi$-independent.

\subsection{Corrections to the Higgs cubic vertex}\label{Corrections to the three-point vertex in Rxi}

Triangle diagrams with Higgs external legs yield corrections to the Higgs cubic vertex.
Such corrections will play a crucial role in the definition of the one-loop scalar potential.
The external momenta are taken to be all inflowing, thus satisfying $p_1+p_2+p_3=0$.
Triangle diagrams can be split in "reducible" and "irreducible" kinds.
Reducible are those that are expressible in terms of 2-point function diagrams and irreducible are those that are not.

\subsubsection{Reducible Triangles}\label{Reducible Triangles}

The first reducible Triangle diagram is:
\vskip .5cm
\begin{center}
\begin{tikzpicture}[scale=0.7]
\draw [dashed,thick] (0.9,0)--(2.3,1.3);
\draw [dashed,thick] (0.9,0)--(2.3,-1.3);
\draw [dashed,thick] (-2.5,0)--(-0.9,0);
\draw [dashed,thick] (0,0) circle [radius=0.9];
\draw [<-]  (0.45,0.6) arc [start angle=45, end angle=135, radius=0.6cm];
\node at (0,1.2) {$k+P_1$};
\draw [->]  (0.45,-0.6) arc [start angle=-45, end angle=-135, radius=0.6cm];
\node at (0,-1.2) {$k$};
\draw [<-]  (1.3,0.6)--(1.9,1.2);
\node at (2.7,1.3) {$p_3$};
\draw [<-]  (1.3,-0.6)--(1.9,-1.2);
\node at (2.7,-1.3) {$p_2$};
\draw [<-] (-1.5,0.2)--(-2.2,0.2);
\node at (-2.5,0.2) {$p_1$};
\node at (4.5,0) {$=\, \, i{\cal K}^{R_\xi ,\phi\phi}_H$};
\end{tikzpicture}
\end{center}
As anticipated, it is not an independent diagram.
It is the same loop-diagram as in \eq{full MH4} (with the same symmetry factor), divided by $v_0$. We can therefore write directly the result:
\bea\label{RedTriangle1}
{\cal K}^{R_\xi, \phi\phi}_H &=& 3\cdot 18 \frac{\l^{3/2}_0}{\sqrt{2}} m_{H_0}{\mu ^{\ve}}B_0(P_1,m_{H_0},m_{H_0})\, .
\eea
The factor of 3 is due to two additional diagrams, obtained from the above by cyclically permuting the external momenta.
These contributions however, evaluated at either $p_1=p_2=p_3=0$ or $p_1^2=p_2^2=p_3^2=m_H^2$ give an identical result.
 
The next diagram is one with a Goldstone in the loop: 
\vskip .5cm
\begin{center}
\begin{tikzpicture}[scale=0.7]
\draw [dashed,thick] (0.9,0)--(2.3,1.3);
\draw [dashed,thick] (0.9,0)--(2.3,-1.3);
\draw [dashed,thick] (-2.5,0)--(-0.9,0);
\draw [] (0,0) circle [radius=0.9];
\draw [<-]  (0.45,0.6) arc [start angle=45, end angle=135, radius=0.6cm];
\node at (0,1.2) {$k+P_1$};
\draw [->]  (0.45,-0.6) arc [start angle=-45, end angle=-135, radius=0.6cm];
\node at (0,-1.2) {$k$};
\draw [<-]  (1.3,0.6)--(1.9,1.2);
\node at (2.7,1.3) {$p_3$};
\draw [<-]  (1.3,-0.6)--(1.9,-1.2);
\node at (2.7,-1.3) {$p_2$};
\draw [<-] (-1.5,0.2)--(-2.2,0.2);
\node at (-2.5,0.2) {$p_1$};
\node at (4.5,0) {$=\, \, i{\cal K}^{R_\xi, \chi\chi}_H$};
\end{tikzpicture}
\end{center}
and it is equal to \eq{full MH5} divided by $v_0$:
\bea\label{RedTriangle2}
&& {\cal K}^{R_\xi, \chi\chi}_H= \nonumber\\
&& 3\cdot 8 \frac{\l^{3/2}_0 }{m^3_{H_0}} {\mu ^{\ve}}\Biggl \{ m_{\chi_0}^2 A_0(m_{\chi_0}) +(m_{\chi_0}^2 - P_1^2)
g_{\m\n} B^{\m\n}(P_1,m_{\chi_0},m_{\chi_0}) + P_{1\m} P_{1\n} B^{\m\n}(P_1,m_{\chi_0},m_{\chi_0}) \Biggr \}\, . \nonumber\\
\eea
The factor of 3 has a similar origin as before.

The diagram with a gauge boson loop
\vskip .5cm
\begin{center}
\begin{tikzpicture}[scale=0.7]
\draw [dashed,thick] (0.9,0)--(2.3,1.3);
\draw [dashed,thick] (0.9,0)--(2.3,-1.3);
\draw [dashed,thick] (-2.5,0)--(-0.9,0);
\draw [photon] (0,0) circle [radius=0.9];
\draw [<-]  (0.45,0.6) arc [start angle=45, end angle=135, radius=0.6cm];
\node at (0,1.4) {$k+P_1$};
\draw [->]  (0.45,-0.4) arc [start angle=-45, end angle=-135, radius=0.6cm];
\node at (0,-1) {$k$};
\draw [<-]  (1.3,0.6)--(1.9,1.2);
\node at (2.7,1.3) {$p_3$};
\draw [<-]  (1.3,-0.6)--(1.9,-1.2);
\node at (2.7,-1.3) {$p_2$};
\draw [<-] (-1.5,0.2)--(-2.2,0.2);
\node at (-2.5,0.2) {$p_1$};
\node at (4.5,0) {$=\, \, i{\cal K}^{R_\xi, ZZ}_H$};
\end{tikzpicture}
\end{center}
is equal to \eq{full MH7} divided by $v_0$:
\bea\label{RedTriangle3}
{\cal K}^{R_\xi, ZZ}_H&= &3\cdot \frac{8}{\sqrt{2}} \frac{m^4_{Z_{0}}}{m^3_{H_0}}\l^{3/2}_0  {\mu ^{\ve}} \Biggl \{d B_0(P_1,m_{Z_0},m_{Z_0}) - (1-\xi)  \Bigl \{g_{\m\n} C^{1\m\n}(p,m_{Z_0},m_{Z_0},m_{\chi_0}) \nonumber\\
&+& g_{\m\n} C^{1\m\n}(P_1,m_{Z_0},m_{Z_0},m_{\chi_0})  \Bigr \} \nonumber\\ 
&+& (1-\xi)^2 \Bigl \{ g_{\m\n} C^{1\m\n}(P_1,m_{Z_0},m_{Z_0},m_{\chi_0})  \nonumber\\
&+& (m_{Z_0}^2 - P_1^2) g_{\m\n}D^{\m\n}(P_1,m_{Z_0},m_{Z_0},m_{\chi_0},m_{\chi_0}) + P_{1\m} P_{1\n} D^{\m\n}(P_1,m_{Z_0},m_{Z_0},m_{\chi_0},m_{\chi_0})\, ,
\Bigr \}   \Biggr \} \, . \nonumber\\
\eea
The last reducible Triangle is:
\vskip .5cm
\begin{center}
\begin{tikzpicture}[scale=0.7]
\draw [dashed,thick] (1,0)--(2.3,1.3);
\draw [dashed,thick] (1,0)--(2.3,-1.3);
\draw [dashed,thick] (-2.5,0)--(-1,0);
\draw [] (-1,0)--(1,0);
\draw [<-]  (1.3,0.6)--(1.9,1.2);
\node at (2.7,1.3) {$p_3$};
\draw [<-]  (1.3,-0.6)--(1.9,-1.2);
\node at (2.7,-1.3) {$p_2$};
\draw [<-] (-1.5,0.2)--(-2.2,0.2);
\node at (-2.5,0.2) {$p_1$};
\draw [<-]  (0.45,0.7) arc [start angle=45, end angle=135, radius=0.6cm];
\node at (0,1.6) {$k+P_1$};
\draw [<-] (-0.4,0.2)--(0.4,0.2);
\node at (0,-0.5) {$k$};
\node at (4.5,0) {$=\,\, i{\cal K}^{R_\xi, \chi Z}_H$ };
\draw  [photon] (-1,0) .. controls (-1,0.555) and (-0.555,1) .. (0,1)
.. controls (0.555,1) and (1,0.555) .. (1,0);
\end{tikzpicture}
\end{center}
It is the same as \eq{full MH8} divided by $v_0$:
\bea\label{RedTriangle4}
{\cal K}^{R_\xi, \chi Z}_H  &=& 3\cdot \frac{16}{\sqrt{2}} \l^{3/2}_0 \frac{m^2_{Z_{0}}}{m^3_{H_0}}  {\mu ^{\ve}}   \Biggl \{ - g_{\m\n} B^{\m\n}(P_1,m_{\chi_0},m_{Z_0}) \nonumber\\
&+& (1-\xi )\Bigl \{ g_{\m\n} B^{\m\n}(P_1,m_{\chi_0},m_{\chi_0}) + (m_{Z_0}^2 - P_1^2) g_{\m\n} C^{1\m\n}(P_1,m_{Z_0},m_{\chi_0},m_{Z_0}) \nonumber\\
&+&P_{1\m} P_{1\n} C^{1\m\n}(P_1,m_{\chi_0},m_{Z_0},m_{\chi_0}) \Bigr \} \Biggr \}\, . 
\eea
Let us collect all reducible Triangle contributions by adding \eq{RedTriangle1}, \eq{RedTriangle2}, \eq{RedTriangle3} and \eq{RedTriangle4}. 
At zero external momenta, we obtain
\bea\label{fullKHRD}
(4\pi)^{d/2} {\cal K}^{R_\xi, {\rm red.}}_{H} &=& \m^\ve \left( [{\cal K}^{R_\xi, {\rm red.}}_{H} ]_{\ve} + 
\left\{ {\cal K}^{R_\xi, {\rm red.}}_{H}\right\}_{\ve} +[ {\cal K}^{R_\xi, {\rm red.}}_{H} ]_{f} + \left\{ {\cal K}^{R_\xi, {\rm red.}}_{H}\right\}_{f} \right)\, ,
\eea
with 
\bea\label{KHRxiRDnotation}
\ve[{\cal K}^{R_\xi, {\rm red.}}_{H}]_{\ve}  &=& \frac{m_{H_0}}{\sqrt{2 \l_0}}  \left( 108  \l_0^2 + 144 \frac{ \l_0^2 m_{Z_0}^4}{m_{H_0}^4} \right)  \nonumber\\
\left\{ {\cal K}^{R_\xi, {\rm red.}}_{H} \right\}_{\varepsilon} &=& 0.
\eea
and 
\bea
[{\cal K}^{R_\xi, {\rm red.}}_{H}]_f &=& \frac{m_{H_0}}{\sqrt{2 \l_0}} \Bigl ( 54 \l_0^2 \ln \frac{\m^2}{m_{H_0}^2} + 72 \frac{ \l^2_0 m^4_{Z_0}}{m_{H_0}^4} \ln \frac{\m^2}{m_{Z_0}^2}  \Bigr )\nonumber\\
\left\{ {\cal K}^{R_\xi, {\rm red.}}_{H} \right\}_{f} &=& 0 \, .
\eea

\subsubsection{Irreducible Triangles}\label{Irreducible Triangles}

We now turn to the irreducible Triangles. 
All irreducible Triangle diagrams can be labelled by the momenta $P_1=p_1$ and $P_2=p_1 + p_3$.
A simplifying consequence of renormalizing at $p_i=0$ is that we can set $P_i = 0$, hence
we can extract both divergent and finite parts, using
\bea
\lim_{P_i\to 0} {\cal K}^{R_\xi,\cdots}_H(P_1,P_2) \equiv {\cal K}^{R_\xi,\cdots}_H (0,0). 
\eea
The limit should be carefully taken, as explained in Sect. \ref{basics}. 
Regarding denominators, from now on, we start following the notation of \eq{denom.}.
Here, apart from finite integrals of the $E$-type, we will also see the appearance of several divergent $U$-integrals. 
All finite integrals and $U$-integrals here and in the following are computed (sometimes without further notice) in Appendices \ref{FiniteParts} and \ref{Uint} respectively. 
Finite diagrams do not play a role in the renormalization program but they contribute to the scalar potential.

The first contribution to the irreducible Triangle class involves a Higgs loop and it is finite. 
It is the diagram
\vskip .5cm
\begin{center}
\begin{tikzpicture}[scale=0.7]
\draw [dashed,thick] (-1,0) -- (1,1);
\draw [dashed,thick] (-1,0) -- (1,-1);
\draw [dashed,thick] (1,1) -- (1,-1);
\draw [dashed,thick] (-2,0) -- (-1,0);
\draw [dashed,thick] (1,1) -- (2.5,1);
\draw [dashed,thick] (1,-1) -- (2.5,-1);
\node at (1.7,1.3) {$p_3$};
\draw [<-] (1.3,0.8) -- (1.9,0.8);
\node at (1.7,-1.3) {$p_2$};
\draw [<-] (1.3,-0.8) -- (1.9,-0.8);
\node at (-1.9,0.2) {$p_1$};
\draw [->] (-1.6,0.2) -- (-1.1,0.2);
\draw [->] (-0.4,0.1) -- (0.2,0.4);
\node at (-0.3,1) {$k+P_1$};
\draw [<-] (-0.5,-0.1) -- (0.2,-0.4);
\node at (-0.1,-1) {$k$};
\draw [<-] (0.8,-0.3) -- (0.8,0.3);
\node at (2.2,0) {$k+P_2$};
\node at (5.5,0) {$=\,\,  i{\cal K}^{R_\xi, \phi\phi\phi}_H.$};.
\end{tikzpicture}
\end{center}
It is equal to\footnote{We thank A. Chatziagapiou for pointing out a factor of 2 in this diagram that was missing 
in the previous version of the paper.}
\bea
{\cal K}^{R_\xi,\phi\phi\phi}_H &=& \frac{108}{\sqrt{2}} \l_0^{3/2} m^3_{H_0} \int {\frac{{{d^4}k}}{{{{\left( {2\pi } \right)}^4}}}} (-i)\frac{1}{D_1 D_2 D_3}
\eea
with a symmetry factor $S^{5}_{K_H} = 1$. 
In DR it can be expressed as
\bea\label{full KH5}
{\cal K}^{R_\xi,\phi\phi\phi}_H (P_1,P_2) &=& \frac{108}{\sqrt{2}} \l_0^{3/2} m^3_{H_0} \m^\ve C_0(P_1,P_2,m_{H_0},m_{H_0},m_{H_0}).
\eea
As explained,
\bea
\lim_{P_i\to 0} {\cal K}^{R_\xi,\phi\phi\phi}_H(P_1,P_2) \equiv {\cal K}^{R_\xi,\phi\phi\phi}_H (0,0).
\eea
There is another finite diagram, the one with a gauge boson loop:
\vskip .5cm
\begin{center}
\begin{tikzpicture}[scale=0.7]
\draw [photon] (-1,0) -- (1,1);
\draw [photon] (-1,0) -- (1,-1);
\draw [photon] (1,1) -- (1,-1);
\draw [dashed,thick] (-2,0) -- (-1,0);
\draw [dashed,thick] (1,1) -- (2.5,1);
\draw [dashed,thick] (1,-1) -- (2.5,-1);
\node at (1.7,1.3) {$p_3$};
\draw [<-] (1.3,0.8) -- (1.9,0.8);
\node at (1.7,-1.3) {$p_2$};
\draw [<-] (1.3,-0.8) -- (1.9,-0.8);
\node at (-1.9,0.2) {$p_1$};
\draw [->] (-1.6,0.2) -- (-1.1,0.2);
\draw [->] (-0.4,0.1) -- (0.2,0.4);
\node at (-0.5,1) {$k+P_1$};
\draw [<-] (-0.5,-0.1) -- (0.2,-0.4);
\node at (-0.1,-1) {$k$};
\draw [<-] (0.8,-0.3) -- (0.8,0.3);
\node at (2.2,0) {$k+P_2$};
\node at (5.5,0) {$=\,\,  i{\cal K}^{R_\xi, ZZZ}_H$};
\end{tikzpicture}
\end{center}
It is equal to
\bea\label{KH6}
{\cal K}^{R_\xi, ZZZ}_H&=&- 16\sqrt{2}\frac{m^6_{Z_0} \l_0^{3/2}}{m^3_{H_0}}  {g^{\mu \nu }}{g^{\alpha \beta }}g^{\gamma \d }  
\int {\frac{{{d^4}k}}{{{{\left( {2\pi } \right)}^4}}}
\frac{{-i\left( { - {g_{\mu \gamma }} + \frac{{{(1-\xi)k_\mu }{k_\gamma }}}{{{k^2 - \xi m_{Z_0}^2}}}} \right)}}{{D_1}}
\frac{{\left( { - {g_{\nu \a }} + \frac{{{{(1-\xi)\left( {k + P_1} \right)}_\nu }{{\left( {k + P_1} \right)}_\a }}}{{{(k+P_1)^2 - \xi m_{Z_0}^2}}}} \right)}}{{D_2}}}\nonumber\\
&\times& \frac{{\left( { - {g_{\d \b }} + \frac{{{{(1-\xi)\left( {k + P_2} \right)}_\d }{{\left( {k + P_2} \right)}_\b }}}{{{(k+P_2)^2 - \xi m_{Z_0}^2}}}} \right)}}{{D_3}}
\eea
with a symmetry factor ${\cal S}^{6}_{{\cal {K}}_H } = 1$. Its expression in DR is
\bea\label{full KH6}
{\cal K}^{R_\xi, ZZZ}_H (P_1,P_2)&=& (K^{R_\xi, ZZZ}_{H}(P_1,P_2))_f  \, ,
\eea
where we have used the notation for finite (parts of) diagrams, explained in \eq{entirefinite}.
Following again previous arguments, at zero momentum we obtain 
\bea
\lim_{P_i\to 0} {\cal K}^{R_\xi, ZZZ}_H (P_1,P_2) \equiv {\cal K}^{R_\xi, ZZZ}_H (0,0) \, .
\eea
This is the first of several irreducible diagrams whose explicit form is not particularly illuminating, so we directly transfer
it to Appendix \ref{FiniteParts}.

Now, let us move on to diagrams that have both an infinite and a finite part. The first such diagram is 
\vskip .5cm
\begin{center}
\begin{tikzpicture}[scale=0.7]
\draw [photon] (-1,0) -- (1,1);
\draw [photon] (-1,0) -- (1,-1);
\draw [] (1,1) -- (1,-1);
\draw [dashed,thick] (-2,0) -- (-1,0);
\draw [dashed,thick] (1,1) -- (2.5,1);
\draw [dashed,thick] (1,-1) -- (2.5,-1);
\node at (1.7,1.3) {$p_3$};
\draw [<-] (1.3,0.8) -- (1.9,0.8);
\node at (1.7,-1.3) {$p_2$};
\draw [<-] (1.3,-0.8) -- (1.9,-0.8);
\node at (-1.9,0.2) {$p_1$};
\draw [->] (-1.6,0.2) -- (-1.1,0.2);
\draw [->] (-0.4,0.1) -- (0.2,0.4);
\node at (-0.5,1) {$k+P_1$};
\draw [<-] (-0.4,-0.1) -- (0.2,-0.4);
\node at (-0.1,-1) {$k$};
\draw [<-] (0.8,-0.3) -- (0.8,0.3);
\node at (2.2,0) {$k+P_2$};
\node at (5.5,0) {$=\,\,  i{\cal K}^{R_\xi, Z\chi Z}_{H}$};
\end{tikzpicture}
\end{center}
It is equal to
\bea
{\cal K}^{R_\xi, Z\chi Z}_{H}&=& 32 v_0 \l_0^2 \frac{m^4_{Z_0} }{m^4_{H_0}} {g^{\mu \nu }}  \int {\frac{{{d^4}k}}{{{{\left( {2\pi } \right)}^4}}}\frac{{-i\left( { - {g_{\mu \a }} + \frac{{{(1-\xi)k_\mu }{k_\a }}}{{{k^2 - \xi m_{Z_0}^2}}}} \right)}}{{D_1}}\frac{{\left( { - {g_{\nu \b }} + \frac{{{{(1-\xi)\left( {k + P_1} \right)}_\nu }{{\left( {k + P_1} \right)}_\b }}}{{{(k+P_1)^2 - \xi m_{Z_0}^2}}}} \right)}}{{D_2}}} \nonumber \\
&\times& \frac{(k + P_2)^\a (k + P_2)^\b}{D_3} \, ,
\eea
with a symmetry factor ${\cal S}^{7}_{{\cal {K}}_H } = 1$.
In DR it becomes
\bea\label{KH7}
{\cal K}^{R_\xi, Z\chi Z}_{H} (P_1,P_2) &=& - 32 v_0 \l_0^2 \frac{m^4_{Z_0} }{m^4_{H_0}}  \m^{\ve} \Biggl \{ B_0(P_1,m_{Z_0},m_{Z_0}) + m^2_{\chi_0} C_0(P_1,m_{Z_0},m_{Z_0},m_{\chi_0}) \nonumber\\
&-& (1-\xi) \Bigl \{ 2 B_0(P_1,m_{Z_0},m_{Z_0}) + 2 m^2_{\chi_0} C_0(P_1,m_{Z_0},m_{Z_0},m_{\chi_0})   \nonumber\\
&+& m^4_{\chi_0}D_0(P_1,P_2,m_{Z_0},m_{Z_0},m_{\chi_0},m_{\chi_0}) + P_2^\m P_2^\n  D_{\m\n}(P_1,P_2,m_{Z_0},m_{Z_0},m_{\chi_0},m_{\chi_0})      \Bigr \}\nonumber\\
&+&(1-\xi)^2 \Bigl \{ D_{{\cal B} 4}(P_1,P_2,m_{Z_0},m_{Z_0},m_{\chi_0},m_{\chi_0}) \nonumber\\
&+& m^2_{\chi_0}E_4(D_1,D_2,D_3,D_4(0,m_{\chi_0}),D_5(P_1,m_{\chi_0})) \nonumber\\
&+&2 P_{2\m} E^\m_5(D_1,D_2,D_3,D_4(0,m_{\chi_0}),D_5(P_1,m_{\chi_0}))\nonumber\\  
&+& P_{2\m}P_{2\n}E^{\m\n}_4(D_1,D_2,D_3,D_4(0,m_{\chi_0}),D_5(P_1,m_{\chi_0})) \Bigr \}   \Biggr \}\, ,
\eea
where the mass arguments of the $D_{1,2,3}$ denominators are easily recovered from the $Z\chi Z$ superscript structure of the diagram: $D_1(m_Z)$, $D_2(m_{\chi})$ and $D_3(m_Z)$.
We are not done yet since there are three different ways to insert the Goldstone propagator in the loop. Therefore, there are two more 
contributing diagrams of the same kind as ${\cal K}^{R_\xi, Z\chi Z}_{H}$. These are the diagrams  
\be
\begin{tikzpicture}[scale=0.5]
\draw [photon] (-1,0) -- (1,1);
\draw [] (-1,0) -- (1,-1);
\draw [photon] (1,1) -- (1,-1);
\draw [dashed,thick] (-2,0) -- (-1,0);
\draw [dashed,thick] (1,1) -- (2.5,1);
\draw [dashed,thick] (1,-1) -- (2.5,-1);
\node at (6.5,0) {$=\,\,  i{\cal K}^{R_\xi, Z Z \chi }_{H}(P_1,P_2)$};
\end{tikzpicture}
\hskip .85cm
\begin{tikzpicture}[scale=0.5]
\draw [] (-1,0) -- (1,1);
\draw [photon] (-1,0) -- (1,-1);
\draw [photon] (1,1) -- (1,-1);
\draw [dashed,thick] (-2,0) -- (-1,0);
\draw [dashed,thick] (1,1) -- (2.5,1);
\draw [dashed,thick] (1,-1) -- (2.5,-1);
\node at (6.5,0) {$=\,\,  i{\cal K}^{R_\xi, \chi Z Z}_{H}(P_1,P_2)$};
\end{tikzpicture}
\hskip .5cm
\nonumber
\ee
Performing the calculations at zero external momenta, the above diagrams have identical divergent and finite parts with ${\cal K}^{R_\xi, Z \chi Z}_{H}$, that is
\bea
{\cal K}^{R_\xi, Z \chi Z}_{H}(0,0) = {\cal K}^{R_\xi, Z Z \chi }_{H}(0,0) = {\cal K}^{R_\xi, \chi Z Z}_{H}(0,0). 
\eea  
Next, we have the diagram with two Goldstones and one gauge boson in the loop:
\vskip .5cm
\begin{center}
\begin{tikzpicture}[scale=0.7]
\draw [] (-1,0) -- (1,1);
\draw [] (-1,0) -- (1,-1);
\draw [photon] (1,1) -- (1,-1);
\draw [dashed,thick] (-2,0) -- (-1,0);
\draw [dashed,thick] (1,1) -- (2.5,1);
\draw [dashed,thick] (1,-1) -- (2.5,-1);
\node at (1.7,1.3) {$p_3$};
\draw [<-] (1.3,0.8) -- (1.9,0.8);
\node at (1.7,-1.3) {$p_2$};
\draw [<-] (1.3,-0.8) -- (1.9,-0.8);
\node at (-1.9,0.2) {$p_1$};
\draw [->] (-1.6,0.2) -- (-1.1,0.2);
\draw [->] (-0.4,0.1) -- (0.2,0.4);
\node at (-0.5,1) {$k+P_1$};
\draw [<-] (-0.4,-0.1) -- (0.2,-0.4);
\node at (-0.1,-1) {$k$};
\draw [<-] (0.8,-0.3) -- (0.8,0.3);
\node at (2.2,0) {$k+P_2$};
\node at (5.5,0) {$=\,\,  i{\cal K}^{R_\xi,\chi Z\chi}_{H}$};
\end{tikzpicture}
\end{center}
It is given by the relation  
\bea
{\cal K}^{R_\xi,\chi Z \chi}_H&=& - 32 v_0 \l_0^2 \frac{m^2_{Z_0} }{m^4_{H_0}}  \int {\frac{{{d^4}k}}{{{{\left( {2\pi } \right)}^4}}}\frac{{-i\left( { - {g_{\mu \n }} + \frac{{{(1-\xi)(k+P_2)_\mu }{(k+P_2)_\n }}}{{{\left( {k + P_2} \right)^2 - \xi m_{Z_0}^2}}}} \right)}}{{D_1 D_2 D_3}} k^\m (k+ P_1)^\n }  (k+P_1) \cdot k \nonumber\\
\eea
with symmetry factor ${\cal S}^{8}_{{\cal {K}}_H } = 1$. In DR it reads
\bea\label{KH8}
{\cal K}^{R_\xi,\chi Z\chi}_{H}(P_1,P_2) &=& 32 v_0 \l_0^2 \frac{m^2_{Z_0} }{m^4_{H_0}}  \m^{\ve} \Biggl \{ U_{K4}(P_1,P_2,m_{Z_0},m_{\chi_0},m_{\chi_0}) \nonumber\\
&+&2 (P_1 + P_2)_\m C_{{\cal K}3}^\m(P_1,P_2,m_{Z_0},m_{\chi_0},m_{\chi_0}) \nonumber\\
&+& (P_{1\m}P_{1\n} +2P_{1\m}P_{2\n} +P_{2\m}P_{2\n} )C^{\m\n}(P_1,P_2,m_{Z_0},m_{\chi_0},m_{\chi_0})\nonumber\\
&+& 2P_1 \cdot P_2 g_{\m\n} C^{\m\n} (P_1,P_2,m_{Z_0},m_{\chi_0},m_{\chi_0})        \nonumber\\
&-& (1-\xi) \Bigl \{ U_{K4}(P_1,P_2,m_{Z_0},m_{\chi_0},m_{\chi_0}) + m^2_{\chi_0}g_{\m\n} C^{\m\n} (P_1,P_2,m_{Z_0},m_{\chi_0},m_{\chi_0}) \nonumber\\
&+& 2 (P_1 + P_2)_\m C_{{\cal K}3}^\m(P_1,P_2,m_{Z_0},m_{\chi_0},m_{\chi_0}) \nonumber\\
&+& (P_{1\m}P_{1\n} +3P_{1\m}P_{2\n} +P_{2\m}P_{2\n} )C^{\m\n}(P_1,P_2,m_{Z_0},m_{\chi_0},m_{\chi_0})\nonumber\\ 
&+& P_1 \cdot P_2 g_{\m\n} C^{\m\n} (P_1,P_2,m_{Z_0},m_{\chi_0},m_{\chi_0}) \nonumber\\
&+& P_1 \cdot P_2 P_{1\m}P_{2\n} D^{\m\n}(P_1,P_2,P_3m_{Z_0},m_{\chi_0},m_{\chi_0},m_{\chi_0})    \Bigr \}    \Biggr \}\, .
\eea
Now, similarly to the previous case there are three ways to insert the gauge boson in the loop,
which means that there are two more diagrams of the same kind as ${\cal K}^{R_\xi,\chi Z\chi}_{H}$:
\be
\begin{tikzpicture}[scale=0.5]
\draw [] (-1,0) -- (1,1);
\draw [photon] (-1,0) -- (1,-1);
\draw [] (1,1) -- (1,-1);
\draw [dashed,thick] (-2,0) -- (-1,0);
\draw [dashed,thick] (1,1) -- (2.5,1);
\draw [dashed,thick] (1,-1) -- (2.5,-1);
\node at (6.5,0) {$=\,\,  i{\cal K}^{R_\xi,\chi \chi Z}_{H}(P_1,P_2)$};
\end{tikzpicture}
\hskip .85cm
\begin{tikzpicture}[scale=0.5]
\draw [photon] (-1,0) -- (1,1);
\draw [] (-1,0) -- (1,-1);
\draw [] (1,1) -- (1,-1);
\draw [dashed,thick] (-2,0) -- (-1,0);
\draw [dashed,thick] (1,1) -- (2.5,1);
\draw [dashed,thick] (1,-1) -- (2.5,-1);
\node at (6.5,0) {$=\,\,  i{\cal K}^{R_\xi,Z \chi \chi}_{H}(P_1,P_2).$};
\end{tikzpicture}
\hskip .5cm
\nonumber
\ee
Again here, at zero external momenta all three diagrams have the same divergent and finite parts:
\bea
{\cal K}^{R_\xi, \chi Z \chi}_{H}(0,0) = {\cal K}^{R_\xi, \chi \chi Z}_{H}(0,0) = {\cal K}^{R_\xi,Z \chi \chi}_{H}(0,0).
\eea
The last one-loop correction to the three-point vertex comes from the irreducible Triangle
\vskip .5cm
\begin{center}
\begin{tikzpicture}[scale=0.7]
\draw [] (-1,0) -- (1,1);
\draw [] (-1,0) -- (1,-1);
\draw [] (1,1) -- (1,-1);
\draw [dashed,thick] (-2,0) -- (-1,0);
\draw [dashed,thick] (1,1) -- (2.5,1);
\draw [dashed,thick] (1,-1) -- (2.5,-1);
\node at (1.7,1.3) {$p_3$};
\draw [<-] (1.3,0.8) -- (1.9,0.8);
\node at (1.7,-1.3) {$p_2$};
\draw [<-] (1.3,-0.8) -- (1.9,-0.8);
\node at (-1.9,0.2) {$p_1$};
\draw [->] (-1.6,0.2) -- (-1.1,0.2);
\draw [->] (-0.4,0.1) -- (0.2,0.4);
\node at (-0.5,1) {$k+P_1$};
\draw [<-] (-0.4,-0.1) -- (0.2,-0.4);
\node at (-0.1,-1) {$k$};
\draw [<-] (0.8,-0.3) -- (0.8,0.3);
\node at (2.2,0) {$k+P_2$};
\node at (5.5,0) {$=\,\,  i{\cal K}^{R_\xi,\chi\chi\chi}_{H}$};.
\end{tikzpicture}
\end{center}
given by the expression
\bea
{\cal K}^{R_\xi,\chi\chi\chi}_H&=&- 32 v_0  \frac{\l_0^2 }{m^4_{H_0}}  \int {\frac{{{d^4}k}}{{{{\left( {2\pi } \right)}^4}}}\frac{-i k \cdot (k+ P_1)}{D_1 D_2 D_3} (k+P_1) \cdot (k+ P_2) }  (k+P_2) \cdot k\, , \nonumber\\
\eea
with symmetry factor ${\cal S}^{9}_{{\cal {K}}_H } = 1$. There is only one diagram of this kind and its explicit form in DR reads 
\bea\label{full KH9}
{\cal K}^{R_\xi,\chi\chi\chi}_H(P_1,P_2) &=& - 32 v_0  \frac{\l_0^2 }{m^4_{H_0}}  \m^{\ve} \Biggl \{U_{K6}(P_1,P_2,m_{\chi_0},m_{\chi_0},m_{\chi_0}) \nonumber\\
&+& 2(P_1 + P_2)_\m U^\m_{K5}(P_1,P_2,m_{\chi_0},m_{\chi_0},m_{\chi_0})  \nonumber\\
&+& (P_{1\m}P_{1\n} +3P_{1\m}P_{2\n} +P_{2\m}P_{2\n} ) U^{\m\n}_{K4}(P_1,P_2,m_{\chi_0},m_{\chi_0},m_{\chi_0}) \nonumber\\
&+& P_1 \cdot P_2 U_{K4}(P_1,P_2,m_{\chi_0},m_{\chi_0},m_{\chi_0}) \nonumber\\
&+& (P_{1\m}P_{1\n}P_{2\a} + P_{1\m}P_{2\n}P_{2\a} )   C^{\m\n\a}(P_1,P_2,m_{\chi_0},m_{\chi_0},m_{\chi_0}) \nonumber\\
&+& P_1 \cdot P_2 (P_{1\m} + P_{2\m} ) C^\m_{{\cal K}3}(P_1,P_2,m_{\chi_0},m_{\chi_0},m_{\chi_0})\nonumber\\
&+& P_1 \cdot P_2 P_{1\m}P_{2\n} C^{\m\n} (P_1,P_2,m_{\chi_0},m_{\chi_0},m_{\chi_0})        \Biggr \}\, .
\eea
Summing up all the irreducible Triangles, we find that at zero external momenta
\bea\label{fullKHIRDRxi}
(4\pi)^{d/2} {\cal K}^{R_\xi, {\rm irred.}}_{H} &=& \mu ^{\ve} \left( [{\cal K}^{R_\xi, {\rm irred.}}_{H} ]_{\ve} + 
\left\{{\cal K}^{R_\xi, {\rm irred.}}_{H} \right\}_{\ve} +[ {\cal K}^{R_\xi, {\rm irred.}}_{H} ]_{f} + \left\{ {\cal K}^{R_\xi, {\rm irred.}}_{H} \right\}_{f} \right)\nonumber\\
\eea
with 
\bea\label{KHIRDRxinotation}
[{\cal K}^{R_\xi, {\rm irred.}}_{H} ]_{\ve}  &=& 0 
\nonumber\\
\left\{ {\cal K}^{R_\xi, {\rm irred.}}_{H} \right\}_{\ve} &=& 0
\eea
and
\bea
[{\cal K}^{R_\xi, {\rm irred.}}_{H}]_f &=& - \frac{m_{H_0}}{\sqrt{2 \l_0}} \Bigl ( 
54 \l_0^2 + 48 \frac{ \l^2_0 m^4_{Z_0}}{m_{H_0}^4}   \Bigr )\nonumber\\
\left\{ {\cal K}^{R_\xi, {\rm irred.}}_{H} \right\}_{f} &=& 0 \, .
\eea
We see that the irreducible Triangles do not contribute to the divergent part of the 3-point function. 

It is worth looking a bit closer at the cancellation of the gauge fixing parameter $\xi$ from the finite part of this sector.
The finite parts, collected according to the loop propagators, are
\be
\begin{tikzpicture}[scale=0.5]
\draw [photon] (-1,0) -- (1,1);
\draw [photon] (-1,0) -- (1,-1);
\draw [photon] (1,1) -- (1,-1);
\draw [dashed,thick] (-2,0) -- (-1,0);
\draw [dashed,thick] (1,1) -- (2.5,1);
\draw [dashed,thick] (1,-1) -- (2.5,-1);
\node at (10,0) {$=\,\,  \frac{m_{H_0}}{\sqrt{2\l_0}} \Bigl ( - \frac{48 \l_0^2 m_{Z_0}^4}{ m_{H_0}^4} - \frac{16 \l_0^2 m_{Z_0}^4 \xi^2}{ m_{H_0}^4} \Bigr ) $};.
\end{tikzpicture}
\nonumber
\ee
\be
\begin{tikzpicture}[scale=0.5]
\draw [photon] (-1,0) -- (1,1);
\draw [photon] (-1,0) -- (1,-1);
\draw [dashed,thick] (1,1) -- (1,-1);
\draw [dashed,thick] (-2,0) -- (-1,0);
\draw [dashed,thick] (1,1) -- (2.5,1);
\draw [dashed,thick] (1,-1) -- (2.5,-1);
\node at (10.5,0) {$=\,\,  \frac{m_{H_0}}{\sqrt{2\l_0}} \Bigl (\frac{48 \l_0^2 m_{Z_0}^4 \xi^2}{ m_{H_0}^4} - \frac{96 \l_0^2 m_{Z_0}^4 \xi^2}{ m_{H_0}^4} \ln \frac{\m^2}{m_{Z_0}^2 \xi}  \Bigr )  $};.
\end{tikzpicture}
\hskip .75cm
\nonumber
\ee
\be
\begin{tikzpicture}[scale=0.5]
\draw [dashed,thick] (-1,0) -- (1,1);
\draw [dashed,thick] (-1,0) -- (1,-1);
\draw [photon] (1,1) -- (1,-1);
\draw [dashed,thick] (-2,0) -- (-1,0);
\draw [dashed,thick] (1,1) -- (2.5,1);
\draw [dashed,thick] (1,-1) -- (2.5,-1);
\node at (10,0) {$=\,\, \frac{m_{H_0}}{\sqrt{2\l_0}} \Bigl (\frac{48 \l_0^2 m_{Z_0}^4 \xi^2}{ m_{H_0}^4} + \frac{288 \l_0^2 m_{Z_0}^4 \xi^2}{ m_{H_0}^4} \ln \frac{\m^2}{m_{Z_0}^2 \xi}  \Bigr ) $};.
\end{tikzpicture}
\nonumber
\ee
\be
\begin{tikzpicture}[scale=0.5]
\draw [] (-1,0) -- (1,1);
\draw [] (-1,0) -- (1,-1);
\draw [] (1,1) -- (1,-1);
\draw [dashed,thick] (-2,0) -- (-1,0);
\draw [dashed,thick] (1,1) -- (2.5,1);
\draw [dashed,thick] (1,-1) -- (2.5,-1);
\node at (10.5,0) {$=\,\,  \frac{m_{H_0}}{\sqrt{2\l_0}} \Bigl (\frac{-80 \l_0^2 m_{Z_0}^4 \xi^2}{ m_{H_0}^4} - \frac{192 \l_0^2 m_{Z_0}^4 \xi^2}{ m_{H_0}^4} \ln \frac{\m^2}{m_{Z_0}^2 \xi}  \Bigr ) $};.
\end{tikzpicture}
\nonumber
\ee
It is easy to see the cancellation of $\xi$ in the sum.

We now add reducible and irreducible contributions into ${\cal K}^{R_\xi}_H = {\cal K}^{R_\xi, {\rm red.}}_H + {\cal K}^{R_\xi, {\rm irred.}}_H$ and we have
\bea\label{fullKHRxi}
(4\pi)^{d/2} {\cal K}^{R_\xi}_H = \mu ^{\ve} \left( [{\cal K}^{R_\xi}_H ]_{\varepsilon} + \left\{{\cal K}^{R_\xi}_H \right\}_{\varepsilon} +[ {\cal K}^{R_\xi}_H ]_{f} + \left\{ {\cal K}^{R_\xi}_H  \right\}_{f} \right)
\eea
where 
\bea\label{KHRxinotation}
\ve [{\cal K}^{R_\xi}_H ]_{\ve}  &=& \frac{m_{H_0}}{\sqrt{2 \l_0}}  \left( 108  \l_0^2 + 144 \frac{ \l_0^2 m_{Z_0}^4}{m_{H_0}^4} \right)
\nonumber\\
\left\{ {\cal K}^{R_\xi}_H \right\}_{\varepsilon} &=& 0
\eea
and
\bea\label{KHRxifinite}
[{\cal K}^{R_\xi}_H]_f &=& \frac{m_{H_0}}{\sqrt{2 \l_0}} \Bigl (
-54 \l_0^2 -  24 \frac{ \l^2_0 m^4_{Z_0}}{m_{H_0}^4} + 54 \l_0^2 \ln \frac{\m^2}{m_{H_0}^2} + 72 \frac{ \l^2_0 m^4_{Z_0}}{m_{H_0}^4} \ln \frac{\m^2}{m_{Z_0}^2}  \Bigr )\nonumber\\
\left\{ {\cal K}^{R_\xi}_H \right\}_{f} &=& 0 \, .
\eea

\subsection{Corrections to the quartic coupling}\label{Corrections to the quartic coupling in Rxi}

Diagrams with four external Higgs fields contribute through their divergent parts to the running of the Higgs quartic self coupling $\l$ and 
through their finite parts they contribute to the one-loop scalar potential. 
They are collectively called "Box diagrams", denoted as ${\cal B}_H$ and come in three classes. The first two of these classes contain reducible diagrams and the third
class contains irreducible Box diagrams, called "Square ($S$)-Boxes". The reducible class is further split in two subclasses, called "Candy ($C$)-Boxes" and "Triangular ($T$)-Boxes".
They have the following structure:
\be
\begin{tikzpicture}[scale=0.5]
\draw [dashed,thick] (1,0)--(2.5,1.5);
\draw [dashed,thick] (1,0)--(2.5,-1.5);
\draw [dashed,thick] (-2.5,1.5)--(-1,0);
\draw [dashed,thick] (-2.5,-1.5)--(-1,0);
\draw [thick] [fill=gray] (0,0) circle [radius=1];
\node at (4,0) {$=\, \, {\cal B}^C_H,$};
\end{tikzpicture}
\hskip .85cm
\begin{tikzpicture}[scale=0.5]
\draw [dashed,thick] (0.8,0.5)--(2.3,1.3);
\draw [dashed,thick] (0.8,-0.5)--(2.3,-1.3);
\draw [dashed,thick] (-2.5,1.5)--(-1,0);
\draw [dashed,thick] (-2.5,-1.5)--(-1,0);
\draw [thick] [fill=gray] (0,0) circle [radius=1];
\node at (4,0) {$=\,\, {\cal B}^T_H,$};
\end{tikzpicture}
\hskip .5cm
\begin{tikzpicture}[scale=0.5]
\draw [dashed,thick] (0.8,0.5)--(2.3,1.3);
\draw [dashed,thick] (0.8,-0.5)--(2.3,-1.3);
\draw [dashed,thick] (-0.8,0.5)--(-2.3,1.3);
\draw [dashed,thick] (-0.8,-0.5)--(-2.3,-1.3);
\draw [thick] [fill=gray] (0,0) circle [radius=1];
\node at (4,0) {$ =\,\, {\cal B}^S_H $ };
\end{tikzpicture}
\nonumber
\ee
Regarding the momentum flow, we take all four external momenta $p_i=1,2,3,4$ to be inflowing and satisfying $p_1+p_2+p_3+p_4=0$.
Candies and $S$-Boxes come in three versions, corresponding to the usual $s$, $t$ and $u$ channels, where
\be
s=(p_1+p_2)^2,\hskip 1cm t=(p_1+p_3)^2,\hskip 1cm u=(p_1+p_4)^2
\ee
$T$-Boxes come in six versions instead because they are not invariant under a reflection with respect to the axis passing through the centre of the loop in the diagram.
There are two inequivalent topologies and each topology comes with $s$, $t$ and $u$ channels.
Any $U$-integral that may appear is dealt with in Appendix \ref{Uint} and finite integrals of the $E$, $F$, $G$ and $H$-type in Appendix \ref{FiniteParts}.

\subsubsection{Reducible Boxes}\label{Reducible Boxes}

Candies have the generic momentum dependence ${\cal B}^{R_\xi, C}_{H}(P_1)$, where $P_1=\sqrt{s}$, $\sqrt{t}$ and $\sqrt{u}$ for the three channels.
Their total contribution is then a sum over $P_1$.

The first Candy is the famous diagram
\vskip .5cm
\begin{center}
\begin{tikzpicture}[scale=0.7]
\draw [dashed,thick] (0.9,0)--(2.5,1.5);
\draw [dashed,thick] (0.9,0)--(2.5,-1.5);
\draw [dashed,thick] (-2.5,1.5)--(-0.9,0);
\draw [dashed,thick] (-2.5,-1.5)--(-0.9,0);
\draw [<-]  (1.7,1)--(2.2,1.5);
\node at (2.7,1.1) {$p_3$};
\draw [<-]  (1.7,-1)--(2.2,-1.5);
\node at (2.7,-1.1) {$p_4$};
\draw [<-] (-1.7,1)--(-2.2,1.5);
\node at (-2.7,1.1) {$p_1$};
\draw [<-] (-1.7,-1)--(-2.2,-1.5);
\node at (-2.7,-1.1) {$p_2$};
\draw [dashed,thick] (0,0) circle [radius=0.9];
\draw [<-]  (0.45,0.6) arc [start angle=45, end angle=135, radius=0.6cm];
\node at (0,1.2) {$k+P_1$};
\draw [->]  (0.45,-0.6) arc [start angle=-45, end angle=-135, radius=0.6cm];
\node at (0,-1.2) {$k$};
\node at (6,0) {$=\, \, i{\cal B}^{R_\xi,\phi\phi}_{H}, \,\,\, S^1_{{\cal B}^{C}_H} = \frac{1}{2} $};
\end{tikzpicture}
\end{center}
which solely determines the $\b$-function in pure scalar theories, in the case where the Higgs is expressed in the Cartesian basis,
all other diagrams being finite. Here in the Polar basis we will see that this is still the case alas in a non-trivial way.
The result for this diagram can be obtained from \eq{RedTriangle1} (without the factor of 3) divided by $v_0$ and evaluated at $P_1$:
\bea\label{full BHC1}
{\cal B}^{R_\xi,\phi\phi}_{H}&= &18 \l^2_0 {\mu ^{\ve}}B_0(P_1,m_{H_0},m_{H_0})\, .
\eea
The Goldstone Candy
\vskip .5cm
\begin{center}
\begin{tikzpicture}[scale=0.7]
\draw [dashed,thick] (0.9,0)--(2.5,1.5);
\draw [dashed,thick] (0.9,0)--(2.5,-1.5);
\draw [dashed,thick] (-2.5,1.5)--(-0.9,0);
\draw [dashed,thick] (-2.5,-1.5)--(-0.9,0);
\draw [<-]  (1.7,1)--(2.2,1.5);
\node at (2.7,1.1) {$p_3$};
\draw [<-]  (1.7,-1)--(2.2,-1.5);
\node at (2.7,-1.1) {$p_4$};
\draw [<-] (-1.7,1)--(-2.2,1.5);
\node at (-2.7,1.1) {$p_1$};
\draw [<-] (-1.7,-1)--(-2.2,-1.5);
\node at (-2.7,-1.1) {$p_2$};
\draw [] (0,0) circle [radius=0.9];
\draw [<-]  (0.45,0.6) arc [start angle=45, end angle=135, radius=0.6cm];
\node at (0,1.2) {$k+P_1$};
\draw [->]  (0.45,-0.6) arc [start angle=-45, end angle=-135, radius=0.6cm];
\node at (0,-1.2) {$k$};
\node at (6,0) {$=\, \, i{\cal B}^{R_\xi,\chi\chi}_{H}, \,\,\,  S^2_{{\cal B}^C_H} = \frac{1}{2} $};
\end{tikzpicture}
\end{center}
is analogously equal to \eq{RedTriangle2} divided by $v_0$ (without the factor of 3), evaluated at $P_1$:
\bea\label{full BHC2}
{\cal B}^{R_\xi,\chi\chi}_{H} &= & 8 \frac{\l^2_0 }{m^4_{H_0}} {\mu ^{\ve}}\Biggl \{ m_{\chi_0}^2 A_0(m_{\chi_0}) + (m_{\chi_0}^2 - P_1^2) g_{\m\n} B^{\m\n}(P_1,m_{\chi_0},m_{\chi_0}) \nonumber\\
&+& P_{1\m} P_{1\n} B^{\m\n}(P_1,m_{\chi_0},m_{\chi_0}) \Biggr \}\, .
\eea
The gauge Candy 
\vskip .5cm
\begin{center}
\begin{tikzpicture}[scale=0.7]
\draw [dashed,thick] (0.9,0)--(2.5,1.5);
\draw [dashed,thick] (0.9,0)--(2.5,-1.5);
\draw [dashed,thick] (-2.5,1.5)--(-0.9,0);
\draw [dashed,thick] (-2.5,-1.5)--(-0.9,0);
\draw [<-]  (1.7,1)--(2.2,1.5);
\node at (2.7,1.1) {$p_3$};
\draw [<-]  (1.7,-1)--(2.2,-1.5);
\node at (2.7,-1.1) {$p_4$};
\draw [<-] (-1.7,1)--(-2.2,1.5);
\node at (-2.7,1.1) {$p_1$};
\draw [<-] (-1.7,-1)--(-2.2,-1.5);
\node at (-2.7,-1.1) {$p_2$};
\draw [photon] (0,0) circle [radius=0.9];
\draw [<-]  (0.45,0.6) arc [start angle=45, end angle=135, radius=0.6cm];
\node at (0,1.3) {$k+P_1$};
\draw [->]  (0.45,-0.4) arc [start angle=-45, end angle=-135, radius=0.6cm];
\node at (0,-1.1) {$k$};
\node at (6,0) {$=\, \, i{\cal B}^{R_\xi,ZZ}_{H}, \,\,\, S^3_{{\cal B}^C_H} = \frac{1}{2}$};
\end{tikzpicture}
\end{center}
is obtained from \eq{RedTriangle3}:
\bea\label{full BHC3}
{\cal B}^{R_\xi,ZZ}_{H} &=& 8 \frac{m^4_{Z_{0}}}{m^4_{H_0}}\l^2_0  {\mu ^{\ve}}\Biggl \{d B_0(P_1,m_{Z_0},m_{Z_0}) - (1-\xi)  \Bigl \{g_{\m\n} C^{1\m\n}(P_1,m_{Z_0},m_{Z_0},m_{\chi_0}) \nonumber\\
&+& g_{\m\n} C^{1\m\n}(P_1,m_{Z_0},m_{Z_0},m_{\chi_0})  \Bigr \} \nonumber\\ 
&+& (1-\xi)^2 \Bigl \{ g_{\m\n} C^{1\m\n}(P_1,m_{Z_0},m_{Z_0},m_{\chi_0})  \nonumber\\
&+& (m_{Z_0}^2 - P_1^2) g_{\m\n}D^{\m\n}(P_1,m_{Z_0},m_{Z_0},m_{\chi_0},m_{\chi_0}) + P_{1\m} P_{1\n} D^{\m\n}(P_1,m_{Z_0},m_{Z_0},m_{\chi_0},m_{\chi_0})\, ,
\Bigr \}   \Biggr \} \, .\nonumber\\
\eea
Finally, there is a mixed Candy
\vskip .5cm
\begin{center}
\begin{tikzpicture}[scale=0.7]
\draw [dashed,thick] (1,0)--(2.5,1.5);
\draw [dashed,thick] (1,0)--(2.5,-1.5);
\draw [dashed,thick] (-2.5,1.5)--(-1,0);
\draw [dashed,thick] (-2.5,-1.5)--(-1,0);
\draw [<-]  (1.5,0.8)--(2.2,1.5);
\node at (2.7,1.1) {$p_3$};
\draw [<-]  (1.5,-0.8)--(2.2,-1.5);
\node at (2.7,-1.1) {$p_4$};
\draw [<-] (-1.5,0.8)--(-2.2,1.5);
\node at (-2.7,1.1) {$p_1$};
\draw [<-] (-1.5,-0.8)--(-2.2,-1.5);
\node at (-2.7,-1.1) {$p_2$};
\draw  [photon] (-1,0) .. controls (-1,0.555) and (-0.555,1) .. (0,1)
.. controls (0.555,1) and (1,0.555) .. (1,0);
\draw  [] (-1,0) .. controls (-1,-0.555) and (-0.555,-1) .. (0,-1)
.. controls (0.555,-1) and (1,-0.555) .. (1,0);
\draw [<-]  (0.45,0.7) arc [start angle=45, end angle=135, radius=0.6cm];
\node at (0,1.45) {$k+P_1$};
\draw [->]  (0.45,-0.6) arc [start angle=-45, end angle=-135, radius=0.6cm];
\node at (0,-1.25) {$k$};
\node at (6,0) {$=\, \, i{\cal B}^{R_\xi, Z\chi}_{H}, \,\,\, S^4_{{\cal B}^C_H} = 1$};
\end{tikzpicture}
\end{center}
obtained easily from \eq{RedTriangle4}:
\bea\label{full BHC4}
{\cal B}^{R_\xi, Z\chi}_{H} &=& 16  \l^2_0 \frac{m^2_{Z_{0}}}{m^4_{H_0}}  {\mu ^{\ve}}   \Biggl \{ - g_{\m\n} B^{\m\n}(P_1,m_{\chi_0},m_{Z_0}) \nonumber\\
&+& (1-\xi )\Bigl \{ g_{\m\n} B^{\m\n}(P_1,m_{\chi_0},m_{\chi_0}) + (m_{Z_0}^2 - P_1^2) g_{\m\n} C^{1\m\n}(P_1,m_{Z_0},m_{\chi_0},m_{Z_0}) \nonumber\\
&+&P_{1\m} P_{1\n} C^{1\m\n}(P_1,m_{\chi_0},m_{Z_0},m_{\chi_0}) \Bigr \} \Biggr \}\, .
\eea
The full contribution of the Candies is obtained by adding \eq{full BHC1}, \eq{full BHC2}, \eq{full BHC3} and \eq{full BHC4} (summed over the three channels)
and can be expressed as
\bea\label{fullBHCRxi}
(4\pi)^{d/2} {\cal B}^{R_\xi, C}_{H} = \mu ^{\ve} \left( [{\cal B}^{R_\xi, C}_{H}]_{\ve} + \left\{{\cal B}^{R_\xi, C}_{H}\right\}_{\ve} +[ {\cal B}^{R_\xi, C}_{H}]_{f} + \left\{ {\cal B}^{R_\xi, C}_{H}\right\}_{f} \right)
\eea
where 
\bea\label{BHCRxinotation}
\ve [{\cal B}^{R_\xi, C}_{H} ]_{\ve} &=& 108 \l_0^2 + 144  \frac{\l_0^2 m^4_{Z_0}}{m^4_{H_0}} +  4\frac{s^2 \l_0^2 }{m^4_{H_0}} + 4 \frac{t^2 \l_0^2 }{m^4_{H_0}} +  4\frac{u^2 \l_0^2 }{m^4_{H_0}} \nonumber\\
\left\{ {\cal B}^{R_\xi,C}_{H} \right\}_{\ve}& =& 0.
\eea
and
\bea
[{\cal B}^{R_\xi, C}_{H}]_f &=& 54 \l_0^2 \ln \frac{\m^2}{m_{H_0}^2} + 72 \frac{ \l^2_0 m^4_{Z_0}}{m_{H_0}^4} \ln \frac{\m^2}{m_{Z_0}^2} \nonumber\\
\left\{ {\cal B}^{R_\xi, C}_{H} \right\}_{f} &=& 0 
\eea
for $p_i=0$.
We now turn to the $T$-Boxes that have three propagators in the loop.
Each of the six channels of a given $T$-Box is determined by two linear combinations of the external momenta that we call $P_1$ and $P_2$.
A consistent choice for $(P_1,P_2)$ for the channels $T_{1,\cdots ,6}$ is for example $T_1:(\sqrt{s},p_1+p_2+p_3)$,
$T_2:(\sqrt{t},p_1+p_3+p_4)$, $T_3:(\sqrt{u},p_1+p_3+p_4)$, $T_4:(\sqrt{s},p_2+p_3+p_4)$, $T_5:(p_1,\sqrt{t})$, $T_6:(p_1,\sqrt{u})$.
The generic form of a $T$-Box is therefore ${\cal B}^{R_\xi,T}_{H}(P_1,P_2)$ and the total contribution is obtained by a sum over the six different pairs $(P_1,P_2)$.
Note that the Mandelstam variables enter also via the inner products
\bea\label{stuMan}
p_1 \cdot p_2 &=& p_3 \cdot p_4 = \frac{s}{2} \nonumber\\
p_1 \cdot p_3 &=& p_2 \cdot p_4 = \frac{t}{2}  \nonumber\\
p_1 \cdot p_4 &=& p_2 \cdot p_3 = \frac{u}{2} 
\eea 
All $T$-Boxes can be obtained from their corresponding irreducible Triangles.
They share a common symmetry factor and analytical structure with all the difference encoded in the values that the pair $(P_1,P_2)$ can take.
Hence, in this sector we give directly the formal expression in order to avoid unnecessary repetitions.
To begin, we have two finite diagrams:
\vskip .5cm
\begin{center}
\begin{tikzpicture}[scale=0.7]
\draw [dashed,thick] (-1,0) -- (1,1);
\draw [dashed,thick] (-1,0) -- (1,-1);
\draw [dashed,thick] (1,1) -- (1,-1);
\draw [dashed,thick] (-2,1) -- (-1,0);
\draw [dashed,thick] (-2,-1) -- (-1,0);
\draw [dashed,thick] (1,1) -- (2.5,1);
\draw [dashed,thick] (1,-1) -- (2.5,-1);
\node at (1.7,1.3) {$p_3$};
\draw [<-] (1.3,0.8) -- (1.9,0.8);
\node at (1.7,-1.3) {$p_4$};
\draw [<-] (1.3,-0.8) -- (1.9,-0.8);
\node at (-1.9,1.3) {$p_1$};
\draw [->] (-1.6,0.9) -- (-1.1,0.4);
\node at (-1.9,-1.30) {$p_2$};
\draw [->] (-1.6,-0.9) -- (-1.1,-0.4);
\draw [->] (-0.1,0.2) -- (0.5,0.5);
\node at (-0.4,1) {$k+P_1$ };
\draw [<-] (-0.1,-0.2) -- (0.5,-0.5);
\node at (0,-0.9) {$k$};
\draw [<-] (0.8,-0.3) -- (0.8,0.3);
\node at (2.0,0) {$k+P_2$};
\node at (8.50,0) {$=\,\, i{\cal B}^{R_\xi,\phi\phi\phi}_{H}=\sum_{(P_1,P_2)} \frac{1}{v_0} i{\cal K}_H^{R_\xi,\phi\phi\phi}(P_1,P_2)$};
\end{tikzpicture}
\end{center}
and 
\vskip .5cm
\begin{center}
\begin{tikzpicture}[scale=0.7]
\draw [photon] (-1,0) -- (1,1);
\draw [photon] (-1,0) -- (1,-1);
\draw [photon] (1,1) -- (1,-1);
\draw [dashed,thick] (-2,1) -- (-1,0);
\draw [dashed,thick] (-2,-1) -- (-1,0);
\draw [dashed,thick] (1,1) -- (2.5,1);
\draw [dashed,thick] (1,-1) -- (2.5,-1);
\node at (1.7,1.3) {$p_3$};
\draw [<-] (1.3,0.8) -- (1.9,0.8);
\node at (1.7,-1.3) {$p_4$};
\draw [<-] (1.3,-0.8) -- (1.9,-0.8);
\node at (-1.9,1.3) {$p_1$};
\draw [->] (-1.6,0.9) -- (-1.1,0.4);
\node at (-1.9,-1.30) {$p_2$};
\draw [->] (-1.6,-0.9) -- (-1.1,-0.4);
\draw [->] (-0.1,0.2) -- (0.5,0.5);
\node at (-0.4,1) {$k+P_1$ };
\draw [<-] (-0.1,-0.2) -- (0.5,-0.5);
\node at (0,-0.9) {$k$};
\draw [<-] (0.8,-0.3) -- (0.8,0.3);
\node at (2.0,0) {$k+P_2$};
\node at (8.50,0) {$=\,\, i{\cal B}^{R_\xi,ZZZ}_{H}=\sum_{(P_1,P_2)} \frac{1}{v_0} i{\cal K}_H^{R_\xi,ZZZ}(P_1,P_2)$};
\end{tikzpicture}
\end{center}
There are diagrams that have both divergent and finite parts. There are three with one Goldstone and two gauge bosons in the loop:
\be
\begin{tikzpicture}[scale=0.7]
\draw [photon] (-1,0) -- (1,1);
\draw [photon] (-1,0) -- (1,-1);
\draw [] (1,1) -- (1,-1);
\draw [dashed,thick] (-2,1) -- (-1,0);
\draw [dashed,thick] (-2,-1) -- (-1,0);
\draw [dashed,thick] (1,1) -- (2.5,1);
\draw [dashed,thick] (1,-1) -- (2.5,-1);
\node at (1.7,1.3) {$p_3$};
\draw [<-] (1.3,0.8) -- (1.9,0.8);
\node at (1.7,-1.3) {$p_4$};
\draw [<-] (1.3,-0.8) -- (1.9,-0.8);
\node at (-1.9,1.3) {$p_1$};
\draw [->] (-1.6,0.9) -- (-1.1,0.4);
\node at (-1.9,-1.30) {$p_2$};
\draw [->] (-1.6,-0.9) -- (-1.1,-0.4);
\draw [->] (-0.1,0.2) -- (0.5,0.5);
\node at (-0.4,1) {$k+P_1$ };
\draw [<-] (-0.1,-0.2) -- (0.5,-0.5);
\node at (0,-0.9) {$k$};
\draw [<-] (0.8,-0.3) -- (0.8,0.3);
\node at (2.0,0) {$k+P_2$};
\end{tikzpicture}
\hskip .75cm
\begin{tikzpicture}[scale=0.7]
\draw [] (-1,0) -- (1,1);
\draw [photon] (-1,0) -- (1,-1);
\draw [photon] (1,1) -- (1,-1);
\draw [dashed,thick] (-2,1) -- (-1,0);
\draw [dashed,thick] (-2,-1) -- (-1,0);
\draw [dashed,thick] (1,1) -- (2.5,1);
\draw [dashed,thick] (1,-1) -- (2.5,-1);
\node at (1.7,1.3) {$p_3$};
\draw [<-] (1.3,0.8) -- (1.9,0.8);
\node at (1.7,-1.3) {$p_4$};
\draw [<-] (1.3,-0.8) -- (1.9,-0.8);
\node at (-1.9,1.3) {$p_1$};
\draw [->] (-1.6,0.9) -- (-1.1,0.4);
\node at (-1.9,-1.30) {$p_2$};
\draw [->] (-1.6,-0.9) -- (-1.1,-0.4);
\draw [->] (-0.1,0.2) -- (0.5,0.5);
\node at (-0.4,1) {$k+P_1$ };
\draw [<-] (-0.1,-0.2) -- (0.5,-0.5);
\node at (0,-0.9) {$k$};
\draw [<-] (0.8,-0.3) -- (0.8,0.3);
\node at (2.0,0) {$k+P_2$};
\end{tikzpicture}
\hskip .75cm
\begin{tikzpicture}[scale=0.7]
\draw [photon] (-1,0) -- (1,1);
\draw [] (-1,0) -- (1,-1);
\draw [photon] (1,1) -- (1,-1);
\draw [dashed,thick] (-2,1) -- (-1,0);
\draw [dashed,thick] (-2,-1) -- (-1,0);
\draw [dashed,thick] (1,1) -- (2.5,1);
\draw [dashed,thick] (1,-1) -- (2.5,-1);
\node at (1.7,1.3) {$p_3$};
\draw [<-] (1.3,0.8) -- (1.9,0.8);
\node at (1.7,-1.3) {$p_4$};
\draw [<-] (1.3,-0.8) -- (1.9,-0.8);
\node at (-1.9,1.3) {$p_1$};
\draw [->] (-1.6,0.9) -- (-1.1,0.4);
\node at (-1.9,-1.30) {$p_2$};
\draw [->] (-1.6,-0.9) -- (-1.1,-0.4);
\draw [->] (-0.1,0.2) -- (0.5,0.5);
\node at (-0.4,1) {$k+P_1$ };
\draw [<-] (-0.1,-0.2) -- (0.5,-0.5);
\node at (0,-0.9) {$k$};
\draw [<-] (0.8,-0.3) -- (0.8,0.3);
\node at (2.0,0) {$k+P_2$};
\end{tikzpicture}
\nonumber
\ee
which are equal to $i{\cal B}^{R_\xi,Z\chi Z}_{H} = \sum_{(P_1,P_2)} \frac{1}{v_0} i{\cal K}_H^{R_\xi,Z\chi Z}(P_1,P_2)$, 
$i{\cal B}^{R_\xi,\chi ZZ}_{H} = \sum_{(P_1,P_2)} \frac{1}{v_0} i{\cal K}_H^{R_\xi,\chi ZZ}(P_1,P_2)$ and 
$i{\cal B}^{R_\xi,ZZ\chi}_{H} = \sum_{(P_1,P_2)} \frac{1}{v_0} i{\cal K}_H^{R_\xi,ZZ\chi}(P_1,P_2)$ respectively.\\
And there are three with one gauge boson and two Goldstones in the loop:
\be
\begin{tikzpicture}[scale=0.7]
\draw [photon] (-1,0) -- (1,1);
\draw [] (-1,0) -- (1,-1);
\draw [] (1,1) -- (1,-1);
\draw [dashed,thick] (-2,1) -- (-1,0);
\draw [dashed,thick] (-2,-1) -- (-1,0);
\draw [dashed,thick] (1,1) -- (2.5,1);
\draw [dashed,thick] (1,-1) -- (2.5,-1);
\node at (1.7,1.3) {$p_3$};
\draw [<-] (1.3,0.8) -- (1.9,0.8);
\node at (1.7,-1.3) {$p_4$};
\draw [<-] (1.3,-0.8) -- (1.9,-0.8);
\node at (-1.9,1.3) {$p_1$};
\draw [->] (-1.6,0.9) -- (-1.1,0.4);
\node at (-1.9,-1.30) {$p_2$};
\draw [->] (-1.6,-0.9) -- (-1.1,-0.4);
\draw [->] (-0.1,0.2) -- (0.5,0.5);
\node at (-0.4,1) {$k+P_1$ };
\draw [<-] (-0.1,-0.2) -- (0.5,-0.5);
\node at (0,-0.9) {$k$};
\draw [<-] (0.8,-0.3) -- (0.8,0.3);
\node at (2.0,0) {$k+P_2$};
\end{tikzpicture}
\hskip .75cm
\begin{tikzpicture}[scale=0.7]
\draw [] (-1,0) -- (1,1);
\draw [] (-1,0) -- (1,-1);
\draw [photon] (1,1) -- (1,-1);
\draw [dashed,thick] (-2,1) -- (-1,0);
\draw [dashed,thick] (-2,-1) -- (-1,0);
\draw [dashed,thick] (1,1) -- (2.5,1);
\draw [dashed,thick] (1,-1) -- (2.5,-1);
\node at (1.7,1.3) {$p_3$};
\draw [<-] (1.3,0.8) -- (1.9,0.8);
\node at (1.7,-1.3) {$p_4$};
\draw [<-] (1.3,-0.8) -- (1.9,-0.8);
\node at (-1.9,1.3) {$p_1$};
\draw [->] (-1.6,0.9) -- (-1.1,0.4);
\node at (-1.9,-1.30) {$p_2$};
\draw [->] (-1.6,-0.9) -- (-1.1,-0.4);
\draw [->] (-0.1,0.2) -- (0.5,0.5);
\node at (-0.4,1) {$k+P_1$ };
\draw [<-] (-0.1,-0.2) -- (0.5,-0.5);
\node at (0,-0.9) {$k$};
\draw [<-] (0.8,-0.3) -- (0.8,0.3);
\node at (2.0,0) {$k+P_2$};
\end{tikzpicture}
\hskip .75cm
\begin{tikzpicture}[scale=0.7]
\draw [] (-1,0) -- (1,1);
\draw [photon] (-1,0) -- (1,-1);
\draw [] (1,1) -- (1,-1);
\draw [dashed,thick] (-2,1) -- (-1,0);
\draw [dashed,thick] (-2,-1) -- (-1,0);
\draw [dashed,thick] (1,1) -- (2.5,1);
\draw [dashed,thick] (1,-1) -- (2.5,-1);
\node at (1.7,1.3) {$p_3$};
\draw [<-] (1.3,0.8) -- (1.9,0.8);
\node at (1.7,-1.3) {$p_4$};
\draw [<-] (1.3,-0.8) -- (1.9,-0.8);
\node at (-1.9,1.3) {$p_1$};
\draw [->] (-1.6,0.9) -- (-1.1,0.4);
\node at (-1.9,-1.30) {$p_2$};
\draw [->] (-1.6,-0.9) -- (-1.1,-0.4);
\draw [->] (-0.1,0.2) -- (0.5,0.5);
\node at (-0.4,1) {$k+P_1$ };
\draw [<-] (-0.1,-0.2) -- (0.5,-0.5);
\node at (0,-0.9) {$k$};
\draw [<-] (0.8,-0.3) -- (0.8,0.3);
\node at (2.0,0) {$k+P_2$};
\end{tikzpicture}
\nonumber
\ee
which are equal to $i{\cal B}^{R_\xi,Z\chi\chi}_{H}=\sum_{(P_1,P_2)} \frac{1}{v_0} i{\cal K}_H^{R_\xi,Z\chi\chi}(P_1,P_2)$, 
$i{\cal B}^{R_\xi,\chi Z\chi}_{H}=\sum_{(P_1,P_2)} \frac{1}{v_0} i{\cal K}_H^{R_\xi,\chi Z\chi}(P_1,P_2)$ and 
$ i{\cal B}^{R_\xi,\chi\chi Z}_{H}=\sum_{(P_1,P_2)} \frac{1}{v_0} i{\cal K}_H^{R_\xi,\chi\chi Z}(P_1,P_2)$ respectively.\\
We finally have the Goldstone $T$-Box
\vskip .5cm
\begin{center}
\begin{tikzpicture}[scale=0.7]
\draw [] (-1,0) -- (1,1);
\draw [] (-1,0) -- (1,-1);
\draw [] (1,1) -- (1,-1);
\draw [dashed,thick] (-2,1) -- (-1,0);
\draw [dashed,thick] (-2,-1) -- (-1,0);
\draw [dashed,thick] (1,1) -- (2.5,1);
\draw [dashed,thick] (1,-1) -- (2.5,-1);
\node at (1.7,1.3) {$p_3$};
\draw [<-] (1.3,0.8) -- (1.9,0.8);
\node at (1.7,-1.3) {$p_4$};
\draw [<-] (1.3,-0.8) -- (1.9,-0.8);
\node at (-1.9,1.3) {$p_1$};
\draw [->] (-1.6,0.9) -- (-1.1,0.4);
\node at (-1.9,-1.30) {$p_2$};
\draw [->] (-1.6,-0.9) -- (-1.1,-0.4);
\draw [->] (-0.1,0.2) -- (0.5,0.5);
\node at (-0.4,1) {$k+P_1$ };
\draw [<-] (-0.1,-0.2) -- (0.5,-0.5);
\node at (0,-0.9) {$k$};
\draw [<-] (0.8,-0.3) -- (0.8,0.3);
\node at (2.0,0) {$k+P_2$};
\node at (9,0) {$=\,\, i{\cal B}^{R_\xi,\chi\chi\chi}_{H}=\sum_{(P_1,P_2)} \frac{1}{v_0} i{\cal K}_H^{R_\xi,\chi\chi\chi}(P_1,P_2)$};
\end{tikzpicture}
\end{center}
Adding up all the $T$-Boxes we find that at zero external momenta
\bea\label{fullBHTRxi}
(4\pi)^{d/2} {\cal B}^{R_\xi,T}_{H} = \mu ^{\ve} \left( [ {\cal B}^{T}_{HR_\xi} ]_{\ve} + \left\{ {\cal B}^{T}_{HR_\xi} \right\}_{\ve} +[  {\cal B}^{T}_{HR_\xi} ]_{f} + \left\{ {\cal B}^{T}_{HR_\xi}  \right\}_{f} \right)
\eea
where 
\bea\label{BHTRxinotation}
\ve [ {\cal B}^{R_\xi,T}_{H} ]_{\ve} &=& -  28\frac{s^2 \l_0^2 }{m^4_{H_0}} - 40 \frac{s \cdot t \l_0^2 }{m^4_{H_0}} - 28 \frac{t^2 \l_0^2 }{m^4_{H_0}} - 40 \frac{s \cdot u \l_0^2 }{m^4_{H_0}} - 40 \frac{t\cdot u \l_0^2 }{m^4_{H_0}} - 28 \frac{u^2 \l_0^2 }{m^4_{H_0}} \nonumber\\
\left\{  {\cal B}^{T}_{HR_\xi} \right\}_{\ve}& =& 0.
\eea
and
\bea
[{\cal B}^{R_\xi,T}_{H}]_f &=& 
-162 \l_0^2 - 288 \frac{ \l^2_0 m^4_{Z_0}}{m_{H_0}^4} \nonumber\\
\left\{ {\cal B}^{R_\xi,T}_{H} \right\}_{f} &=& 0 \, .
\eea
\subsubsection{Irreducible Boxes}\label{Irreducible Boxes}

Irreducible, or $S$-Boxes, have a channel structure determined by three linear combinations of momenta, called $P_1$, $P_2$ and $P_3$.
The resulting three independent channels are the usual $s$, $t$ and $u$ channels
\be
\begin{tikzpicture}[scale=0.65]
\node at (-1.3,0) {$k$ };
\node at (0,1.5) {$k+P_1$ };
\node at (0,-1.5) {$k+P_2$ };
\node at (2.0,0) {$k+P_3$ };
\node at (0,0) {${\bf s}$ };
\draw [->] [thick] (-1,1) -- (1,1);
\draw [->] [thick] (1,1) -- (1,-1);
\draw [->] [thick] (1,-1) -- (-1,-1);
\draw [->] [thick] (-1,-1) -- (-1,1);
\draw [dashed,thick] (-2,2) -- (-1,1);
\draw [<-]  (-1.8,1.3)--(-2.3,1.8);
\node at (-2.3,2.2) {$p_1$};
\draw [dashed,thick] (2,2) -- (1,1);
\draw [<-]  (1.8,1.3)--(2.3,1.8);
\node at (2.3,2.2) {$p_3$};
\draw [dashed,thick] (2,-2) -- (1,-1);
\draw [<-]  (1.8,-1.3)--(2.3,-1.8);
\node at (2.3,-2.2) {$p_4$};
\draw [dashed,thick] (-2,-2) -- (-1,-1);
\draw [<-]  (-1.8,-1.3)--(-2.3,-1.8);
\node at (-2.3,-2.2) {$p_2$};
\end{tikzpicture}
\hskip .75cm
\begin{tikzpicture}[scale=0.65]
\node at (-1.3,0) {$k$ };
\node at (0,1.5) {$k+P_1$ };
\node at (0,-1.5) {$k+P_2$ };
\node at (2.0,0) {$k+P_3$ };
\node at (0,0) {${\bf t}$ };
\draw [->] [thick] (-1,1) -- (1,1);
\draw [->] [thick] (1,1) -- (1,-1);
\draw [->] [thick] (1,-1) -- (-1,-1);
\draw [->] [thick] (-1,-1) -- (-1,1);
\draw [dashed,thick] (-1,1) -- (2,2);
\draw [<-]  (-1.5,1.5)--(-2.3,1.8);
\node at (-2.3,2.2) {$p_1$};
\draw [dashed,thick] (1,1) -- (-2,2);
\draw [<-]  (1.5,1.5)--(2.3,1.8);
\node at (2.3,2.2) {$p_3$};
\draw [dashed,thick] (2,-2) -- (1,-1);
\draw [<-]  (1.8,-1.3)--(2.3,-1.8);
\node at (2.3,-2.2) {$p_4$};
\draw [dashed,thick] (-2,-2) -- (-1,-1);
\draw [<-]  (-1.8,-1.3)--(-2.3,-1.8);
\node at (-2.3,-2.2) {$p_2$};
\end{tikzpicture}
\hskip .75cm
\begin{tikzpicture}[scale=0.65]
\node at (-1.3,0) {$k$ };
\node at (0,1.5) {$k+P_1$ };
\node at (0,-1.5) {$k+P_2$ };
\node at (2.0,0) {$k+P_3$ };
\node at (0,0) {${\bf u}$ };
\draw [->] [thick] (-1,1) -- (1,1);
\draw [->] [thick] (1,1) -- (1,-1);
\draw [->] [thick] (1,-1) -- (-1,-1);
\draw [->] [thick] (-1,-1) -- (-1,1);
\draw [dashed,thick] (-2,2) -- (-1,1);
\draw [<-]  (-1.8,1.3)--(-2.3,1.8);
\node at (-2.3,2.2) {$p_1$};
\draw [dashed,thick] (1,1) -- (2,-2);
\draw [<-]  (2.1,1.2)--(2.3,1.8);
\node at (2.3,2.2) {$p_3$};
\draw [dashed,thick] (1,-1) -- (2,2);
\draw [<-]  (2.1,-1.2)--(2.3,-1.8);
\node at (2.3,-2.2) {$p_4$};
\draw [dashed,thick] (-2,-2) -- (-1,-1);
\draw [<-]  (-1.8,-1.3)--(-2.3,-1.8);
\node at (-2.3,-2.2) {$p_2$};
\end{tikzpicture}
\nonumber
\ee
with the momenta $P_{1,2,3}$ determined easily from the above figure:
\bea\label{Pstu}
s\,\, {\rm -channel}: && P_1=p_1,\,\, P_2=p_1+p_3+p_4 ,\,\, P_3=p_1+p_3  \nonumber\\
t\,\, {\rm  -channel}: && P_1=p_3,\,\,  P_2=p_1+p_3+p_4 ,\,\,  P_3=p_1+p_3 \nonumber\\
u\,\, {\rm -channel}: && P_1=p_1, \,\, P_2=p_1+p_3+p_4 ,\,\,  P_3=p_1+p_4  
\eea
Irreducible Boxes have therefore the generic form ${\cal B}^{R_\xi,S}_{H}(P_1,P_2,P_3)$ and receive a contribution from the $s$, $t$ and $u$ channels. In order to take into account
all channels, after considering the different diagram topologies, we sum over the $(P_1,P_2,P_3)$ according to the above rule.
All $S$-Boxes have symmetry factor 1.
Several annoying for the eye expressions for finite terms $( B^{R_\xi, \cdots}_{H})_f$ are moved to Appendix \ref{FiniteParts}.

We start the computation of the $S$-Boxes with the finite Higgs loop diagram
\vskip .5cm
\begin{center}
\begin{tikzpicture}[scale=0.7]
\draw [dashed,thick] (-1,1) -- (1,1);
\draw [dashed,thick] (1,1) -- (1,-1);
\draw [dashed,thick] (1,-1) -- (-1,-1);
\draw [dashed,thick] (-1,-1) -- (-1,1);
\draw [dashed,thick] (-2,2) -- (-1,1);
\draw [dashed,thick] (-2,-2) -- (-1,-1);
\draw [dashed,thick] (2,2) -- (1,1);
\draw [dashed,thick] (2,-2) -- (1,-1);
\draw [<-]  (-1.8,1.3)--(-2.3,1.8);
\node at (-2.3,2.2) {$p_1$};
\draw [<-]  (-1.8,-1.3)--(-2.3,-1.8);
\node at (-2.3,-2.2) {$p_2$};
\draw [<-]  (1.8,1.3)--(2.3,1.8);
\node at (2.3,2.2) {$p_3$};
\draw [<-]  (1.8,-1.3)--(2.3,-1.8);
\node at (2.3,-2.2) {$p_4$};
%
\draw [->] (-0.3,0.9) -- (0.5,0.9);
\node at (0,1.3) {$k+P_1$ };
\draw [->] (-0.9,-0.2) -- (-0.9,0.5);
\node at (-1.3,0) {$k $};
\draw [<-] (-0.3,-0.9) -- (0.5,-0.9);
\node at (0,-1.4) {$k +P_2$};
\draw [<-] (0.9,-0.2) -- (0.9,0.5);
\node at (2.0,0) {$k+ P_3$};
\node at (5,0) {$=\,\, i{\cal B}^{R_\xi,\phi\phi\phi\phi}_{H}$};
\end{tikzpicture}
\end{center}
given by the expression
\bea
{\cal B}^{R_\xi,\phi\phi\phi\phi}_{H} &=& 324 \l_0^2 m^2_{H_0} \int {\frac{{{d^4}k}}{{{{\left( {2\pi } \right)}^4}}}}
\frac{ (-i)}{D_1 D_2 D_3 D_4}\, ,
\eea
with the $D_i$ in the denominator defined in Appendix \ref{FiniteIntegrals}.
This is just a finite $D_0$ PV integral:
\be\label{full BHRxiR1}
{\cal B}^{R_\xi,\phi\phi\phi\phi}_{H} =
324 \l_0^{2} m^2_{H_0} \m^\ve D_0(P_1,P_2,P_3,m_{H_0},m_{H_0},m_{H_0},m_{H_0})\, .
\ee
The next diagram is also finite. It is
\vskip .5cm
\begin{center}
\begin{tikzpicture}[scale=0.7]
\draw [photon] (-1,1) -- (1,1);
\draw [photon] (1,1) -- (1,-1);
\draw [photon] (1,-1) -- (-1,-1);
\draw [photon] (-1,-1) -- (-1,1);
\draw [dashed,thick] (-2,2) -- (-1,1);
\draw [dashed,thick] (-2,-2) -- (-1,-1);
\draw [dashed,thick] (2,2) -- (1,1);
\draw [dashed,thick] (2,-2) -- (1,-1);
\draw [<-]  (-1.8,1.3)--(-2.3,1.8);
\node at (-2.3,2.2) {$p_1$};
\draw [<-]  (-1.8,-1.3)--(-2.3,-1.8);
\node at (-2.3,-2.2) {$p_2$};
\draw [<-]  (1.8,1.3)--(2.3,1.8);
\node at (2.3,2.2) {$p_3$};
\draw [<-]  (1.8,-1.3)--(2.3,-1.8);
\node at (2.3,-2.2) {$p_4$};
%
\draw [->] (-0.3,0.8) -- (0.5,0.8);
\node at (0,1.5) {$k+P_1$ };
\draw [->] (-0.8,-0.2) -- (-0.8,0.5);
\node at (-1.3,0) {$k $};
\draw [<-] (-0.3,-0.8) -- (0.5,-0.8);
\node at (0,-1.5) {$k +P_2$};
\draw [<-] (0.8,-0.2) -- (0.8,0.5);
\node at (2.0,0) {$k+ P_3$};
\node at (5,0) {$=\,\, i{\cal B}^{R_\xi,ZZZZ}_{H}$};
\end{tikzpicture}
\end{center}
and is equal to 
\bea
{\cal B}^{R_\xi,ZZZZ}_{H}  &=& 64  \frac{m_{Z_0}^8 \l_0^2}{m_{H_0}^4}  {g^{\mu \nu }}{g^{\alpha \beta }}g^{\gamma \d } g^{\epsilon \zeta } 
\int {\frac{{{d^4}k}}{{{{\left( {2\pi } \right)}^4}}}\frac{{-i\left( { - {g_{\mu \zeta }} + \frac{{{(1-\xi)k_\mu }{k_\zeta }}}{{{k^2 - \xi m_{Z_0}^2}}}} \right)}}{{D_1}}
\frac{{\left( { - {g_{\nu \a }} + \frac{{{{(1-\xi)\left( {k + P_1} \right)}_\nu }{{\left( {k + P_1} \right)}_\a }}}{{{(k+P_1)^2 - \xi m_{Z_0}^2}}}} \right)}}{{D_2}}} \nonumber \\
&\times& \frac{{\left( { - {g_{\d \b }} + \frac{{{{(1-\xi)\left( {k + P_2} \right)}_\d }{{\left( {k + P_2} \right)}_\b }}}{{{(k+P_2)^2 - \xi m_{Z_0}^2}}}} \right)}}{{D_3}}
\frac{{\left( { - {g_{\epsilon \gamma }} + \frac{{{{(1-\xi)\left( {k + P_3} \right)}_\epsilon }{{\left( {k + P_3} \right)}_\gamma }}}{{{(k+P_3)^2 - \xi m_{Z_0}^2}}}} \right)}}{{D_4}}\, .
\eea
In DR it takes the form
\bea
{\cal B}^{R_\xi,ZZZZ}_{H} &=& ( B^{R_\xi, ZZZZ}_{H})_f  \, .
\eea
Another finite diagram is
\vskip .5cm
\begin{center}
\begin{tikzpicture}[scale=0.7]
\draw [] (-1,1) -- (1,1);
\draw [photon] (1,1) -- (1,-1);
\draw [photon] (1,-1) -- (-1,-1);
\draw [photon] (-1,-1) -- (-1,1);
\draw [dashed,thick] (-2,2) -- (-1,1);
\draw [dashed,thick] (-2,-2) -- (-1,-1);
\draw [dashed,thick] (2,2) -- (1,1);
\draw [dashed,thick] (2,-2) -- (1,-1);
\draw [<-]  (-1.8,1.3)--(-2.3,1.8);
\node at (-2.3,2.2) {$p_1$};
\draw [<-]  (-1.8,-1.3)--(-2.3,-1.8);
\node at (-2.3,-2.2) {$p_2$};
\draw [<-]  (1.8,1.3)--(2.3,1.8);
\node at (2.3,2.2) {$p_3$};
\draw [<-]  (1.8,-1.3)--(2.3,-1.8);
\node at (2.3,-2.2) {$p_4$};
%
\draw [->] (-0.3,0.8) -- (0.5,0.8);
\node at (0,1.5) {$k+P_1$ };
\draw [->] (-0.8,-0.2) -- (-0.8,0.5);
\node at (-1.3,0) {$k $};
\draw [<-] (-0.3,-0.8) -- (0.5,-0.8);
\node at (0,-1.5) {$k +P_2$};
\draw [<-] (0.8,-0.2) -- (0.8,0.5);
\node at (2.0,0) {$k+ P_3$};
\node at (5,0) {$=\,\, i{\cal B}^{R_\xi, \chi ZZZ}_{H}$};
\end{tikzpicture}
\end{center}
which is equal to 
\bea
{\cal B}^{R_\xi, \chi ZZZ}_{H} &=& 64  \frac{m_{Z_0}^6 \l_0^2}{m_{H_0}^4}  {g^{\mu \nu }}g^{\gamma \d }  \int {\frac{{{d^4}k}}{{{{\left( {2\pi } \right)}^4}}}\frac{{-i\left( { - {g_{\mu \d }} 
+ \frac{{{(1-\xi)k_\mu }{k_\d }}}{{{k^2 - \xi m_{Z_0}^2}}}} \right)}}{{D_1}}\frac{{\left( { - {g_{\nu \a }} + \frac{{{{(1-\xi)\left( {k + P_2} \right)}_\nu }
{{\left( {k + P_2} \right)}_\a }}}{{{(k+P_2)^2 - \xi m_{Z_0}^2}}}} \right)}}{{D_3}}} \nonumber \\
&\times&  \frac{(k+P_1)^\a (k+P_1)^\b}{D_2}\frac{{\left( { - {g_{\gamma \b }} + \frac{{{{(1-\xi)\left( {k + P_3} \right)}_\gamma }
{{\left( {k + P_3} \right)}_\b }}}{{{(k+P_3)^2 - \xi m_{Z_0}^2}}}} \right)}}{{D_4}}
\eea
and in DR to
\bea
&& {\cal B}^{R_\xi, \chi ZZZ}_{H} =  ( B^{R_\xi, \chi ZZZ}_{H})_f \, .
\eea
Now, there are three additional diagrams of this type, ${\cal B}^{R_\xi, Z\chi ZZ}_{H}$, ${\cal B}^{R_\xi, ZZ\chi Z}_{H}$ and ${\cal B}^{R_\xi, ZZZ\chi }_{H}$, 
giving different, but finite result. The total contribution of these diagrams too is obtained by summing over 
the three possible channels $(P_1,P_2,P_3)$, i.e. the $s$, $t$ and $u$ channels, according to \eq{Pstu}.

We turn to the $S$-Boxes which have both infinite and finite parts. The first one of this type is
\vskip .5cm
\begin{center}
\begin{tikzpicture}[scale=0.7]
\draw [photon] (-1,1) -- (1,1);
\draw [photon] (1,1) -- (1,-1);
\draw [] (1,-1) -- (-1,-1);
\draw [] (-1,-1) -- (-1,1);
\draw [dashed,thick] (-2,2) -- (-1,1);
\draw [dashed,thick] (-2,-2) -- (-1,-1);
\draw [dashed,thick] (2,2) -- (1,1);
\draw [dashed,thick] (2,-2) -- (1,-1);
\draw [<-]  (-1.8,1.3)--(-2.3,1.8);
\node at (-2.3,2.2) {$p_1$};
\draw [<-]  (-1.8,-1.3)--(-2.3,-1.8);
\node at (-2.3,-2.2) {$p_2$};
\draw [<-]  (1.8,1.3)--(2.3,1.8);
\node at (2.3,2.2) {$p_3$};
\draw [<-]  (1.8,-1.3)--(2.3,-1.8);
\node at (2.3,-2.2) {$p_4$};
%
\draw [->] (-0.3,0.8) -- (0.5,0.8);
\node at (0,1.5) {$k+P_1$ };
\draw [->] (-0.8,-0.2) -- (-0.8,0.5);
\node at (-1.3,0) {$k $};
\draw [<-] (-0.3,-0.8) -- (0.5,-0.8);
\node at (0,-1.5) {$k +P_2$};
\draw [<-] (0.8,-0.2) -- (0.8,0.5);
\node at (2.0,0) {$k+ P_3$};
\node at (5,0) {$=\,\, i{\cal B}^{R_\xi,ZZ\chi\chi}_{H}$};
\end{tikzpicture}
\end{center}
and is given by the expression 
\bea
{\cal B}^{R_\xi,ZZ\chi\chi}_H &=& 64  \frac{m_{Z_0}^4 \l_0^2}{m_{H_0}^4} 
g^{\m\b} \int \frac{d^4 k}{\left( 2\pi \right)^4}
\frac{-i
\left( 
-g_{\n \b} + 
\frac{(1-\xi)(k+P_1)_\n (k+P_1)_\b}{(k+P_1)^2 - \xi m_{Z_0}^2} 
\right)}{D_1 D_2 D_3 D_4}\cdot\nonumber\\
&& \left(- g_{\m \a} + 
\frac{(1-\xi)\left( k + P_3 \right)_\m \left( k + P_3 \right)_\a }
{(k+P_3)^2 - \xi m_{Z_0}^2}
\right)
k^\n (k+P_2)^\a k \cdot (k+P_2).\nonumber\\
\eea
Its DR form is
\bea\label{BHS4}
{\cal B}^{R_\xi,ZZ\chi\chi}_{H} &=& 64 \frac{\l_0^2 m_{Z_0}^4}{m_{H_0}^4} \m^\ve \Biggr \{D_{{\cal B}4}(P_1,P_2,P_3,m_{Z_0},m_{Z_0},m_{\chi_0},m_{\chi_0})  \nonumber\\
&+&(1-\xi) \Bigr \{ 2 D_{{\cal B}4}(P_1,P_2,P_3,m_{\chi_0},m_{Z_0},m_{\chi_0},m_{\chi_0})   \Bigl \}\nonumber\\
&+&(1-\xi)^2 \Bigr \{ D_{{\cal B}4}(P_1,P_2,P_3,m_{\chi_0},m_{Z_0},m_{\chi_0},m_{\chi_0})    \Bigl \} \Biggl \} \nonumber\\
&+&  ( B^{R_\xi, ZZ \chi \chi }_{H})_{f,1} + (1-\xi) ( B^{R_\xi, ZZ \chi \chi }_{H})_{f,2}  + (1-\xi)^2 ( B^{R_\xi, ZZ \chi \chi }_{H})_{f,3} \, . \nonumber\\
\eea
A portion of the finite part of the above diagram is actually built in the $D_{{\cal B}4}$-integrals (defined at the end of Sect. \ref{TensorPV})
while the rest of it is given in Appendix \ref{FiniteParts}. 
There are in total six different topologies for this diagram corresponding to the different ways the two Goldstones can be distributed in the loop.
All these diagrams have exactly the same divergent part as ${\cal B}^{R_\xi,ZZ\chi\chi}_{H}$ but not the same finite part. 

Next, we have
\vskip .5cm
\begin{center}
\begin{tikzpicture}[scale=0.7]
\draw [] (-1,1) -- (1,1);
\draw [] (1,1) -- (1,-1);
\draw [] (1,-1) -- (-1,-1);
\draw [photon] (-1,-1) -- (-1,1);
\draw [dashed,thick] (-2,2) -- (-1,1);
\draw [dashed,thick] (-2,-2) -- (-1,-1);
\draw [dashed,thick] (2,2) -- (1,1);
\draw [dashed,thick] (2,-2) -- (1,-1);
\draw [<-]  (-1.8,1.3)--(-2.3,1.8);
\node at (-2.3,2.2) {$p_1$};
\draw [<-]  (-1.8,-1.3)--(-2.3,-1.8);
\node at (-2.3,-2.2) {$p_2$};
\draw [<-]  (1.8,1.3)--(2.3,1.8);
\node at (2.3,2.2) {$p_3$};
\draw [<-]  (1.8,-1.3)--(2.3,-1.8);
\node at (2.3,-2.2) {$p_4$};
%
\draw [->] (-0.3,0.8) -- (0.5,0.8);
\node at (0,1.5) {$k+P_1$ };
\draw [->] (-0.8,-0.2) -- (-0.8,0.5);
\node at (-1.3,0) {$k $};
\draw [<-] (-0.3,-0.8) -- (0.5,-0.8);
\node at (0,-1.5) {$k +P_2$};
\draw [<-] (0.8,-0.2) -- (0.8,0.5);
\node at (2,0) {$k+ P_3$};
\node at (5,0) {$=\,\, i{\cal B}^{R_\xi, \chi\chi\chi Z}_{H}$};
\end{tikzpicture}
\end{center}
together with ${\cal B}^{R_\xi, Z\chi\chi\chi }_{H}$, ${\cal B}^{R_\xi, \chi Z \chi\chi }_{H}$ and ${\cal B}^{R_\xi, \chi\chi Z \chi}_{H}$ that have also identical divergent parts but different finite parts.
${\cal B}^{R_\xi, \chi\chi\chi Z}_{H}$ is equal to 
\bea
{\cal B}^{R_\xi, \chi\chi\chi Z}_{H} &=&  64  \frac{m_{Z_0}^2 \l_0^2}{m_{H_0}^4} \int {\frac{{{d^4}k}}{{{{\left( {2\pi } \right)}^4}}}\frac{{-i\left( { - {g_{\mu \n }} 
+ \frac{{{(1-\xi)k_\mu }{k_\n }}}{{{k^2 - \xi m_{Z_0}^2}}}} \right)}}{{D_1 D_2 D_3 D_4}} (k+P_1)^\m (k+P_2)^\n }\nonumber \\
&\times&  (k+P_1)\cdot (k+P_3)  (k+P_3)\cdot (k+P_2) 
\eea
and in DR evaluates to
\bea\label{BHS5}
{\cal B}^{R_\xi, \chi\chi\chi Z}_{H} &=& - 64 \frac{\l_0^2 m_{Z_0}^2}{m_{H_0}^4} \m^\ve \Biggr \{U_{B6}(P_1,P_2,P_3,m_{Z_0},m_{\chi_0},m_{\chi_0},m_{\chi_0}) \nonumber\\
&+&2 (P_1 + P_2 +P_3 )_\m U^\m_{B5}(P_1,P_2,P_3,m_{Z_0},m_{\chi_0},m_{\chi_0},m_{\chi_0})  \nonumber\\
& +& (P_{1\m} P_{1\n} + 3P_{1\m} P_{2\n} + P_{2\m} P_{2\n} \nonumber\\
&+& 3P_{1\m} P_{3\n} +3P_{2\m} P_{3\n} + P_{3\m} P_{3\n} ) D^{\m\n}_{B4}(P_1,P_2,P_3,m_{Z_0},m_{\chi_0},m_{\chi_0},m_{\chi_0}) \nonumber\\
&+& ( P_1 \cdot P_2 + P_1 \cdot P_3 + P_2 \cdot P_3 ) D_{B4}(P_1,P_2,P_3,m_{Z_0},m_{\chi_0},m_{\chi_0},m_{\chi_0}) \nonumber\\
&-&(1-\xi) \Bigr \{U_{B6}(P_1,P_2,P_3,m_{Z_0},m_{\chi_0},m_{\chi_0},m_{\chi_0}) \nonumber\\
&+& m_{\chi_0}^2 U_{B4}(P_1,P_2,P_3,m_{Z_0},m_{\chi_0},m_{\chi_0},m_{\chi_0})\nonumber\\
& +&2 (P_1 + P_2 +P_3 )_\m U^\m_{B5}(P_1,P_2,P_3,m_{Z_0},m_{\chi_0},m_{\chi_0},m_{\chi_0})  \nonumber\\
& +&(P_{1\m} P_{1\n} + 3P_{1\m} P_{2\n} + P_{2\m} P_{2\n} \nonumber\\
&+& 4P_{1\m} P_{3\n} +3P_{2\m} P_{3\n} + P_{3\m} P_{3\n} ) D^{\m\n}_{B4}(P_1,P_2,P_3,m_{Z_0},m_{\chi_0},m_{\chi_0},m_{\chi_0}) \nonumber\\
&+&( P_1 \cdot P_2 + P_2 \cdot P_3 ) D_{B4}(P_1,P_2,P_3,m_{Z_0},m_{\chi_0},m_{\chi_0},m_{\chi_0})      \Bigl \}       \Biggl \} 
\nonumber\\
&+& ( B^{R_\xi, \chi \chi \chi Z}_{H})_{f,1} + (1-\xi) ( B^{R_\xi, \chi \chi \chi Z}_{H})_{f,2} \, ,
\eea
where $( B^{R_\xi, \chi \chi \chi Z}_{H})_{f,1}$ and $( B^{R_\xi, \chi \chi \chi Z}_{H})_{f,2}$ are entirely finite, given in Appendix \ref{FiniteParts}. 
The rest of the finite part is encoded in the PV and $U$-integrals.

The last divergent $S$-Box is the one with only Goldstone bosons inside the loop: 
\vskip .5cm
\begin{center}
\begin{tikzpicture}[scale=0.7]
\draw [] (-1,1) -- (1,1);
\draw [] (1,1) -- (1,-1);
\draw [] (1,-1) -- (-1,-1);
\draw [] (-1,-1) -- (-1,1);
\draw [dashed,thick] (-2,2) -- (-1,1);
\draw [dashed,thick] (-2,-2) -- (-1,-1);
\draw [dashed,thick] (2,2) -- (1,1);
\draw [dashed,thick] (2,-2) -- (1,-1);
\draw [<-]  (-1.8,1.3)--(-2.3,1.8);
\node at (-2.3,2.2) {$p_1$};
\draw [<-]  (-1.8,-1.3)--(-2.3,-1.8);
\node at (-2.3,-2.2) {$p_2$};
\draw [<-]  (1.8,1.3)--(2.3,1.8);
\node at (2.3,2.2) {$p_3$};
\draw [<-]  (1.8,-1.3)--(2.3,-1.8);
\node at (2.3,-2.2) {$p_4$};
%
\draw [->] (-0.3,0.8) -- (0.5,0.8);
\node at (0,1.5) {$k+P_1$ };
\draw [->] (-0.8,-0.2) -- (-0.8,0.5);
\node at (-1.3,0) {$k $};
\draw [<-] (-0.3,-0.8) -- (0.5,-0.8);
\node at (0,-1.5) {$k +P_2$};
\draw [<-] (0.8,-0.2) -- (0.8,0.5);
\node at (2,0) {$k+ P_3$};
\node at (5,0) {$=\,\, i{\cal B}^{R_\xi, \chi\chi\chi\chi}_{H}$};
\end{tikzpicture}
\end{center}
and is equal to 
\bea
{\cal B}^{R_\xi, \chi\chi\chi\chi}_{H} &=&  64  \frac{ \l_0^2}{m_{H_0}^4}  \int {\frac{{{d^4}k}}{{{{\left( {2\pi } \right)}^4}}}\frac{-i k \cdot (k+P_1)}{D_1 D_2 D_3 D_4}  (k+P_1) \cdot (k+P_3) } \times   (k+P_3)\cdot (k+P_2) \times (k+P_2) \cdot k .\nonumber\\
\eea
Its DR form is
\bea\label{BS6Rxi}
{\cal B}^{R_\xi, \chi\chi\chi\chi}_{H} &=& 64 \frac{\l_0^2 m_{Z_0}^2}{m_{H_0}^4} \m^\ve \Biggr \{U_{B8}(P_1,P_2,P_3,m_{\chi_0},m_{\chi_0},m_{\chi_0},m_{\chi_0}) \nonumber\\
&+& 2(P_1 + P_2 +P_3 )_\m U^\m_{B7}(P_1,P_2,P_3,m_{\chi_0},m_{\chi_0},m_{\chi_0},m_{\chi_0}) \nonumber\\
&+& \Bigr \{ P_{1\m} P_{1\n} + 3P_{1\m} P_{2\n} + P_{2\m} P_{2\n} \nonumber\\
&+& 4 P_{1\m} P_{3\n} + 3 P_{2\m} P_{3\n} + P_{3\m} P_{3\n}   \Bigl \} U^{\m\n}_{B6}(P_1,P_2,P_3,m_{\chi_0},m_{\chi_0},m_{\chi_0},m_{\chi_0}) \nonumber\\
&+& ( P_1 \cdot P_2 + P_2 \cdot P_3) U_{B6}(P_1,P_2,P_3,m_{\chi_0},m_{\chi_0},m_{\chi_0},m_{\chi_0}) \nonumber\\
&+& \Bigr \{ P_{1\m} P_{1\n}P_{2\a} +P_{1\m} P_{2\n}P_{2\a} +2P_{1\m} P_{1\n}P_{3\a} + 4P_{1\m} P_{2\n}P_{3\a} \nonumber\\
&+& P_{2\m} P_{2\n}P_{3\a} + 2P_{1\m} P_{3\n}P_{3\a} + P_{2\m} P_{3\n}P_{3\a} \Bigl \} U^{\m\n\a}_{B5}(P_1,P_2,P_3,m_{\chi_0},m_{\chi_0},m_{\chi_0},m_{\chi_0})\nonumber\\
&+&\Bigr \{  P_1 \cdot P_2 (P_1 + P_2 +2 P_3 )_\m   \nonumber\\
&+&  P_2 \cdot P_3 (2P_1 + P_2 + P_3 )_\m \Bigl \} U^\m_{B5}(P_1,P_2,P_3,m_{\chi_0},m_{\chi_0},m_{\chi_0},m_{\chi_0})  \nonumber\\
& +& \Bigr \{ P_{1\m} P_{1\n}P_{2\a}P_{3\b} + P_{1\m} P_{2\n}P_{2\a}P_{3\b} \nonumber\\
&+& P_{1\m} P_{1\n}P_{3\a}P_{3\b} + P_{1\m} P_{2\n}P_{3\a}P_{3\b}    \Bigl \} D^{\m\n\a\b}(P_1,P_2,P_3,m_{\chi_0},m_{\chi_0},m_{\chi_0},m_{\chi_0}) \nonumber\\
&+& \Bigr \{ P_1 \cdot P_2 [P_{1\m} P_{1\n}  +2 P_{1\m} P_{3\n} +  P_{2\m} P_{3\n} + P_{3\m} P_{3\n} ] \nonumber\\
&+& P_2 \cdot P_3 [P_{1\m} P_{1\n}  + P_{1\m} P_{2\n} + 2 P_{1\m} P_{3\n} + P_{2\m} P_{3\n}  ]   \Bigl \} D^{\m\n}_{B4}(P_1,P_2,P_3,m_{\chi_0},m_{\chi_0},m_{\chi_0},m_{\chi_0})\nonumber\\
&+& ( P_1 \cdot P_2 \times P_2 \cdot P_3 ) D_{B4}(P_1,P_2,P_3,m_{\chi_0},m_{\chi_0},m_{\chi_0},m_{\chi_0})  \Biggl \}\, .
\eea
Summing up all topologies and channels, at zero external momenta, we find for the $S$-Boxes the relations
\bea\label{BHRRxinotation}
\ve [  {\cal B}^{R_\xi, S}_{H} ]_{\varepsilon} &=& 24\frac{s^2 \l_0^2 }{m^4_{H_0}} + 40 \frac{s \cdot t \l_0^2 }{m^4_{H_0}} + 24 \frac{t^2 \l_0^2 }{m^4_{H_0}} 
+ 40 \frac{s \cdot u \l_0^2 }{m^4_{H_0}} + 40 \frac{t\cdot u \l_0^2 }{m^4_{H_0}} + 24 \frac{u^2 \l_0^2 }{m^4_{H_0}}  \nonumber\\
\left\{  {\cal B}^{R_\xi, S}_{H} \right\}_{\ve}& =& 0
\eea
and
\bea
[{\cal B}^{R_\xi, S}_{H} ]_f &=& 162 \l_0^2 + 96 \frac{ \l^2_0 m^4_{Z_0}}{m_{H_0}^4} \nonumber\\
\left\{ {\cal B}^{R_\xi, S}_{H}  \right\}_{f} &=& 0 
\eea
We note the interesting facts that ${\cal B}^{R_\xi, ZZ\chi\chi}_{H}$ has only a $\{  \cdots \}_{\ve}$ part and that
\bea
\Bigl \{ {\cal B}^{R_\xi,ZZ\chi\chi}_{H} +{\cal B}^{R_\xi,\chi\chi\chi Z}_{H} +{\cal B}^{R_\xi,\chi\chi\chi\chi}_{H}  \Bigr \}_{\ve} = 0\, ,
\eea
while $[ {\cal B}^{R_\xi,\chi\chi\chi Z}_{H} ]_{\ve} \sim \frac{\l_0^2 m^2_{Z_0}}{m^2_{H_0}}$ and $[ {\cal B}^{R_\xi,\chi\chi\chi\chi}_{H}  ]_{\ve}  \sim  \l_0^2$.

The cancellation of $\xi$ from the finite parts is more illuminating when results from individual loop structures are shown:
\be
\begin{tikzpicture}[scale=0.5]
\draw [photon] (-1,1) -- (1,1);
\draw [photon] (1,1) -- (1,-1);
\draw [photon] (1,-1) -- (-1,-1);
\draw [photon] (-1,-1) -- (-1,1);
\draw [dashed,thick] (-2,2) -- (-1,1);
\draw [dashed,thick] (-2,-2) -- (-1,-1);
\draw [dashed,thick] (2,2) -- (1,1);
\draw [dashed,thick] (2,-2) -- (1,-1);
\node at (7,0) {$=\,\, \frac{96 \l_0^2 m_{Z_0}^4}{ m_{H_0}^4} + \frac{32 \l_0^2 m_{Z_0}^4 \xi^2}{ m_{H_0}^4}  $};
\end{tikzpicture}
\nonumber
\ee
\be
\begin{tikzpicture}[scale=0.5]
\draw [] (-1,1) -- (1,1);
\draw [photon] (1,1) -- (1,-1);
\draw [photon] (1,-1) -- (-1,-1);
\draw [photon] (-1,-1) -- (-1,1);
\draw [dashed,thick] (-2,2) -- (-1,1);
\draw [dashed,thick] (-2,-2) -- (-1,-1);
\draw [dashed,thick] (2,2) -- (1,1);
\draw [dashed,thick] (2,-2) -- (1,-1);
\node at (5.5,0) {$=\,\, \frac{256 \l_0^2 m_{Z_0}^4 \xi^2}{ m_{H_0}^4}  $};
\end{tikzpicture}
\nonumber
\ee
\be
\begin{tikzpicture}[scale=0.5]
\draw [] (-1,1) -- (1,1);
\draw [] (1,1) -- (1,-1);
\draw [photon] (1,-1) -- (-1,-1);
\draw [photon] (-1,-1) -- (-1,1);
\draw [dashed,thick] (-2,2) -- (-1,1);
\draw [dashed,thick] (-2,-2) -- (-1,-1);
\draw [dashed,thick] (2,2) -- (1,1);
\draw [dashed,thick] (2,-2) -- (1,-1);
\node at (10,0) {$=\,\, - \frac{960 \l_0^2 m_{Z_0}^4 \xi^2}{ m_{H_0}^4} + \frac{1152 \l_0^2 m_{Z_0}^4 \xi^2}{ m_{H_0}^4}\ln \frac{\m^2}{m_{Z_0}^2 \xi}  $};
\end{tikzpicture}
\nonumber
\ee
\be
\begin{tikzpicture}[scale=0.5]
\draw [] (-1,1) -- (1,1);
\draw [] (1,1) -- (1,-1);
\draw [] (1,-1) -- (-1,-1);
\draw [photon] (-1,-1) -- (-1,1);
\draw [dashed,thick] (-2,2) -- (-1,1);
\draw [dashed,thick] (-2,-2) -- (-1,-1);
\draw [dashed,thick] (2,2) -- (1,1);
\draw [dashed,thick] (2,-2) -- (1,-1);
\node at (10,0) {$=\,\, \frac{256 \l_0^2 m_{Z_0}^4 \xi^2}{ m_{H_0}^4} - \frac{3072 \l_0^2 m_{Z_0}^4 \xi^2}{ m_{H_0}^4}\ln \frac{\m^2}{m_{Z_0}^2 \xi}   $};
\end{tikzpicture}
\nonumber
\ee
\be
\begin{tikzpicture}[scale=0.5]
\draw [] (-1,1) -- (1,1);
\draw [] (1,1) -- (1,-1);
\draw [] (1,-1) -- (-1,-1);
\draw [] (-1,-1) -- (-1,1);
\draw [dashed,thick] (-2,2) -- (-1,1);
\draw [dashed,thick] (-2,-2) -- (-1,-1);
\draw [dashed,thick] (2,2) -- (1,1);
\draw [dashed,thick] (2,-2) -- (1,-1);
\node at (10,0) {$=\,\,  \frac{416 \l_0^2 m_{Z_0}^4 \xi^2}{ m_{H_0}^4} + \frac{1920 \l_0^2 m_{Z_0}^4 \xi^2}{ m_{H_0}^4}\ln \frac{\m^2}{m_{Z_0}^2 \xi}.  $};
\end{tikzpicture}
\nonumber
\ee
The cancellation of $\xi$ is now evident.

The final step here is to collect all Boxes and sum them up.
Adding Eq.\eqref{BHCRxinotation}, Eq.\eqref{BHTRxinotation} and Eq.\eqref{BHRRxinotation}, we obtain at $p_i=0$:
\bea\label{BHRxinotation}
\ve [  {\cal B}^{R_\xi}_H ]_{\varepsilon} &=&  108 \l_0^2 + 144 \frac{ \l_0^2  m^4_{Z_0}}{m^4_{H_0}}  \nonumber\\
\left\{  {\cal B}^{R_\xi}_H  \right\}_{\ve}& =& 0.
\eea
and
\bea
[{\cal B}^{R_\xi}_H ]_f &=&  -168 \frac{ \l^2_0 m^4_{Z_0}}{m_{H_0}^4} + 54 \l_0^2 \ln \frac{\m^2}{m_{H_0}^2} + 72 \frac{ \l^2_0 m^4_{Z_0}}{m_{H_0}^4} \ln \frac{\m^2}{m_{Z_0}^2} \nonumber\\
\left\{ {\cal B}^{R_\xi}_H  \right\}_{f} &=& 0 \, .
\eea
A couple of final comments are in order. First, each block of box diagrams, Candies, $T$ and $S$-Boxes is by itself $\xi$-independent. 
Moreover, looking at the results from the three sectors, 
one notices that they have an explicit $s$, $t$ and $u$-dependence. So, one would expect that $[{\cal B}^{R_\xi}_H ]_{\varepsilon} $ 
could also be channel dependent. Nevertheless, Eq.\eqref{BHRxinotation} shows that the $s$, $t$ and $u$-dependence cancels
when the full contribution of the box diagrams is taken into account.

\tikzset{
photon/.style={decorate, decoration={snake}, draw=black},
electron/.style={draw=black, postaction={decorate},
decoration={markings,mark=at position .55 with {\arrow[draw=black]{>}}}},
gluon/.style={decorate, draw=black,
decoration={coil,amplitude=4pt, segment length=5pt}} 
}

\section{Unitary Gauge}\label{Abelian Higgs Model in the Unitary Gauge}

In the previous section, we computed one-loop processes in the Abelian Higgs model when the gauge symmetry is broken using an $R_\xi$ gauge fixing term. 
However one of our main goals here is to investigate this model in the Unitary gauge. This is interesting since, in the Unitary gauge only physical degrees of freedom are present, 
in contrast to the $R_\xi$ gauge. 
Moreover, there are statements in the literature that argue that the Unitary gauge may be problematic at the quantum level, 
so by comparing it to the $R_\xi$ gauge, we will try to clarify the correctness of these arguments.

In the Unitary gauge, no gauge fixing is needed, therefore there is no need for ghosts.
The Unitary gauge Lagrangean can be simply obtained from \eq{LRx} by removing gauge fixing and ghost terms
and setting $\chi_0=0$ in the remaining. Doing so, we obtain the Unitary gauge Lagrangean
\bea\label{Unitary Lagrangian}
{\cal L}_{AH} &= & - \frac{1}{4}F_{0,\mu \nu }^2 + \frac{1}{2}\left( {{\partial _\mu }\phi_0 } \right)\left( {{\partial ^\mu }\phi_0 } \right) 
+ \frac{1}{2}m_{Z_0}^2{A^0_\mu }{A^{0\mu} } + {{{g^{\mu \nu }}}}\frac{\l_0 m_{Z_0}^2}{m_{H_0}^2}{\phi_0^2}{A^0_\mu }{A^0_\nu } \nonumber\\
&+ & {g^{\mu \nu }} \frac{\sqrt{2 \l_0} m_{Z_0}^2}{m_{H_0}} \phi_0 {A^0_\mu }{A^0_\nu } - \frac{{1}}{2}m_{H_0}^2{\phi_0^2} -  \sqrt{\frac{\lambda_0}{2}}m_{H_0}{\phi_0^3} - \frac{\lambda_0 }{{4}}{\phi_0^4} + const.
\eea 
from which the Feynman rules can be derived:
\begin{itemize}
\item Gauge boson propagator
\begin{center}
\begin{tikzpicture}[scale=0.8]
\draw[photon] (-1,-0.19)--(1.5,-0.19) ;
\node at (5,0) {$=\displaystyle
  \frac{{i\left( { - {g^{\mu \nu }} +  \frac{{{k^\mu }{k^\nu }}}{{{ m_{Z_0}^2}}}} \right)}}{{{k^2} - m_{Z_0}^2 + i\varepsilon }} $};
\end{tikzpicture}
\end{center}
\item Higgs propagator 
\begin{center}
\begin{tikzpicture}[scale=0.8]
\draw[dashed,thick] (-1,0)--(1.5,0) ;
\node at (5,0) {$=\displaystyle \frac{i}{{{k^2} - m_{H_0}^2 + i\varepsilon }} $};
\end{tikzpicture}
\end{center}
\item $\phi$-$Z$-$Z$ vertex
\begin{center}
\begin{tikzpicture}[scale=0.7]
\draw [photon] (-2.5,1.5)--(-1,0);
\draw [photon] (-2.5,-1.5)--(-1,0);
\draw [<-] (-1.5,0.8)--(-2.2,1.5);
\node at (-2.7,1.1) {$p_1$};
\draw [<-] (-1.5,-0.8)--(-2.2,-1.5);
\node at (-2.7,-1.1) {$p_2$};
\draw [<-] (-0.5,0.2)--(0.3,0.2);
\node at (0,-0.3) {$p_3$};
\draw[dashed,thick] (-1,0)--(1,0) ;
\node at (4,0) {$=  \displaystyle 2i{g^{\mu \nu }}{\frac{m^2_{Z_0}}{m_{H_0}}\sqrt{2 \l_0}}$};
\end{tikzpicture}
\end{center}
\item $\phi$-$\phi$-$\phi$ vertex
\begin{center}
\begin{tikzpicture}[scale=0.7]
\draw [dashed,thick] (-2.5,1.5)--(-1,0);
\draw [dashed,thick] (-2.5,-1.5)--(-1,0);
\draw [<-] (-1.5,0.8)--(-2.2,1.5);
\node at (-2.7,1.1) {$p_a$};
\draw [<-] (-1.5,-0.8)--(-2.2,-1.5);
\node at (-2.7,-1.1) {$p_2$};
\draw [<-] (-0.5,0.2)--(0.3,0.2);
\node at (0,-0.3) {$p_3$};
\draw[dashed,thick] (-1,0)--(1,0) ;
\node at (4,0) {$= \displaystyle  - 6i\sqrt{\frac{\lambda_0}{2}} {m_{H_0}}$};
\end{tikzpicture}
\end{center}
\item $\phi$-$\phi$-$Z$-$Z$ vertex
\begin{center}
\begin{tikzpicture}[scale=0.5]
\draw [dashed,thick] (0,0)--(1.5,1.4);
\draw [dashed,thick] (0,0)--(1.5,-1.4);
\draw [photon] (-1.5,1.4)--(0,0);
\draw [photon] (-1.5,-1.4)--(0,0);
\draw [<-]  (0.7,0.3)--(1.3,0.9);
\node at (1.7,1.7) {$p_3$};
\draw [<-]  (0.9,-0.4)--(1.5,-1);
\node at (1.7,-1.7) {$p_4$};
\draw [<-] (-0.9,0.3)--(-1.6,1);
\node at (-1.7,1.7) {$p_1$};
\draw [<-] (-0.8,-0.4)--(-1.5,-1);
\node at (-1.7,-1.7) {$p_2$};
\node at (6,0) {$=\displaystyle 4i{{{} \frac{ \l_0 m^2_{Z_0}}{m^2_{H_0}}g^{\mu \nu }}}$};
\end{tikzpicture}
\end{center}
\item $\phi$-$\phi$-$\phi$-$\phi$ vertex
\begin{center}
\begin{tikzpicture}[scale=0.5]
\draw [dashed,thick] (0,0)--(1.5,1.4);
\draw [dashed,thick] (0,0)--(1.5,-1.4);
\draw [dashed,thick] (-1.5,1.4)--(0,0);
\draw [dashed,thick] (-1.5,-1.4)--(0,0);
\draw [<-]  (0.7,0.3)--(1.3,0.9);
\node at (1.7,1.7) {$p_3$};
\draw [<-]  (0.9,-0.4)--(1.5,-1);
\node at (1.7,-1.7) {$p_4$};
\draw [<-] (-0.9,0.3)--(-1.6,1);
\node at (-1.7,1.7) {$p_1$};
\draw [<-] (-0.8,-0.4)--(-1.5,-1);
\node at (-1.7,-1.7) {$p_2$};
\node at (5,0) {$= \displaystyle -6i \lambda_0$};
\end{tikzpicture}
\end{center}
\end{itemize}
In the Unitary gauge, integrals of the $U$-type are ubiquitous.
But such integrals we have already seen in the $R_\xi$ gauge due to the momentum dependent vertices of the Polar basis for the Higgs field.
In the Unitary gauge we do not have momentum dependent vertices, the $U$-integrals arise only because of the form of the propagators.

In the following sections we compute one-loop diagrams, setting the external momenta to zero at the end, as in the $R_\xi$ gauge.
Again for completeness, we present some on-shell results in Appendix \ref{Onshell}.
We will directly insert symmetry factors here since they are the same as in the corresponding $R_\xi$ calculation.
In the Unitary gauge, since there is no gauge fixing parameter, we have trivially $\{\star\}_\ve = \{\star\}_f\equiv 0$.
As before, we will consistently move nasty expressions for finite parts to Appendix \ref{FiniteParts}.

\subsection{Tadpoles}\label{Tadpoles in Unitary}

The Higgs tadpole is
\vskip .5cm
\begin{center}
\begin{tikzpicture}[scale=0.7]
\draw [dashed,thick] (0,0) circle [radius=1];
\draw [dashed,thick] (-2.5,0)--(-1,0);
\draw [->]  (-2.2,0.2)--(-1.5,0.2);
\node at (-2.9,0) {$p$};
\draw [->]  (0.45,-0.6) arc [start angle=-45, end angle=-135, radius=0.6cm];
\node at (0,-0.4) {$k$};
\node at (3,0) {$=\, \, i{\cal T}^{U,\phi}_H$};
\end{tikzpicture}
\end{center}
\begin{eqnarray}\label{Th1}
 {\cal T}^{U,\phi}_H &=& -  6 \frac{1}{2} \sqrt{\frac{\lambda_0}{2}}m_{H_0}\int {\frac{{{d^4}k}}{{{{\left( {2\pi } \right)}^4}}}\frac{i}{{\left( {{k^2} - m_{H_0}^2} \right)}}} 
\end{eqnarray}
and in DR reads
\begin{equation}\label{final Tad1}
{\cal T}^{U,\phi}_H={3  \sqrt{\frac{\lambda_0}{2}}m_{H_0}{\mu ^{{\ve}}}}A_0(m_{H_0}).
\end{equation}
The gauge tadpole is
\vskip .5cm
\begin{center}
\begin{tikzpicture}[scale=0.7]
\draw [photon] (0,0) circle [radius=1];
\draw [dashed,thick] (-2.5,0)--(-1,0);
\draw [->]  (-2.2,0.2)--(-1.5,0.2);
\node at (-2.9,0) {$p$};
\draw [->]  (0.45,-0.5) arc [start angle=-45, end angle=-135, radius=0.6cm];
\node at (0,-0.3) {$k$};
\node at (3,0) {$=\, \, i{\cal T}^{U,Z}_H$};
\end{tikzpicture}
\end{center}
and following Appendix \ref{UGD}, this becomes
\be\label{final TadH}
{\cal T}^U_H={\mu^{\ve}} \Biggl ( { 3\sqrt{\frac{\lambda_0}{2}}m_{H_0}{}{ }}A_0(m_{H_0}) +  3\frac{\sqrt{2 \l_0} m_{Z_0}^2}{m_{H_0}}A_0(m_{Z_0})  \Biggr ).
\ee
Adding up the two results, we have
\begin{equation}\label{full TadU}
(4\pi)^{d/2}{\cal T}^{U}_H= \mu ^{\ve} \left( [{\cal T}^{U}_H ]_{\varepsilon}  +[ {\cal T}^{U}_H ]_{f}   \right)\, 
\end{equation}
with 
\bea\label{TadUnotation}
\ve [{\cal T}^{U}_H ]_{\varepsilon}  &=& 6 \sqrt{\frac{\l_0}{2}}m^3_{H_0} + 6 \frac{\sqrt{2 \l_0}m^4_{Z_0}}{m_{H_0}}
\eea
and
\bea\label{THUF}
[{\cal T}^{U}_H ]_{f}  &=&  3 \sqrt{\frac{\l_0}{2}}m^3_{H_0} + 3 \frac{\sqrt{2 \l_0}m^4_{Z_0}}{m_{H_0}} + 
3 \sqrt{\frac{\l_0}{2}}m^3_{H_0} \ln \frac{\m^2}{m_{H_0}^2} + 3\frac{\sqrt{2 \l_0}m^4_{Z_0}}{m_{H_0}} \ln \frac{\m^2}{m_{Z_0}^2} \, .
\eea
The above expressions are identical to the $R_\xi$ expressions \eq{Tade} and \eq{Tadf}.

\subsection{Corrections to the gauge boson mass}\label{Corrections to the Z mass in Unitary gauge}

The $Z$ boson mass receives its first correction from
\vskip .5cm
\begin{center}
\begin{tikzpicture}[scale=0.7]
\draw [photon] (0,0)--(1.8,0);
\draw [dashed,thick] (0,0.9) circle [radius=0.9];
\draw [photon] (-1.8,0)--(-0,0);
\draw [->]  (-1.7,0.2)--(-1,0.2);
\node at (-2.2,0.2) {$p$};
\draw [->]  (0.45,0.3) arc [start angle=-45, end angle=-135, radius=0.6cm];
\node at (0,0.4) {$k$};
\draw [->]  (1.7,0.2)--(1,0.2);
\node at (4.3,0) {$=\, \, i{\cal M}^{U,\phi}_{Z,\mu\nu}$};
\end{tikzpicture}
\end{center}
which is equal to
\begin{eqnarray}\label{M1Z}
{\cal M}^{U,\phi}_{Z,\mu\nu} &=& -2{{g_{\mu \nu }}}\frac{m^2_{Z_0}}{m^2_{H_0}} \l_0 \mu^{\ve} A_0(m_{H_0})\, .
\end{eqnarray}
Next is the Higgs sunset
\vskip .5cm
\begin{center}
\begin{tikzpicture}[scale=0.7]
\draw [photon] (-2.3,0)--(2.3,0);
\draw [->]  (-1.9,0.2)--(-1.2,0.2);
\node at (-2.5,0.2) {$p$};
\draw [<-]  (0.45,0.7) arc [start angle=45, end angle=135, radius=0.6cm];
\node at (0,1.3) {$k+p$};
\draw [<-] (-0.4,0.2)--(0.4,0.2);
\node at (0,-0.3) {$k$};
\draw [->]  (1.9,0.2)--(1.2,0.2);
\node at (4,0) {$=\,\, i{\cal M}^{U,\phi Z}_{Z,\mu\nu}$ };
\draw  [dashed,thick] (-1,0) .. controls (-1,0.555) and (-0.555,1) .. (0,1)
.. controls (0.555,1) and (1,0.555) .. (1,0);
\end{tikzpicture}
\end{center}
translating in DR to
\bea\label{final M2Z}
{\cal M}^{U,\phi Z}_{Z,\mu\nu} & =&  - 8{g_{\mu \nu }}{{}\frac{m^4_{Z_0}}{m^2_{H_0}} \l_0 \mu^{\ve}}B_0(p,m_{Z_0},m_{H_0}) + 8 \frac{m^2_{Z_0}}{m^2_{H_0}} \l_0\mu^{\ve}B_{\m\n}(p,m_{Z_0},m_{H_0}) \, . 
\eea
The sum of these two corrections is
\bea\label{full MZ}
{\cal M}^{U}_{Z,\mu\nu}  &=& \frac{m^2_{Z_0}}{m^2_{H_0}} \l_0\mu^{\ve}\Biggl \{- 8g_{\m\n} {}{{}m_{Z_0}^2}B_0(p,m_{Z_0},m_{H_0}) - 2 g_{\m\n} A_0(m_{H_0}) \nonumber\\
&+& 8 B_{\m\n}(p,m_{Z_0},m_{H_0})  \Biggr \}\, . 
\eea
As before, the proper contraction we are after is 
\bea
{\cal M}^{U}_Z = \frac{1}{3} \left(-g^{\m\n} + \frac{p^\m p^\n}{p^2}\right) {\cal M}^{U}_{Z,{\m\n}}(p)
\eea
which can be easily shown to be equal to
\bea\label{reduced MZ}
{\cal M}^{U}_{Z}(p) &=& -\frac{1}{3} \frac{m^2_{Z_0}}{m^2_{H_0}} \l_0 \mu^{\ve} \Biggl \{  \Bigl \{ - 8{(d+\varepsilon)}{{}m_{Z_0}^2}B_0(p,m_{Z_0},m_{H_0}) - 2 (d+\varepsilon)A_0(m_{H_0}) \nonumber\\
&+& 8 g^{\m\n}B_{\m\n}(p,m_{Z_0},m_{H_0}) \Bigr \}  - 8{}{{}m_{Z_0}^2}B_0(p,m_{Z_0},m_{H_0}) - 2 A_0(m_{H_0}) \nonumber\\
&+&8 \frac{p^\m p^\n}{p^2} B_{\m\n}(p,m_{Z_0},m_{H_0})  \Biggr \} \, .
\eea
At $p = 0$, we have
\begin{equation}\label{full MZUnot}
(4\pi)^{d/2}{\cal M}^{U}_Z= \mu ^{\ve} \left( [{\cal M}^{U}_Z ]_{\varepsilon}  +[ {\cal M}^{U}_Z ]_{f}   \right)\, 
\end{equation}
with 
\bea\label{MZUnotation}
\ve [{\cal M}^{U}_Z ]_{\varepsilon}  &=&  12 \frac{\l_0 m_{Z_0}^4}{m_{H_0}^2}\, .
\eea
The wave function renormalization factor is now straightforward to compute:
\bea\label{adMZUn}
\d A^U & =&- {\left. {\frac{d {\cal M}^U_Z(p) }{d p^2}} \right|_{{p^2} = 0}} = \frac{\m^\ve}{(4\pi)^{d/2}} \left([\d A^U ]_{\varepsilon} + [\d A^U]_{f} \right)
\eea
with 
\bea\label{dAUnotation}
\ve [\d A^U ]_{\varepsilon}  &=& - \frac{4}{3} \l_0 \frac{m_{Z_0}^2}{m_{H_0}^2} \, .
\eea
Glancing back on Sect.\ref{Corrections to the Z mass in Rxi} we see the first sign of the consistency of the calculation by noticing
that ${\cal M}^U_Z $ and ${\cal M}^{R_\xi}_Z $ have the same gauge-independent divergent parts and divergent parts of their respective anomalous dimensions.

Regarding the finite parts we have
\bea\label{MZUfinite}
 [{\cal M}^{U}_Z ]_{f}  &=& -\frac{4}{3}\frac{m_{H_0}^2 m_{Z_0}^2 \l_0}{( m_{H_0}^2 -  m_{Z_0}^2  )} + \frac{10}{3}\frac{ m_{Z_0}^4 \l_0}{( m_{H_0}^2 -  m_{Z_0}^2  )} 
 - \frac{26}{3}\frac{ m_{Z_0}^6 \l_0}{m_{H_0}^2( m_{H_0}^2 -  m_{Z_0}^2  )} \nonumber\\
 &+& \frac{16}{3} \frac{ m_{Z_0}^4 \l_0}{m_{H_0}^2( m_{H_0}^2 -  m_{Z_0}^2  )} \ln \frac{\m^2}{m_{H_0}^2}+ 
 \frac{2}{3} \frac{ m_{Z_0}^4 \l_0}{( m_{H_0}^2 -  m_{Z_0}^2  )} \ln \frac{\m^2}{m_{Z_0}^2} - 6 \frac{ m_{Z_0}^6 \l_0}{m_{H_0}^2( m_{H_0}^2 -  m_{Z_0}^2  )} \ln \frac{\m^2}{m_{Z_0}^2}\, \nonumber\\
\eea
and
\bea\label{dAUfinite}
 [\d A^U ]_{f}  &=& -\frac{2}{3}\frac{ m_{Z_0}^2 \l_0}{( m_{H_0}^2 -  m_{Z_0}^2 )} 
 \left(\left(1+\ln \frac{\m^2}{m_{H_0}^2}\right) - \frac{m_{Z_0}^2}{m_{H_0}^4} \left(1+ \ln \frac{\m^2}{m_{Z_0}^2} \right)\right)\, .
\eea
It is interesting to notice that the last two expressions for the finite parts can be obtained from the corresponding
$R_\xi$ gauge expressions in \eq{MZRxfinite} and \eq{dARxfinite} for $\xi=1$ and not for $\xi\to \infty$.

\subsection{Corrections to the Higgs mass}\label{Corrections to the Higgs mass in Unitary gauge}

The first correction to the Higgs mass comes from
\vskip .5cm
\begin{center}
\begin{tikzpicture}[scale=0.7]
\draw [dashed,thick] (0,0)--(1.8,0);
\draw [photon] (0,0.9) circle [radius=0.9];
\draw [dashed,thick] (-1.8,0)--(-0,0);
\draw [->]  (-1.7,0.2)--(-1,0.2);
\node at (-2.2,0.2) {$p$};
\draw [->]  (0.45,0.5) arc [start angle=-45, end angle=-135, radius=0.6cm];
\node at (0,0.75) {$k$};
\draw [->]  (1.7,0.2)--(1,0.2);
\node at (4.5,0) {$=\, \,i{\cal M}^{U,Z}_H$};
\end{tikzpicture}
\end{center}
and in DR it becomes
\bea\label{final M1H}
{\cal M}_H^{U,Z}&=& \m^{\ve}\Bigl \{ {2 d \frac{ \l_0 m_{Z_0}^2}{m_{H_0}^2}}A_0(m_{Z_0}) - 2 \frac{ \l_0}{m_{H_0}^2} U_{\cal T}(1, m_{Z_0}) \Bigr \} \nonumber\\
&=& \m^{\ve} {6\frac{ \l_0 m_{Z_0}^2}{m_{H_0}^2}}A_0(m_{Z_0}) \, .
\eea
Next is
\vskip .5cm
\begin{center}
\begin{tikzpicture}[scale=0.7]
\draw [dashed,thick] (0,0)--(1.8,0);
\draw [dashed,thick] (0,0.9) circle [radius=0.9];
\draw [dashed,thick] (-1.8,0)--(-0,0);
\draw [->]  (-1.7,0.2)--(-1,0.2);
\node at (-2.2,0.2) {$p$};
\draw [->]  (0.45,0.4) arc [start angle=-45, end angle=-135, radius=0.6cm];
\node at (0,0.5) {$k$};
\draw [->]  (1.7,0.2)--(1,0.2);
\node at (4.5,0) {$=\, \, i{\cal M}^{U,\phi}_H$};
\end{tikzpicture}
\end{center}
In DR it reads,
\be\label{final M2H}
{\cal M}_H^{U,\phi}={3\l_0  {\mu^{\ve}}{}}A_0(m_{H_0}).
\ee 
The Higgs vacuum polarization diagram
\vskip .5cm
\begin{center}
\begin{tikzpicture}[scale=0.7]
\draw [dashed,thick] (0.9,0)--(2.2,0);
\draw [dashed,thick] (0,0) circle [radius=0.9];
\draw [dashed,thick] (-2.2,0)--(-0.9,0);
\draw [->]  (-1.9,0.2)--(-1.2,0.2);
\node at (-2.6,0.2) {$p$};
\draw [<-]  (0.45,0.6) arc [start angle=45, end angle=135, radius=0.6cm];
\node at (0,1.2) {$k+p$};
\draw [->]  (0.45,-0.6) arc [start angle=-45, end angle=-135, radius=0.6cm];
\node at (0,-1.2) {$k$};
\draw [->]  (1.9,0.2)--(1.2,0.2);
\node at (4.5,0) {$=\, \, i{\cal M}^{U,\phi\phi}_H$};
\end{tikzpicture}
\end{center}
which in DR is equal to
\bea\label{final M3H}
{\cal M}^{U,\phi\phi}_H&= &{9 \l_0}m_{H_0}^2{\mu ^{\ve}}B_0(p,m_{H_0},m_{H_0})\, .
\eea
The corresponding gauge loop is
\vskip .5cm
\begin{center}
\begin{tikzpicture}[scale=0.7]
\draw [dashed,thick] (0.9,0)--(2.2,0);
\draw [photon] (0,0) circle [radius=0.9];
\draw [dashed,thick] (-2.2,0)--(-0.9,0);
\draw [->]  (-1.9,0.2)--(-1.2,0.2);
\node at (-2.6,0.2) {$p$};
\draw [<-]  (0.45,0.6) arc [start angle=45, end angle=135, radius=0.6cm];
\node at (0,1.3) {$k+p$};
\draw [->]  (0.45,-0.4) arc [start angle=-45, end angle=-135, radius=0.6cm];
\node at (0,-1.1) {$k$};
\draw [->]  (1.9,0.2)--(1.2,0.2);
\node at (4.5,0) {$=\, \, i{\cal M}^{U,ZZ}_H$};
\end{tikzpicture}
\end{center}
and in DR
\bea\label{final M4H}
{\cal M}^{U,ZZ}_H &=& \mu ^{\ve}{}\Biggl \{ 4 d \frac{\l_0 m^4_{Z_0}}{m_{H_0}^2} B_0(p,m_{Z_0},m_{Z_0}) \nonumber\\
&-& 4 \frac{\l_0 m^2_{Z_0}}{m_{H_0}^2}{g_{\mu\nu}}B_{k+p}^{\mu\nu}(p,m_{Z_0},m_{Z_0})
+ 4 \frac{ \l_0 }{m_{H_0}^2}m_{Z_0}^2 A_0(m_{Z_0}) + 4 \frac{ \l_0 }{m_{H_0}^2}m_{Z_0}^4  \nonumber\\
&-&4 \frac{ \l_0 }{m_{H_0}^2} p^2 g_{\m\n}B^{\mu\nu}(p,m_{Z_0},m_{Z_0}) + 4 \frac{ \l_0 }{m_{H_0}^2} p_\m p_\n B^{\mu\nu}(p,m_{Z_0},m_{Z_0})    \Biggr \} \, ,
\eea
where we have defined
\bea
g_{\m\n}B_{k+p}^{\m\n}(p,m_{Z_0},m_{Z_0}) = \int {\frac{{{d^4}k}}{{{{\left( {2\pi } \right)}^4}}}\frac{-i{{(k  +  p)^2 } } }{\left({{k^2} - m_{Z_0}^2}\right){\left( {{{\left( {k + p} \right)}^2} - m_{Z_0}^2} \right)}}}\, .
\eea
Adding up \eq{final M1H}, \eq{final M2H}, \eq{final M3H} and \eq{final M4H} we obtain
\begin{eqnarray}\label{full MHUn}
{\cal M}^U_H(p)& = &\mu ^{\ve} \Biggl \{ {6}\frac{\l m^2_{Z_0}}{m_{H_0}^2}A_0(m_{Z_0}) + \frac{2\l m^4_{Z_0}}{m_{H_0}^2} + {3\l_0{{}}}{{{}}}A_0(m_{H_0}) \nonumber \\
 &+& 9\l_0 m_{H_0}^2{}B_0(p,m_{H_0},m_{H_0}) + \frac{\l_0 m^2_{Z_0}}{m_{H_0}^2}\Big \{ 4 d m_{Z_0}^2 B_0(p,m_{Z_0},m_{Z_0}) \nonumber\\
&-& 4{ {}}{{}}{g_{\mu\nu}}B_{k+p}^{\mu\nu}(p,m_{Z_0},m_{Z_0}) + 4 A_0(m_{Z_0})  - 4 \frac{p^2}{m_{Z_0}^2}{ {}}{{}}{g_{\mu\nu}}B^{\mu\nu}(p,m_{Z_0},m_{Z_0}) \nonumber\\
&+& 4 m_{Z_0}^2 + 4 \frac{p_\m p_\n}{m_{Z_0}^2} B^{\mu\nu}(p,m_{Z_0},m_{Z_0})    \Big \} \Biggr \}.
\end{eqnarray}
Using the reduction formulae in Appendix \ref{PassarinoVeltman} and summing up all contributions we can extract
\bea\label{fullMHRxi2}
(4\pi)^{d/2} {\cal M}^U_H &=& \mu ^{\ve} \left( [{\cal M}^U_H ]_{\varepsilon} + [ {\cal M}^U_H ]_{f} \right)
\eea
with 
\bea\label{MHUnotation}
\ve [{\cal M}^U_H ]_{\varepsilon}  &=&24 \l_0 m_{H_0}^2 + 36 \frac{\l_0 m_{Z_0}^4}{m_{H_0}^2} 
\eea
and
\bea\label{MHUF}
[{\cal M}^U_H ]_{f}  &=&  3 \l_0 m^2_{H_0} + 6 \frac{ \l_0 m^4_{Z_0}}{m_{H_0}^2} + 12 \l_0 m^2_{H_0} \ln \frac{\m^2}{m_{H_0}^2} + 18 \frac{ \l_0 m^4_{Z_0}}{m_{H_0}^2} \ln \frac{\m^2}{m_{Z_0}^2} \, .
\eea
The anomalous dimension is then
\bea\label{adMHUn}
\d \phi^U & =&- {\left. {\frac{d  {\cal M}^U_H(p) }{d p^2}} \right|_{{p^2} = 0}} = \frac{\m^\ve}{(4\pi)^{d/2}} \left( [\d \phi^U ]_{\varepsilon} + [\d \phi^U ]_{f} \right)
\eea
with 
\bea\label{dphiUnotation}
\ve [\d \phi^U ]_{\varepsilon}  &=& 12 \l_0 \frac{m_{Z_0}^2}{m_{H_0}^2} \, .
\eea
and
\bea\label{dphiUfinite}
[\d \phi^U ]_{f}  &=& 2 \l_0 \frac{m_{Z_0}^2}{m_{H_0}^2} + 6\l_0 \frac{m^4_{Z_0}}{m_{H_0}^2} \ln \frac{\m^2}{m_{Z_0}^2}\, .
\eea
Comparing to the result of Sect. \ref{Corrections to the Higgs mass in Rx} we can see that \eq{MHUnotation} and \eq{MHUF} of ${\cal M}^{U}_H$ 
are identical to \eq{MHRxinotation} and \eq{MHRxifinite} of ${\cal M}^{R_\xi}_H$ and (the $\xi$-independent) anomalous dimension of the $R_\xi$ gauge is equal to $\d \phi^U$.

\subsection{Corrections to the Higgs cubic vertex}\label{Corrections to the three-point vertex in Unitary gauge}

As before we split the diagrams in two classes. The first class
consists of the reducible Triangle diagrams, while the second one consists of the irreducible Triangles. 
The contribution of the various topologies and channels is contained in their $P_1$ and $(P_1,P_2)$ dependence
respectively, exactly as in the $R_\xi$-gauge calculation.

\subsubsection{Reducible Triangles}\label{Reducible Triangles in Unitary}

The first reducible Triangle is
\vskip .5cm
\begin{center}
\begin{tikzpicture}[scale=0.7]
\draw [dashed,thick] (0.9,0)--(2.3,1.3);
\draw [dashed,thick] (0.9,0)--(2.3,-1.3);
\draw [dashed,thick] (-2.5,0)--(-0.9,0);
\draw [dashed,thick] (0,0) circle [radius=0.9];
\draw [<-]  (0.45,0.6) arc [start angle=45, end angle=135, radius=0.6cm];
\node at (0,1.2) {$k+P_1$};
\draw [->]  (0.45,-0.6) arc [start angle=-45, end angle=-135, radius=0.6cm];
\node at (0,-1.2) {$k$};
\draw [<-]  (1.3,0.6)--(1.9,1.2);
\node at (2.7,1.3) {$p_3$};
\draw [<-]  (1.3,-0.6)--(1.9,-1.2);
\node at (2.7,-1.3) {$p_2$};
\draw [<-] (-1.5,0.2)--(-2.2,0.2);
\node at (-2.5,0.2) {$p_1$};
\node at (6,0) {$=\, \, i{\cal K}^{U,\phi\phi}_{H}$};
\end{tikzpicture}
\end{center}
and it is the same loop as in \eq{final M3H} with $p\to P_1$, with the same symmetry factor, divided by $v_0$:
\bea
{\cal K}^{U,\phi\phi}_{H} &=& 18 \frac{\l^{3/2}_0}{\sqrt{2}} m_{H_0}{\mu ^{\ve}}B_0(P_1,m_{H_0},m_{H_0})\, ,
\eea
where the three different channels are obtained by summing over $P_1$, with $P_1=p_1, p_2, p_3$
(resulting to an overall factor of 3, as in the $R_\xi$-gauge).

Next is the gauge loop
\vskip .5cm
\begin{center}
\begin{tikzpicture}[scale=0.7]
\draw [dashed,thick] (0.9,0)--(2.3,1.3);
\draw [dashed,thick] (0.9,0)--(2.3,-1.3);
\draw [dashed,thick] (-2.5,0)--(-0.9,0);
\draw [photon] (0,0) circle [radius=0.9];
\draw [<-]  (0.45,0.6) arc [start angle=45, end angle=135, radius=0.6cm];
\node at (0,1.4) {$k+P_1$};
\draw [->]  (0.45,-0.4) arc [start angle=-45, end angle=-135, radius=0.6cm];
\node at (0,-1) {$k$};
\draw [<-]  (1.3,0.6)--(1.9,1.2);
\node at (2.7,1.3) {$p_3$};
\draw [<-]  (1.3,-0.6)--(1.9,-1.2);
\node at (2.7,-1.3) {$p_2$};
\draw [<-] (-1.5,0.2)--(-2.2,0.2);
\node at (-2.5,0.2) {$p_1$};
\node at (6,0) {$=\, \, i{\cal K}^{U,ZZ}_{H}$};
\end{tikzpicture}
\end{center}
which is \eq{final M4H}, divided by $v_0$:
\bea
{\cal K}^{U,ZZ}_{H}&= & \frac{1}{v_0}\frac{4 \l_0 }{m_{H_0}^2} \mu ^{\ve}{}\Biggl \{ d m^4_{Z_0} B_0(P_1,m_{Z_0},m_{Z_0}) \nonumber\\
&-& m^2_{Z_0}{g_{\mu\nu}}B_{k+P_1}^{\mu\nu}(P_1,m_{Z_0},m_{Z_0}) +  m_{Z_0}^2 A_0(m_{Z_0}) + 4 m_{Z_0}^2\nonumber\\
&-& P_1^2 g_{\m\n}B^{\mu\nu}(P_1,m_{Z_0},m_{Z_0}) + P_{1\m} P_{1\n} B^{\mu\nu}(P_1,m_{Z_0},m_{Z_0})    \Biggr \}.
\eea
There are again two more channels with $P_1=p_2$ and $P_1=p_3$, yielding identical contributions.

Adding up all the reducible Triangles, we have
\bea\label{fullKHRDU}
(4\pi)^{d/2} {\cal K}^{U,{\rm red.}}_H  &=& \mu ^{\ve} \left( [{\cal K}^{U,{\rm red.}}_H ]_{\ve} +[ {\cal K}^{U,{\rm red.}}_H]_{f}  \right)
\eea
with 
\bea\label{KHRxiRDnotation}
\ve [{\cal K}^{U,{\rm red.}}_H]_{\ve}  &=&\frac{m_{H_0}}{\sqrt{2 \l_0}}  \left( 108  \l_0^2 + 144 \frac{ \l_0^2 m_{Z_0}^4}{m_{H_0}^4} \right)
\eea
and
\bea\label{KHredUF}
[{\cal K}^{U,{\rm red.}}_H ]_{f}  &=& \frac{m_{H_0}}{\sqrt{2 \l }}\Bigl ( 54 \l^2_0 \ln \frac{\m^2}{m_{H_0}^2} + 72 \frac{ \l^2_0 m^4_{Z_0}}{m_{H_0}^4} \ln \frac{\m^2}{m_{Z_0}^2} \Big )\,  \nonumber\\
\eea
at $p_i=0$.

\subsubsection{Irreducible Triangles}\label{Irreducible Triangles in Unitary}

The first irreducible Triangle is
\vskip .5cm
\begin{center}
\begin{tikzpicture}[scale=0.7]
\draw [dashed,thick] (-1,0) -- (1,1);
\draw [dashed,thick] (-1,0) -- (1,-1);
\draw [dashed,thick] (1,1) -- (1,-1);
\draw [dashed,thick] (-2,0) -- (-1,0);
\draw [dashed,thick] (1,1) -- (2.5,1);
\draw [dashed,thick] (1,-1) -- (2.5,-1);
\node at (1.7,1.3) {$p_3$};
\draw [<-] (1.3,0.8) -- (1.9,0.8);
\node at (1.7,-1.3) {$p_2$};
\draw [<-] (1.3,-0.8) -- (1.9,-0.8);
\node at (-1.9,0.2) {$p_1$};
\draw [->] (-1.6,0.2) -- (-1.1,0.2);
\draw [->] (-0.4,0.1) -- (0.2,0.4);
\node at (-0.3,1) {$k+P_1$};
\draw [<-] (-0.4,-0.1) -- (0.2,-0.4);
\node at (-0.1,-1) {$k$};
\draw [<-] (0.8,-0.3) -- (0.8,0.3);
\node at (2.2,0) {$k+P_2$};
\node at (6,0) {$=\,\,  i{\cal K}^{U,\phi\phi\phi}_H$};.
\end{tikzpicture}
\end{center}
\bea
{\cal K}^{U,\phi\phi\phi}_H &=& \frac{%
108}{\sqrt{2}} \l_0^{3/2} m^3_{H_0} \int {\frac{{{d^4}k}}{{{{\left( {2\pi } \right)}^4}}}} (-i)\frac{1}{D_1 D_2 D_3}.
\eea
and in DR
\bea\label{full KH3U}
{\cal K}^{U\phi\phi\phi}_H &=& \frac{%
108}{\sqrt{2}} \l_0^{3/2} m^3_{H_0} \m^\ve C_0(P_1,P_2,m_{H_0},m_{H_0},m_{H_0}).
\eea
In the Unitary gauge there is only one more divergent irreducible Triangle, which is
\vskip .5cm
\begin{center}
\begin{tikzpicture}[scale=0.7]
\draw [photon] (-1,0) -- (1,1);
\draw [photon] (-1,0) -- (1,-1);
\draw [photon] (1,1) -- (1,-1);
\draw [dashed,thick] (-2,0) -- (-1,0);
\draw [dashed,thick] (1,1) -- (2.5,1);
\draw [dashed,thick] (1,-1) -- (2.5,-1);
\node at (1.7,1.3) {$p_3$};
\draw [<-] (1.3,0.8) -- (1.9,0.8);
\node at (1.7,-1.3) {$p_2$};
\draw [<-] (1.3,-0.8) -- (1.9,-0.8);
\node at (-1.9,0.2) {$p_1$};
\draw [->] (-1.6,0.2) -- (-1.1,0.2);
\draw [->] (-0.4,0.1) -- (0.2,0.4);
\node at (-0.4,1) {$k + P_1$};
\draw [<-] (-0.4,-0.1) -- (0.2,-0.4);
\node at (-0.1,-1) {$k$};
\draw [<-] (0.8,-0.3) -- (0.8,0.3);
\node at (2.2,0) {$k + P_2$};
\node at (6,0) {$=\,\,  i{\cal K}^{U,ZZZ}_H$};.
\end{tikzpicture}
\end{center}
\bea\label{KH4U}
{\cal K}^{U,ZZZ}_H&=&- 16\sqrt{2}\frac{m^6_{Z_0} \l_0^{3/2}}{m^3_{H_0}}  {g^{\mu \nu }}{g^{\alpha \beta }}g^{\gamma \d }  \int {\frac{{{d^4}k}}{{{{\left( {2\pi } \right)}^4}}}
\frac{{-i\left( { - {g_{\mu \gamma }} + \frac{{{k_\mu }{k_\gamma }}}{{{ m_{Z_0}^2}}}} \right)}}{{D_1 D_2 D_3}}{{\left( { - {g_{\nu \a }} + \frac{{{{\left( {k + P_1} \right)}_\nu }{{\left( {k + P_1} \right)}_\a }}}{{{m_{Z_0}^2}}}} \right)}}} \nonumber \\
&\times& {{\left( { - {g_{\d \b }} + \frac{{{{\left( {k + P_2} \right)}_\d }{{\left( {k + P_2} \right)}_\b }}}{{{ m_{Z_0}^2}}}} \right)}}
\eea
and in DR
\bea\label{full KH4U}
{\cal K}^{U,ZZZ}_H&=& -32 v_0 \frac{m^6_{Z_0} \l_0^2}{m^4_{H_0}} \m^\ve \Biggl \{ -d C_0(P_1,P_2,m_{Z_0},m_{Z_0},m_{Z_0}) \nonumber\\
&+&\frac{1}{m_{Z_0}^2}\Bigl \{ (3 m_{Z_0}^2 + P_1 \cdot P_1 + P_2 \cdot P_2 ) C_0(P_1,P_2,m_{Z_0},m_{Z_0},m_{Z_0}) + 3 B_0(P_1,m_{Z_0},m_{Z_0})  \Bigr \}\nonumber\\
&+&\frac{1}{m_{Z_0}^4}\Bigl \{3 U_{K4}(P_1,P_2,m_{Z_0},m_{Z_0},m_{Z_0}) +4(P_1 + P_2)_{\m}C_{{\cal K}3}^\m(P_1,P_2,m_{Z_0},m_{Z_0},m_{Z_0})\nonumber\\
&+& 2(P_{1\m}P_{1\n} + P_{1\m}P_{2\n} +P_{2\m}P_{2\n})C^{\m\n}(P_1,P_2,m_{Z_0},m_{Z_0},m_{Z_0}) + 2P_1 \cdot P_2 B_0(p_1,m_{Z_0},m_{Z_0})  \nonumber\\
&+&2 P_1 \cdot P_2 (P_1 + P_2)_\m C^{\m}(P_1,P_2,m_{Z_0},m_{Z_0},m_{Z_0}) + P_1 \cdot P_2 C_0(P_1,P_2,m_{Z_0},m_{Z_0},m_{Z_0})     \Bigr \}\nonumber\\
&+&\frac{1}{m_{Z_0}^6}\Bigl \{U_{K6}(P_1,P_2,m_{Z_0},m_{Z_0},m_{Z_0}) + 2(P_1 + P_2)_\m U^\m_{K5}(P_1,P_2,m_{Z_0},m_{Z_0},m_{Z_0})  \nonumber\\
&+& (P_{1\m}P_{1\n} +3P_{1\m}P_{2\n} +P_{2\m}P_{2\n} ) U^{\m\n}_{K4}(P_1,P_2,m_{Z_0},m_{Z_0},m_{Z_0}) \nonumber\\
&+& P_1 \cdot P_2 U_{K4}(P_1,P_2,m_{Z_0},m_{Z_0},m_{Z_0}) \nonumber\\
&+& (P_{1\m}P_{1\n}P_{2\a} + P_{1\m}P_{2\n}P_{2\a} )   C^{\m\n\a}(P_1,P_2,m_{Z_0},m_{Z_0},m_{Z_0}) \nonumber\\
&+& P_1 \cdot P_2 (P_{1\m} + P_{2\m} )C_{{\cal K}3}^\m(P_1,P_2,m_{Z_0},m_{Z_0},m_{Z_0}) \nonumber\\
&+& P_1 \cdot P_2 P_{1\m}P_{2\n} C^{\m\n} (P_1,P_2,m_{Z_0},m_{Z_0},m_{Z_0})       \Bigr \}     \Biggr \}\, ,
\eea
where in this case $P_1 = p_1$ and $P_2 = p_1 + p_3$. The contracted PV integral $C_{{\cal K}3}^\m$ is defined in Appendix \ref{TensorPV}.

In total, we find for the irreducible Triangles
\bea\label{fullKHIRDUn}
(4\pi)^{d/2} {\cal K}^{U,{}\rm irred.}_{H} &=& \mu ^{\ve} \left( [{\cal K}^{U,{}\rm irred.}_{H}]_{\ve} + [ {\cal K}^{U,{}\rm irred.}_{H}]_{f} \right)
\eea
with 
\bea\label{KHIRDUnotation}
[{\cal K}^{U,{}\rm irred.}_{H} ]_{\ve}  &=& 0
\eea
and
\bea\label{KHirredUF}
[{\cal K}^{U,{}\rm irred.}_{H} ]_{f}  &=& - \frac{m_{H_0}}{\sqrt{2 \l }}\Bigl ( 
54 \l^2_0  + 48 \frac{ \l^2_0 m^4_{Z_0}}{m_{H_0}^4}  \Big )\, .
\eea
Finally, adding reducible and irreducible contributions, we have
\bea\label{fullKHU}
(4\pi)^{d/2} {\cal K}^U_H = \mu ^{\ve} \left( [{\cal K}^U_H ]_{\ve} +[ {\cal K}^U_H ]_{f} \right)
\eea
where 
\bea\label{KHUnotation}
\ve [{\cal K}^U_H ]_{\varepsilon} &=& \frac{m_{H_0}}{\sqrt{2 \l_0}}  \left( 108  \l_0^2 + 144 \frac{ \l_0^2 m_{Z_0}^4}{m_{H_0}^4} \right)
\eea
and
\bea\label{KHUF}
[{\cal K}^{U}_H ]_{f}  &=& \frac{m_{H_0}}{\sqrt{2 \l }}\Bigl ( 
-54 \l^2_0  -24 \frac{ \l^2_0 m^4_{Z_0}}{m_{H_0}^4} + 
54 \l^2_0 \ln \frac{\m^2}{m_{H_0}^2} + 72 \frac{ \l^2_0 m^4_{Z_0}}{m_{H_0}^4} \ln \frac{\m^2}{m_{Z_0}^2} \Big ).
\eea
These are the same results as the ones found in \eq{KHRxinotation} and \eq{KHRxifinite} in the $R_\xi$-gauge.

\subsection{Corrections to the quartic coupling}\label{Corrections to the quartic coupling in Unitary gauge}

The separation of the Box diagrams into $C$, $T$ and $S$-Boxes in the Unitary gauge holds exactly like in the $R_\xi$ gauge.
The same goes for the labelling of the various channels by the momenta $P_1$ (for Candies), $(P_1,P_2)$ for $T$-Boxes
and $(P_1,P_2,P_3)$ for $S$-Boxes. 

\subsubsection{Reducible Boxes}\label{Reducible Boxes in Unitary}

The Higgs Candy 
\vskip .5cm
\begin{center}
\begin{tikzpicture}[scale=0.7]
\draw [dashed,thick] (0.9,0)--(2.5,1.5);
\draw [dashed,thick] (0.9,0)--(2.5,-1.5);
\draw [dashed,thick] (-2.5,1.5)--(-0.9,0);
\draw [dashed,thick] (-2.5,-1.5)--(-0.9,0);
\draw [<-]  (1.7,1)--(2.2,1.5);
\node at (2.7,1.1) {$p_3$};
\draw [<-]  (1.7,-1)--(2.2,-1.5);
\node at (2.7,-1.1) {$p_4$};
\draw [<-] (-1.7,1)--(-2.2,1.5);
\node at (-2.5,1.1) {$p_1$};
\draw [<-] (-1.7,-1)--(-2.2,-1.5);
\node at (-2.5,-1.1) {$p_2$};
\draw [dashed,thick] (0,0) circle [radius=0.9];
\draw [<-]  (0.45,0.6) arc [start angle=45, end angle=135, radius=0.6cm];
\node at (0,1.2) {$k+P_1$};
\draw [->]  (0.45,-0.6) arc [start angle=-45, end angle=-135, radius=0.6cm];
\node at (0,-1.2) {$k$};
\node at (5,0) {$=\, \, i{\cal B}^{U,\phi\phi}_{H}$};
\end{tikzpicture}
\end{center}
has been computed already in the $R_\xi$ gauge with the result
\bea\label{full BHCU1}
{\cal B}^{U,\phi\phi}_{H}&= &18 \l^2_0 {\mu ^{\ve}}B_0(P_1,m_{H_0},m_{H_0})\, .
\eea
The gauge Candy is slightly different due to the different gauge boson propagator. The diagram is
\vskip .5cm
\begin{center}
\begin{tikzpicture}[scale=0.7]
\draw [dashed,thick] (0.9,0)--(2.5,1.5);
\draw [dashed,thick] (0.9,0)--(2.5,-1.5);
\draw [dashed,thick] (-2.5,1.5)--(-0.9,0);
\draw [dashed,thick] (-2.5,-1.5)--(-0.9,0);
\draw [<-]  (1.7,1)--(2.2,1.5);
\node at (2.7,1.1) {$p_3$};
\draw [<-]  (1.7,-1)--(2.2,-1.5);
\node at (2.7,-1.1) {$p_4$};
\draw [<-] (-1.7,1)--(-2.2,1.5);
\node at (-2.5,1.1) {$p_1$};
\draw [<-] (-1.7,-1)--(-2.2,-1.5);
\node at (-2.5,-1.1) {$p_2$};
\draw [photon] (0,0) circle [radius=0.9];
\draw [<-]  (0.45,0.6) arc [start angle=45, end angle=135, radius=0.6cm];
\node at (0,1.3) {$k+P_1$};
\draw [->]  (0.45,-0.4) arc [start angle=-45, end angle=-135, radius=0.6cm];
\node at (0,-1.1) {$k$};
\node at (5,0) {$=\, \, i{\cal B}^{U,ZZ}_{H}$};
\end{tikzpicture}
\end{center}
It evaluates in DR to
\bea\label{full BHCU2}
(4\pi)^{d/2}{\cal B}^{U,ZZ}_{H} &=& 8 \frac{m^2_{Z_{0}}}{m^4_{H_0}}\l^2_0  {\mu ^{\ve}} \Biggl \{ d m^2_{Z_0} B_0(P_1,m_{Z_0},m_{Z_0}) \nonumber\\
&-& {g_{\mu\nu}}B_{k+P_1}^{\mu\nu}(P_1,m_{Z_0},m_{Z_0}) + A_0(m_{Z_0}) +2 m_{Z_0}^2\nonumber\\
&-& \frac{P_1^2}{m_{Z_0}^2} g_{\m\n}B^{\mu\nu}(P_1,m_{Z_0},m_{Z_0}) + \frac{P_{1\m} P_{1\n}}{m_{Z_0}^2} B^{\mu\nu}(P_1,m_{Z_0},m_{Z_0})    \Biggr \} \, .
\eea
These are the only two contributions in this sector and adding them up, we have
\bea\label{fullBHCU}
(4\pi)^{d/2} {\cal B}^{U,C}_{H} = \mu ^{\ve} \left( [{\cal B}^{U,C}_{H} ]_{\ve} +[ {\cal B}^{U,C}_{H}  ]_{f}  \right)
\eea
where 
\bea\label{BHCUnotation}
\ve [{\cal B}^{U,C}_{H}  ]_{\ve} &=& 108 \l_0^2 + 144  \frac{\l_0^2 m^4_{Z_0}}{m^4_{H_0}} +  4\frac{s^2 \l_0^2 }{m^4_{H_0}} + 4 \frac{t^2 \l_0^2 }{m^4_{H_0}} +  4\frac{u^2 \l_0^2 }{m^4_{H_0}} 
\eea
and
\bea\label{BCUF}
[{\cal B}^{U,C}_{H}  ]_{f}  &=& 54 \l^2_0 \ln \frac{\m^2}{m_{H_0}^2} + 72 \frac{ \l^2_0 m^4_{Z_0}}{m_{H_0}^4} \ln \frac{\m^2}{m_{Z_0}^2} .
\eea
We turn to the $T$-Boxes.
In the Unitary gauge we have only two contributing diagrams, one finite and one divergent. 
The finite diagram is
\vskip .5cm
\begin{center}
\begin{tikzpicture}[scale=0.7]
\draw [dashed,thick] (-1,0) -- (1,1);
\draw [dashed,thick] (-1,0) -- (1,-1);
\draw [dashed,thick] (1,1) -- (1,-1);
\draw [dashed,thick] (-2,1) -- (-1,0);
\draw [dashed,thick] (-2,-1) -- (-1,0);
\draw [dashed,thick] (1,1) -- (2.5,1);
\draw [dashed,thick] (1,-1) -- (2.5,-1);
\node at (1.7,1.3) {$p_3$};
\draw [<-] (1.3,0.8) -- (1.9,0.8);
\node at (1.7,-1.3) {$p_4$};
\draw [<-] (1.3,-0.8) -- (1.9,-0.8);
\node at (-1.9,1.3) {$p_1$};
\draw [->] (-1.6,0.9) -- (-1.1,0.4);
\node at (-1.9,-1.30) {$p_2$};
\draw [->] (-1.6,-0.9) -- (-1.1,-0.4);
\draw [->] (-0.1,0.2) -- (0.5,0.5);
\node at (-0.3,1) {$k+P_1$ };
\draw [<-] (-0.1,-0.2) -- (0.5,-0.5);
\node at (0,-0.9) {$k$};
\draw [<-] (0.8,-0.3) -- (0.8,0.3);
\node at (2,0) {$k+P_2$};
\node at (5,0) {$=\,\, i{\cal B}^{U,\phi\phi\phi}_{H}$};
\end{tikzpicture}
\end{center}
The $s$ channel contribution is equal to
\bea\label{full BHUT1}
{\cal B}^{U,\phi\phi\phi}_{H}&=& 108 \l_0^{2} m^2_{H_0} \m^\ve C_0(P_1,m_{H_0},m_{H_0},m_{H_0}) .
\eea
The $t$ and $u$ channels can be obtained as explained in the $R_\xi$-gauge calculation.
The divergent diagram in this sector is
\vskip .5cm
\begin{center}
\begin{tikzpicture}[scale=0.7]
\draw [photon] (-1,0) -- (1,1);
\draw [photon] (-1,0) -- (1,-1);
\draw [photon] (1,1) -- (1,-1);
\draw [dashed,thick] (-2,1) -- (-1,0);
\draw [dashed,thick] (-2,-1) -- (-1,0);
\draw [dashed,thick] (1,1) -- (2.5,1);
\draw [dashed,thick] (1,-1) -- (2.5,-1);
\node at (1.7,1.3) {$p_3$};
\draw [<-] (1.3,0.8) -- (1.9,0.8);
\node at (1.7,-1.3) {$p_4$};
\draw [<-] (1.3,-0.8) -- (1.9,-0.8);
\node at (-1.9,1.3) {$p_1$};
\draw [->] (-1.6,0.9) -- (-1.1,0.4);
\node at (-1.9,-1.30) {$p_2$};
\draw [->] (-1.6,-0.9) -- (-1.1,-0.4);
\draw [->] (-0.1,0.2) -- (0.5,0.5);
\node at (-0.35,1) {$k+P_1$ };
\draw [<-] (-0.1,-0.2) -- (0.5,-0.5);
\node at (0,-0.9) {$k$};
\draw [<-] (0.8,-0.3) -- (0.8,0.3);
\node at (2,0) {$k+P_2$};
\node at (5,0) {$=\,\, i{\cal B}^{U,ZZZ}_{H}$};
\end{tikzpicture}
\end{center}
and can be easily be obtained from Eq.\eqref{full KH4U} by dividing by $v_0$:
\bea\label{full BHUT2}
{\cal B}^{U,ZZZ}_{H}&=& -32 \frac{m^6_{Z_0} \l_0^2}{m^4_{H_0}} \m^\ve \Biggl \{ - d C_0(P_1,P_2,m_{Z_0},m_{Z_0},m_{Z_0}) \nonumber\\
&+&\frac{1}{m_{Z_0}^2}\Bigl \{ (3 m_{Z_0}^2 + P_1 \cdot P_1 + P_2 \cdot P_2 ) C_0(P_1,P_2,m_{Z_0},m_{Z_0},m_{Z_0}) + 3 B_0(P_1,m_{Z_0},m_{Z_0})  \Bigr \}\nonumber\\
&+&\frac{1}{m_{Z_0}^4}\Bigl \{3 U_{K4}(P_1,P_2,m_{Z_0},m_{Z_0},m_{Z_0}) +4(P_1 + P_2)_{\m}C_{{\cal K}3}^\m(P_1,P_2,m_{Z_0},m_{Z_0},m_{Z_0})\nonumber\\
&+& 2(P_{1\m}P_{1\n} + P_{1\m}P_{2\n} +P_{2\m}P_{2\n})C^{\m\n}(P_1,P_2,m_{Z_0},m_{Z_0},m_{Z_0}) + 2P_1 \cdot P_2 B_0(p_1,m_{Z_0},m_{Z_0})  \nonumber\\
&+&2 P_1 \cdot P_2 (P_1 + P_2)_\m C^{\m}(P_1,P_2,m_{Z_0},m_{Z_0},m_{Z_0}) + P_1 \cdot P_2 C_0(P_1,P_2,m_{Z_0},m_{Z_0},m_{Z_0})     \Bigr \}\nonumber\\
&+&\frac{1}{m_{Z_0}^6}\Bigl \{U_{K6}(P_1,P_2,m_{Z_0},m_{Z_0},m_{Z_0}) + 2(P_1 + P_2)_\m U^\m_{K5}(P_1,P_2,m_{Z_0},m_{Z_0},m_{Z_0})  \nonumber\\
&+& (P_{1\m}P_{1\n} +3P_{1\m}P_{2\n} +P_{2\m}P_{2\n} ) U^{\m\n}_{K4}(P_1,P_2,m_{Z_0},m_{Z_0},m_{Z_0}) \nonumber\\
&+& P_1 \cdot P_2 U_{K4}(P_1,P_2,m_{Z_0},m_{Z_0},m_{Z_0}) \nonumber\\
&+& (P_{1\m}P_{1\n}P_{2\a} + P_{1\m}P_{2\n}P_{2\a} )   C^{\m\n\a}(P_1,P_2,m_{Z_0},m_{Z_0},m_{Z_0}) \nonumber\\
&+& P_1 \cdot P_2 (P_{1\m} + P_{2\m} )C_{{\cal K}3}^\m(P_1,P_2,m_{Z_0},m_{Z_0},m_{Z_0}) \nonumber\\
&+& P_1 \cdot P_2 P_{1\m}P_{2\n} C^{\m\n} (P_1,P_2,m_{Z_0},m_{Z_0},m_{Z_0})       \Bigr \}     \Biggr \}\, .
\eea
After summing over all channels by summing over all $(P_1,P_2)$, we obtain
\bea\label{fullBHTU}
(4\pi)^{d/2} {\cal B}^{U,T}_{H} = \mu ^{\ve} \left( [ {\cal B}^{U,T}_{H} ]_{\ve} +[  {\cal B}^{U,T}_{H} ]_{f}  \right)
\eea
where 
\bea\label{BHTUnotation}
\ve [ {\cal B}^{U,T}_{H} ]_{\ve} &=& -  28\frac{s^2 \l_0^2 }{m^4_{H_0}} - 40 \frac{s \cdot t \l_0^2 }{m^4_{H_0}} - 28 
\frac{t^2 \l_0^2 }{m^4_{H_0}} - 40 \frac{s \cdot u \l_0^2 }{m^4_{H_0}} - 40 \frac{t\cdot u \l_0^2 }{m^4_{H_0}} - 28 \frac{u^2 \l_0^2 }{m^4_{H_0}} \nonumber\\
\eea
and
\bea\label{BTUF}
[{\cal B}^{U,T}_{H}  ]_{f}  &=& 
-162 \l^2_0  - 288 \frac{ \l^2_0 m^4_{Z_0}}{m_{H_0}^4}\,
\eea
at $p_i=0$.

\subsubsection{Irreducible Boxes}\label{Irreducible Boxes}

The irreducible, $S$-Boxes in the Unitary gauge are only two, one finite and one divergent.
The finite diagram is the Higgs $S$-Box
\vskip .5cm
\begin{center}
\begin{tikzpicture}[scale=0.7]
\draw [dashed,thick] (-1,1) -- (1,1);
\draw [dashed,thick] (1,1) -- (1,-1);
\draw [dashed,thick] (1,-1) -- (-1,-1);
\draw [dashed,thick] (-1,-1) -- (-1,1);
\draw [dashed,thick] (-2,2) -- (-1,1);
\draw [dashed,thick] (-2,-2) -- (-1,-1);
\draw [dashed,thick] (2,2) -- (1,1);
\draw [dashed,thick] (2,-2) -- (1,-1);
\draw [<-]  (-1.8,1.3)--(-2.3,1.8);
\node at (-2.3,2.2) {$p_1$};
\draw [<-]  (-1.8,-1.3)--(-2.3,-1.8);
\node at (-2.3,-2.2) {$p_2$};
\draw [<-]  (1.8,1.3)--(2.3,1.8);
\node at (2.3,2.2) {$p_3$};
\draw [<-]  (1.8,-1.3)--(2.3,-1.8);
\node at (2.3,-2.2) {$p_4$};
%
\draw [->] (-0.3,0.9) -- (0.5,0.9);
\node at (0,1.3) {$k+P_1$ };
\draw [->] (-0.9,-0.2) -- (-0.9,0.5);
\node at (-1.3,0) {$k $};
\draw [<-] (-0.3,-0.9) -- (0.5,-0.9);
\node at (0,-1.3) {$k +P_2$};
\draw [<-] (0.9,-0.2) -- (0.9,0.5);
\node at (2,0) {$k+ P_3$};
\node at (5,0) {$=\,\, i{\cal B}^{U,\phi\phi\phi\phi}_{H}$};
\end{tikzpicture}
\end{center}
\bea
{\cal B}^{U,\phi\phi\phi\phi}_{H} &=& 324 \l_0^2 m^2_{H_0} \int {\frac{{{d^4}k}}{{{{\left( {2\pi } \right)}^4}}}}\frac{-i}{D_1 D_2 D_3 D_4}.
\eea
In DR
\bea\label{full BHRxiR1}
{\cal B}^{U,\phi\phi\phi\phi}_{H}&=& 324 \l_0^{2} m^2_{H_0} \m^\ve D_0(P_1,P_2,P_3,m_{H_0},m_{H_0},m_{H_0},m_{H_0}) \, .
\eea
The divergent $S$-Box is the one with a gauge loop
\vskip .5cm
\begin{center}
\begin{tikzpicture}[scale=0.7]
\draw [photon] (-1,1) -- (1,1);
\draw [photon] (1,1) -- (1,-1);
\draw [photon] (1,-1) -- (-1,-1);
\draw [photon] (-1,-1) -- (-1,1);
\draw [dashed,thick] (-2,2) -- (-1,1);
\draw [dashed,thick] (-2,-2) -- (-1,-1);
\draw [dashed,thick] (2,2) -- (1,1);
\draw [dashed,thick] (2,-2) -- (1,-1);
\draw [<-]  (-1.8,1.3)--(-2.3,1.8);
\node at (-2.3,2.2) {$p_1$};
\draw [<-]  (-1.8,-1.3)--(-2.3,-1.8);
\node at (-2.3,-2.2) {$p_2$};
\draw [<-]  (1.8,1.3)--(2.3,1.8);
\node at (2.3,2.2) {$p_3$};
\draw [<-]  (1.8,-1.3)--(2.3,-1.8);
\node at (2.3,-2.2) {$p_4$};
%
\draw [->] (-0.3,0.8) -- (0.5,0.8);
\node at (0,1.5) {$k+P_1$ };
\draw [->] (-0.8,-0.2) -- (-0.8,0.5);
\node at (-1.3,0) {$k $};
\draw [<-] (-0.3,-0.8) -- (0.5,-0.8);
\node at (0,-1.4) {$k +P_2$};
\draw [<-] (0.8,-0.2) -- (0.8,0.5);
\node at (2,0) {$k+ P_3$};
\node at (5,0) {$=\,\, i{\cal B}^{U,ZZZZ}_{H}$};
\end{tikzpicture}
\end{center}
which is equal to
\bea
{\cal B}^{U,ZZZZ}_{H} &=& 64  \frac{m_{Z_0}^8 \l_0^2}{m_{H_0}^4}  {g^{\mu \nu }}{g^{\alpha \beta }}g^{\gamma \d } g^{\epsilon \zeta } \int {\frac{{{d^4}k}}{{{{\left( {2\pi } \right)}^4}}}
\frac{{-i\left( { - {g_{ \zeta \m }} + \frac{{{k_\zeta }{k_\m }}}{{{m_{Z_0}^2}}}} \right)}}{{D_1 D_2 D_3 D_4}}{{\left( { - {g_{\nu \a }} + 
\frac{{{{\left( {k + P_1} \right)}_\nu }{{\left( {k + P_1} \right)}_\a }}}{{{m_{Z_0}^2}}}} \right)}}} \nonumber \\
&\times& {{\left( { - {g_{ \b \gamma }} + \frac{{{{\left( {k + P_2} \right)}_\b }{{\left( {k + P_2} \right)}_\gamma }}}{{{m_{Z_0}^2}}}} \right)}}
{{\left( { - {g_{\d \epsilon }} + \frac{{{{\left( {k + P_3} \right)}_\d }{{\left( {k + P_3} \right)}_\epsilon }}}{{{m_{Z_0}^2}}}} \right)}}\, ,
\eea
and in DR
\bea
{\cal B}^{U,ZZZZ}_{H}&=& 64 \frac{\l_0^2 m_{Z_0}^8}{m_{H_0}^4} \m^\ve \Biggr \{d D_0(P_1,P_2,P_3,m_{Z_0},m_{Z_0},m_{Z_0},m_{Z_0})\nonumber\\
&-& \frac{1}{m_{Z_0}^2}\Bigl \{4g_{\m\n}D^{\m\n}(P_1,P_2,P_3,m_{Z_0},m_{Z_0},m_{Z_0},m_{Z_0}) \nonumber\\
&+& 2(P_1 +P_2 + P_3)_\m D^{\m}(P_1,P_2,P_3,m_{Z_0},m_{Z_0},m_{Z_0},m_{Z_0}) \nonumber\\
&+& (P^2_1 +P^2_2 +P^2_3)D_0(P_1,P_2,P_3,m_{Z_0},m_{Z_0},m_{Z_0},m_{Z_0})       \Bigr \}      \nonumber\\
&+& \frac{1}{m_{Z_0}^4}\Bigl \{6 D_{{\cal B}4}(P_1,P_2,P_3,m_{Z_0},m_{Z_0},m_{Z_0},m_{Z_0}) \Bigr \}      \nonumber\\
&-& \frac{1}{m_{Z_0}^6}\Bigl \{ 4U_{B_6}(P_1,P_2,P_3,m_{Z_0},m_{Z_0},m_{Z_0},m_{Z_0}) \nonumber\\
&+&6(P_1 +P_2 +P_3)_\m  U^\m_{B_5}(P_1,P_2,P_3,m_{Z_0},m_{Z_0},m_{Z_0},m_{Z_0})  \nonumber\\
&+& 3 ( 2P_{1\m} P_{2\n} + P_{2\m} P_{2\n} + 2P_{1\m} P_{3\n} + 2P_{2\m} P_{3\n} \nonumber\\
&+& P_{3\m} P_{3\n} + P_{1\m} P_{1\n}) D^{\m\n}_{{\cal B}4}(P_1,P_2,P_3,m_{Z_0},m_{Z_0},m_{Z_0},m_{Z_0}) \nonumber\\
&+&2(P_1 \cdot P_2 + P_1 \cdot P_3 +P_2 \cdot P_3)D_{{\cal B}4}(P_1,P_2,P_3,m_{Z_0},m_{Z_0},m_{Z_0},m_{Z_0})   \Bigr \}\nonumber\\
&+& \frac{1}{m_{Z_0}^8}\Bigl \{ U_{B8}(P_1,P_2,P_3,m_{Z_0},m_{Z_0},m_{Z_0},m_{Z_0}) \nonumber\\
&+& 2(P_1 + P_2 +P_3 )_\m U^\m_{B7}(P_1,P_2,P_3,m_{Z_0},m_{Z_0},m_{Z_0},m_{Z_0}) \nonumber\\
&+& \Bigr \{ P_{1\m} P_{1\n} + 3P_{1\m} P_{2\n} + P_{2\m} P_{2\n} \nonumber\\
&+& 4 P_{1\m} P_{3\n} + 3 P_{2\m} P_{3\n} + P_{3\m} P_{3\n}   \Bigl \} U^{\m\n}_{B6}(P_1,P_2,P_3,m_{Z_0},m_{\chi_0},m_{Z_0},m_{Z_0}) \nonumber\\
&+& ( P_1 \cdot P_2 + P_2 \cdot P_3) U_{B6}(P_1,P_2,P_3,m_{Z_0},m_{Z_0},m_{Z_0},m_{Z_0}) \nonumber\\
&+& \Bigr \{ P_{1\m} P_{1\n}P_{2\a} +P_{1\m} P_{2\n}P_{2\a} +2P_{1\m} P_{1\n}P_{3\a} + 4P_{1\m} P_{2\n}P_{3\a} \nonumber\\
&+& P_{2\m} P_{2\n}P_{3\a} + 2P_{1\m} P_{3\n}P_{3\a} + P_{2\m} P_{3\n}P_{3\a} \Bigl \} U^{\m\n\a}_{B5}(P_1,P_2,P_3,m_{Z_0},m_{Z_0},m_{Z_0},m_{Z_0})\nonumber\\
&+&\Bigr \{  P_1 \cdot P_2 [P_1 + P_2 +2 P_3 ]_\m   +  P_2 \cdot P_3 [2P_1 + P_2 + P_3 ]_\m \Bigl \} U^\m_{B5}(P_1,P_2,P_3,m_{Z_0},m_{Z_0},m_{Z_0},m_{Z_0})  \nonumber\\
& +& \Bigr \{ P_{1\m} P_{1\n}P_{2\a}P_{3\b} + P_{1\m} P_{2\n}P_{2\a}P_{3\b} \nonumber\\
&+& P_{1\m} P_{1\n}P_{3\a}P_{3\b} + P_{1\m} P_{2\n}P_{3\a}P_{3\b}    \Bigl \} D^{\m\n\a\b}(P_1,P_2,P_3,m_{Z_0},m_{Z_0},m_{Z_0},m_{Z_0}) \nonumber\\
&+& \Bigr \{ P_1 \cdot P_2 [P_{1\m} P_{1\n}  +2 P_{1\m} P_{3\n} +  P_{2\m} P_{3\n} + P_{3\m} P_{3\n} ] \nonumber\\
&+& P_2 \cdot P_3 [P_{1\m} P_{1\n}  + P_{1\m} P_{2\n} + 2 P_{1\m} P_{3\n} + P_{2\m} P_{3\n}  ]   \Bigl \} D^{\m\n}_{{\cal B}4}(P_1,P_2,P_3,m_{Z_0},m_{Z_0},m_{Z_0},m_{Z_0})\nonumber\\
&+& ( P_1 \cdot P_2 \times P_2 \cdot P_3 ) D_{{\cal B}4}(P_1,P_2,P_3,m_{Z_0},m_{Z_0},m_{Z_0},m_{Z_0})  \Bigr \} \Biggl \} \nonumber\\
&+& (B^{U,ZZZZ}_{H})_{f,1} + (B^{U,ZZZZ}_{H})_{f,2}  + (B^{U,ZZZZ}_{H})_{f,3}  \, .
\eea
The $(B^{U,ZZZZ}_{H})_{f,1}$, $(B^{U,ZZZZ}_{H})_{f,2}$ and $(B^{U,ZZZZ}_{H})_{f,3}$ are finite integrals, moved to Appendix \ref{FiniteParts}. 
The $U$-integrals are dealt with in Appendix \ref{Uint}.
The $s$, $t$ and $u$ channels are taken into account by summing over the $(P_1,P_2,P_3)$ as in the $R_\xi$ gauge.
The divergence structure of the sum of the $S$-Boxes at zero external momenta is revealed through the relation
\bea\label{fullBHRUn}
(4\pi)^{d/2} {\cal B}^{U,S}_{H} &=&  \mu ^{\ve} \left( [  {\cal B}^{U,S}_{H}  ]_{\ve}  + [ {\cal B}^{U,S}_{H} ]_{f} \right)
\eea
where 
\bea\label{BHRUnotation}
\ve [  {\cal B}^{U,S}_{H}]_{\ve} &=&  24\frac{s^2 \l_0^2 }{m^4_{H_0}} + 40 \frac{s \cdot t \l_0^2 }{m^4_{H_0}} + 
24 \frac{t^2 \l_0^2 }{m^4_{H_0}} + 40 \frac{s \cdot u \l_0^2 }{m^4_{H_0}} + 40 \frac{t\cdot u \l_0^2 }{m^4_{H_0}} + 24 \frac{u^2 \l_0^2 }{m^4_{H_0}} \nonumber\\
\eea
and
\bea\label{BSUF}
[ {\cal B}^{U,S}_{H} ]_{f}  &=& 162 \l^2_0  + 96 \frac{ \l^2_0 m^4_{Z_0}}{m_{H_0}^4}  . 
\eea
The total sum of the Boxes then satisfies
\bea\label{BHUnotation}
\ve [{\cal B}^{U}_H ]_{\ve} &=&  108 \l_0^2 + 144 \frac{ \l_0^2  m^4_{Z_0}}{m^4_{H_0}}
\eea
and
\bea\label{BHUF}
[{\cal B}^{U}_H ]_{f}  &=& -168 \frac{ \l^2_0 m^4_{Z_0}}{m_{H_0}^4} + 54 \l^2_0 \ln \frac{\m^2}{m_{H_0}^2} + 72 \frac{ \l^2_0 m^4_{Z_0}}{m_{H_0}^4} \ln \frac{\m^2}{m_{Z_0}^2} .
\eea
In the 4-point function as well, we observe a sector by sector agreement between the $R_\xi$ and Unitary gauges.

\tikzset{
photon/.style={decorate, decoration={snake}, draw=black},
electron/.style={draw=black, postaction={decorate},
decoration={markings,mark=at position .55 with {\arrow[draw=black]{>}}}},
gluon/.style={decorate, draw=black,
decoration={coil,amplitude=4pt, segment length=5pt}} 
}

\section{Renormalization}\label{Renormalization}

In this section we will renormalize the Abelian-Higgs model in both the $R_\xi$ and Unitary gauges.
In fact, since we will be concerned here mainly with the $Z$-mass and scalar potential we will be able to
perform the renormalization program simultaneously for both. 

One of the results of the calculation of the previous two sections is that the one-loop corrections to the $Z$-mass and the scalar potential
have identical divergent parts in the $R_\xi$ and Unitary gauges. This means that the two gauges have the same $\b$-functions for the masses and quartic coupling.
Now the Lagrangean in the Unitary gauge can be obtained from the $R_\xi$-gauge Lagrangean by dropping the gauge fixing and ghost terms
and simply setting the Goldstone field to zero. Since we are interested here only in the common subsector that consists of the 
$Z$-mass term and the scalar potential, we can carry out the renormalization program on the Unitary gauge Lagrangean 
and only when we arrive at the stage where we analyse the finite, renormalized scalar potential where finite corrections become relevant, 
we may have to distinguish between the two gauges if necessary. 
Thus for now and until further notice, we drop the "U" superscript that denotes Unitary gauge. 

\subsection{Counter-terms}\label{Counterterms}

We introduce the counter-terms 
\bea\label{renorm.}
m_0^2 &=& m^2 + \d m \nonumber\\
g_0 &=& g + \d g \nonumber\\ 
\l_0 &=& \l + \d \l 
\eea
In the Abelian-Higgs model there is a non-zero anomalous dimension for the scalar (as opposed to $\phi^4$ theory and the linear sigma model)
and likewise a non-zero anomalous dimension for the gauge boson. 
Therefore, we also introduce
\bea\label{ren.fields}
H_0 &=& \sqrt{Z_\phi}\, \phi = \sqrt{1+\d \phi} \, \phi \nonumber\\
A_0 &=& \sqrt{Z_A}A = \sqrt{1+\d A} \, A\, .
\eea
Substituting the above in \eq{clas.Lag.1} and then expressing the Higgs in the Polar basis, we obtain 
\bea
{\cal L} = {\cal L}^{{\rm tree}} + {\cal L}^{{\rm count.}}\, ,
\eea
with ${\cal L}^{{\rm tree}}$ the renormalized tree-level Lagrangean
\bea\label{RIN.U}
{\cal L}^{{\rm R,tree}}&=&- \frac{1}{4}F_{\mu \nu }^2 + \frac{1}{2}\left( {{\partial _\mu }\phi } \right)\left( {{\partial ^\mu }\phi } \right)  + 
\frac{1}{2}m_{Z}^2{A_\mu }{A^{\mu} } +  {{{g^{\mu \nu }}}}{}{\frac{ \l m^2_{Z}}{m^2_{H}}}{A_\mu }{A_\nu } {{\phi ^2} {}} \nonumber\\
&+& {g^{\mu \nu }}{\frac{m^2_{Z}}{m_{H}}\sqrt{2 \l}}\phi {A_\mu }{A_\nu }   - \frac{1}{2}{m_{H}^2}{\phi^2} -  \sqrt{\frac{\lambda}{2}} {m_{H}}{\phi ^3}- \frac{\lambda }{{4}}{\phi ^4} \eea
and ${\cal L}^{{\rm count.}}$ the counterterm Lagrangian
\bea\label{COUNT.U}
{\cal L}^{{\rm count.}} &=& \frac{1}{2} \Biggl \{ -( p^2g_{\m\n} -p_\m p_\n - g_{\m\n} m_{Z}^2 )\d A + g_{\m\n}m_{Z}^2 \d \phi + g_{\m\n} \frac{2 m_H m_Z}{\sqrt{2 \l}} \d g +{\cal M}_{Z,\m\n} \Biggr \}{A^\m A^\n} \nonumber\\
&+& g^{\mu \nu } \Biggl \{ \frac{\sqrt{2 \l }m_Z}{m_H} \d g + \frac{\l m_Z^2}{m_H^2}\d A  + \frac{\l m_Z^2}{m_H^2}\d \phi + {\cal M}_{Z4} \Biggr \} {A_\mu }{A_\nu } {{\phi ^2} {}} \nonumber\\
&+&g^{\mu \nu } \Biggl \{ 2 m_Z \d g + \frac{\sqrt{2 \l} m_Z^2}{m_H}\d A  + \frac{\sqrt{2 \l}  m_Z^2}{m_H}\d \phi + {\cal M}_{Z3}  \Biggr \} \phi {A_\mu }{A_\nu } \nonumber\\
&+&\frac{m_H}{\sqrt{2 \l}} \Biggr \{ -\frac{1}{2} m_H^2 \d \phi + \d m - \frac{m_H^2}{2 \l}\d \l + {\cal T}_H \Biggr \} \phi \nonumber\\
&+&\frac{1}{2}\Biggl \{ (p^2 - \frac{5}{2}m_H^2)\d \phi + \d m - \frac{3m_H^2}{2 \l}\d \l + {\cal M}_H  \Biggr \} {\phi ^2} \nonumber\\
&-&\frac{m_H}{\sqrt{2 \l}} \Biggl \{ \d \l + 2 \l \d \phi - \frac{\sqrt{2 \l}}{m_H}\frac{{\cal K}_H}{6} \Bigr \} {\phi ^3} - \frac{1}{4} \Biggl \{ \d \l + 2 \l \d \phi - \frac{{\cal B}_H}{6} \Bigr \} {\phi ^4} \, ,
\eea
with the computed one-loop corrections also added with appropriate factors.
Corrections to interaction terms between the Higgs and the $Z$ ($ {\cal M}_{Z3}$ and $ {\cal M}_{Z4}$) we have not computed but we are not concerned with those here.

The Feynman rules for the counterterms deriving from the above expression are
\begin{itemize}
\item Higgs 1-point function
\begin{center}
\begin{tikzpicture}
\draw[dashed,thick] (-2.2,0)--(0.3,0) ;
\draw [thick] [fill=black] (0.25,0) circle [radius=0.1];
\node at (6,0) {$=\displaystyle i \frac{m_H}{\sqrt{2 \l}} \left[ -\frac{1}{2} m_H^2 \d \phi + \d m - \frac{m_H^2}{2 \l}\d \l     \right] $};
\end{tikzpicture}
\end{center}
\item Gauge boson 2-point function
\begin{center}
\begin{tikzpicture}
\draw[photon] (-3,0)--(-0.5,0) ;
\draw [thick] [fill=black] (-1.7,0) circle [radius=0.1];
\node at (6,0) {$=\displaystyle
 i\left[ -( p^2g_{\m\n} -p_\m p_\n - g_{\m\n}m_{Z}^2 )\d A + g_{\m\n}m_{Z}^2 \d \phi + g_{\m\n} \frac{\sqrt{2}m_H m_Z}{\sqrt{ \l}} \d g  \right]$};
\end{tikzpicture}
\end{center}
\item Higgs 2-point function
\begin{center}
\begin{tikzpicture}
\draw[dashed,thick] (-1,0)--(1.5,0) ;
\draw [thick] [fill=black] (0.25,0) circle [radius=0.1];
\node at (6,0) {$=\displaystyle i\left[ (p^2 - \frac{5}{2}m_H^2)\d \phi + \d m - \frac{3m_H^2}{2 \l}\d \l    \right] $};
\end{tikzpicture}
\end{center}
\item For completeness, the $\phi$-$Z$-$Z$ counterterm vertex
\begin{center}
\begin{tikzpicture}[scale=0.7]
\draw [photon] (-2.5,1.5)--(-1,0);
\draw [photon] (-2.5,-1.5)--(-1,0);
\draw[dashed,thick] (-1,0)--(1,0);
\draw [thick] [fill=black] (-1,0) circle [radius=0.1];
\node at (8,0) {$=  \displaystyle i g^{\mu \nu }\left[ 4 m_Z \d g + 2\frac{\sqrt{2 \l} m_Z^2}{m_H}\d A  +2 \frac{\sqrt{2 \l}  m_Z^2}{m_H}\d \phi    \right]$};
\end{tikzpicture}
\end{center}
\item $\phi$-$\phi$-$\phi$ vertex counterterm
\begin{center}
\begin{tikzpicture}[scale=0.7]
\draw [dashed,thick] (-2.5,1.5)--(-1,0);
\draw [dashed,thick] (-2.5,-1.5)--(-1,0);
\draw[dashed,thick] (-1,0)--(1,0);
\draw [thick] [fill=black] (-1,0) circle [radius=0.1];
\node at (5,0) {$= \displaystyle  - 6i\frac{m_H}{\sqrt{2 \l}} \left[ \d \l + 2 \l \d \phi    \right]$};
\end{tikzpicture}
\end{center}
\item For completeness, the $\phi$-$\phi$-$Z$-$Z$ vertex counterterm
\begin{center}
\begin{tikzpicture}[scale=0.7]
\draw [dashed,thick] (0,0)--(1.5,1.4);
\draw [dashed,thick] (0,0)--(1.5,-1.4);
\draw [photon] (-1.5,1.4)--(0,0);
\draw [photon] (-1.5,-1.4)--(0,0);
\draw [thick] [fill=black] (0,0) circle [radius=0.1];
\node at (8,0) {$=\displaystyle ig^{\mu \nu }\left[ 4\frac{\sqrt{2 \l }m_Z}{m_H} \d g + \frac{4 \l m_Z^2}{m_H^2}\d A  + \frac{4 \l m_Z^2}{m_H^2}\d \phi  \right] $};
\end{tikzpicture}
\end{center}
\item $\phi$-$\phi$-$\phi$-$\phi$ vertex counterterm
\begin{center}
\begin{tikzpicture}[scale=0.7]
\draw [dashed,thick] (0,0)--(1.5,1.4);
\draw [dashed,thick] (0,0)--(1.5,-1.4);
\draw [dashed,thick] (-1.5,1.4)--(0,0);
\draw [dashed,thick] (-1.5,-1.4)--(0,0);
\draw [thick] [fill=black] (0,0) circle [radius=0.1];
\node at (5,0) {$= \displaystyle -6i \left[ \d \l + 2 \l \d \phi    \right] $};
\end{tikzpicture}
\end{center}
\end{itemize}
The renormalization conditions are in order. All conditions are imposed at zero external momenta.
Regarding the gauge boson sector, our renormalization condition is that the mass of the gauge boson be $m_Z = g v_0$.
Diagrammatically this condition is
\be
\begin{tikzpicture} [scale=0.9]
\draw [photon,thick] (-2.3,0)--(-1.2,0);
\draw [thick] [fill=gray] (-0.5,0) circle [radius=0.8];
\draw [photon,thick] (0.3,0)--(1.3,0);
\node at (2,0) {$+$};
\draw [photon,thick] (2.5,0)--(3.6,0);
\draw [thick] [fill=black] (3.7,0) circle [radius=0.1];
\draw [photon,thick] (3.8,0)--(4.9,0);
\node at (6,0) {$=\, \, 0$};
\end{tikzpicture}
\nonumber
\ee
and as an equation
\bea
{\cal M}_Z  - \frac{1}{3} \left(g_{\m\n} -\frac{p_\m p_\n}{p^2}   \right)  \left( (- p^2g_{\m\n} + p_\m p_\n +  g_{\m\n} m_{Z}^2 ) \d A  + g_{\m\n}m_{Z}^2 \d \phi 
 + g_{\m\n} \frac{2 m_H m_Z}{\sqrt{2 \l}} \d g \right) = 0\, .\nonumber\\
 \eea
This is an independent condition from the rest that can be directly solved for $\d g$:
\bea\label{deltaZmass}
2 v_0 \d g  &=& \frac{{\cal M}_Z}{m_Z}  - \left( m_{Z} - \frac{p^2}{m_Z} \right) \d A  - m_{Z} \d \phi
\eea
At zero external momentum, this becomes
\be\label{soldZ}
v_0 \d g \equiv \d m_Z = \frac{\m^\ve}{(4\pi)^{2}} \left[ \frac{2}{3} \frac{\l m_Z^3}{m_H^2} \frac{1}{\ve} + \frac{1}{2} \left( \frac{{\cal M}_Z}{m_Z}  - m_{Z} \d A  - m_{Z} \d \phi   \right)_f   \right]\, ,
\ee
where we have absorbed in $\d g$ all the finite parts.
Just one comment on the Higgs-$Z$ interaction terms that we have not computed here: they become finite provided that
\be
\frac{{\cal M}_Z}{m_Z^2} + {\cal M}_{Z3, Z4} = \, {\rm finite}\, . 
\ee

Now we turn to the scalar sector.
In order to avoid the tadpoles contaminating the one-loop vev for the Higgs field, we impose a vanishing tadpole condition.
Diagrammatically it is
\be
\begin{tikzpicture}[scale = 0.9]
\draw [dashed,thick] (-2.5,0)--(-0.8,0);
\draw [thick] [fill=gray] (0,0) circle [radius=0.8];
\node at (1.4,0) {$+$};
\draw [dashed,thick] (2,0)--(3.7,0);
\draw [thick] [fill=black] (3.8,0) circle [radius=0.1];
\node at (5.0,0.05) {$=\, \,0$};
\end{tikzpicture}
\nonumber
\ee
and as an equation
\be\label{Rencond1}
 {\cal T}_H + \frac{m_H}{\sqrt{2 \l}} \left( -\frac{1}{2} m_H^2 \d\phi  + \d m  - \frac{m_H^2}{2 \l} \d \l \right)  = 0\, .
\ee
The second condition is the requirement that the only term that remains in the quadratic part of the potential, is $m_H$.
This means that in the quadratic term of the potential
we absorb, together with the divergent part, the entire finite term as well:
\be
\begin{tikzpicture} [scale=0.9]
\draw [dashed,thick] (-2.3,0)--(-1.2,0);
\draw [thick] [fill=gray] (-0.5,0) circle [radius=0.8];
\draw [dashed,thick] (0.3,0)--(1.3,0);
\node at (2,0) {$+$};
\draw [dashed,thick] (2.5,0)--(3.6,0);
\draw [thick] [fill=black] (3.7,0) circle [radius=0.1];
\draw [dashed,thick] (3.8,0)--(4.9,0);
\node at (5.8,0) {$=\, \, 0 $};
\end{tikzpicture}
\nonumber
\ee
or, in equation at $p = 0$,
\be\label{Rencond2}
{\cal M}_H - \frac{5}{2}m_H^2 \d \phi  + \d m  - \frac{3m_H^2}{2 \l}  \d \l = 0\, .
\ee
These two conditions fix completely our freedom. The solution to the system of \eq{Rencond1} and \eq{Rencond2} is
\bea
\label{soldm}
\d m &=& \frac{\m^\ve}{(4\pi)^{2}} \left[ \frac{1}{\ve} \frac{1}{2} \left(6 \l m_H^2 - 12 \l m_Z^2 \right) + \frac{1}{2} \left({\cal M}_{H} - m_H^2 \d\phi - 3 \frac{\sqrt{2\l}}{m_H} {{\cal T}_H}\right)_f\right]\\
\label{soldl}
\d \l &=& \frac{\m^\ve}{(4\pi)^{2}} \left[ \frac{1}{\ve} \left(18 \l^2 - 24\l^2 \frac{m_Z^2}{m_H^2} + 24 \l^2 \frac{m_Z^4}{m_H^4}\right) 
+ \frac{\l }{m_H^2} \left({{\cal M}_H} - 2 m_H^2 \d\phi - \frac{\sqrt{2\l}}{m_H} {{\cal T}_H} \right)_f \right]\nonumber\\
\eea
For later reference note that
\be\label{dmH}
\d m_H = 2 \d m\, .
\ee
Divergences must cancel automatically from the rest. 
Absence of divergences in the cubic coupling means 
\be\label{cond3}
\begin{tikzpicture} [scale=0.9]
\draw [dashed,thick] (0.3,0.4)--(1.1,0.9);
\draw [dashed,thick] (0.3,-0.4)--(1.1,-0.9);
\draw [dashed,thick] (-1.3,0)--(-2.3,0);
\draw [thick] [fill=gray] (-0.5,0) circle [radius=0.8];
\node at (1.9,0) {$+$};
\draw [dashed,thick] (3.8,0.0)--(5.0,0.8);
\draw [dashed,thick] (3.8,0.0)--(5.0,-0.8);
\draw [dashed,thick] (3.7,0)--(2.7,0);
\draw [thick] [fill=black] (3.8,0) circle [radius=0.1];
\node at (6,0) {$=\,\,  {\rm finite}$};
\end{tikzpicture}
\nonumber
\ee
or
\be\label{Rencond3}
\frac{{\cal K}_{H}}{6} -  \frac{m_H}{\sqrt{2 \l}} \left(2 \l \d \phi + \d \l \right)  = {\rm finite}\, .
\ee
Substituting \eq{soldm} and \eq{soldl} in the above, we find that the divergent part cancels as expected and we are left with the finite terms
\be\label{CF3}
C_{\phi^3} = \frac{1}{16\pi^2}  \Bigl( \frac{{\cal K}_{Hf}}{6} - \frac{1}{2} \frac{\sqrt{2 \l}}{m_H} {\cal M}_{Hf} + \frac{\l }{m_H^2} {\cal T}_{Hf} \Bigr)\, .
\ee
Absence of divergences from the quartic coupling on the other hand requires
\be\label{cond4}
\begin{tikzpicture} [scale=0.9]
\draw [dashed,thick] (0.3,0.4)--(1.1,0.9);
\draw [dashed,thick] (0.3,-0.4)--(1.1,-0.9);
\draw [dashed,thick] (-1.9,0.8)--(-1,0);
\draw [dashed,thick] (-1.9,-0.8)--(-1,0);
\draw [thick] [fill=gray] (-0.5,0) circle [radius=0.8];
\node at (1.9,0) {$+$};
\draw [dashed,thick] (3.8,0.0)--(5.0,0.8);
\draw [dashed,thick] (3.8,0.0)--(5.0,-0.8);
\draw [dashed,thick] (3.8,0.0)--(2.6,0.8);
\draw [dashed,thick] (3.8,0.0)--(2.6,-0.8);
\draw [thick] [fill=black] (3.8,0) circle [radius=0.1];
\node at (6,0) {$=\,\,  {\rm finite}$};
\end{tikzpicture}
\nonumber
\ee
or
\be\label{Rencond4}
\frac{{\cal B}_H}{6} -  (2 \l \d \phi + \d \l)  = {\rm finite}\, .
\ee
Substituting again \eq{soldm} and \eq{soldl} in the above, we again observe the cancellation of the divergent part
and we collect the finite piece
\be\label{CF4}
C_{\phi^4} = \frac{1}{16\pi^2} \Bigl( \frac{{\cal B}_{Hf}}{6} - \frac{{\l}}{m_H^2} {\cal M}_{Hf} + \frac{\l \sqrt{2\l}}{m_H^3} {\cal T}_{Hf} \Bigr)\, .
\ee
The subscript $f$ in the various one-loop quantities denotes finite part, in either the $R_\xi$ or the Unitary gauge.
For example, ${\cal T}_{Hf} = [{\cal T}^{R_\xi}_H]_f + \{{\cal T}^{R_\xi}_H\}_f $ in the $R_\xi$ gauge ${\cal T}_{Hf} = [{\cal T}^{U}_H]_f$ in the Unitary gauge.

The finite one-loop Higgs potential we are left with after renormalization is then
\bea\label{fullVrenU2}
V_1(\phi) &=& \frac{1}{2} m_H^2 \phi^2 + \left[ \sqrt{\frac{\l}{2}} m_H + C_{\phi^3} \right]{\phi ^3} + \frac{1}{4}\left[ \l +C_{\phi^4} \right]  \phi^4\, .
\eea
The Higgs field anomalous dimension has apparently cancelled from the renormalized potential.
We are now ready to minimize this potential. There are three extrema. Our preferred "physical" solution for the global minimum is 
\be
\langle \phi \rangle \equiv v = 0\, ,
\ee
for which the potential satisfies $V_1''(v) = m_H^2$.
The quantities $C_{\phi^3}$ and $C_{\phi^4}$ are examples of the $\star$ quantities in \eq{starcomm} of the Introduction.
Note finally that $V_1(v)=0$.

\subsection{Physical quantities and the $\b$-functions}\label{Physical quantities and the b-functions}

Let us denote by $\a_0$ a generic bare coupling. Its counter-term $\d \a(\m)$ is introduced via the relation
\bea\label{a0aR}
\a_0 = \a(\m) + \d \a(\m)
\eea
where $\a(\m)$ is the renormalized running coupling. At one-loop the counter-term has the form
\bea\label{counterterm}
\d \a (\m) &=& \frac{\mu^{\varepsilon}}{(4\pi)^2} \left(\frac{C_\a}{\varepsilon} + \sum_k f_{A_0}^{k} \ln \frac{\m^2}{m_{k}^2}  + 
\sum_{k,i}  f_{B_0}^{k}\int\limits_0^1 {dx}  \ln \left(\frac{\m^2}{\Delta^i_{k}(m_k,m_i)} \right)  + \sum_k f^k_{A_0} \right)\, .\nonumber\\
\eea
The indices $i$ and $k$ are counting the fields running in the loop. The entire divergent part has been collected in the term proportional to $C_a$.
In the notation of the previous sections, we can identify $(\m^\ve / 16\pi^2) C_\a = \ve ({\d \a})_\ve = \ve [\d \a]_\ve + \ve \{\d \a \}_\ve$ and $({\d \a})_f=[\d \a]_f + \{  \d \a \}_f$ 
containing the finite logarithms and the non-logarithmic finite term.
As already noted, in the Unitary gauge we have $\{\d \a \}_\ve = \{  \d \a \}_f = 0$ by definition.

It is useful to review the calculations of $\b$-functions in the presence of multiple couplings.
We introduce the boundary condition
\bea
\a(\m = m_{\rm phys.}) \equiv \a\, .
\eea
We also use the following standard definitions
\bea\label{beta function}
\d_ \a &\equiv& \frac{\d \a(\m)}{a(\m)} \nonumber\\
\b_\a &\equiv  & \m \frac{d}{d \m} \a(\m) \nonumber\\
\tilde {\b_\a} &\equiv&  \frac{ \b_\a}{\a} 
\eea
For a general coupling we recall the successive relations
\bea
0&=&\m \frac{d}{d \m} \a_0 = \m \frac{d}{d \m}\Bigl \{ \m^{\varepsilon}  \a(\m) (1 + \d_\a )  \Bigr \} = \m \frac{d}{d \m}\Bigl \{ \m^{\varepsilon} \a(\m) + \m^{\varepsilon}\d\a(\m)\Bigr \} \Leftrightarrow \nonumber\\
0&=& \m \Biggl \{ \varepsilon \a (1+ \d_ \a) +(1+ \d_ \a) \m \frac{\partial  \a}{\partial \m} + \a \m \frac{\partial \d_ \a}{\partial \m}   \Biggr \} \Leftrightarrow \nonumber\\
\b_\a(1 + \d_\a) &=& - \varepsilon \a (1+ \d_ \a)  - \a \m \frac{\partial \d_ \a}{\partial \m} \Leftrightarrow \nonumber\\
\b_\a &=& - \varepsilon \a  - \a \m \frac{\partial \d_ \a}{\partial \m} (1 + \d_\a)^{-1} \Leftrightarrow \nonumber\\
\b_\a &=& - \varepsilon \a  - \a \m \frac{\partial \d_ \a}{\partial \m} \Leftrightarrow \, ,
\eea
where, since $\d_\a \sim O(\hbar)$, we have performed an expansion in $\hbar$ in order to get rid of terms of ${\cal O}(\hbar^2)$ like $\d_ \a \cdot \frac{\partial \d_ \a}{\partial \m}$. 
In the case of the AH model where we have three couplings, we will have a system of equations:
\bea
\b_\l &=& - \varepsilon \l  - \l \Biggl \{ \b_\l \frac{\partial \d_ \l}{\partial \l} + \b_{m_H^2}\frac{\partial \d_ \l}{\partial m_H^2} + \b_{m_Z^2}\frac{\partial \d_ \l}{\partial m_Z^2}     \Biggr \} \nonumber\\
\b_{m_H^2} &=& - \varepsilon {m_H^2}  - {m_H^2} \Biggl \{ \b_\l \frac{\partial \d_{m_H}}{\partial \l} + \b_{m_H^2}\frac{\partial \d_{m_H}}{\partial m_H^2} + \b_{m_Z^2}\frac{\partial \d_{m_H}}{\partial m_Z^2}     \Biggr \} \nonumber\\
\b_{m_Z^2} &=& - \varepsilon {m_Z^2}  - {m_Z^2} \Biggl \{ \b_\l \frac{\partial \d_{m_Z}}{\partial \l} + \b_{m_H^2}\frac{\partial \d_{m_Z}}{\partial m_H^2} + \b_{m_Z^2}\frac{\partial \d_{m_Z}}{\partial m_Z^2}     \Biggr \} \, .
\eea
This system can be rewritten as
\bea
\b_\l (1 + \frac{\partial \d_ \l}{\partial \l}  ) + \l \Biggl \{ \b_{m_H^2}\frac{\partial \d_ \l}{\partial m_H^2} + \b_{m_Z^2}\frac{\partial \d_ \l}{\partial m_Z^2}     \Biggr \} \   &=& - \varepsilon \l   \nonumber\\
\b_{m_H^2}(1+  \frac{\partial \d_{m_H}}{\partial m_H^2}  ) + {m_H^2} \Biggl \{ \b_\l \frac{\partial \d_{m_H}}{\partial \l} + \b_{m_Z^2}\frac{\partial \d_{m_H}}{\partial m_Z^2}     \Biggr \}  &=& - \varepsilon {m_H^2}  \nonumber\\
\b_{m_Z^2}(1 + \frac{\partial \d_{m_Z}}{\partial m_Z^2}  ) +  {m_Z^2} \Biggl \{ \b_\l \frac{\partial \d_{m_Z}}{\partial \l} + \b_{m_H^2}\frac{\partial \d_{m_Z}}{\partial m_H^2}  \Biggr \}&=& - \varepsilon {m_Z^2} \, ,
\eea
or in matrix form as
\bea
\left( {\begin{array}{*{20}{c}}
{1 + \frac{\partial \d_ \l}{\partial \l} }&{\l \frac{\partial \d_ \l}{\partial m_H^2} }&{\l \frac{\partial \d_ \l}{\partial m_Z^2}}\\
{m_H^2 \frac{\partial \d_{m_H}}{\partial \l}}&{1+  \frac{\partial \d_{m_H}}{\partial m_H^2} }&{m_H^2 \frac{\partial \d_{m_H}}{\partial m_Z^2} }\\
{m_Z^2\frac{\partial \d_{m_Z}}{\partial \l}}&{m_Z^2 \frac{\partial \d_{m_Z}}{\partial m_H^2}}&{1 + \frac{\partial \d_{m_Z}}{\partial m_Z^2}}
\end{array}} \right) \cdot \left( {\begin{array}{*{20}{c}}
{\b_\l}\\
{\b_{m_H^2}}\\
{\b_{m_Z^2}}
\end{array}} \right) =- \varepsilon \left( {\begin{array}{*{20}{c}}
{ \l}\\
{ {m_H^2}}\\
{ {m_Z^2} }
\end{array}} \right)
\eea
Inverting the matrix we obtain
\bea\label{betasystem}
16\pi^2 \b_\l &=& \l^2 \frac{\partial \frac{C_\l}{\l}}{\partial \l} + \l m_H^2 \frac{\partial \frac{C_\l}{\l}}{\partial m_H} + \l m_Z^2 \frac{\partial \frac{C_\l}{\l}}{\partial m_Z}\nonumber\\
16\pi^2 \b_{m_H^2} &=& \l m_H^2 \frac{\partial \frac{C_{m_H^2}}{{m_H^2}}}{\partial \l} + \frac{1}{2}m_H^2 \frac{\partial \frac{C_{m_H^2}}{{m_H^2}}}{\partial m_H} + \frac{1}{2} m_H^2 \frac{\partial \frac{C_{m_H^2}}{{m_H^2}}}{\partial m_Z}\nonumber\\
16\pi^2 \b_{m_Z^2} &=&  \l m_Z^2 \frac{\partial \frac{C_{m_Z^2}}{{m_Z^2}}}{\partial \l} + \frac{1}{2}m_Z^2 \frac{\partial \frac{C_{m_Z^2}}{{m_Z^2}}}{\partial m_H} + \frac{1}{2} m_Z^2 \frac{\partial \frac{C_{m_Z^2}}{{m_Z^2}}}{\partial m_Z}\, .
\eea
Thus, all that we need to do in order to obtain the various $\b$-functions, is to identify from the explicit form of the counterterms 
the quantities $C_\a$ defined in Eq.\eqref{counterterm} and build its $\b$-function, according to Eq.\eqref{betasystem}. 

Moreover, solving the differential equation for the running coupling yields the Renormalization Group flow of the coupling $\a$:
\bea\label{phy.quant.2}
\a(\m) &=& \frac{\a}{1+\tilde{\b_\a}{}\, \ln \left(\frac{m_{\rm phys.}}{\m}\right) } 
\eea
The Landau pole associated with the coupling $\a$ is 
\bea\label{Landau pole}
\m_{L}^\a &=& m_{\rm phys.}\, e^{\frac{\a}{\b_\a}} \, .
\eea
In the AH model we found from our one-loop calculation that
$C_{m_Z} = \frac{2}{3} \frac{\l m_Z^4}{m_H^2}$, $C_{m_H} = 6 \l m^2_H  - 12 \l m_Z^2$ and $C_{\l} = 18\l^2 -24 \l^2 \frac{m_Z^2}{m_H^2} + 24 \l^2 \frac{m_Z^4}{m_H^4}$
that immediately determine
\bea\label{beta}
\b_{m_Z^2} &=&  \frac{1}{16\pi^2} \left(\frac{2}{3} \frac{\l m_Z^4}{m_H^2}\right)\nonumber\\
\b_{m_H^2} &=&  \frac{1}{16\pi^2} \left( 6 \l m^2_H  - 12 \l m_Z^2 \right)\nonumber\\
\b_{\l} &=&  \frac{1}{16\pi^2} \left( 18\l^2 -24 \l^2 \frac{m_Z^2}{m_H^2} + 24 \l^2 \frac{m_Z^4}{m_H^4} \right)\, .
\eea
The above expressions hold for both $R_\xi$ and Unitary gauges. 
It is interesting to notice that these $\b$-functions are not identically the same as those that one would compute in a Cartesian basis
for the Higgs. For a comparison see the next section.

\section{The one-loop Higgs potential}
\label{Potential}

To summarize our result regarding the one-loop Higgs potential, in \eq{fullVrenU2} we found that it is determined by the
quantities $C_{\phi^3}$ and $C_{\phi^4}$ in \eq{CF3} and in \eq{CF4} respectively, yielding the finite expression
\bea\label{oneloopV}
V_1(\phi) &=& \frac{1}{2} m_H^2 \phi^2 + \left[ \sqrt{\frac{\l}{2}} m_H - \frac{m_H}{16 \pi^2\sqrt{2\l}} \Biggl ( 
9 \l^2 + \frac{8 \l^2 m_Z^4}{ m_H^4} \Biggl ) \right]{\phi ^3} \nonumber\\
&+& \frac{1}{4}\left[ \l - \frac{1}{16 \pi^2} \Biggl ( \frac{32 \l^2 m_Z^4}{ m_H^4} \Biggl ) \right]  \phi^4\, .
\eea
The form of the potential is such that the $\xi$-independent part of its $R_\xi$ gauge expression is the same as the Unitary gauge expression.
Moreover, using the standard prescription to compute the $U_{\cal T}$ integral, see \eq{gmnJmn1}, the $\xi$-dependent part is made to vanish.
In addition, the potential has no explicit $\m$-dependence. All the $\m$-dependence is implicit, through the dependence on $\m$ of
the renormalized quantities $\l$, $m_H$ and $m_Z$ via their RG evolution.
In the following numerical analysis one can either interpret the potential in \eq{oneloopV} as a $\xi$-independent $R_\xi$ gauge potential or 
simply as a Unitary gauge potential, as these are the same independently of prescriptions.

Before we proceed with the study of the gauge invariant Higgs potential \eq{oneloopV}, we recall the standard result used to extract physics from the Higgs potential.
The usual method to compute the Higgs potential is in the context of the background field method,
where a functional integration yields the so called Higgs effective potential. We will not review the details of
this well known calculation; we assume that the reader has some familiarity with it.
The derivation of an effective scalar potential via the background field method at one-loop is much simpler than computing
Feynman diagrams. The simplicity is related among other reasons to the Cartesian basis representation of the 
Higgs field because in the Cartesian basis the Gaussian path integral involved, is almost trivial.
The result of the calculation for the scalar potential in the Abelian-Higgs model (see for example \cite{Schwartz}) with the same normalization of the mass and quartic coupling
as in \eq{clas.Lag.1}, in ${\overline {\rm MS}}$ scheme with Fermi gauge fixing, is ($H_0\to \phi/\sqrt{2}$)
\bea\label{Veff}
V^{\overline {\rm MS}}_{1,\rm eff.} &=& - \frac{1}{2} m^2 \phi^2 + \frac{\l}{4} \phi^4\nonumber\\
&-& \frac{1}{64\pi^2} \left\{ 3m_A^4   \left( \ln\frac{\m^2}{m_A^2} + \frac{5}{6}\right) + m_B^4   \left( \ln\frac{\m^2}{m_B^2} + \frac{3}{2}\right)
+ \sum_{\pm }m_{C_\pm}^4   \left( \ln\frac{\m^2}{m_{C_\pm}^2} + \frac{3}{2}\right)\right\} \nonumber\\
\eea
with
\bea
m_A^2 &=& g^2 \phi^2\nonumber\\
m_B^2 &=& 3\l \phi^2 - m^2\nonumber\\
m_{C_\pm}^2 &=& \frac{1}{2} \left[\left( \l \phi^2 - m^2 \right) \pm \sqrt{(\l \phi^2 - m^2)^2 - 4\xi g^2 \phi^2 (\l \phi^2 - m^2)} \right]\, .
\eea
The derivation of the $\b$-function is also simple, provided that a separate diagram calculation has yielded the also well known result
\be
16\pi^2 \g = g^2  (-3 + \xi)
\ee
for the anomalous dimension $\g$. Then, one finds
\bea\label{betaeff}
16\pi^2 \b_{m^2} &=& 8\l m^2 - 6g^2 m^2\nonumber\\
16\pi^2 \b_{\l} &=& 20\l^2 - 12 \l g^2 + 6g^4\nonumber\\
16\pi^2 \b_{g} &=& \frac{1}{3} g^4
\eea
for the $\b$-functions. These Cartesian basis results are not identically the same as the Polar basis results in \eq{beta}.
Despite however the fact that some coefficients are not the same, their physical content is similar.

\begin{figure}[t]
\centerline{\includegraphics[width=70mm]{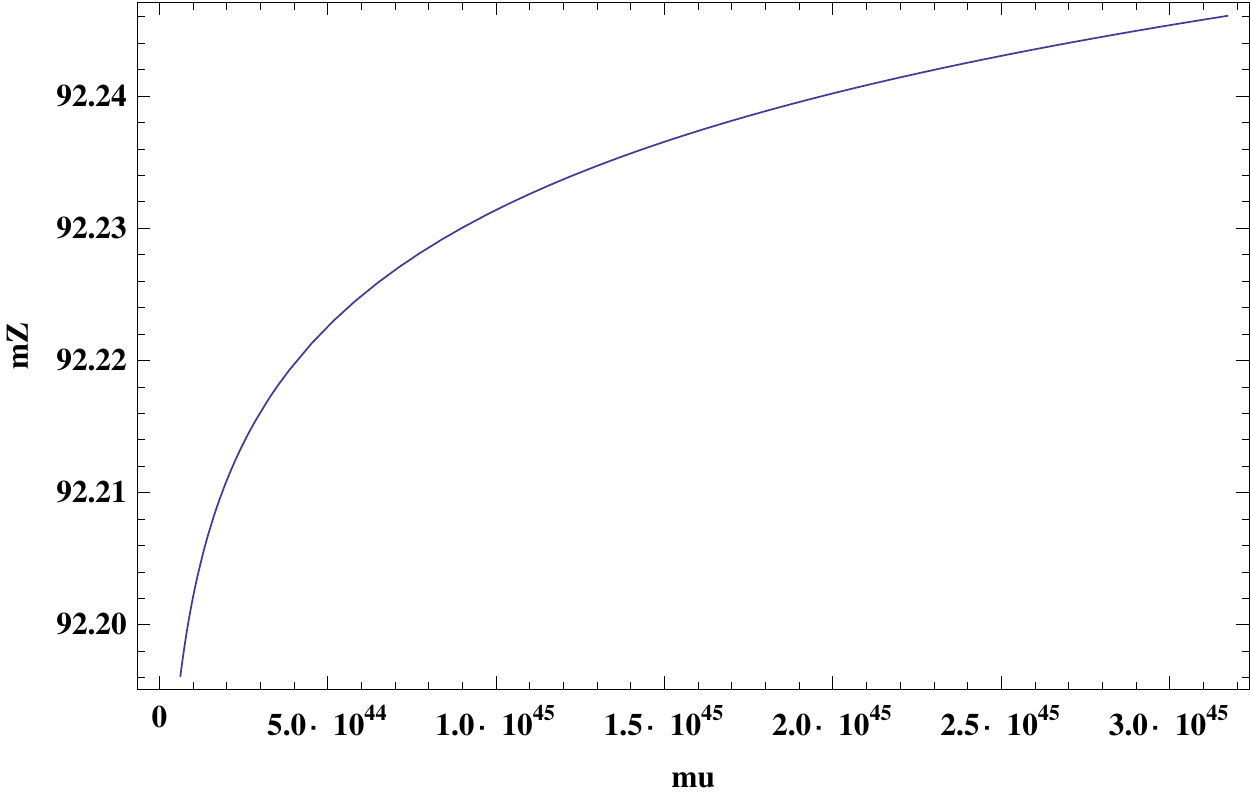}\quad\includegraphics[width=70mm]{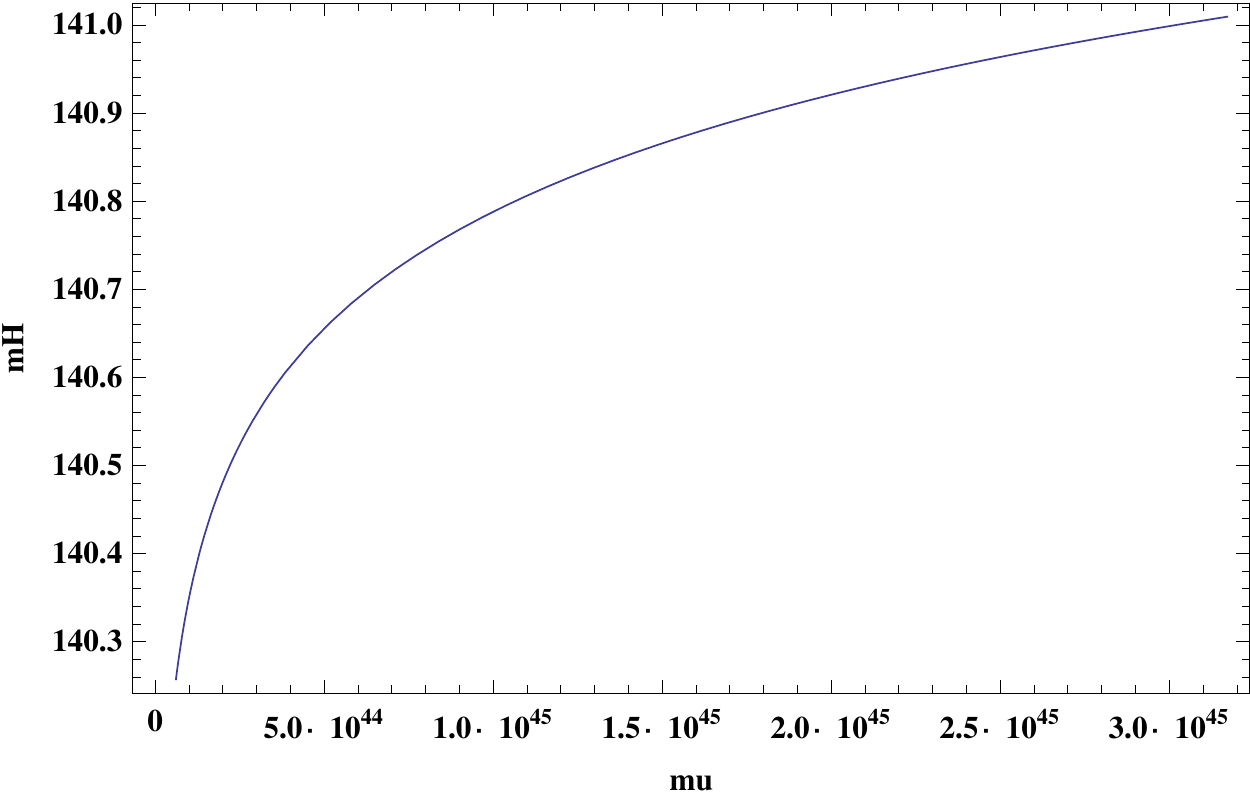}}
\caption{\small{The RG evolution of the $Z$ (left) and Higgs (right) mass.}}
\label{mRG}
\end{figure}

\begin{figure}[t!]
\begin{center}
\includegraphics[width=0.5\textwidth]{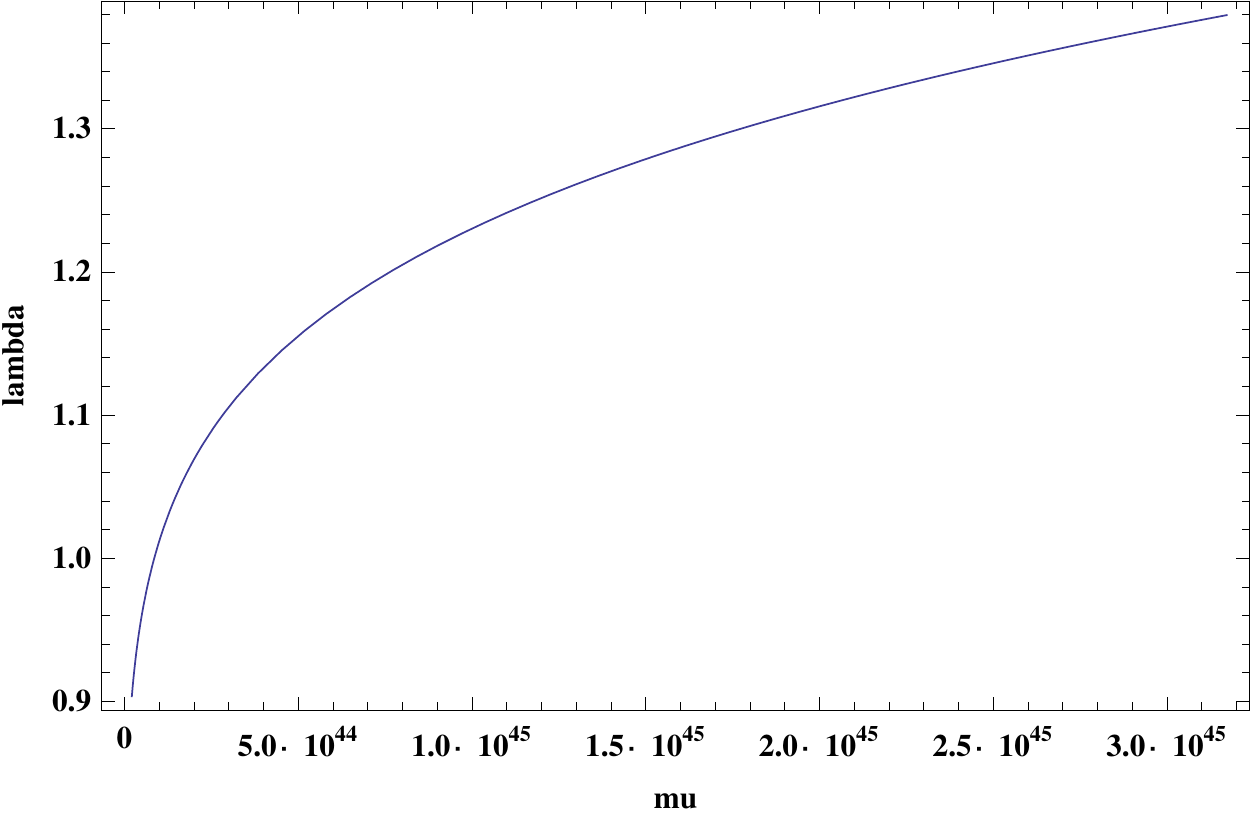}
\end{center}
\caption{\footnotesize The RG evolution of $\l$.}
\label{lambdaRG}
\end{figure}

In Figs. \ref{mRG} and \ref{lambdaRG} we plot the RG evolution of the Higgs and the $Z$ mass as well as that of the coupling $\l$,
as a function of the Renormalization scale $\m$, as determined in \eq{beta}.
We do not produce separate figures for \eq{betaeff} because the numerical differences are quite small.
The physical values at $\m=125$ GeV we use are 125 GeV for the Higgs mass, 91 GeV for the $Z$-mass and 0.12 for $\l$. 
We stop the evolution at a certain scale
\be\label{Iscale}
\m_I \simeq 3.03 \cdot 10^{46} \,\, {\rm GeV}
\ee
whose physical meaning will be discussed below. We observe the usual perturbative, logarithmic evolution of the couplings.

\begin{figure}[t!]
\begin{center}
\includegraphics[width=1.0\textwidth]{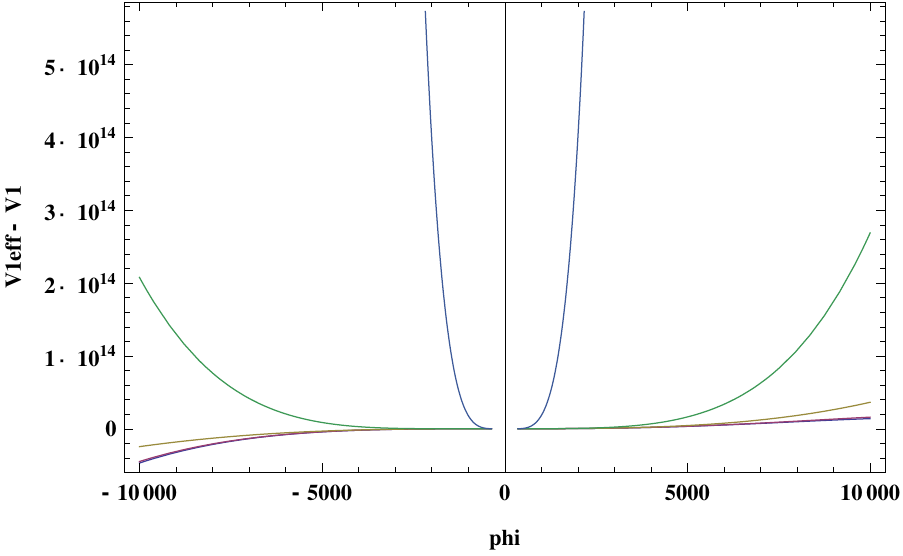}
\end{center}
\caption{\footnotesize The difference $V_1(\phi)- V^{\overline {\rm MS}}_{1,\rm eff.}(\phi)$ between 
the 1-loop and effective Higgs potentials at $\m=125$ GeV for various values of $\xi=10^{-3}, 1, 10, 100, 10^{5}$. 
Curves taking larger values on the vertical axis correspond to larger $\xi$.}
\label{Vdifference}
\end{figure}

The potential in \eq{oneloopV} is gauge invariant and the one in \eq{Veff} is manifestly gauge dependent. 
To quantify the effect of $\xi$, the different basis for the Higgs and the different subtraction schemes, we compare numerically the two results.
The comparison that follows should be clearly taken with a grain of salt, as the two objects are quite different. 
In Fig. \ref{Vdifference} we plot the difference between the one-loop Higgs potential \eq{oneloopV} and the effective potential \eq{Veff}
at $\m=125$ GeV, for $\xi=0.001,1,10,100,100000$. 
The running values of $m_H$, $m_Z$ and $\l$ in the potential $V_1$ are determined from \eq{beta},
while those of $m$, $g$ and $\l$ in $V^{\overline {\rm MS}}_{1,\rm eff.}$ are determined using \eq{betaeff}.
A regime around $\phi=0$ in the plot is missing because there the effective potential is imaginary. 
There seem to be ways to fix this \cite{Martin} but this is not our concern here. $V_1$ remains instead always real.
What we observe is that as $\xi$ increases, the regime where the two expressions agree shrinks.
There seems to be a value of $\xi$ for which the best agreement is achieved, which from the figure is around $\xi < O(10)$.
The limit $\xi\to\infty$ of the effective potential is clearly singular: in \eq{starcomm} we have $\lim_{\xi\to\infty} g (\xi) = \infty$.

\begin{figure}[t!]
\begin{center}
\includegraphics[width=0.75\textwidth]{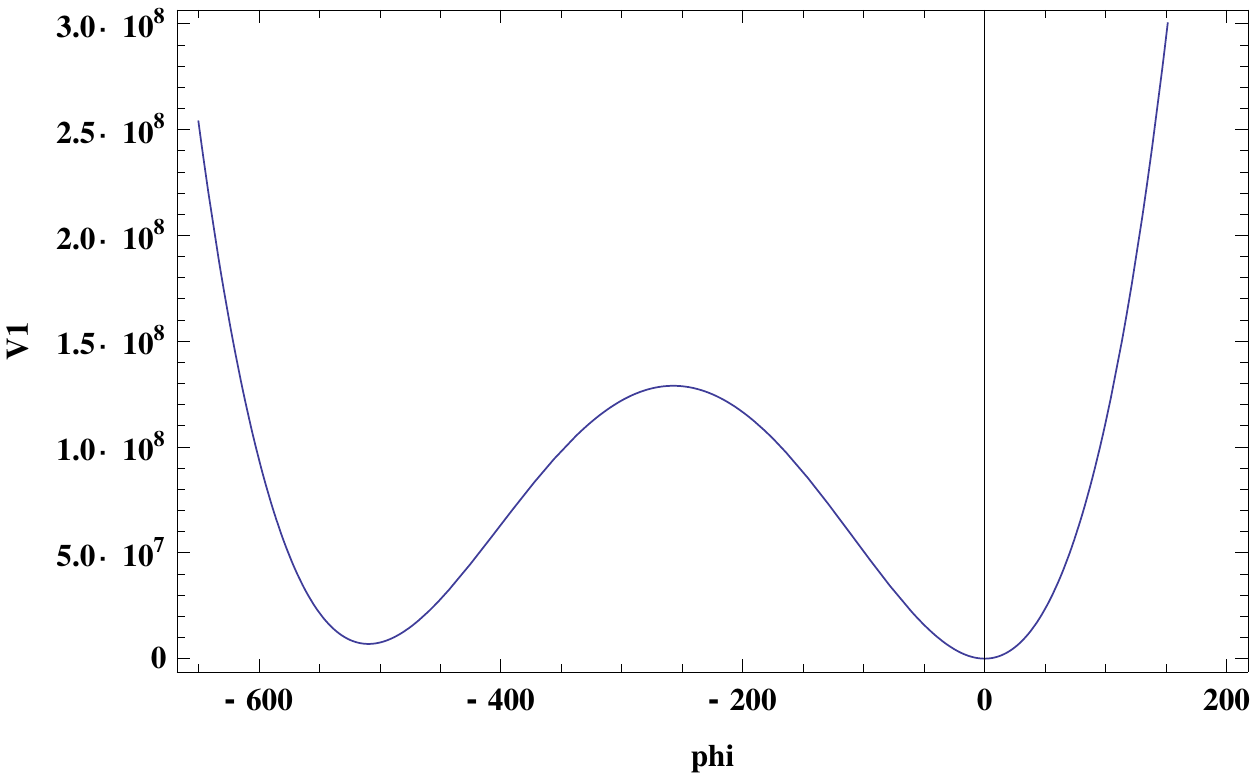}
\end{center}
\caption{\footnotesize The gauge invariant 1-loop Higgs potential $V_1(\phi)$ at $\m=125$ GeV.}
\label{V1mu125}
\end{figure}

\begin{figure}[t]
\centerline{\includegraphics[width=90mm]{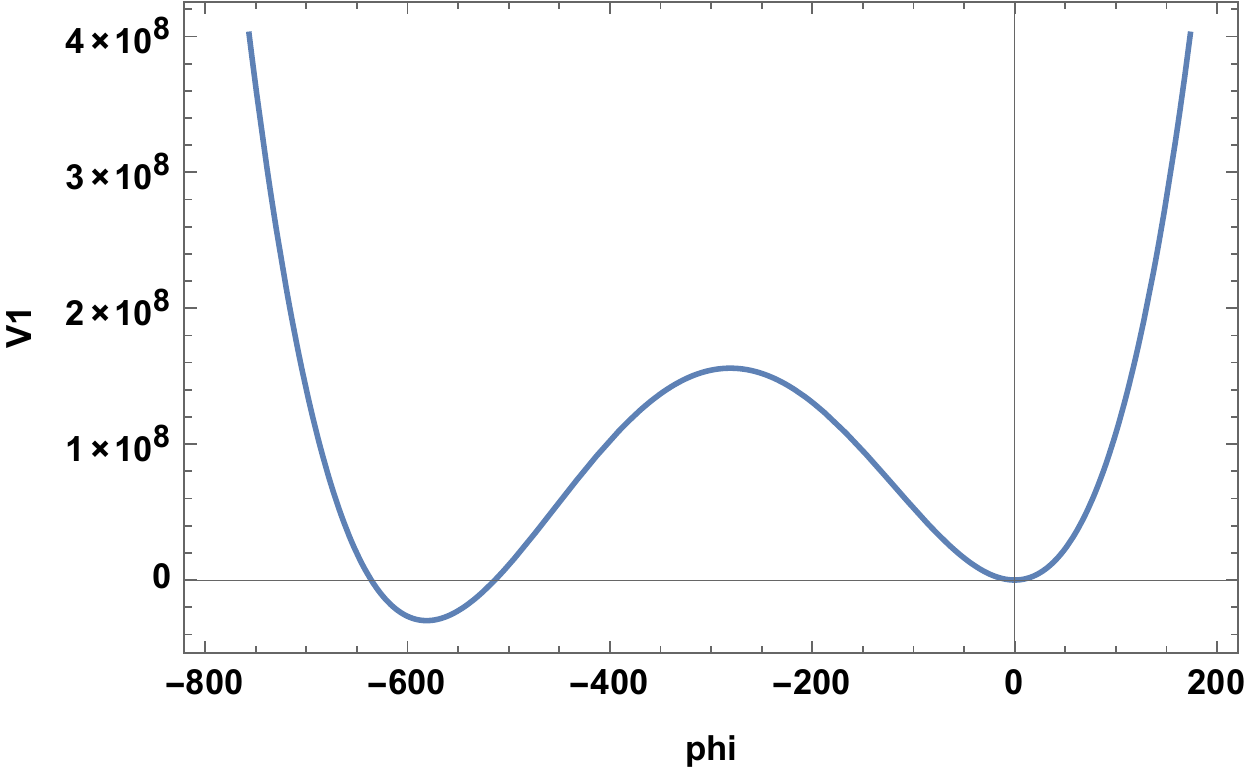}\quad\includegraphics[width=94mm]{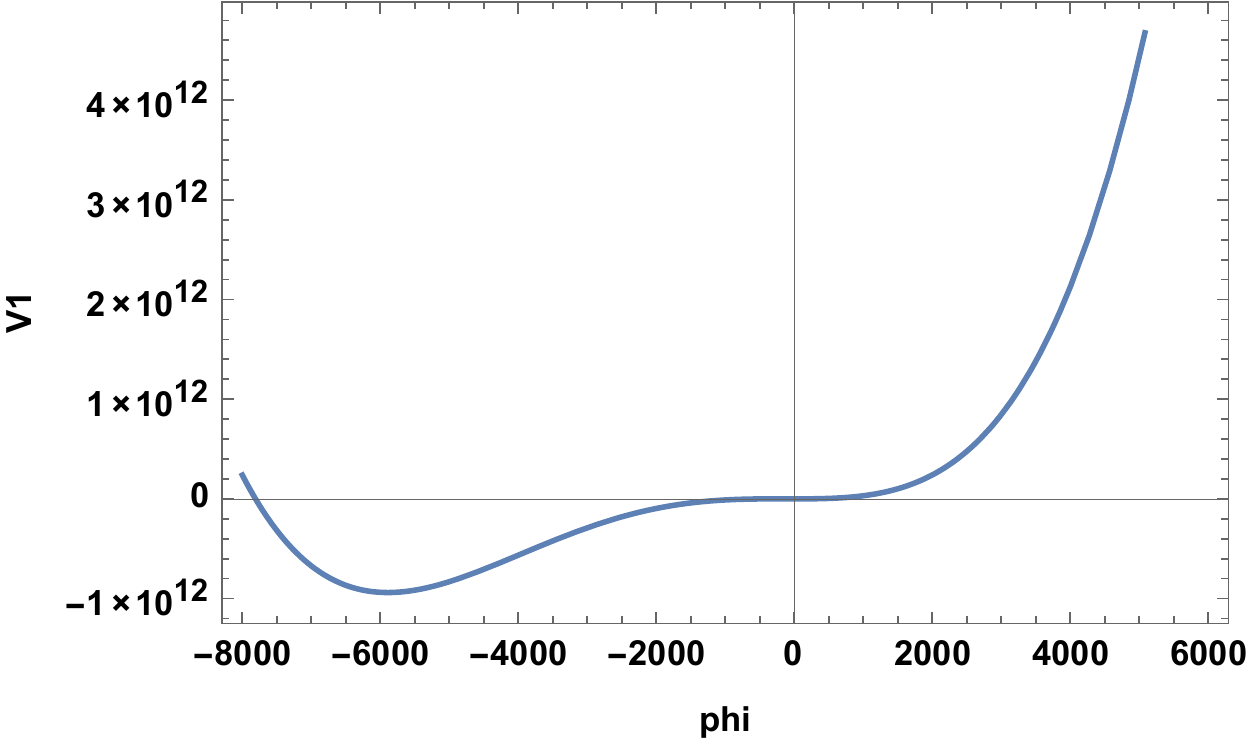}}
\caption{\small{The potential $V_1(\phi)$ at $\m\simeq 10^{40}$ GeV (left) and just below the instability scale $\m_I$, at $\m\simeq 2.0 \cdot 10^{45}$ 
 GeV (right).}}
\label{V1muInst1}
\end{figure}

From now on we concentrate on $V_1(\phi)$ in \eq{oneloopV} and we analyze it numerically.
First, in Fig. \ref{V1mu125} we plot it at the physical scale $\m=125$ GeV. 
We observe that it has two minima, of which the one at $\phi=0$ is the global minimum, as claimed. 
The tilt in the mexican hat shape implies that the global $Z_2$ symmetry $H_0\to -H_0$ of the classical potential
has been spontaneously broken by quantum effects. The breaking is small and the vacuum in the interior of the Higgs phase is stable.
Going to higher scales, we observe a big desert. 
\begin{figure}[t]
\centerline{\includegraphics[width=90mm]{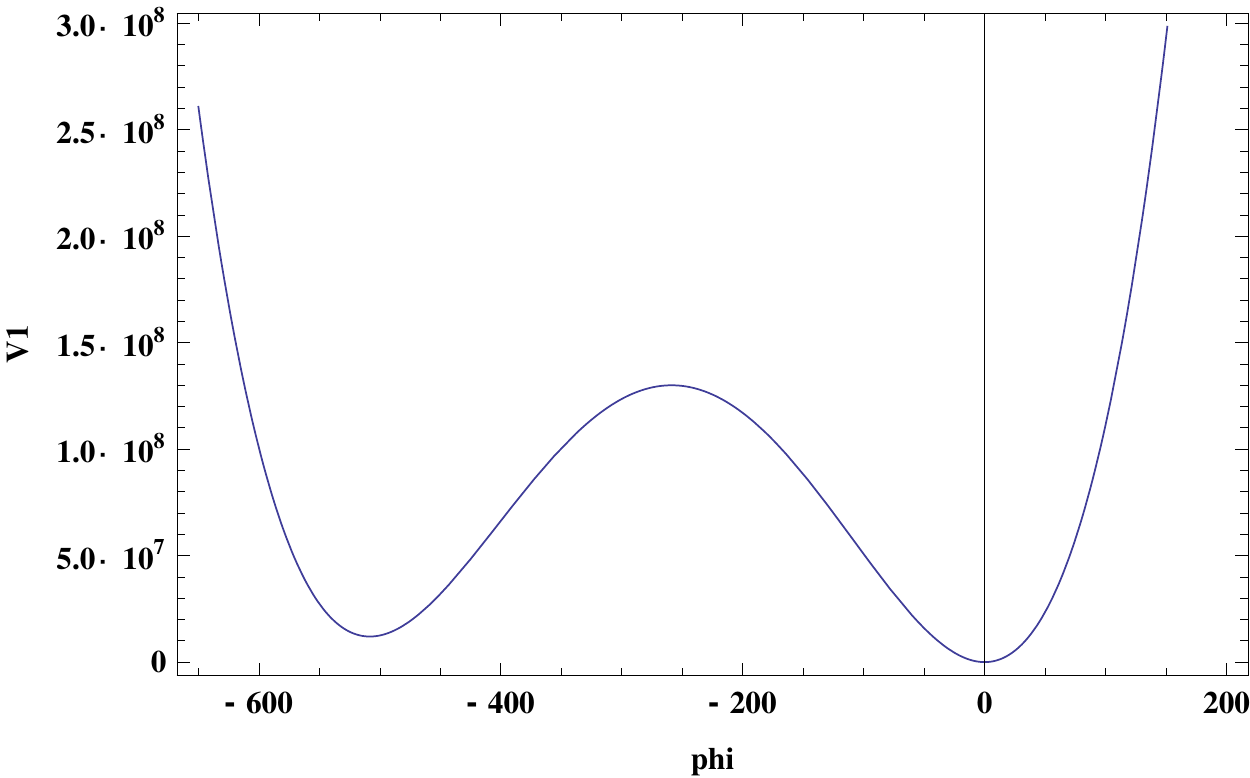}\quad\includegraphics[width=90mm]{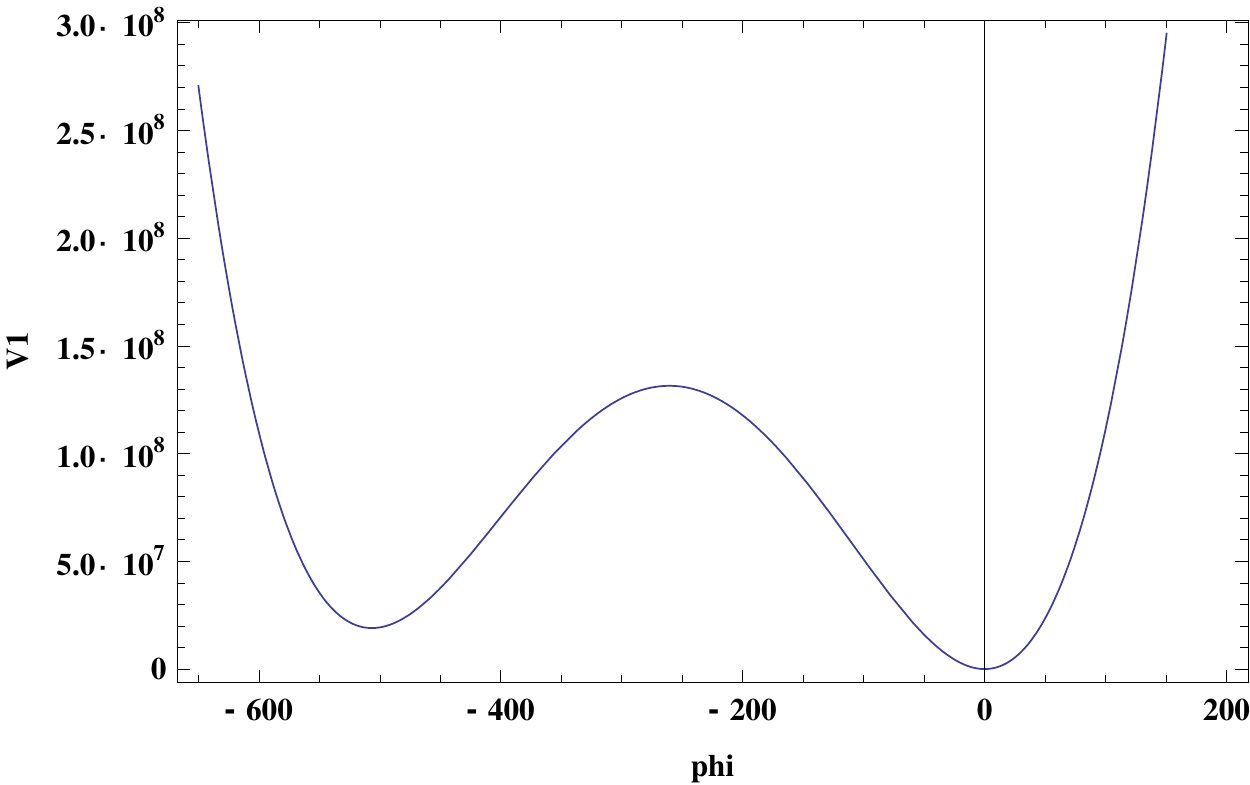}}
\caption{\small{The 1-loop Higgs potential $V_1(\phi)$ at $\m=10^{12}$ GeV (left) and $\m=10^{19}$ GeV (right).}}
\label{V1mu12Planck}
\end{figure}
To illustrate this, in Fig. \ref{V1mu12Planck} we plot the potential for $\m=10^{12}$ GeV and $\m=10^{19}$ GeV respectively.
The Abelian-Higgs model does not care about 'low' intermediate scales including the Planck scale.
It roams through them perturbatively to much higher scales.
The first qualitative change observed is around $\m\simeq 10^{40}$ GeV where the local minimum at negative values of $\phi$ becomes the global one, see the left of \fig{V1muInst1}.
For the next five or so orders of magnitude in $\m$, the new global minimum becomes deeper while the local one, at $\phi=0$, becomes shallower. Nevertheless, if we zoom in near $\phi=0$ we will see that the local minimum is still there.
Just below $\m_I$ the picture of the potential remains the same, the local minimum becomes even deeper and there is no restoration of the global $Z_2$ symmetry, see right of \fig{V1muInst1}.
\begin{figure}[t]
\centerline{\includegraphics[width=110mm]{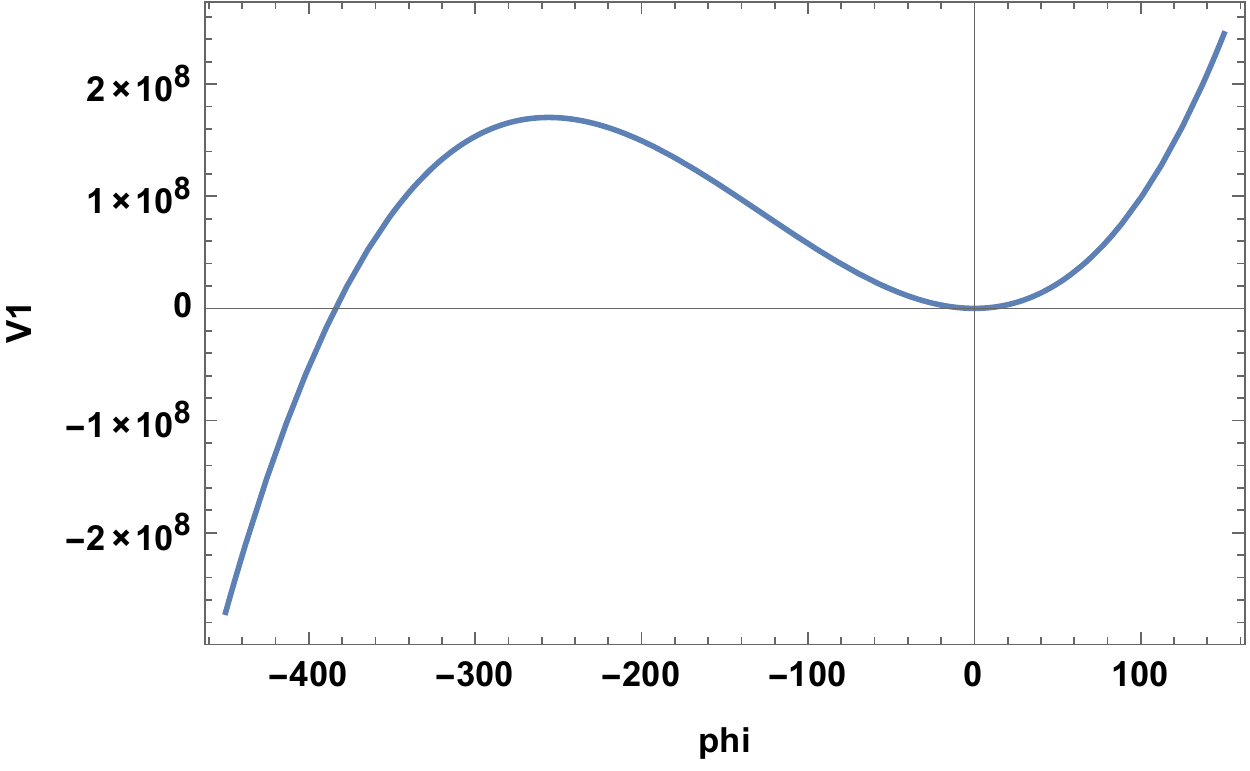}}
\caption{\small{The potential $V_1(\phi)$ just above the instability scale $\m_I$, at $\m\simeq 3.0 \cdot 10^{45}$GeV.}}
\label{V1muInst2}
\end{figure}
At this scale, $\l\simeq 1.8$, i.e. still perturbative. In Fig. \ref{V1muInst2} we plot the potential just above $\m_I$, at $3.0\cdot 10^{45}$ GeV; 
 we observe that $\m_I$ is the approximate scale where 
the potential develops an instability. 
A related question that arises is what is the scale where perturbation theory breaks down, that is which scale produces a quartic coupling of $\l\simeq 4\pi$
(according to the usual perturbative argument based on the fact that each loop introduces an additional factor of $\frac{\l^2}{16\pi^2}$).
This scale turns out to be 
\be
\m_{NP}\simeq 1.5 \cdot 10^{49}\,\, {\rm GeV}\, ,
\ee
a scale slightly larger than the instability scale and remarkably close to the Landau pole of $\l$, $\m_L^\l \simeq 4.3 \cdot 10^{49}$ GeV.
The three orders of magnitude in $\m$ between $\m_I$ and $\m_{\rm NP}$ is just an order of magnitude or less on a logarithmic scale.
In a gauge invariant cut-off regularization such as the lattice, this is typical in the vicinity of a phase transition 
and can be achieved by a moderate change of the bare couplings.
We could say that an interval of scale evolution, long in perturbative time can be short in (cut-off) non-perturbative time.
More on this in the next section.

\begin{figure}[t]
\centerline{\includegraphics[width=90mm]{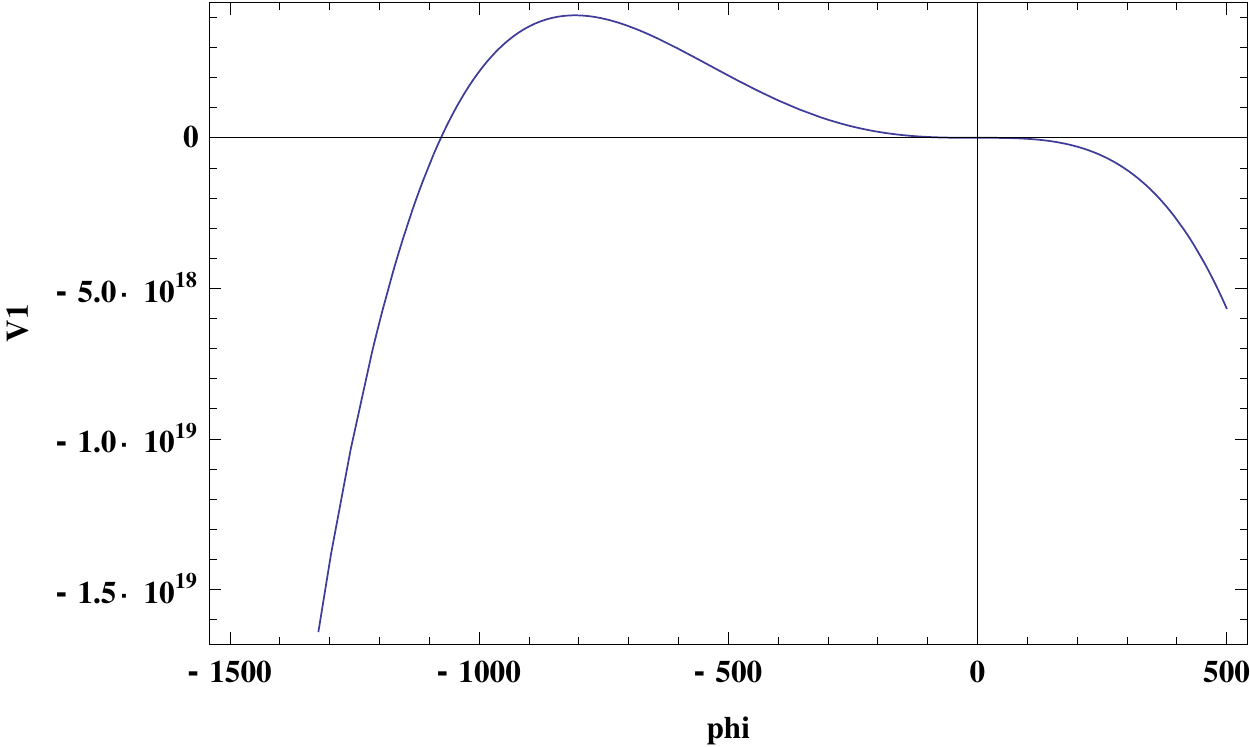}\quad\includegraphics[width=90mm]{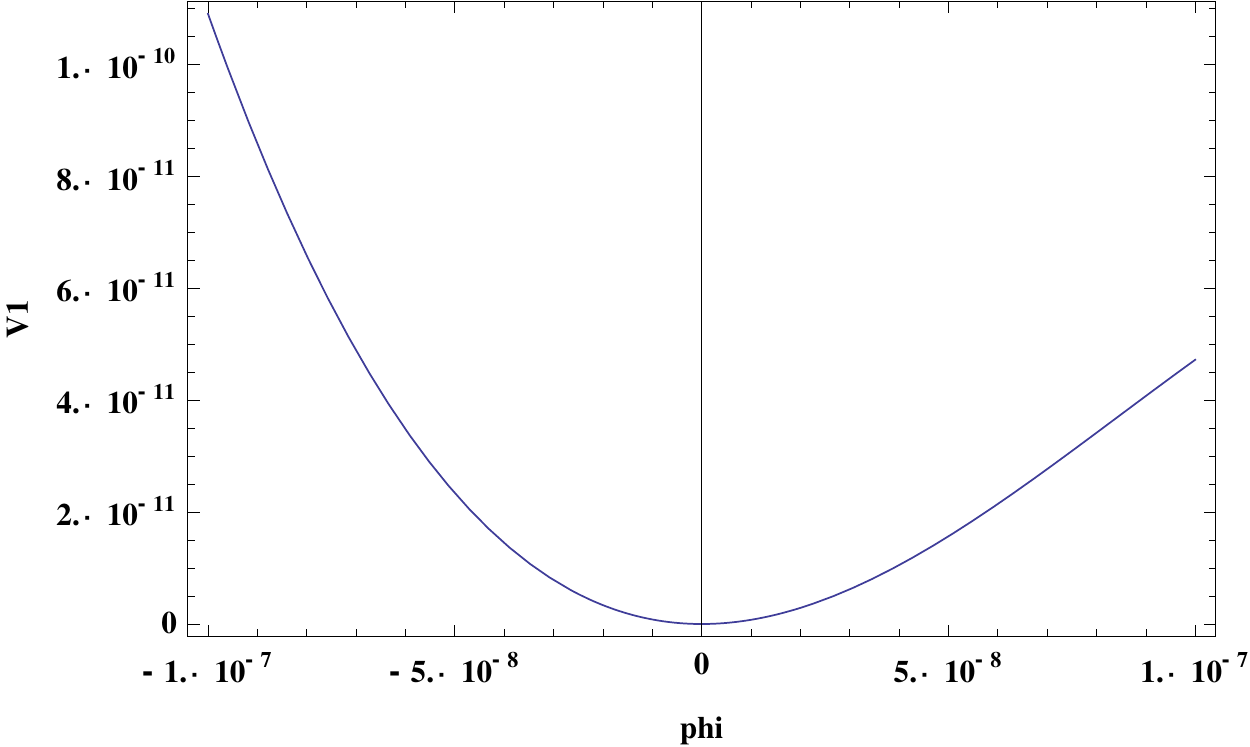}}
\caption{\small{The potential $V_1(\phi)$ near the Landau pole $\m_L^\l$ (left) and its $\phi=0$ regime zoomed in (right). }}
\label{V1muLandau}
\end{figure}

In Fig. \ref{V1muLandau} we plot once more the potential in the vicinity of the Landau pole.
On the left, we show the global form of the potential, where we can see the flattening of the $\phi=0$ regime.
On the right, we zoom in the $\phi=0$ regime and observe that the local minimum is still there, even though
the closer one gets to the Landau pole the more one must zoom in to see it, as the flattening gets closer to forming a saddle point.
Note that the gauge coupling $g$ changes very little during these scale changes thus remains perturbative.
This means that beyond the Landau pole where $\l$ turns negative lays the Coulomb phase. 

The physics that we extract from the numerical analysis is that the Abelian-Higgs model remains perturbative 
almost all the way to the Landau pole $\m_L^\l$ of $\l$.
At a scale approximately $\m_I \simeq 7 \cdot 10^{-4}\, \m_L^\l$ an instability develops, the onset of the system trying
to pass from the Higgs phase into the Coulomb phase.  
Above $\m_I$ there are two vacua, one is the old global minimum that has turned into a local minimum and the
unstable vacuum, corresponding to the Coulomb phase.
Up to the scale of $0.35\, \m_L^\l$ the system is trying perturbatively to stabilize in the Coulomb phase and this
is finally achieved at the Landau pole where in the last few tiny scale seconds, approximately between $[0.35-1.00]\cdot\m_L^\l$,
non-perturbative evolution takes over. The breadth of the unstable regime reflects perhaps the fact that the
phase transition between the Higgs and the Coulomb phase is a strong first order phase transition \cite{AHlattice}.
It is likely that in a Monte Carlo simulation, as the phase transition is crossed, 
one would observe the system tunelling back and forth between the two vacua. 

\section{Lines of Constant Physics}
\label{LCP}

The following section has a more speculative character in comparison to the previous ones in an attempt to make a connection
to non-perturbative properties of the Abelian-Higgs model.
In a non-perturbative approach one of the first thing one does is to map the phase diagram of a model via Monte Carlo simulations.
For the Abelian-Higgs model the phase diagram is defined by three axes in the space of the three bare parameters $g_0$, $\l_0$ and $m_0$.
For instance in a lattice regularization, the three axes can be chosen to be 
\be\label{axes}
\b\equiv \frac{1}{g_0^2},\hskip 1cm \kappa \equiv \frac{1-2\l_0}{1+\frac{1}{2} a^2 m_{H_0}^2}, \hskip 1cm \l_0\, ,
\ee 
with $a$ the lattice spacing. The $\b$ above should not be confused with a beta-function, it is the conventional notation for a lattice gauge coupling.
The mapping of the phase diagram has been done for the Abelian-Higgs model to some extent in the past \cite{AHlattice} and more recently,
for $\l_0=0.15$, in \cite{Sheida}. It seems to depend weakly on $\l_0$. Here we reproduce it (semi) qualitatively:
\begin{center}
\begin{tikzpicture}
\draw[help lines] (0,0) grid (10,6);
\node at (0.1,-0.25) {0};
\node at (-0.20,0.20) {0};
\node at (5,-0.25) {1};
\node at (5,-0.75) {$\b$};
\node at (10,-0.25) {2};
\node at (-0.50,3) {0.15};
\node at (-1.25,3) {$\kappa$};
\node at (-0.50,6) {0.30};
\draw [cyan,thick] (5.05, 0.0) -- (5.05,3.00) -- (4.75,3.30) -- (4.55,3.80) -- (2.75,4.80);
\draw [cyan,thick] (4.55, 3.80) -- (5.50,3.35) -- (10.00,3.20);
\node at (7.5,4.5) {Higgs};
\node at (7.5,1.5) {Coulomb};
\node at (1.5,2.5) {Confined};
\end{tikzpicture}
\end{center}
We also note that the above phase diagram has been constructed using $\kappa \equiv \frac{1-2\l_0}{8+a^2 m_{0}^2}$, a slightly different normalization for $\kappa$ in \eq{axes}. 
This just corresponds to choosing a different normalization for the bare vev parameter $v_0$.
Here we will use the normalization of \eq{axes} and $v_0=246$ GeV.

A Line of Constant Physics (LCP) of the AH model is defined by the curve on its phase diagram all of whose points satisfy
the constraint that $m_H$, $m_Z$ and $\l$ are some chosen constants. We can choose for example
\be
m_H = 125\, {\rm GeV}, \hskip 1cm m_Z = 91\, {\rm GeV}, \hskip 1cm \l=0.12\, ,
\label{LCPparams}
\ee
and then the points of the LCP can be defined perturbatively to be determined by the equations
\bea
m_{H_0}^2(\m) &=& 125^2 + (\d m_{H})_f (m_H=125,m_Z=91,\l=0.12; \m)\nonumber\\
m_{Z_0}^2(\m) &=& 91^2 + (\d m_{Z})_f (m_H=125,m_Z=91,\l=0.12; \m)\nonumber\\
\l_0(\m) &=& 0.12 + (\d \l)_f (m_H=125,m_Z=91,\l=0.12; \m)
\label{LCPeqs}
\eea
where
\bea
(\d m_{H})_f (m_H=125,m_Z=91,\l=0.12; \m) &=& -19557.3 + 5625 \ln \frac{\m^2}{15625} - 5962.32 \ln \frac{\m^2}{8281} \nonumber\\
(\d \l)_f (m_H=125,m_Z=91,\l=0.12; \m) &=& -0.0305 + 0.01296 \ln \frac{\m^2}{15625} - 0.043 \ln \frac{\m^2}{8281} \nonumber\\
(\d m_{Z})_f (m_H=125,m_Z=91,\l=0.12; \m) &=& 2537.18 + 6723.0.4 \ln \frac{\m^2}{15625} - 52.0632 \ln \frac{\m^2}{8281}\nonumber\\
\eea
are the finite parts of the Unitary gauge counter-terms, according to \eq{soldm}, \eq{soldl} and \eq{soldZ}. 
We recall that $m_{H_0} = 2 m_0$, $m_{Z_0}=g_0 v_0$ and correspondingly $\d m_H = 2 \d m$, $\d m_Z = v_0 \d g$.

We now have all the ingredients to plot a first "perturbative LCP".
\begin{figure}[t]
\centerline{\includegraphics[width=120mm]{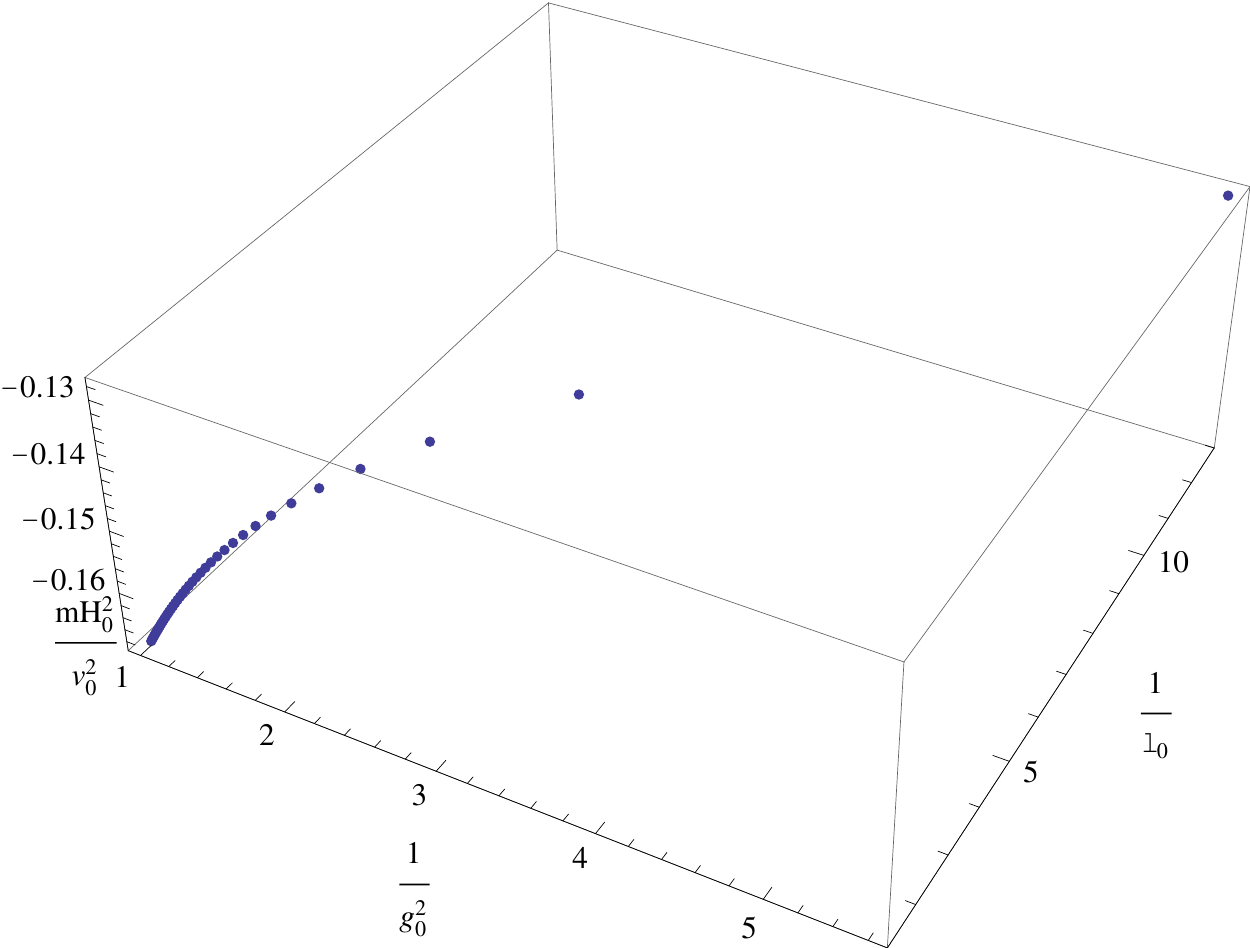}}
\caption{\small{The perturbative LCP defined by \eq{LCPparams} for $v_0=246$ GeV.}}
\label{LCPpert}
\end{figure}
In \fig{LCPpert} we plot the LCP, defined by \eq{LCPeqs} and \eq{LCPparams} with $v_0=246$ GeV, for the range
$\m \in [125, 5000]$ GeV. The part (lower left) of the line where the points are denser corresponds to larger values of $\m$.
There, the parameter $\b=1/g_0^2$ (that decreases quite rapidly as $\m$ increases) approaches 1, which means that 
most likely we are about to enter into either the Confined or the Coulomb phase.
There is potentially a non-trivial message in this so it is worth redoing it this time using \eq{axes}, i.e. to try to actually put the perturbative LCP
on the non-perturbative phase diagram. 
The only obstacle is to relate the lattice spacing to some scale parameter in the perturbative calculation.
This can be quite an involved operation, depending on our desired level of precision.
For a discussion of this issue, see \cite{LuscherWeisz}.
Here we will simplify the discussion and make the naive identification $a=1/\Lambda$ with $\Lambda$ a momentum cut-off.
\begin{figure}[t]
\centerline{\includegraphics[width=90mm]{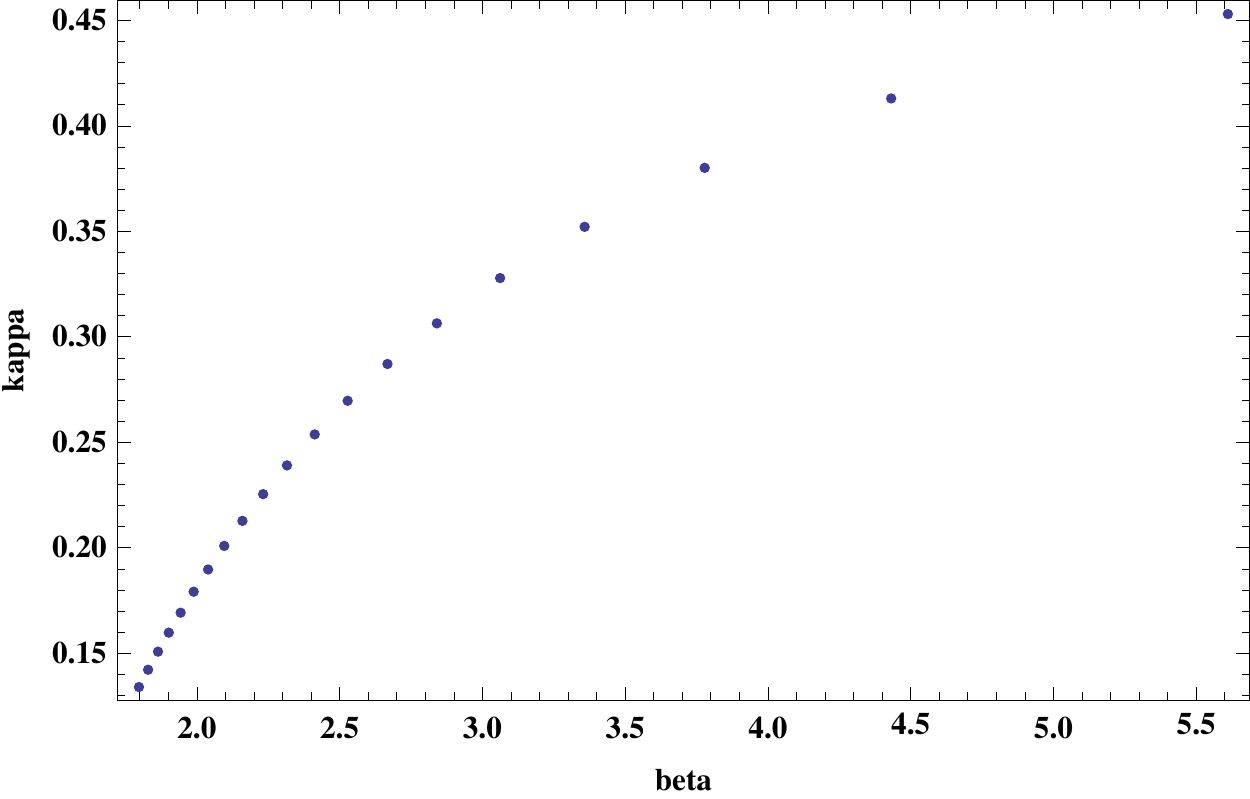}\quad\includegraphics[width=90mm]{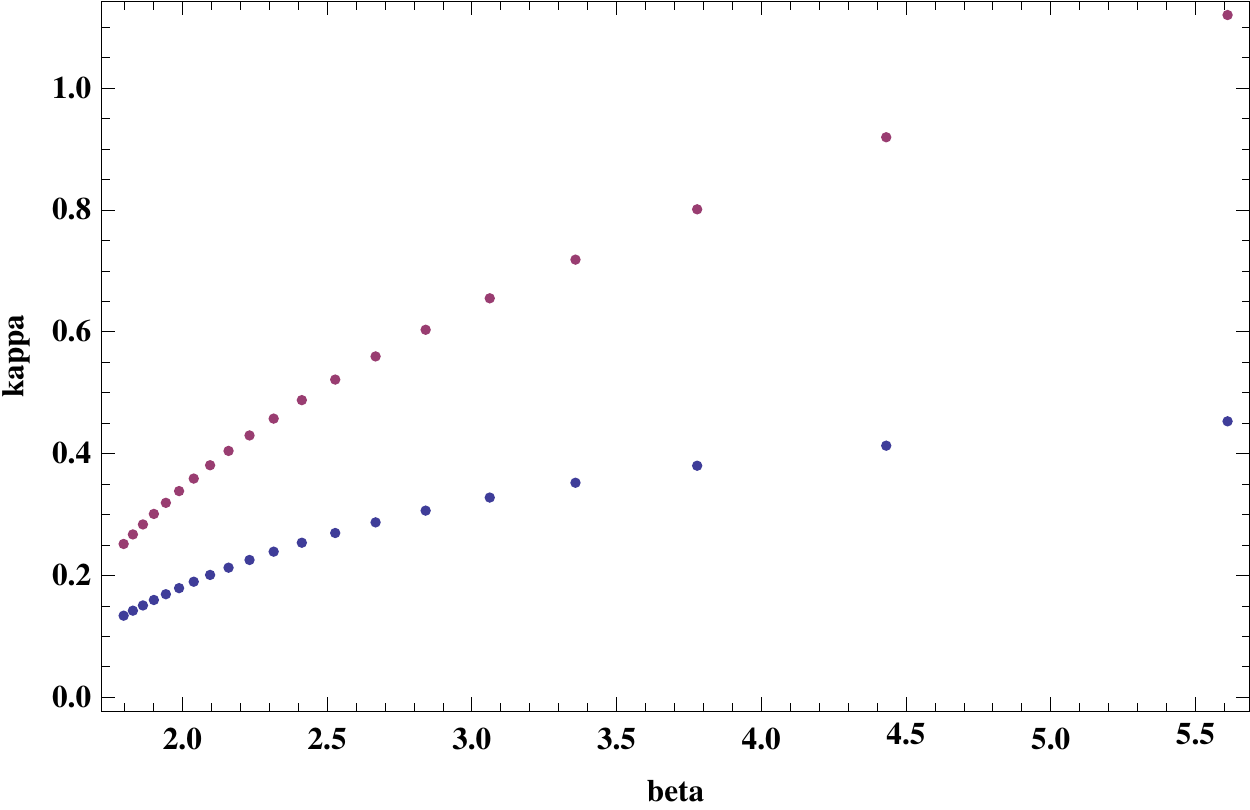}}
\caption{\small{The cut-off LCP that identifies $\Lambda=1/a$ for $v_0=246$ GeV projected on the $\b-\kappa$ plane (left) and a comparison with the corresponding LCP 
with the identification $\m=1/a$ (right). The lower curve is the cut-off LCP.}}
\label{LCPnonpert}
\end{figure}
\begin{table}
\begin{center}
\begin{tabular}{c|c|c|c|c}
$\Lambda$ & $\b$ & $\kappa$ & $\l_0$ & $q$ \\
\hline
125 & 5.6 & 0.45 & 0.07 & 1.7 \\
250 & 3.0 & 0.32 & 0.19 & 3.8 \\
375 & 2.4 & 0.25 & 0.26 & 7.3 \\
500 & 2.0 & 0.2 & 0.31 & 12.1 \\
625 & 1.9 & 0.15 & 0.35 & 18.3\\
750 & 1.7 & 0.12 & 0.38 & 25.9\\
\end{tabular}
\end{center}
\caption{Scale and corresponding bare parameters of the LCP defined by
$m_Z = 91$ GeV, $m_H=125$ GeV and $\l=0.12$ with the lattice spacing identified as $a=1/\Lambda$.}
\label{table_LCPparams1}
\end{table}
We need to do some extra work though, as we have to estimate 
$\d m_H$ in a cut-off regularization. Fortunately this is not too hard, it yields to leading order
\be
m_{H_0}^2(\Lambda) = 125^2 + \frac{6\l}{16\pi^2} \Lambda^2 + \cdots
\ee
that should replace the first of \eq{LCPeqs} in \eq{axes}. The dots stand for small corrections.
For $m_Z$ and $\l$ on the other hand that have a logarithmic
cut-off dependence we keep the same relations as before. This is not a completely well defined identification but it is good enough for the information we want to extract.
We call the resulting line on the $\b-\kappa$ plane, the "cut-off LCP".
On the left of \fig{LCPnonpert} we plot the cut-off LCP defined by \eq{LCPparams} for $v_0=246$ GeV and
projecting the $\l_0$ dependence on the $\b-\kappa$ plane. The values of the cut-off range in $\Lambda \in [125,700]$ GeV
and they increase as we move down and left along the curve.
By comparing to the Monte Carlo phase diagram we can now clearly see  that around the upper limit of the $\Lambda$-range
the system hits the Higgs-Coulomb phase transition.
Some representative values of the parameters can be seen on Table \ref{table_LCPparams1}.
On the right of \fig{LCPnonpert} we compare the cut-off LCP to the one we would have obtained by the identification $\m=1/a$.
Some representative values of the parameters for this identification can be seen on Table \ref{table_LCPparams2}.
\begin{table}
\begin{center}
\begin{tabular}{c|c|c|c|c}
$\m$ & $\b$ & $\kappa$ & $\l_0$ & $q$ \\
\hline
125 & 5.6 & 1.11 & 0.07 & 0\\
250 & 3.0 & 0.65 & 0.19 & 0.0014\\
375 & 2.4 & 0.48 & 0.26 & 0.0023\\
500 & 2.0 & 0.38 & 0.31 & 0.0029\\
625 & 1.9 & 0.30 & 0.35 & 0.0033\\
750 & 1.7 & 0.23 & 0.38 & 0.0037\\
\end{tabular}
\end{center}
\caption{Scale and corresponding bare parameters of the LCP defined by
$m_Z = 91$ GeV, $m_H=125$ GeV and $\l=0.12$ with the lattice spacing identified as $a=1/\m$.}
\label{table_LCPparams2}
\end{table}
The numbers change but the conclusion remains: If we want to construct an LCP that keeps the values of the physical quantities
close to their "experimentally measured values" (we are not in the Standard Model but the point we would like to make should be clear), 
we can not take the cut-off too high, since around 1 TeV or so
we are forced to cross into the Coulomb phase. Notice that the analysis of the RG trajectories and of the potential in the previous section did not warn us
about this. This has potentially consequences for the fine tuning of the Higgs mass. In Tables \ref{table_LCPparams1} and \ref{table_LCPparams2} 
we introduced the unsophisticated fine tuning parameter
\be
q=\frac{\left|(\d m_H)_f\right|}{m_H}
\ee
from which we can see the increase of fine tuning as the scale increases, with the $\Lambda=1/a$ identification naturally worse.
Even so though, since the cut-off stops at a maximum value of about 1 TeV, it can not take very large values.

\section{Conclusions}
\label{Conclusions}

We carried out the renormalization of the Higgs potential and of the $Z$-mass in the Abelian-Higgs model
in the $R_\xi$ and Unitary gauges. The fact that this is a sector of the action not containing derivatives, suggested
us to renormalize at zero external momenta. The renormalization conditions we used imply subtractions that are necessarily different than ${\overline {\rm MS}}$.
In this context, we showed that these two gauges are completely equivalent both at the level of the $\b$-functions
and at the level of the finite remnants after renormalization, that determine the one-loop Higgs potential.
Moreover, we showed that after the renormalization procedure we obtain automatically the gauge independence of the potential.
The cancellation of the gauge fixing parameter $\xi$ from the scalar potential
obtained via Feynman diagram calculations in an $R_\xi$ gauge fixing scheme is intriguing as the 
corresponding effective potential obtained via the background field method is manifestly $\xi$-dependent.
Its equivalence to the Unitary gauge on the other hand could open a new direction in Standard Model calculations,
as the number of diagrams in the Unitary gauge is significantly smaller. 
Apart however from simplifying already known results, computations such as done here could also render currently ambiguous results
based on the dynamics of the scalar potential away from its extrema, more robust. 
Since this work is just a demonstration of gauge independence of the Higgs potential by construction
 it would be certainly worth trying to construct some sort of a more formal 'proof'.

We have analyzed numerically the Higgs potential and saw that the system generically behaves perturbatively until quite close to the Landau pole of $\l$.
It becomes metastable already in the perturbative regime and only during its last steps towards the phase transition into the Coulomb phase
turns non-perturbative. Finally, we constructed a couple of Lines of Constant Physics in the Unitary gauge and placed them on the Abelian-Higgs phase diagram
in order to enhance the perturbative analysis by some non-perturbative input.
This suggested us that in a Higgs-scalar system the cut-off can not be generically driven to the highest possible scales that the 
RG flow or the stability of the potential allows, if we want to keep the physics constant. This is because it is possible that way before those scales
are reached, a phase transition may be encountered.

Regarding generalizations of this work, simple bosonic extensions should work in a similar way, without any surprises.
In the presence of fermions some extra care may be needed to handle the $U$-integrals but this is expected to be
a straightforward extra technical step. 
In this case though it is also expected that a new instability will appear at some intermediate scale
and it would be interesting to work out in our framework how this modifies the bosonic system.
Finally, if the observed consistency between the two gauges breaks down at higher loops
then there is something about the quantum internal structure of spontaneously broken gauge theories that needs to be understood better.

{\bf Acknowledgements.}

We would like to thank C. Corian\'o, A. Kehagias, A. Mariano and C. Marzo for discussions.
F. K. thanks C. Corian\'o and the U. del Salento, Lecce for warm hospitality during the initial stage of this work. 
N. I. would like to thank F. Knechtli for discussions on the lattice AH model.

\begin{appendices}

\section{One-loop integrals}
\label{FiniteIntegrals}

The general form of a Feynman diagram ${F}^{G,L}_{E}$ (see \eq{general FD} in the text for the explanation of the notation) is
\bea
(4\pi)^{d/2} {F}^{G,L}_{E} = \m^\ve \left([ F^{G,L}_{E} ]_{\ve} + \left\{F^{G,L}_{E} \right\}_{\ve} +[ F^{G,L}_{E} ]_{f} + \left\{ F^{G,L}_{E} \right\}_{f}\right) \, .
\eea
Here we are interested in the finite part of this expression, the general form of which is
\be\label{genfin1}
\frac{1}{(4\pi)^{d/2}} \left(F^{G,L}_{E}\right)_f = \frac{1}{(4\pi)^{d/2}} \left( [F^{G,L}_{E} ]_{f} + \left\{ F^{G,L}_{E} \right\}_{f}\right) = 
\int \frac{d^d k}{i(4\pi)^{d/2}} \sum_{n}\frac{\left(N_{F^{G,L}_{E}}^{(n)}\right)_f(k)}{D_1D_2\cdots D_n}\, ,
\ee
where the range of the sum over $n$ depends on the diagram. The $n$'th term of the numerator has a denominator with $n$ factors,
yields a finite result after integration over the loop momentum $k$ and it is denoted as $\left(N_{F^{G,L}_{E}}^{(n)}\right)_f(k)$. It is also diagram dependent.
In the Unitary gauge, even though this notation is redundant, we will still use it for extra clarity.
The denominators are defined as
\be\label{denom.}
D_i(P_{i-1},m_i) = (k+P_{i-1})^2 - m_i^2, \hskip 1cm i=1,\cdots, n
\ee
and $P_0\equiv 0$. Notice that the proper argument of the denominators is $D_i(P_{i-1},m_i)$. Nevertheless, we do not always show both arguments systematically in the main text, 
apart from cases where otherwise their absence could cause ambiguities. Such a case is some finite diagrams, which generally have more than four denominators.
In other cases where there can be no confusion, we show none, or only one of the arguments.

Feynman parametrization is implemented using
\be\label{Feynman parameters n}
\frac{1}{{{D_1}{D_2} \cdot  \cdot  \cdot {D_n}}} = \left( {n - 1} \right)! \int_0^1 {\cal D}^n x \frac{1}{{{{\left( {{D_1}{x_1} + {D_2}{x_2} +  \cdot  \cdot  \cdot  + {D_n}{x_n}} \right)}^n}}}\, ,
\ee
where we defined
\be
 \int_{0}^{1} dx_1\cdots \int_0^1 dx_n \, \d(1-\sum_{i=1}^{n} x_i) \equiv \int_0^1 {\cal D}^n x\, .
\ee
Passing now to Dimensional Regularization, we introduce the useful shorthand notation
\be
\int {\frac{{{d^d}k}}{{{{i \left( {4\pi } \right)}^{d/2}}}}} (\cdots) \equiv \left\langle \cdots \right\rangle \, .
\ee
We can then express \eq{genfin1} in DR as
\be
\frac{1}{(4\pi)^{d/2}} \left(F^{G,L}_{E}\right)_f = \sum_n \Gamma(n) \int_0^1 {\cal D}^n x 
\left\langle\frac{\left(N_{F^{G,L}_{E}}^{(n)}\right)_f(k\to k - \sum_{i=1}^{n-1}P_i x_{i+1} )}{\left(k^2 - \Delta_{F_{0,n}}\right)^n}\right\rangle\, ,
\ee
with
\be
\Delta_{F_{0,n}}(x_1,\cdots ,x_n) = - \sum_{i=1}^{n-1} P_i^2 x_{i+1} + \left( \sum_{i=1}^{n-1} P_i x_{i+1} \right)^2 +  \sum_{i=1}^{n} x_i m_i^2\, .
\ee
In a few simple cases we use the conventional notation
\bea
\Delta_{F_{0,2}} \equiv \Delta_{B_0}\nonumber\\
\Delta_{F_{0,3}} \equiv \Delta_{C_0}\nonumber\\
\Delta_{F_{0,4}} \equiv \Delta_{D_0}
\eea

\section{Passarino -Veltman}
\label{PassarinoVeltman}

In this Appendix we collect some standard integrals of the Passarino-Veltman type, encountered in the text.

\subsection{Scalars}

The simplest PV integral is the scalar tadpole integral
\be
A_0(m)=\int {\frac{{{d^4}k}}{{{{i \left( {2\pi } \right)}^4}}}\frac{1}{{{k^2} - m^2}}}\, ,
\ee
naively quadratically divergent with a cut-off. In Dimensional Regularization, it can be expressed as (attaching the factor $\m^\ve$)
\be
\m^\ve A_0(m) = -\frac{\m^{\ve}}{(4\pi)^{d/2}} \Gamma \left(1-\frac{d}{2} \right) \frac{1}{m^{1-d/2}}\, 
\ee
and by expanding in $\ve$ finally as
\bea\label{A0ma}
\m^\ve  A_0(m)&=& \frac{1}{(4 \pi)^2} m^2 \left(  \frac{2}{\varepsilon} + \ln \frac{\m^2}{m^2} +1 \right) \, . 
\eea
The scalar, naively logarithmically diveregent PV integral is
\be
B_0(P_1,m_1,m_2)=\int \frac{d^4k}{i \left(2\pi\right)^4}\frac{1}{D_1D_2}\, .
\ee
It can be computed explicitly using the formulation of Appendix \ref{FiniteIntegrals}. It corresponds to $n=2$ and a numerator equal to 1.
In $d$-dimensions, this integral is
\bea\label{B0m1m2}
B_0(P_1,m_1,m_2)&=& \int\limits_0^1 {\cal D}^2 x \left\langle \frac{1}{( k^2 - \Delta_{B_0})^2} \right\rangle \nonumber \\
&=& \frac{1}{(4\pi)^{d/2}}\int\limits_0^1 {dx}{\Gamma \left( {2 - \frac{d}{2}} \right){{\left( {\frac{1}{\Delta_{B_0} }} \right)}^{2 - {d \mathord{\left/
 {\vphantom {d 2}} \right. \kern-\nulldelimiterspace} 2}}}}\, .
\eea
Expanding in $\ve$ we get
\bea
B_0(P_1,m_1,m_2)&=& \frac{1}{(4\pi)^{2}}\left( {\frac{2}{\varepsilon } + \int\limits_0^1 {dx} \ln \frac{{4\pi {e^{ - {\gamma _E}}}}}{\Delta_{B_0} }} \right) \, 
\eea
and finally
\be\label{B0ma}
\m^\ve B_0(P_1,m_1,m_2) = \frac{1}{(4\pi)^{2}}\left( {\frac{2}{\varepsilon } + \int\limits_0^1 {dx} \ln \frac{\m^2}{\Delta_{B_0} }} \right)\, .
\ee
We note that this integral is symmetric under $m_1\leftrightarrow m_2$.
An important special case is when $P_1=0$ in which case
\be\label{B0p0}
 B^1_0(m_1,m_2) \equiv B_0(P_1=0,m_1,m_2)  = \frac{A_0(m_1) - A_0(m_2)}{m_1^2 - m_2^2} \, .
\ee
The finite scalar integral that appears in Triangles and Boxes is of the form
\bea\label{U1Th}
C_0(P_1,P_2,m_1,m_2,m_3) = \int {\frac{{{d^4}k}}{{{{i \left( {2\pi } \right)}^4}}}} \frac{{ 1{}}}{D_1 D_2 D_3}\, .
\eea
In DR, this integral becomes
\bea\label{U1Tha}
C_0(P_1,P_2,m_1,m_2,m_3) &=& 2 \int_0^1 {\cal D}^3 x \left\langle \frac{1}{\left(k^2 - \Delta_{C_0} \right)^3} \right\rangle\, .
\eea 
Expanding in $\ve$ it reduces to
\bea\label{c0m123}
\m^\ve C_0(P_1,P_2,m_1,m_2,m_3)&=& - \frac{\m^\ve}{(4\pi)^{d/2}} \int_0^1 {\cal D}^3 x {{{ {\frac{\Gamma \left( {3 - \frac{d}{2}} \right)}{\Delta_{C_0}^{3-d/2} }}}}} \nonumber \\
 &=& - \frac{\m^\ve}{16\pi^2} \int_0^1 {\cal D}^3 x \frac{1}{\Delta_{C_0}}\, .
\eea
We also define some special cases of $C_0$ integrals that appear in the main text in various places:
\bea\label{C012ab}
C_0^1(p,m_{Z_0},m_{Z_0},m_{\chi_0}) &\equiv&C_0 (P_1=p,P_2=0,m_{Z_0},m_{Z_0},m_{\chi_0}) \nonumber\\
C_0^2(p,m_{Z_0},m_{Z_0},m_{\chi_0}) &\equiv& C_0(P_1=p,P_2=p,m_{Z_0},m_{Z_0},m_{\chi_0}) \nonumber\\
C_0^3(p,m_{Z_0},m_{\chi_0},m_{\chi_0}) &\equiv&  C_0(P_1=0,P_2=p,m_{Z_0},m_{\chi_0},m_{\chi_0})  \, . 
\eea
The last scalar is the finite Box integral
\be\label{U0Box}
D_0(P_1,P_2,P_3,m_1,m_2,m_3,m_4) = \int {\frac{{{d^4}k}}{{{{i \left( {2\pi } \right)}^4}}}} \frac{{ 1{}}}{D_1D_2D_3D_4}\, . 
\ee
The integral in $d$-dimensions becomes
\bea
\m^\ve D_0(P_1,P_2,P_3,m_1,m_2,m_3,m_4) &=&6 \m^\ve \int_0^1 {\cal D}^4 x \left\langle \frac{1}{\left( k^2 - \Delta_{D_0} \right)^4}\right\rangle\nonumber\\
&=& \frac{\m^\ve}{(4\pi)^{d/2}}  \int_0^1 {\cal D}^4 x {{{ {\frac{\Gamma \left( {4 - \frac{d}{2}} \right)}{\Delta_{D_0}^{4-d/2} }}}}}  \nonumber \\
&=&  \frac{\m^\ve}{16\pi^2}  \int_0^1 {\cal D}^4 x  \frac{1}{{\Delta}_{D_0}^2 }\, .
\eea 

\subsection{Tensors}
\label{TensorPV}

Standard tensor PV integrals can be algebraically reduced to scalar integrals.
Actually in one-loop diagrams only contractions of tensors with the metric and external momenta occur.
The tadpole integral has no standard PV extension. It has an extension of the $U$-type, naively
quartically divergent with a cut-off and it will be computed in Appendix \ref{Uint}.

Let us introduce some shorthand notation. First, define for any $P$:
\be
B_0(1,2) \equiv B_0(P,m_1,m_2)
\ee
and for any $P_{1}$ and $P_{2}$
\bea
B_0(1,3) &\equiv& B_0(P_2,m_1,m_3)\nonumber\\
B_0(2,3) &\equiv& B_0(P_2-P_1,m_2,m_3)\, .
\eea
The simplest $B$-tensor is the linearly divergent
\be
B^{\m}(P,m_1,m_2) = \left\langle  \frac{k^\m}{(k^2-m_1^2)((k+P)^2-m_2^2)}  \right\rangle
\ee
and it can be contracted only by a momentum
\be\label{PmBm}
P_\m B^\m(P,m_1,m_2) = \frac{1}{2} \left( f_1(P) B_0(1,2) + A_0(m_1) - A_0(m_2) \right) \, ,
\ee
where
\be
f_1(P) = m_2^2 - m_1^2 - P^2\, .
\ee
The other PV tensor extension of the $B$-type is of the form
\be
B^{\m\n} = \left\langle  \frac{k^\m k^\n}{(k^2-m_1^2)((k+P)^2-m_2^2)}  \right\rangle
\ee
and it is quadratically divergent. Its contraction with the metric is
\be\label{gmnBmn}
g_{\m\n}{B^{\mu \nu }}(P,m_1,m_2) = m_1^2 B_0(P,m_1,m_2) + A_0(m_2)
\ee
while its contraction with $P_{\m}P_{\n}$ is
\bea\label{pmpnBmn1}
P_{\m}P_{\n}{B^{\mu \nu }}(P,m_1,m_2)  &=& \frac{m_2^2- m_1^2 - P^2}{4} A_0(m_1) + \frac{m_1^2- m_2^2 + 3P^2}{4} A_0(m_2) \nonumber\\
&+&\left( \frac{m_1^4 +  m_2^4 - 2 m_1^2 m_2^2}{4} + \frac{P^2 (2 m_1^2 -  2m_2^2 +P^2)}{4}     \right) B_0(P,m_1,m_2)\, .  \nonumber\\
\eea
The simplest $C$-tensor is 
\be
C^{\m}(P_1,P_2,m_1,m_2,m_3) = \left\langle  \frac{k^\m }{D_1D_2D_3} \right\rangle \, .
\ee
It contracts either as
\be
R_1^{\left[ c \right]}\equiv P_{1\m} C^\m = \frac{1}{2} \left( f_1(P_1)C_0(P_1,P_2,m_1,m_2,m_3) + B_0(1,3) - B_0(2,3) \right)
\ee
or as
\be
R_2^{\left[ c \right]}\equiv (P_2-P_1)_\m C^\m = \frac{1}{2} \left( f_2(P_1,P_2)C_0(P_1,P_2,m_,m_2,m_3) + B_0(1,2) - B_0(1,3) \right)
\ee
where
\be
f_2(P_1,P_2) = m_3^2 - m_2^2 - P_2^2 + P_1^2\, .
\ee
The second $C$-tensor, logarithmically divergent, integral is
\be
C^{\m\n}(P_1,P_2,m_1,m_2,m_3) = \left\langle  \frac{k^\m k^\n}{D_1D_2D_3} \right\rangle\, .
\ee
Its contraction with the metric is 
\be\label{gmnCmn}
g_{\m\n} C^{\m\n} = m_1^2 C_0(P_1,P_2,m_1,m_2,m_3) + B_0(P_2-P_1,m_2,m_3)\, .
\ee
There are three different momentum contractions that can appear.
Following \cite{Kunszt}, we define the matrix 
\begin{equation*}
G_2^{ - 1} = \frac{1}{{\det {G_2}}}\left( {\begin{array}{*{20}{c}}
{{(P_2-P_1)} \cdot {(P_2-P_1)}}&{ - {P_1} \cdot {(P_2-P_1)}}\\
{ - {P_1} \cdot {(P_2-P_1)}}&{{P_1} \cdot {P_1}}
\end{array}} \right)
\end{equation*} 
and using this matrix the quantities
\be\label{system2}
\left( {\begin{array}{*{20}{c}}
{{C_1}}\\
{{C_2}}
\end{array}} \right)  = G_2^{-1}\left( {\begin{array}{*{20}{c}}
{R_1^{\left[ c \right]}}\\
{R_2^{\left[ c \right]}}
\end{array}} \right).
\ee
We also define 
\bea
B_1(1,2)\equiv  B_1(P_1,m_1,m_2) &=& \frac{1}{2P_1^2}\left( {{f_1(P_1)}{B_0}\left( {1,2} \right) + {A_0}\left( {m_1} \right) - {A_0}\left( {m_2} \right)} \right)\nonumber\\
B_1(1,3)\equiv  B_1(P_2,m_1,m_3) &=& \frac{1}{2 P_2^2}\left( {{f_1(P_2)}{B_0}\left( {1,3} \right) + {A_0}\left( {m_1} \right) - {A_0}\left( {m_3} \right)} \right)\nonumber\\
B_1(2,3)\equiv  B_1(P_2-P_1,m_2,m_3) &=& \frac{1}{2(P_2-P_1)^2}\left( {{f_1(P_2-P_1)}{B_0}\left( {2,3} \right) + {A_0}\left( {m_2} \right) - {A_0}\left( {m_3} \right)} \right)\nonumber\\
{C_{00}}(1,2,3) &=& \frac{1}{{2\left( {d - 2} \right)}}\left( {2m_1^2{C_0} - {f_2(P_2,P_1)}{C_2} - {f_1(P_1)}{C_1} + {B_0}\left( {2,3} \right)} \right)\nonumber\\
\eea
out of which we construct the four quantities
\bea
R^{[c_1]}_{1}& =& \frac{1}{2}\left( f_1(P_1)C_1+ B_1\left( 1,3 \right) + B_0\left( 2,3 \right) - 2C_{00}(1,2,3) \right) \nonumber\\
R^{[c_1]}_{2}& =&\frac{1}{2}\left( f_2(P_2,P_1)C_1  + B_1\left( 1,2 \right) - B_1\left( 1,3 \right) \right) \nonumber\\
R^{[c_2]}_{1}& =& \ \frac{1}{2}\left( f_1(P_1)C_2 + B_1\left(1,3 \right) - B_1\left( 2,3 \right) \right) \nonumber\\
R^{[c_2]}_{2}& =&\frac{1}{2}\left( f_2(P_2,P_1)C_2 - B_1\left(1,3\right) - 2C_{00}(1,2,3) \right).
\eea
This data determines the quantities
\bea
C_{11} &=& \frac{1}{\det G_2} \Bigl \{ (P_2-P_1)^2 R^{[c_1]}_{1} - P_1 \cdot (P_2-P_1) R^{[c_1]}_{2}    \Bigr \} \nonumber\\
C_{12} &=& C_{21} = \frac{1}{\det G_2} \Bigl \{  P_1^2 R^{[c_1]}_{2} - P_1 \cdot (P_2-P_1) R^{[c_1]}_{1}  \Bigr \} \nonumber\\
C_{22} &=& \frac{1}{\det G_2} \Bigl \{  P_1^2 R^{[c_2]}_{2} - P_1 \cdot (P_2-P_1) R^{[c_2]}_{1}  \Bigr \}
\eea
and in terms of these we can express $C_{\m\n}$ itself:
\bea\label{f.Cmn}
C_{\m\n} &=& \frac{g_{\m\n}}{{2\left( {d - 2} \right)}}\left( {2m_1^2{C_0}(P_1,P_2,m_1,m_2,m_3) - {f_2(P_2,P_1)}{C_2} - {f_1(P_1)}{C_1} + {B_0}\left( {2,3} \right)} \right) \nonumber\\
&+& P_{1\m}P_{1\n} C_{11} + \left( P_{1\m}(P_2-P_1)_{\n} + (P_2-P_1)_{\m}P_{1\n} \right)C_{12} + (P_2-P_1)_{\m}(P_2-P_1)_{\n} C_{22} \nonumber\\
\eea
and from the above expression it is straightforward to compute all its momentum contractions.

We sometimes encounter $C$-integrals with ${\rm det}[G_2]=0$. Such integrals are computed with direct Feynman parametrization and DR (i.e. without algebraic reduction).
To give an example, we may stumble on
\bea\label{explicit Cmn1}
C^{1}_{\m\n}(p,m_1,m_2,m_3) &\equiv& C_{\m\n}(P_1=p,P_2=0,m_1,m_2,m_3) \nonumber\\
&=& \left\langle {{\frac{k_\m k_\n}{(k^2 - m_1^2) ((k+p)^2 - m_2^2 )( k^2 - m_3^2 )}} } \right\rangle \nonumber\\
&=& 2 \int_0^1{\cal D}^3 x \left\langle {{\frac{k_\m k_\n + x_2 k_\m p_\n + x_2 k_\n p_\n + x_2^2 p_\m p_\n}{(k^2 - \Delta_{C_0})^3 }} } \right\rangle \, ,
\eea
which in DR, after expanding in $\ve$, becomes
\bea\label{explicit Cmn2}
16\pi^2 \m^\ve C_{\m\n}^{1}(d=4,\varepsilon \to 0) &=&\frac{g_{\m\n}}{4} \Biggl \{ \frac{2}{\varepsilon} + 2 \int_0^1{\cal D}^3 x \ln \frac{\m^2}{\Delta_{C_0}}  \Biggr \} +2 \int_0^1{\cal D}^3 x  \frac{x_2^2 p^2}{\Delta_{C_0}}  \, .
\eea
Let us now consider the linearly divergent rank 3 integral
\bea
C^{\m\n\a} (P_1,P_2,m_1,m_2,m_3)=\left\langle {\frac{{{k^\m k^\n k^\a}}}{{D_1D_2D_3}}} \right\rangle \, .
\eea
This integral could be reduced in principle algebraically, following \cite{Kunszt}.
Here, as it is only linearly divergent, we will compute it in a brute force way with Feynman parametrization. 
It is a case with $n=3$ and $N^{(3)}_{C^{\m\n\a}} = k^\m k^\n k^\a$ in the language of Appendix \ref{FiniteIntegrals}.
It explicitly evaluates to
\bea\label{UK3mna}
16\pi^2 \m^\ve C^{\m\n\a} (P_1,P_2,m_1,m_2,m_3) &=& - 2 \frac{g^{\m\n}}{d} \int_0^1{\cal D}^3 x  [P_1x_2 + P_2 x_3]^\a B^{K^a}_0 \nonumber\\
&-& 2 \frac{g^{\m\a}}{d} \int_0^1{\cal D}^3 x  [P_1 x_2 + P_2 x_3]^\n B^{K^a}_0 \nonumber\\
&-& 2 \frac{g^{\a\n}}{d} \int_0^1{\cal D}^3 x  [P_1x_2 + P_2 x_3]^\m B^{K^a}_0 \nonumber\\
&+& \int_0^1{\cal D}^3 x \frac{ [P_1x_2 + P_2 x_3]^\m [P_1x_2 + P_2 x_3]^\n [P_1x_2 + P_2 x_3]^\a}{\Delta_{C_0}}\nonumber\\
\eea
where
\bea
B^{K^a}_0 = \frac{2}{\ve} + \ln \frac{\m^2}{\Delta_{C_0}}\, .
\eea
It is now easy to compute any contraction of the above expression with the metric and/or momenta.
A useful contraction is $C_{{\cal K}3}^\m \equiv g_{\n\a} C^{\m\n\a}$.

All Box tensor integrals are computed directly as in Appendix \ref{FiniteIntegrals} since they are at most logarithmically divergent.
$D^{\m}$, $D^{\m\n}$, $D^{\m\n\a}$ and $D^{\m\n\a\b}$ are computed as
\be\label{DPV}
D^{\m,\m\n,\m\n\a,\m\n\a\b} = 6 \int_0^1 {\cal D}^4 x \left\langle  \frac{N_{D^{\m,\m\n,\m\n\a,\m\n\a\b}}^{(4)}(k-\sum_{i=1}^{4} P_ix_{i+1})}{D_1D_2D_3D_4} \right\rangle\, ,
\ee
with $N_{D^{\m}}^{(4)} = k^{\m}$, $N_{D^{\m\n}}^{(4)} = k^{\m} k^{\n}$, $N_{D^{\m\n\a}}^{(4)} = k^{\m}k^\n k^\a$ and $N_{D^{\m\n\a\b}}^{(4)} = k^{\m}k^\n k^\a k^\b$.
Two useful contractions are $D_{{\cal B}4}^{\m\n} \equiv g_{\a\b}D^{\m\n\a\b}$ and $D_{{\cal B}4} \equiv g_{\m\n}D_{{\cal B}4}^{\m\n}$.

\section{$U$-integrals}
\label{Uint}

$U$-integrals are linearly, quadratically, cubically or quartically divergent diagrams that do not appear when the Higgs is expressed in a Cartesian basis
and an $R_\xi$ gauge fixing is performed. If either a Polar basis for the Higgs is used and/or the computation is performed in the Unitary gauge,
these integrals do appear. Clearly, the standard PV reduction formulae must be extended.
At each level of $n$-point functions we meet at least an integral of the $U$-type.

\subsection{$U$-integrals in Tadpoles}
\label{Utadpole}

Indeed, during the Tadpole calculation in section \ref{Tadpoles in Rx} we find the quartically divergent contraction
\bea\label{gmnJmna}
U_{\cal T}^{\mu\nu}(m) &=& \left\langle {{\frac{k^\m k^\n}{k^2 - m^2}} } \right\rangle = \frac{g_{\m\n}}{d}\left\langle {{\frac{k^2}{k^2 - m^2}} } \right\rangle  \Leftrightarrow \nonumber\\
U_{\cal T}(m)={g_{\mu \nu }}U_{\cal T}^{\mu\nu}(m) &=& \frac{d}{d}\left\langle {{\frac{k^2}{k^2 - m^2}} } \right\rangle \nonumber \\
&=& \left\langle {{\frac{m^2}{k^2 - m^2}} } \right\rangle + \left\langle {\frac{k^2 - m^2}{k^2 - m^2} } \right\rangle \nonumber\\
&=& {m^2}A_0(m) + {\cal V}\, .
\eea
where ${\cal V}$ is the DR volume of space-time.
The usual prescription is ${\cal V}=0$,
in which case, the tensor-Tadpole $U$-integral reduces to
\bea\label{gmnJmn1}
U_{\cal T}(m) &=&  {m^2}A_0(m).
\eea

\subsection{$U$-integrals in mass corrections}\label{App.UM4}

The basic $U$-integral with two denominators is the quartically divergent with the cut-off integral
\bea\label{UM4a1}
U_{{\cal M}4} (P_1,m_1,m_2)&=& \left\langle {\frac{{{k^4}}}{{D_1 D_2(P_1)}}} \right\rangle
\eea
with $D_1=k^2-m_1^2$ and $D_2(P_1)=(k+P_1)^2-m_2^2$ as defined in \eq{denom.}. We show only the momentum argument $P_1$
as we would like to follow it.
Adding and subtracting a term, the above integral can be rewritten as
\bea\label{M4redu1}
U_{{\cal M}4}(P_1,m_1,m_2) &=&\left\langle {\frac{{{{\left( {{k^2} - m_1^2} \right)}k^2}}}{{D_1D_2(P_1)}}} \right\rangle  + \left\langle {\frac{{{k^2}m_1^2}}{{D_1D_2(P_1)}}} \right\rangle \nonumber\\
&=& \left\langle {\frac{{{k^2}}}{{D_2(P_1)}}} \right\rangle +  m_1^2 g_{\m\n}{B^{\mu \nu }}\left( {P_1,{m_1},{m_2}} \right). 
\eea 
Now the second term is a standard PV integral, while the first term is still a $U$-integral (it is still quartically divergent) but we can compute it straightforwardly as
\bea\label{M4redu2}
\left\langle {\frac{{{k^2}}}{{D_2(P_1)}}} \right\rangle  &=& \left\langle {\frac{{{k^2}}}{{D_2(0)}}} \right\rangle - \left\langle {\frac{{2{P_1} \cdot k}}{{D_2(0)}}} \right\rangle 
+ \left\langle {\frac{{P_1^2}}{{D_2(0)}}} \right\rangle  \nonumber \\
&=& ( m_2^2 + P_1^2){A_0}\left( {{m_2}} \right) + \frac{1}{16\pi^2} m_2^4 \, ,
\eea
where we have performed the shift $k \to k - P_1$ and we
have dropped the $2k \cdot P_1$ term because it is odd under $k \to -k$. 
Also, we have used \eq{gmnJmn1} for the rational part. 
From \eqref{M4redu1} and \eqref{M4redu2} and using Eq.\eqref{gmnBmn}, we get
\bea\label{UM4redu}
U_{{\cal M}4}(P_1,m_1,m_2) &=& \left ( m^2_1 + m_2^2 + P_1^2  \right ) A_0(m_2) + m_1^4 B_0(P_1,m_1,m_2)  \, . 
\eea
The reduction of $U_{{\cal M}4}$ was carried through using at some point a loop momentum shift.
We know that in DR and in the absence of fermions, momentum shifts are ok, as long as the integral
is up to naively quadratically divergent with the cut-off. $U$-integrals are however often cubically or
quartically divergent. In order to check the validity of momentum shifts, we will now re-compute $U_{{\cal M}4}$
without momentum shift and compare the results.

Starting with \eq{UM4a1}, by adding and subtracting terms, we construct in the numerator $D_2(P_1)$ obtaining 
\bea\label{M4redua}
U_{{\cal M}4}(P_1,m_1,m_2) &=&\left\langle {\frac{{{{\Bigl ( {{(k + P_1)^2} - m_2^2} \Bigr)}k^2}}}{{D_1D_2(P_1)}}} \right\rangle  - 
2 \left\langle {\frac{{{ k^2  P_1 \cdot k }}}{{D_1D_2(P_1)}}} \right\rangle  + (m_2^2 - P_1^2) \left\langle {\frac{{{k^2}}}{{D_1D_2(P_1)}}} \right\rangle \nonumber\\
&=& \left\langle {\frac{{{k^2}}}{{D_1}}} \right\rangle  -  2 \left\langle {\frac{{{ k^2  P_1 \cdot k }}}{{D_1D_2(P_1)}}} \right\rangle +  
( m_2^2 - P_1^2 ) g_{\m\n}{B^{\mu \nu }}\left( {P_1,{m_1},{m_2}} \right). 
\eea 
The second term in the last expression is 
\bea\label{Sec.term.}
-  2 \left\langle {\frac{{{ k^2  P_1 \cdot k }}}{{D_1D_2(P_1)}}} \right\rangle &=& -2 \left\langle {\frac{{{{\Bigl ( {{(k + P_1)^2} - m_2^2} \Bigr)}P_1 \cdot k}}}{{D_1D_2(P_1)}}} \right\rangle \nonumber\\
&+& 4 \left\langle {\frac{{{ P_1 \cdot k  P_1 \cdot k }}}{{D_1D_2(P_1)}}} \right\rangle  -2 (m_2^2 - P_1^2) \left\langle {\frac{{{P_1 \cdot k}}}{{D_1D_2(P_1)}}} \right\rangle \nonumber\\
&=& -2 \left\langle {\frac{{{P_1 \cdot k}}}{{D_1}}} \right\rangle  +  4 P_{1\m} P_{1\n} B^{\mu \nu }\left( {P_1,{m_1},{m_2}} \right) \nonumber\\
&-& 2   ( m_2^2 - P_1^2 ) P_{1\m}{B^{\mu }}\left( {P_1,{m_1},{m_2}} \right).
\eea 
The first term of the above relation is zero since $P_1 \cdot k$ term is odd under $k \to -k$. The third term of \eq{M4redua} 
and the second and third terms of \eq{Sec.term.} are standard PV integrals. The first term of \eq{M4redua} however is still a $U$-integral 
(and still quartically divergent) but we can reduce it easily as
\bea\label{M4redub}
\left\langle {\frac{{{k^2}}}{{D_1 }}} \right\rangle  &=& \left\langle {\frac{{{k^2 - m_1^2 }}}{{D_1(0)}}} \right\rangle 
= m_1^2 {A_0}\left( {{m_1}} \right).
\eea
Combining \eq{M4redua} with \eq{Sec.term.} and \eq{M4redub} and using \eq{PmBm}, \eq{gmnBmn} and \eq{pmpnBmn1} we obtain 
\bea\label{UM4redu.a}
U_{{\cal M}4}(P_1,m_1,m_2) &=& \left ( m^2_1 + m_2^2 + P_1^2  \right ) A_0(m_2) + m_1^4 B_0(P_1,m_1,m_2)  \, ,
\eea
which is the same as \eq{UM4redu}. 
Clearly, we have traded loop momentum shifts in highly divergent integrals for adding and subtracting infinities,
a slightly less disturbing operation. In any case, the final results will justify or not these manipulations.

\subsection{$U$-integrals in Triangles}

In the Triangle sector we first meet
\bea
U^{\m\n}_{{\cal K}4}(P_1,P_2,m_1,m_2,m_3)&=&\left\langle {\frac{{{k^2 k^\m k^\n }}}{{D_1D_2D_3}}} \right\rangle \, .
\eea
It can be reduced easily as
\bea
U^{\m\n}_{{\cal K}4}(P_1,P_2,m_1,m_2,m_3)&=& \left\langle {\frac{{{\left((k + P_2)^2 - m_3^2 \right)k^\m k^\n}}}{{D_1D_2D_3}}} \right\rangle - 2 P_{2\a} C^{\m\n\a}(P_1,P_2,m_1,m_2,m_3) \nonumber\\
&+& (m_3^2 - P_2^2)C^{\m\n}(P_1,P_2,m_1,m_2,m_3)\nonumber\\
&=& B^{\m\n}(P_1,m_1,m_2) - 2 P_{2\a} C^{\m\n\a}(P_1,P_2,m_1,m_2,m_3) \nonumber\\
&+& (m_3^2 - P_2^2)C^{\m\n}(P_1,P_2,m_1,m_2,m_3)
\eea 
and with PV reduction further on.

The next case is the cubically divergent
\bea
U^\m_{{\cal K}5}(P_1,P_2,m_1,m_2,m_3)&=&\left\langle {\frac{{{k^4 k^\m  }}}{{D_1D_2D_3}}} \right\rangle.
\eea
Following similar steps as before
\bea
U^\m_{{\cal K}5}(P_1,P_2,m_1,m_2,m_3)&=&  \left\langle {\frac{{{((k + P_2)^2 - m_3^2 ) k^2 k^\m }}}{{D_1D_2D_3}}} \right\rangle - 2 P_{2\n} U^{\m\n}_{{\cal K}4}(P_1,P_2,m_1,m_2,m_3) \nonumber\\
&+& (m_3^2 - P_2^2)C_{{\cal K}3}^\m(P_1,P_2,m_1,m_2,m_3) \Leftrightarrow \nonumber\\
&=& \left\langle {\frac{{{k^2 k^\m }}}{{D_1D_2}}} \right\rangle - 2 P_{2\n} U^{\m\n}_{{\cal K}4}(P_1,P_2,m_1,m_2,m_3) \nonumber\\
&+& (m_3^2 - P_2^2)C_{{\cal K}3}^\m(P_1,P_2,m_1,m_2,m_3). \nonumber\\
\eea
Only the first term is new (and also a $U$-integral) but it is easy to compute it:
\bea
\left\langle {\frac{{{k^2 k^\m }}}{{D_1D_2}}} \right\rangle &=& \left\langle {\frac{{{(k^2 - m_1^2) k^\m }}}{{D_1D_2}}} \right\rangle + m_1^2 \left\langle {\frac{{{ k^\m }}}{{D_1D_2}}} \right\rangle \nonumber\\
&=& \left\langle {\frac{{{ k^\m }}}{{D_2(P_1)}}} \right\rangle + m_1^2 B^\m(P_1,m_1,m_2) \nonumber\\
&=& \left\langle {\frac{{{ k^\m }}}{{D_2(0)}}} \right\rangle - P_1^\m A_0(m_2) + m_1^2 B^\m(P_1,m_1,m_2) \nonumber\\
&=& - P_1^\m A_0(m_2) + m_1^2 B^\m(P_1,m_1,m_2)
\eea
where in the third line above we have shifted the loop momentum and then neglected the $k$-odd term, as before. It total,
\bea\label{UK5redu}
U^\m_{{\cal K}5}(P_1,P_2,m_1,m_2,m_3) &=& - P_1^\m A_0(m_2) + m_1^2 B^\m(P_1,m_1,m_2) - 2 P_{2\n} U^{\m\n}_{{\cal K}4}(P_1,P_2,m_1,m_2,m_3) \nonumber\\
&+& (m_3^2 - P_2^2)C_{{\cal K}3}^\m(P_1,P_2,m_1,m_2,m_3)\, . 
\eea
The last Triangle $U$-integral is quartically divergent, It is successively reduced as
\bea
U_{{\cal K}6}(P_1,P_2,m_1,m_2,m_3)&=&\left\langle {\frac{{{k^6 }}}{{D_1D_2D_3}}} \right\rangle \nonumber\\
&=&  \left\langle {\frac{{{((k + P_2)^2 - m_3^2 ) k^4 }}}{{D_1D_2D_3}}} \right\rangle  - 2 P_{2\m} U^{\m}_{{\cal K}5}(P_1,P_2,m_1,m_2,m_3) \nonumber\\
&+& (m_3^2 - P_2^2)U_{{\cal K}4}(P_1,P_2,m_1,m_2,m_3) \nonumber\\
&=& \left\langle {\frac{{{k^4 }}}{{D_1D_2}}} \right\rangle - 2 P_{2\m} U^{\m}_{{\cal K}5}(P_1,P_2,m_1,m_2,m_3) \nonumber\\
&+& (m_3^2 - P_2^2)U_{{\cal K}4}(P_1,P_2,m_1,m_2,m_3)\nonumber\\
&=& U_{{\cal M}4}(P_1,m_1,m_2) - 2 P_{2\m} U^{\m}_{{\cal K}5}(P_1,P_2,m_1,m_2,m_3) \nonumber\\
&+& (m_3^2 - P_2^2)U_{{\cal K}4}(P_1,P_2,m_1,m_2,m_3).
\eea
where we have defined $U_{{\cal K}4} = g_{\m\n} U_{{\cal K}4}^{\m\n}$. 

\subsection{$U$-integrals in Boxes}

The lowest Box $U$-integrals have five momenta in their numerator. One is
\bea\label{UB5m}
U^{\m}_{{\cal B}5}(P_1,P_2,P_3,m_1,m_2,m_3,m_4) &=&  \left\langle {\frac{{{k^4 k^\m }}}{{D_1D_2D_3D_4}}} \right\rangle \nonumber\\
 &=& C^{\m}_{{\cal K}3}(P_1,P_2,m_1,m_2,m_3) - 2 P_{3\n} D^{\m\n}_{{\cal B}4}(P_1,P_2,P_3,m_1,m_2,m_3,m_4)\nonumber\\
&+&  (m_4^2 - P_3^2) g_{\n\a}D^{\m\n\a}(P_1,P_2,P_3,m_1,m_2,m_3,m_4)\nonumber\\
\eea
and the other is
\bea
U^{\m\n\a}_{{\cal B}5}(P_1,P_2,P_3,m_1,m_2,m_3,m_4) &=&  \left\langle {\frac{{{k^2 k^\m k^\n k^\a }}}{{D_1D_2D_3D_4}}} \right\rangle  \nonumber\\
 &=& C^{\m\n\a}(P_1,P_2,m_1,m_2,m_3) 
 - 2 P_{3\b} D^{\m\n\a\b}(P_1,P_2,P_3,m_1,m_2,m_3,m_4)\nonumber\\
&+&  (m_4^2 - P_3^2) D^{\m\n\a}(P_1,P_2,P_3,m_1,m_2,m_3,m_4) \, .
\eea
The pattern should start becoming obvious by now. For example, we have
\bea
U^{\m\n}_{{\cal B}6}(P_1,P_2,P_3,m_1,m_2,m_3,m_4)&=&\left\langle {\frac{{{k^4k^\m k^\n}}}{{D_1{{{D_2}} } {{{D_3 D_4}}} }}} \right\rangle\nonumber\\
&=& U^{\m\n}_{{\cal K}4}(P_1,P_2,m_1,m_2,m_3) \nonumber\\
&-& 2 P_{3\a} U^{\m\n\a}_{{\cal B}5}(P_1,P_2,P_3,m_1,m_2,m_3,m_4)\nonumber\\
&+& (m_4^2 - P_3^2) D^{\m\n}_{{\cal B}4}(P_1,P_2,P_3,m_1,m_2,m_3,m_4)\nonumber\\
\eea
and its contracted version $U_{{\cal B}6}\equiv g_{\m\n}U^{\m\n}_{{\cal B}6}$.
The cubically divergent Box $U$-integral is
\bea 
U^\m_{{\cal B}7}(P_1,P_2,P_3,m_1,m_2,m_3,m_4)&=&\left\langle {\frac{{{k^6k^\m }}}{{D_1{{{D_2}} } {{{D_3 D_4}}} }}} \right\rangle \nonumber\\
&=& U^{\m}_{{\cal K}5}(P_1,P_2,m_1,m_2,m_3) \nonumber\\
&-& 2 P_{3\n} U^{\m\n}_{{\cal B}6}(P_1,P_2,P_3,m_1,m_2,m_3,m_4)\nonumber\\
&+& (m_4^2 - P_3^2) U^{\m}_{{\cal B}5}(P_1,P_2,P_3,m_1,m_2,m_3,m_4)\, .\nonumber\\
\eea
Finally we have the quartically divergent  
\bea 
U_{{\cal B}8}(P_1,P_2,P_3,m_1,m_2,m_3,m_4)&=&\left\langle {\frac{{{k^8}}}{{D_1{{{D_2}} } {{{D_3 D_4}}} }}} \right\rangle\nonumber\\
&=& U_{{\cal K}6}(P_1,P_2,m_1,m_2,m_3) \nonumber\\
&-& 2 P_{3\m} U^{\m}_{{\cal B}7}(P_1,P_2,P_3,m_1,m_2,m_3,m_4)\nonumber\\
&+& (m_4^2 - P_3^2) U_{{\cal B}6}(P_1,P_2,P_3,m_1,m_2,m_3,m_4)\, .\nonumber\\
\eea

\section{Explicit calculation of the diagrams}
\label{Explicitcalculation}
In this Appendix we present the explicit calculation of the one-loop Feynman diagrams, concentrating on the Tadpole and the two-point function categories, in both $R_\xi$ and Unitary gauges. 
The Triangle and Box contributions can be straightforwardly computed following similar steps, therefore there is no need to calculate them explicitly here.  

\subsection{$R_\xi$-diagrams}\label{RGD}
In order to be synchronised with our main text, we start form the first set of diagrams which corresponds to the one-point functions. The first such diagram is
\vskip .5cm
\begin{center}
\begin{tikzpicture}[scale=0.7]
\draw [dashed,thick] (0,0) circle [radius=1];
\draw [dashed,thick] (-2.5,0)--(-1,0);
\draw [->]  (-2.2,0.2)--(-1.5,0.2);
\node at (-2.9,0) {$p$};
\draw [->]  (0.45,-0.6) arc [start angle=-45, end angle=-135, radius=0.6cm];
\node at (0,-0.4) {$k$};
\node at (3,0) {$=\, \, i{\cal T}_H^{R_\xi,\phi}$};
\end{tikzpicture}
\end{center}
and analytically evaluates to
\bea
i{\cal T}_H^{R_\xi,\phi} &=& -  6{\cal S}^1_{T}  \sqrt{\frac{\l_0}{2}}m_{H_0}\int {\frac{{{d^4}k}}{{{{\left( {2\pi } \right)}^4}}}\frac{-i}{{\left( {{k^2} - m_{H_0}^2} \right)}}}\, .
\eea
The symmetry factor is ${\cal S}^1_{T}=\frac{1}{2}$ since $(n_O,n_l,\ell_1,v_1)=(3,1,3,1)$.
In DR this integral takes the form 
\be\label{final Tad1Rx}
{\cal T}_H^{R_\xi, \phi} ={3  \sqrt{\frac{\l_0}{2}}m_{H_0}{\mu ^{\ve}}}A_0(m_{H_0}) \, .
\ee
The "tadpole integral" $A_0$ is defined in Appendix \ref{PassarinoVeltman}. It has mass dimension 2 and it is (external) momentum independent.

The next tadpole comes with a gauge boson loop:
\vskip .5cm
\begin{center}
\begin{tikzpicture}[scale=0.7]
\draw [photon] (0,0) circle [radius=1];
\draw [dashed,thick] (-2.5,0)--(-1,0);
\draw [->]  (-2.2,0.2)--(-1.5,0.2);
\node at (-2.9,0) {$p$};
\draw [->]  (0.45,-0.5) arc [start angle=-45, end angle=-135, radius=0.6cm];
\node at (0,-0.3) {$k$};
\node at (3,0) {$=\, \, i{\cal T}_H^{R_\xi, Z}$};
\end{tikzpicture}
\end{center}
It is equal to
\bea
{\cal T}_H^{R_\xi, Z} &=& d \frac{m^2_{Z_0}}{m_{H_0}}\sqrt{2 \l_0}\int {\frac{{{d^4}k}}{{{{\left( {2\pi } \right)}^4}}}\frac{{ - i}}{{{k^2} - m_{Z_0}^2}}} \nonumber\\
&+& {{(1-\xi) \frac{m^2_{Z_0}}{m_{H_0}}\sqrt{2 \l_0}}}{{g_{\m\n}}}\int {\frac{{{d^4}k}}{{{{\left( {2\pi } \right)}^4}}}\frac{{i{k^\m k^\n}}}{{\left({k^2} - m_{Z_0}^2 \right)\left({k^2} - \xi m_{Z_0}^2 \right)}}}
\eea
with symmetry factor ${\cal S}^2_{T} = \frac{1}{2} $ since $(n_O,n_l,\ell_1,v_1)=(1,1,2,1)$.
Using the relation $k^\m k^\n = \frac{g_{\m\n}}{d} k^2$ under the intergral it simplifies to
\bea\label{T2HRx2}
{\cal T}_H^{R_\xi, Z} &=&\frac{m^2_{Z_0}}{m_{H_0}}\sqrt{2 \l_0} {\mu^{\ve}} \Biggl \{  4A_0(m_{Z_0}) - (1-\xi)A_0(\sqrt{\xi} m_{Z_0}) -(1-\xi)m_{Z_0}^2 B_0^1(m_{Z_0},\sqrt{\xi} m_{Z_0}) \Biggr \} \, , \nonumber\\
\eea
where the $B_0$ scalar integral appears at $p^2 = 0$ and we call it $B_0^1$. 
Using Eq.\eqref{B0p0} the gauge tadpole becomes
\bea\label{final Tad2Rx2}
{\cal T}_H^{R_\xi, Z} &=& \frac{m^2_{Z_0}}{m_{H_0}}\sqrt{2 \l_0} {\mu^{\ve}} \Biggl \{  3A_0(m_{Z_0}) + \xi A_0(\sqrt{\xi} m_{Z_0}) \Biggr \}\, . 
\eea
The last tadpole has a Goldstone loop:
\vskip .5cm
\begin{center}
\begin{tikzpicture}[scale=0.7]
\draw [] (0,0) circle [radius=1];
\draw [dashed,thick] (-2.5,0)--(-1,0);
\draw [->]  (-2.2,0.2)--(-1.5,0.2);
\node at (-2.9,0) {$p$};
\draw [->]  (0.45,-0.6) arc [start angle=-45, end angle=-135, radius=0.6cm];
\node at (0,-0.4) {$k$};
\node at (3,0) {$=\, \, i{\cal T}_H^{R_\xi,\chi}$};
\end{tikzpicture}
\end{center}
It is equal to
\bea\label{UT2}
{\cal T}_H^{R_\xi,\chi} &=& - \frac{\sqrt{2 \l_0}}{m_{H_0}} \int {\frac{{{d^4}k}}{{{{\left( {2\pi } \right)}^4}}}\frac{-i k^2}{{\left( {{k^2} - m_{\chi_0}^2} \right)}}} \, ,
\eea
with ${\cal S}^3_{T} = \frac{1}{2}$ as $(n_O,n_l,\ell_1,v_1)=(1,1,2,1)$.
Notice that the integral in \eq{UT2} is a $U$-integral (called $U_{\cal T}$), the first in the class of highly divergent integrals that we will often encounter in the Unitary gauge.
As already mentioned, the origin of such integrals emerging also in the $R_\xi$ gauge can be traced to our Polar basis choice to represent the Higgs field.
Using Eq.\eqref{gmnJmn1} to calculate ${U}_{\cal T}(m_{\chi_0})$ we obtain that in DR, 
\be\label{final Tad3Rx2}
{\cal T}_H^{R_\xi, \chi}= -\frac{\sqrt{2 \l_0}}{m_{H_0}} \m^\ve m_{\chi_0}^2 A_0(m_{\chi_0})  \, .
\ee
Finally, the total tadpole value is the sum of the above three contributions:
\bea\label{full Tad}
{\cal T}_H^{R_\xi} &=& \mu ^{\ve} \Biggl ( 3 \sqrt{\frac{\l_0}{2}}m_{H_0}  A_0(m_H)  + 3 \frac{\sqrt{2 \l_0}m^2_{Z_0}}{m_{H_0}} A_0(m_Z)   \Biggr ) \, .
\eea

Next we have the $Z$ 2-point function. A gauge-boson vacuum polarization amplitude can be Lorentz-covariantly split into a transverse and a longitudinal part
\bea
{\cal M}_{Z,\mu\nu} &=& \left( -g_{\m\n} + \frac{p_\m p_\n}{p^2}   \right) \Pi^T(p^2) + \frac{p_\m p_\n}{p^2} \Pi^L(p^2).
\eea
Contracting with $p^\m p^\n$ both sides fixes
\bea
 \Pi^L(p^2) &=& \frac{p^\m p^\n}{p^2} {\cal M}_{Z,\mu\nu}.
\eea
Contracting with $g^{\m\n}$ gives on the other hand
\bea
g^{\m\n}{\cal M}_{Z,\mu\nu} &=& -(d-1) \Pi^T + \Pi^L
\eea
that can be easily solved for the transverse part in $d = 4$
\bea
\Pi^T &=& \frac{1}{3} \left(-g_{\m\n} + \frac{p_\m p_\n}{p^2}\right) {\cal M}^{\m\n}_{Z}\, .
\eea
Now, the Schwinger-Dyson equation that the dressed $Z$-propagator
\bea
G_{\m\n} &=& - g_{\m\n} G(p^2) + \frac{p_\m p_\n}{m_{Z_0}^2} L(p^2)
\eea
obeys is written as
\bea
G_{\m\n} &=& G_{\m\n} +D_{\m \rho} {\cal M}_Z^{\rho \sigma}G_{\sigma \n}
\eea
with $D_{\m \rho}$ the tree level gauge boson propagator
\bea
D_{\m \rho} = \frac{\left(-g_{\m\rho} + \frac{p_\m p_\rho}{m_{Z_0}^2}\right) }{p^2 - m^2_{Z_0}}.
\eea
So, performing the contractions the Schwinger-Dyson equation becomes
\bea
-g_{\m\n} G + \frac{p_\m p_\n}{m_{Z_0}^2}L = \frac{\left(-g_{\m\n} + \frac{p_\m p_\n}{m_{Z_0}^2}\right) }{p^2 - m^2_{Z_0}} + \frac{\left(-g_{\m\n} + \frac{p_\m p_\n}{p^2}\right) }{p^2 - m^2_{Z_0}}\Pi^T G.
\eea
Contracting again with $p^\m p^\n$ we have that 
\bea
-G +\frac{p^2}{m_{Z_0}^2}L = \frac{1}{m_{Z_0}^2}\left[ 1- \Pi^L(G - L)  \right]
\eea
while contracting with the metric gives
\bea
-d G +\frac{p^2}{m_{Z_0}^2}L = \frac{ -d + \frac{p^2}{m_{Z_0}^2} }{p^2 - m^2_{Z_0}} + \frac{-d + 1}{p^2 - m^2_{Z_0}}\Pi^T G - \frac{1}{m_{Z_0}^2}\Pi^L(G - L).
\eea
The solution of the above equation is
\bea
G(p^2) &=& \frac{1}{p^2 -m_{Z_0}^2 - \Pi^T(p^2)} \nonumber\\
L(p^2) &=&G(p^2) \left[ 1 - \frac{\Pi^T}{p^2 - \Pi^L}  \right]. 
\eea
The quantity that enters in the renormalization of the mass of the $Z$ gauge boson is therefore
\bea\label{calMZRxi2}
{\cal M}_{Z} = -\frac{1}{3} \left( g_{\m\n} - \frac{p_\m p_\n}{p^2} \right) {\cal M}^{\m\n}_{Z}(p) \, .
\eea
We now start computing the one-loop Feynman diagrams contributing to ${\cal M}^{R_\xi}_{Z, \m\n}$.

The first contributing diagram has a Higgs running in the loop:
\vskip .5cm
\begin{center}
\begin{tikzpicture}[scale=0.7]
\draw [photon] (0,0)--(1.8,0);
\draw [dashed,thick] (0,0.9) circle [radius=0.9];
\draw [photon] (-1.8,0)--(-0,0);
\draw [->]  (-1.7,0.2)--(-1,0.2);
\node at (-2.2,0.2) {$p$};
\draw [->]  (0.45,0.3) arc [start angle=-45, end angle=-135, radius=0.6cm];
\node at (0,0.5) {$k$};
\draw [->]  (1.7,0.2)--(1,0.2);
\node at (4.5,0) {$=\, \, i{\cal M}^{R_\xi,\phi}_{Z,\mu\nu}$};
\end{tikzpicture}
\end{center}
and it is equal to
\begin{eqnarray}
{\cal M}^{R_\xi, \phi}_{Z,\mu\nu} &=& -2{{g_{\mu \nu }}}\frac{m^2_{Z_0}}{m^2_{H_0}} \l_0\int {\frac{{{d^4}k}}{{{{\left( {2\pi } \right)}^4}}}\frac{-i}{{{k^2} - m_{H_0}^2}}}\, ,
\end{eqnarray}
where we used that ${\cal S}_{g}^1 =\frac{1}{2}$ because $(n_O,n_l,\ell_1,\ell_2,v_1)=(2,1,2,2,1)$. 
In DR this integral is
\be\label{M1ZRxi2}
{\cal M}^{R_\xi,\phi}_{Z,\mu\nu} = -2{{g_{\mu \nu }}}\frac{m^2_{Z_0}}{m^2_{H_0}} \l_0 \mu^{\ve} A_0(m_{H_0}) \, .
\ee
Next we meet a couple of "sunset" diagrams. The first is 
\vskip .5cm
\begin{center}
\begin{tikzpicture}[scale=0.7]
\draw [photon] (-2.3,0)--(2.3,0);
\draw [->]  (-1.9,0.2)--(-1.2,0.2);
\node at (-2.5,0.2) {$p$};
\draw [<-]  (0.45,0.7) arc [start angle=45, end angle=135, radius=0.6cm];
\node at (0,1.3) {$k+p$};
\draw [<-] (-0.4,0.2)--(0.4,0.2);
\node at (0,-0.3) {$k$};
\draw [->]  (1.9,0.2)--(1.2,0.2);
\node at (4.5,0) {$=\,\,  i{\cal M}^{R_\xi,\phi Z}_{Z,\mu\nu}$ };
\draw  [dashed,thick] (-1,0) .. controls (-1,0.555) and (-0.555,1) .. (0,1)
.. controls (0.555,1) and (1,0.555) .. (1,0);
\end{tikzpicture}
\end{center}
that evaluates to
\bea\label{M2ZRxi2}
{\cal M}^{R_\xi,\phi Z}_{Z,\mu\nu} &=& -8g_{\mu\alpha}}{g_{\nu\b}}\frac{m^4_{Z_0}}{m^2_{H_0}} \l_0\int {\frac{{{d^4}k}}{{{{\left( {2\pi } \right)}^4}}}
\frac{{ - i{g^{\alpha \beta }}}}{{\left( {{k^2} - m_{Z_0}^2} \right)\left( {{{\left( {k + p} \right)}^2} - m_{H_0}^2} \right)}} \nonumber\\
&-&8g_{\mu\alpha}}{g_{\nu\b}}\frac{m^4_{Z_0}}{m^2_{H_0}} \l_0 \int {\frac{{{d^4}k}}{{{{\left( {2\pi } \right)}^4}}}
\frac{{i{( 1-\xi  )k^\alpha }{k^\beta }}}{{\left( {{k^2} - m_{Z_0}^2} \right)\left( {{{\left( {k + p} \right)}^2} - m_{H_0}^2} \right)} \left( k^2 - m_{\chi_0}^2  \right)} \,  \nonumber\\
\eea
with $(n_O,n_l,\ell_1,\ell_2,v_1)=(2 \times 2 \times 2,1,2,2,2)$ and ${\cal S}_{g}^2 = 1$.
In DR and using \eq{explicit Cmn2}, it can be expressed as
\be\label{M2ZRxi2}
{\cal M}^{R_\xi,\phi Z}_{Z,\mu\nu}  = 8{}{{}\frac{m^4_{Z_0}}{m^2_{H_0}} \l_0\mu^{\ve}} \Biggl \{- g_{\mu \nu }B_0(p,m_{Z_0},m_{H_0})
+ ( 1- \xi ) C^{1}_{\m\n}(p,m_{Z_0}, m_{H_0},m_{\chi_0}) \Biggr \}\, .  \nonumber\\
\ee
Notice that the $C$-type integral above is a special PV case, computed in Appendix \ref{PassarinoVeltman} as well.

The next sunset diagram is the last that contributes to the one-loop correction of the gauge boson propagator:
\vskip .5cm
\begin{center}
\begin{tikzpicture}[scale=0.7]
\draw [photon] (-2.3,0)--(-1,0);
\draw [photon] (1,0)--(2.3,0);
\draw [] (-1,0)--(1,0);
\draw [->]  (-1.9,0.2)--(-1.2,0.2);
\node at (-2.5,0.2) {$p$};
\draw [<-]  (0.45,0.7) arc [start angle=45, end angle=135, radius=0.6cm];
\node at (0,1.3) {$k+p$};
\draw [<-] (-0.4,0.2)--(0.4,0.2);
\node at (0,-0.3) {$k$};
\draw [->]  (1.9,0.2)--(1.2,0.2);
\node at (4.5,0) {$=\,\,  i{\cal M}^{R_\xi,\chi\phi}_{Z,\mu\nu}$ };
\draw  [dashed,thick] (-1,0) .. controls (-1,0.555) and (-0.555,1) .. (0,1)
.. controls (0.555,1) and (1,0.555) .. (1,0);
\end{tikzpicture}
\end{center}
It is equal to
\bea
{\cal M}^{R_\xi,\chi\phi }_{Z,\mu\nu} &=& 8{}{}\frac{m^2_{Z_0}}{m^2_{H_0}} \l_0\int {\frac{{{d^4}k}}{{{{\left( {2\pi } \right)}^4}}}\frac{-i k_\m k_\n}{({{{ { k}}^2} -  m_{\chi_0}^2}) ( {{\left(k + p \right)^2} - m_{H_0}^2} )}}\,   
\eea
with ${\cal S}_{g}^3=1$ from $(n_O,n_l,\ell_1,\ell_2,v_1)=(1 \times 1 \times 2,1,1,1,2)$.
In DR it is
\be
{\cal M}^{R_\xi,\chi\phi}_{Z,\mu\nu} = 8 \frac{m^2_{Z_0}}{m^2_{H_0}} \l_0 \mu^{\ve} B_{\m\n}(p,m_{\chi_0},m_{H_0}) \, . 
\ee
Adding up all contributions we obtain
\bea\label{full MZRxi2}
{\cal M}^{R_\xi}_{Z,\mu\nu} &=& \frac{m^2_{Z_0}}{m^2_{H_0}} \l_0 \mu^{\ve} \Biggl \{  -2{{g_{\mu \nu }}}A_0(m_{H_0}) - 8 g_{\mu \nu } m_{Z_0}^2B_0(p,m_{Z_0},m_{H_0}) \nonumber\\
&+& 8( 1- \xi ) m_{Z_0}^2 C^{1}_{\m\n}(m_{Z_0}, m_{H_0},m_{\chi_0}) + 8 B_{\m\n}(p,m_{\chi_0},m_{H_0})   \Biggr \} \, .
\eea 

Next, we consider the one-loop contributions to the Higgs propagator. The first diagram here is
\vskip .5cm
\begin{center}
\begin{tikzpicture}[scale=0.7]
\draw [dashed,thick] (0,0)--(1.8,0);
\draw [photon] (0,0.9) circle [radius=0.9];
\draw [dashed,thick] (-1.8,0)--(-0,0);
\draw [->]  (-1.7,0.2)--(-1,0.2);
\node at (-2.2,0.2) {$p$};
\draw [->]  (0.45,0.50) arc [start angle=-45, end angle=-135, radius=0.6cm];
\node at (0,0.8) {$k$};
\draw [->]  (1.7,0.2)--(1,0.2);
\node at (4.5,0) {$=\, \,i{\cal M}_H^{R_\xi, Z}$};
\end{tikzpicture}
\end{center}
that is,
\bea
{\cal M}_H^{R_\xi, Z} &=& 2 d \frac{m^2_{Z_0}}{m^2_{H_0}} \l_0\int {\frac{{{d^4}k}}{{{{\left( {2\pi } \right)}^4}}}\frac{{ - i}}{{{k^2} - m_{Z_0}^2}}} \nonumber\\
&+&{{\frac{2(1-\xi)m^2_{Z_0}}{m^2_{H_0}} \l_0 }}{{g_{\m\n}}}\int {\frac{{{d^4}k}}{{{{\left( {2\pi } \right)}^4}}}\frac{{i{k^\m k^\n}}}{{\left({k^2} - m_{Z_0}^2 \right)\left({k^2} - \xi m_{Z_0}^2 \right)}}}
\eea
with ${\cal S}^{1}_{{\cal {M}}_H } =  \frac{1}{2} $ from $(n_O,n_l,\ell_1,\ell_2,v_1)=(2,1,2,2,1)$.
In DR it can be written as 
\bea\label{full MH1a}
{\cal M}_H^{R_\xi, Z}&=& {\mu^{\ve}} \Biggl \{2 d \frac{m^2_{Z_0}}{m^2_{H_0}} \l_0 A_0(m_{Z_0}) - 
2 (1-\xi) \frac{m^2_{Z_0}}{m^2_{H_0}} \l_0 g_{\m\n} B^{\m \n}(0,m_{Z_0},\sqrt{\xi} m_{Z_0}) \Biggr \} \nonumber\\
&=& \frac{m^2_{Z_0}}{m^2_{H_0}} \l_0 {\mu^{\ve}} \Biggl \{  8 A_0(m_{Z_0}) - 2(1-\xi)A_0(\sqrt{\xi} m_{Z_0}) -2 (1-\xi)m_{Z_0}^2 B_0^1(m_{Z_0},\sqrt{\xi} m_{Z_0})  \Biggr \}\nonumber\\
\eea 
and after reductions finally as
\bea\label{full MH1b}
{\cal M}_H^{R_\xi, Z}&=&  \frac{m^2_{Z_0}}{m^2_{H_0}} \l_0 {\mu^{\ve}} \Biggl \{  6A_0(m_{Z_0}) +2 \xi A_0( m_{\chi_0})  \Biggr \}\, . 
\eea 
The next contribution comes from the diagram 
\vskip .5cm
\begin{center}
\begin{tikzpicture}[scale=0.7]
\draw [dashed,thick] (0,0)--(1.8,0);
\draw [dashed,thick] (0,0.9) circle [radius=0.9];
\draw [dashed,thick] (-1.8,0)--(-0,0);
\draw [->]  (-1.7,0.2)--(-1,0.2);
\node at (-2.2,0.2) {$p$};
\draw [->]  (0.45,0.4) arc [start angle=-45, end angle=-135, radius=0.6cm];
\node at (0,0.5) {$k$};
\draw [->]  (1.7,0.2)--(1,0.2);
\node at (4.5,0) {$=\, \, i{\cal M}^{R_\xi, \phi}_H$};
\end{tikzpicture}
\end{center}
with explicit form 
\bea\label{M2_H}
{\cal M}^{R_\xi, \phi}_H &=& 3 \l_0 \int {\frac{{{d^4}k}}{{{{\left( {2\pi } \right)}^4}}}\frac{-i}{{{k^2} - m_{H_0}^2}}}\, 
\eea
with $(n_O,n_l,\ell_1,v_1)=(4 \times 3,1,4,1)$ and ${\cal S}^{2}_{{\cal {M}}_H }= \frac{1}{2}$.
This is just
\be\label{full MH2}
{\cal M}_H^{R_\xi,\phi}={3 \l_0  {\mu^{\ve}}{}}A_0(m_{H_0})
\ee 
in DR.

Next comes the Goldstone loop
\vskip .5cm
\begin{center}
\begin{tikzpicture}[scale=0.7]
\draw [dashed,thick] (0,0)--(1.8,0);
\draw [] (0,0.9) circle [radius=0.9];
\draw [dashed,thick] (-1.8,0)--(-0,0);
\draw [->]  (-1.7,0.2)--(-1,0.2);
\node at (-2.2,0.2) {$p$};
\draw [->]  (0.45,0.4) arc [start angle=-45, end angle=-135, radius=0.6cm];
\node at (0,0.5) {$k$};
\draw [->]  (1.7,0.2)--(1,0.2);
\node at (4.5,0) {$=\, \, i{\cal M}^{R_\xi,\chi}_H$};
\end{tikzpicture}
\end{center}
that is equal to
\bea
{\cal M}^{R_\xi,\chi}_{H} &=& - 2  \frac{ \l_0}{m_{H_0}^2}  \int {\frac{{{d^4}k}}{{{{\left( {2\pi } \right)}^4}}}\frac{-i k^2}{{{k^2} - m_{\chi_0}^2}}}
\eea
with $(n_O,n_l,\ell_1, \ell_2, v_1)=(2,1,2,2,1)$ and ${\cal S}^{3}_{{\cal {M}}_H } =  \frac{1}{2}$.
This is
\be\label{full MH3}
{\cal M}_H^{R_\xi,\chi}=-{ \frac{2 \l_0}{m_{H_0}^2} {\mu^{\ve}} m_{\chi_0}^2 } A_0(m_{\chi_0}) \, .
\ee 
A few vacum polarization diagrams are in order. The first is
\vskip .5cm
\begin{center}
\begin{tikzpicture}[scale=0.7]
\draw [dashed,thick] (0.9,0)--(2.2,0);
\draw [dashed,thick] (0,0) circle [radius=0.9];
\draw [dashed,thick] (-2.2,0)--(-0.9,0);
\draw [->]  (-1.9,0.2)--(-1.2,0.2);
\node at (-2.6,0.2) {$p$};
\draw [<-]  (0.45,0.6) arc [start angle=45, end angle=135, radius=0.6cm];
\node at (0,1.2) {$k+p$};
\draw [->]  (0.45,-0.6) arc [start angle=-45, end angle=-135, radius=0.6cm];
\node at (0,-1.2) {$k$};
\draw [->]  (1.9,0.2)--(1.2,0.2);
\node at (4.5,0) {$=\, \, i{\cal M}^{R_\xi,\phi\phi}_H$};
\end{tikzpicture}
\end{center}
and it is equal to 
\bea
{\cal M}^{R_\xi,\phi\phi}_H &= &9  \l_0 m^2_{H_0}\int {\frac{{{d^4}k}}{{{{\left( {2\pi } \right)}^4}}}\frac{-i}{\left({{k^2} - m_{H_0}^2}\right){\left( {{{\left( {k + p} \right)}^2} - m_{H_0}^2} \right)}}}\, 
\eea
and finally in DR to
\bea\label{full MH4}
{\cal M}^{R_\xi,\phi\phi}_H&= &9 \l_0 m^2_{H_0}{\mu ^{\ve}}B_0(p,m_{H_0},m_{H_0})\, ,
\eea
where $(n_O,n_l,\ell_1, \ell_2, v_1)=(3 \times 3 \times 2,2,3,3,2)$ and ${\cal S}^{4}_{{\cal {M}}_H } =  \frac{1}{2}$.

The Goldstone loop contribution
\vskip .5cm
\begin{center}
\begin{tikzpicture}[scale=0.7]
\draw [dashed,thick] (0.9,0)--(2.2,0);
\draw [] (0,0) circle [radius=0.9];
\draw [dashed,thick] (-2.2,0)--(-0.9,0);
\draw [->]  (-1.9,0.2)--(-1.2,0.2);
\node at (-2.6,0.2) {$p$};
\draw [<-]  (0.45,0.6) arc [start angle=45, end angle=135, radius=0.6cm];
\node at (0,1.2) {$k+p$};
\draw [->]  (0.45,-0.6) arc [start angle=-45, end angle=-135, radius=0.6cm];
\node at (0,-1.2) {$k$};
\draw [->]  (1.9,0.2)--(1.2,0.2);
\node at (4.5,0) {$=\, \, i{\cal M}^{R_\xi,\chi\chi}_H$};
\end{tikzpicture}
\end{center}
is equal to 
\bea\label{M5RxUint}
{\cal M}^{R_\xi,\chi\chi}_H &=&- 8{\cal S}^5_{{\cal {M}}_H }  \frac{\l_0 }{m^2_{H_0}} \int {\frac{{{d^4}k}}{{{{\left( {2\pi } \right)}^4}}}
\frac{i(k^4 +  2k^2 k \cdot p + (p \cdot k)^2)}{{\left( {{k^2} - m_{\chi_0}^2} \right)\left( {{(k + p)^2} - m_{\chi_0}^2} \right)}}}
\eea
with symmetry factor ${\cal S}^{5}_{{\cal {M}}_H }= \frac{1}{2}$ from $(n_O,n_l,\ell_1, \ell_2, v_1)=(2 \times 1,2,2,2,2)$.
In DR it becomes
\bea\label{full MH5}
{\cal M}^{R_\xi,\chi\chi}_H&= & 4 \frac{\l_0 }{m^2_{H_0}} {\mu ^{\ve}}\Biggl \{ m_{\chi_0}^2 A_0(m_{\chi_0}) + 
(m_{\chi_0}^2 - p^2 ) g_{\m\n} B^{\m\n}(p,m_{\chi_0},m_{\chi_0}) + p_\m p_\n B^{\m\n}(p,m_{\chi_0},m_{\chi_0}) \Biggr \}\, . \nonumber\\
\eea
Slightly more complicated is the gauge boson loop
\vskip .5cm
\begin{center}
\begin{tikzpicture}[scale=0.7]
\draw [dashed,thick] (0.9,0)--(2.2,0);
\draw [photon] (0,0) circle [radius=0.9];
\draw [dashed,thick] (-2.2,0)--(-0.9,0);
\draw [->]  (-1.9,0.2)--(-1.2,0.2);
\node at (-2.6,0.2) {$p$};
\draw [<-]  (0.45,0.6) arc [start angle=45, end angle=135, radius=0.6cm];
\node at (0,1.3) {$k+p$};
\draw [->]  (0.45,-0.40) arc [start angle=-45, end angle=-135, radius=0.6cm];
\node at (0,-1.1) {$k$};
\draw [->]  (1.9,0.2)--(1.2,0.2);
\node at (4.5,0) {$=\, \, i{\cal M}^{R_\xi,ZZ}_H$};
\end{tikzpicture}
\end{center}
evaluating to
\bea\label{M6_H}
{\cal M}^{R_\xi,ZZ}_H &= &4{g^{\mu \nu }}{g^{\alpha \beta }} \frac{m^4_{Z_{0}}}{m^2_{H_0}}\l_0 \int {\frac{{{d^4}k}}{{{{\left( {2\pi } \right)}^4}}}
\frac{{-i\left( { - {g_{\mu \alpha }} + \frac{{{(1-\xi)k_\mu }{k_\alpha }}}{{{k^2 - \xi m_{Z_0}^2}}}} \right)}}{{\left( {{k^2} - m_{Z_0}^2} \right)}}
\frac{{\left( { - {g_{\nu \b }} + \frac{{{{(1-\xi)\left( {k + p} \right)}_\nu }{{\left( {k + p} \right)}_\b }}}{{{(k+p)^2 - \xi m_{Z_0}^2}}}} \right)}}{{\left( {{{\left( {k + p} \right)}^2} - m_{Z_0}^2} \right)}}}\nonumber\\
\eea
with $(n_O,n_l,\ell_1, \ell_2, v_1)=(2 \times 1,2,2,2,2)$ and ${\cal S}^{6}_{{\cal {M}}_H } =  \frac{1}{2}$.
The numerator of this diagram can be expanded as
\bea
N&=& g^{\m\n} g^{\a\b} \left( { - {g_{\mu \alpha }} + \frac{{{(1-\xi)k_\mu }{k_\alpha }}}{{{k^2 - \xi m_{Z_0}^2}}}} \right) \left( { - {g_{\nu \b }} + \frac{{{{(1-\xi)\left( {k + p} \right)}_\nu }{{\left( {k + p} \right)}_\b }}}{{{(k+p)^2 - \xi m_{Z_0}^2}}}} \right) \nonumber\\
&=&  d - (1-\xi) \left(  \frac{k^2}{k^2 - \xi m_{Z_0}^2} + \frac{k^2 +2 k\cdot p + p^2}{(k+p)^2 - \xi m_{Z_0}^2} \right) \nonumber\\
 &+&(1-\xi)^2 \left( \frac{k^4 + 2k^2 k\cdot p + (k \cdot p)^2}{(k^2 - \xi m_{Z_0}^2)((k+p)^2 - \xi m_{Z_0}^2)} \right)\, 
\eea
and then it is a standard step to express it in terms of PV integrals as
\bea\label{full MH7}
{\cal M}^{R_\xi,ZZ}_H&= &4  \frac{m^4_{Z_{0}}}{m^2_{H_0}}\l_0  {\mu ^{\ve}} \Biggl \{d B_0(p,m_{Z_0},m_{Z_0}) - (1-\xi)  \Bigl \{g_{\m\n} C^{1\m\n}(p,m_{Z_0},m_{Z_0},m_{\chi_0}) \nonumber\\
&+& g_{\m\n} C^{\m\n}(p,p,m_{Z_0},m_{Z_0},m_{\chi_0})  \Bigr \} \nonumber\\ 
&+& (1-\xi)^2 \Bigl \{ g_{\m\n} C^{1\m\n}(p,m_{Z_0},m_{Z_0},m_{\chi_0})  \nonumber\\
&+& (m_{Z_0}^2 - p^2) g_{\m\n}D^{\m\n}(p,m_{Z_0},m_{Z_0},m_{\chi_0},m_{\chi_0}) + p_\m p_\n D^{\m\n}(p,m_{Z_0},m_{Z_0},m_{\chi_0},m_{\chi_0})\, ,
\Bigr \}   \Biggr \} \nonumber\\
\eea
where the $a = 1,2,3$ superscripts on the $C_0$-integrals correspond to the different combinations of the denominators
according to \eq{C012ab} of Appendix \ref{PassarinoVeltman}. The $D^{\m\n}$ integrals are defined in \eq{DPV}.

Finally, the last contribution to the one-loop correction of the Higgs mass comes from the sunset
\vskip .5cm
\begin{center}
\begin{tikzpicture}[scale=0.7]
\draw [dashed,thick] (-2.3,0)--(-1,0);
\draw [dashed,thick] (1,0)--(2.3,0);
\draw [] (-1,0)--(1,0);
\draw [->]  (-1.9,0.2)--(-1.2,0.2);
\node at (-2.5,0.2) {$p$};
\draw [<-]  (0.45,0.7) arc [start angle=45, end angle=135, radius=0.6cm];
\node at (0,1.6) {$k+p$};
\draw [<-] (-0.4,0.2)--(0.4,0.2);
\node at (0,-0.5) {$k$};
\draw [->]  (1.9,0.2)--(1.2,0.2);
\node at (4.5,0) {$=\,\, i{\cal M}^{R_\xi,\chi Z}_H$ };
\draw  [photon] (-1,0) .. controls (-1,0.555) and (-0.555,1) .. (0,1)
.. controls (0.555,1) and (1,0.555) .. (1,0);
\end{tikzpicture}
\end{center}
with explicit form 
\bea
{\cal M}^{R_\xi,\chi Z}_H &=& -8 \l_0 \frac{m^2_{Z_{0}}}{m^2_{H_0}}  \int {\frac{{{d^4}k}}{{{{\left( {2\pi } \right)}^4}}}
\frac{ik^\m k^\n}{{{{ {k} }^2} - m_{\chi_0}^2}}\frac{{\left( { - {g_{\m \n }} + \frac{{{( 1-\xi  )\left( {k + p} \right)_\m }
{\left( {k + p} \right)_\n }}}{{{\left( {k + p} \right)^2 - \xi m_{Z_0}^2}}}} \right)}}{{\left( {{\left( {k + p} \right)^2} - m_{Z_0}^2} \right)}}} \nonumber\\
&=& - 8 \l_0 \frac{m^2_{Z_{0}}}{m^2_{H_0}} g_{\m\n} B^{\m\n}(p,m_{\chi_0},m_{Z_0}) \nonumber\\
&+& 8(1-\xi) \l_0\frac{m^2_{Z_{0}}}{m^2_{H_0}}  \int {\frac{{{d^4}k}}{{{{\left( {2\pi } \right)}^4}}}\frac{-i 
\left( k^4 + 2k^2 k\cdot p + (k \cdot p)^2 \right)}{{{{\left({k^2} - m_{\chi_0}^2\right)\left({\left( {k + p} \right)^2} - m_{Z_0}^2\right)(\left( {k + p} \right)}^2} - m_{\chi_0}^2)}}{{}}{{}}}  \nonumber\\
\eea
where $(n_O,n_l,\ell_1, \ell_2, v_1)=(2 \times 1,1,1,1,2)$ and $ {\cal S}^{7}_{{\cal {M}}_H } = 1$.
Standard steps allow us to write this as
\bea\label{full MH8}
{\cal M}^{R_\xi, \chi Z}_H  &=& 8 \l_0 \frac{m^2_{Z_{0}}}{m^2_{H_0}}  {\mu ^{\ve}}  \Biggl \{ - g_{\m\n} B^{\m\n}(p,m_{\chi_0},m_{Z_0}) \nonumber\\
&+& (1-\xi )\Bigl \{ g_{\m\n} B^{\m\n}(p,m_{\chi_0},m_{\chi_0}) + (m_{Z_0}^2 - p^2) g_{\m\n} C^{1\m\n}(p,m_{Z_0},m_{\chi_0},m_{Z_0}) \nonumber\\
&+&p_\m p_\n C^{1\m\n}(p,m_{\chi_0},m_{Z_0},m_{\chi_0}) \Bigr \} \Biggr \}\, , 
\eea
where $C^1_{\m\n}$ is defined in \eq{explicit Cmn1} in Appendix \ref{PassarinoVeltman}.

\subsection{Unitary gauge diagrams}\label{UGD}
Similarly with the previous subsection, we consider first the Higgs tadpoles starting from
\vskip .5cm
\begin{center}
\begin{tikzpicture}[scale=0.7]
\draw [dashed,thick] (0,0) circle [radius=1];
\draw [dashed,thick] (-2.5,0)--(-1,0);
\draw [->]  (-2.2,0.2)--(-1.5,0.2);
\node at (-2.9,0) {$p$};
\draw [->]  (0.45,-0.6) arc [start angle=-45, end angle=-135, radius=0.6cm];
\node at (0,-0.4) {$k$};
\node at (3,0) {$=\, \, i{\cal T}^{U,\phi}_H$};
\end{tikzpicture}
\end{center}
\begin{eqnarray}\label{Th1}
 {\cal T}^{U,\phi}_H &=& -  6 \frac{1}{2} \sqrt{\frac{\lambda_0}{2}}m_{H_0}\int {\frac{{{d^4}k}}{{{{\left( {2\pi } \right)}^4}}}\frac{i}{{\left( {{k^2} - m_{H_0}^2} \right)}}} 
\end{eqnarray}
and in DR is
\begin{equation*}
{\cal T}^{U,\phi}_H = 3\frac{{\sqrt{\frac{\lambda_0}{2}}m_{H_0}{\mu ^{{\ve}}}}}{{{{\left( {4\pi } \right)}^{d/2}}}}\int {\frac{{{d^d}k}}{{i{\pi ^{d/2}}}}\frac{1}{{\left( {{k^2} - m_{H_0}^2} \right)}}}
\end{equation*}
or simply
\begin{equation}\label{final Tad1}
{\cal T}^{U,\phi}_H={3  \sqrt{\frac{\lambda_0}{2}}m_{H_0}{\mu ^{{\ve}}}}A_0(m_{H_0}).
\end{equation}
Next is the gauge tadpole is
\vskip .5cm
\begin{center}
\begin{tikzpicture}[scale=0.7]
\draw [photon] (0,0) circle [radius=1];
\draw [dashed,thick] (-2.5,0)--(-1,0);
\draw [->]  (-2.2,0.2)--(-1.5,0.2);
\node at (-2.9,0) {$p$};
\draw [->]  (0.45,-0.5) arc [start angle=-45, end angle=-135, radius=0.6cm];
\node at (0,-0.3) {$k$};
\node at (3,0) {$=\, \, i{\cal T}^{U,Z}_H$};
\end{tikzpicture}
\end{center}
\begin{eqnarray}
{\cal T}^{U,Z}_H =d \frac{\sqrt{2 \l_0} m_{Z_0}^2}{m_{H_0}}\int {\frac{{{d^4}k}}{{{{\left( {2\pi } \right)}^4}}}\frac{{ - i}}{{{k^2} - m_{Z_0}^2}}} 
+ \frac{{\sqrt{2 \l_0} }}{{m_{H_0}}}\int {\frac{{{d^4}k}}{{{{\left( {2\pi } \right)}^4}}}\frac{{i{k^2}}}{{{k^2} - m_{Z_0}^2}}} \label{HiggsTad}
\end{eqnarray}
where we have expanded the numerator, used that $g_{\m\n} g^{\m\n} = d $ and that under the integral $k^\m k^\n = \frac{g_{\m\n}}{d} k^2$.
In DR 
\be\label{final Tad2}
{\cal T}^{U,Z}_H = {\mu^{\ve}}\left( d \frac{\sqrt{2 \l_0} m_{Z_0}^2}{m_{H_0}}A_0(m_{Z_0}) 
- { \frac{\sqrt{2 \l_0} }{m_{H_0}}{}}U_{\cal T}(m_{Z_0}) \right)\, .
\ee
Using Eq.\eqref{gmnJmn1} to calculate ${U}_{\cal T}(m_{Z_0})$, this becomes
\be\label{final TadH}
{\cal T}^U_H={\mu^{\ve}} \Biggl ( { 3\sqrt{\frac{\lambda_0}{2}}m_{H_0}{}{ }}A_0(m_{H_0}) +  3\frac{\sqrt{2 \l_0} m_{Z_0}^2}{m_{H_0}}A_0(m_{Z_0})  \Biggr ).
\ee


Next, we present the explicit calculation of the one-loop $Z$ boson mass corrections starting from,
\vskip .5cm
\begin{center}
\begin{tikzpicture}[scale=0.7]
\draw [photon] (0,0)--(1.8,0);
\draw [dashed,thick] (0,0.9) circle [radius=0.9];
\draw [photon] (-1.8,0)--(-0,0);
\draw [->]  (-1.7,0.2)--(-1,0.2);
\node at (-2.2,0.2) {$p$};
\draw [->]  (0.45,0.3) arc [start angle=-45, end angle=-135, radius=0.6cm];
\node at (0,0.5) {$k$};
\draw [->]  (1.7,0.2)--(1,0.2);
\node at (4.3,0) {$=\, \, i{\cal M}^{U,\phi}_{Z,\mu\nu}$};
\end{tikzpicture}
\end{center}
\begin{eqnarray}
{\cal M}^{U,\phi}_{Z,\mu\nu}&=& -2{{g_{\mu \nu }}}\frac{m^2_{Z_0}}{m^2_{H_0}} \l_0 \int {\frac{{{d^4}k}}{{{{\left( {2\pi } \right)}^4}}}\frac{-i}{{{k^2} - m_{H_0}^2}}}
\end{eqnarray}
which is equal to
\begin{eqnarray}\label{M1Z2}
{\cal M}^{U,\phi}_{Z,\mu\nu} &=& -2{{g_{\mu \nu }}}\frac{m^2_{Z_0}}{m^2_{H_0}} \l_0 \mu^{\ve} A_0(m_{H_0})\, .
\end{eqnarray}
Next is the Higgs sunset
\vskip .5cm
\begin{center}
\begin{tikzpicture}[scale=0.7]
\draw [photon] (-2.3,0)--(2.3,0);
\draw [->]  (-1.9,0.2)--(-1.2,0.2);
\node at (-2.5,0.2) {$p$};
\draw [<-]  (0.45,0.7) arc [start angle=45, end angle=135, radius=0.6cm];
\node at (0,1.3) {$k+p$};
\draw [<-] (-0.4,0.2)--(0.4,0.2);
\node at (0,-0.3) {$k$};
\draw [->]  (1.9,0.2)--(1.2,0.2);
\node at (4,0) {$=\,\, i{\cal M}^{U,\phi Z}_{Z,\mu\nu}$ };
\draw  [dashed,thick] (-1,0) .. controls (-1,0.555) and (-0.555,1) .. (0,1)
.. controls (0.555,1) and (1,0.555) .. (1,0);
\end{tikzpicture}
\end{center}
\begin{eqnarray}\label{M2Z2}
{\cal M}^{U,\phi Z}_{Z,\mu\nu}&=& -8{g_{\mu\alpha}}{g_{\nu\b}}\frac{m^4_{Z_0}}{m^2_{H_0}} \l_0\int {\frac{{{d^4}k}}{{{{\left( {2\pi } \right)}^4}}}\frac{i}{{{{\left( {k + p} \right)}^2} - m_{H_0}^2}}
\frac{{ { - {g^{\alpha \beta }} + \frac{{{k^\alpha }{k^\beta }}}{{{m_{Z_0}^2}}}} }}{{\left( {{k^2} - m_{Z_0}^2} \right)}}} 
\end{eqnarray}
translating in DR to
\bea\label{final M2Z}
{\cal M}^{U,\phi Z}_{Z,\mu\nu} & =&  - 8{g_{\mu \nu }}{{}\frac{m^4_{Z_0}}{m^2_{H_0}} \l_0 \mu^{\ve}}B_0(p,m_{Z_0},m_{H_0}) + 8 \frac{m^2_{Z_0}}{m^2_{H_0}} \l_0\mu^{\ve}B_{\m\n}(p,m_{Z_0},m_{H_0}) \, . 
\eea
The sum of these two corrections is
\bea\label{full MZ2}
{\cal M}^{U}_{Z,\mu\nu}  &=& \frac{m^2_{Z_0}}{m^2_{H_0}} \l_0\mu^{\ve}\Biggl \{- 8g_{\m\n} {}{{}m_{Z_0}^2}B_0(p,m_{Z_0},m_{H_0}) - 2 g_{\m\n} A_0(m_{H_0}) \nonumber\\
&+& 8 B_{\m\n}(p,m_{Z_0},m_{H_0})  \Biggr \}\, . 
\eea

Now, we deal with the one-loop corrections of the Higgs mass in Unitary gauge. The first comes from
\vskip .5cm
\begin{center}
\begin{tikzpicture}[scale=0.7]
\draw [dashed,thick] (0,0)--(1.8,0);
\draw [photon] (0,0.9) circle [radius=0.9];
\draw [dashed,thick] (-1.8,0)--(-0,0);
\draw [->]  (-1.7,0.2)--(-1,0.2);
\node at (-2.2,0.2) {$p$};
\draw [->]  (0.45,0.5) arc [start angle=-45, end angle=-135, radius=0.6cm];
\node at (0,0.75) {$k$};
\draw [->]  (1.7,0.2)--(1,0.2);
\node at (4.5,0) {$=\, \,i{\cal M}^{U,Z}_H$};
\end{tikzpicture}
\end{center}
\begin{eqnarray}
{\cal M}_H^{U,Z}=\frac{1}{2}4 d \frac{ \l_0 m_{Z_0}^2}{m_{H_0}^2}\int {\frac{{{d^4}k}}{{{{\left( {2\pi } \right)}^4}}}\frac{{ - i}}{{{k^2} - m_{Z_0}^2}}} 
+ \frac{1}{2} 4 \frac{{\frac{ \l_0 m_{Z_0}^2}{m_{H_0}^2}}}{{m_{Z_0}^2}}\int {\frac{{{d^4}k}}{{{{\left( {2\pi } \right)}^4}}}\frac{{i{k^2}}}{{{k^2} - m_{Z_0}^2}}} . \nonumber\\
\end{eqnarray}
In DR it becomes
\bea\label{final M1H2}
{\cal M}_H^{U,Z}&=& \m^{\ve}\Bigl \{ {2 d \frac{ \l_0 m_{Z_0}^2}{m_{H_0}^2}}A_0(m_{Z_0}) - 2 \frac{ \l_0}{m_{H_0}^2} U_{\cal T}(1, m_{Z_0}) \Bigr \} \nonumber\\
&=& \m^{\ve} {6\frac{ \l_0 m_{Z_0}^2}{m_{H_0}^2}}A_0(m_{Z_0}) \, .
\eea
Next is
\vskip .5cm
\begin{center}
\begin{tikzpicture}[scale=0.7]
\draw [dashed,thick] (0,0)--(1.8,0);
\draw [dashed,thick] (0,0.9) circle [radius=0.9];
\draw [dashed,thick] (-1.8,0)--(-0,0);
\draw [->]  (-1.7,0.2)--(-1,0.2);
\node at (-2.2,0.2) {$p$};
\draw [->]  (0.45,0.4) arc [start angle=-45, end angle=-135, radius=0.6cm];
\node at (0,0.5) {$k$};
\draw [->]  (1.7,0.2)--(1,0.2);
\node at (4.5,0) {$=\, \, i{\cal M}^{U,\phi}_H$};
\end{tikzpicture}
\end{center}
\begin{equation}\label{M2H}
\bal
{\cal M}^{U,\phi}_H &= \frac{1}{2} 6 \l _0\int {\frac{{{d^4}k}}{{{{\left( {2\pi } \right)}^4}}}\frac{-i}{{{k^2} - m_{H_0}^2}}}.
\eal
\end{equation}
In DR,
\be\label{final M2H2}
{\cal M}_H^{U,\phi}={3\l_0  {\mu^{\ve}}{}}A_0(m_{H_0}).
\ee 
The Higgs vacuum polarization diagram
\vskip .5cm
\begin{center}
\begin{tikzpicture}[scale=0.7]
\draw [dashed,thick] (0.9,0)--(2.2,0);
\draw [dashed,thick] (0,0) circle [radius=0.9];
\draw [dashed,thick] (-2.2,0)--(-0.9,0);
\draw [->]  (-1.9,0.2)--(-1.2,0.2);
\node at (-2.6,0.2) {$p$};
\draw [<-]  (0.45,0.6) arc [start angle=45, end angle=135, radius=0.6cm];
\node at (0,1.2) {$k+p$};
\draw [->]  (0.45,-0.6) arc [start angle=-45, end angle=-135, radius=0.6cm];
\node at (0,-1.2) {$k$};
\draw [->]  (1.9,0.2)--(1.2,0.2);
\node at (4.5,0) {$=\, \, i{\cal M}^{U,\phi\phi}_H$};
\end{tikzpicture}
\end{center}
\bea
{\cal M}^{U,\phi\phi}_H&= & \frac{1}{2} 18 {{\lambda }m_{H_0}^2}\int {\frac{{{d^4}k}}{{{{\left( {2\pi } \right)}^4}}}\frac{-i}{\left({{k^2} - m_{H_0}^2}\right){\left( {{{\left( {k + p} \right)}^2} - m_{H_0}^2} \right)}}}
\eea
in DR is equal to
\bea\label{final M3H2}
{\cal M}^{U,\phi\phi}_H&= &{9 \l_0}m_{H_0}^2{\mu ^{\ve}}B_0(p,m_{H_0},m_{H_0})\, .
\eea
The corresponding gauge loop is
\vskip .5cm
\begin{center}
\begin{tikzpicture}[scale=0.7]
\draw [dashed,thick] (0.9,0)--(2.2,0);
\draw [photon] (0,0) circle [radius=0.9];
\draw [dashed,thick] (-2.2,0)--(-0.9,0);
\draw [->]  (-1.9,0.2)--(-1.2,0.2);
\node at (-2.6,0.2) {$p$};
\draw [<-]  (0.45,0.6) arc [start angle=45, end angle=135, radius=0.6cm];
\node at (0,1.3) {$k+p$};
\draw [->]  (0.45,-0.4) arc [start angle=-45, end angle=-135, radius=0.6cm];
\node at (0,-1.1) {$k$};
\draw [->]  (1.9,0.2)--(1.2,0.2);
\node at (4.5,0) {$=\, \, i{\cal M}^{U,ZZ}_H$};
\end{tikzpicture}
\end{center}
\bea\label{M4_H}
{\cal M}^{U,ZZ}_H&= & \frac{1}{2} 8 {g^{\mu \nu }}{g^{\alpha \beta }}\frac{\l_0 m^4_{Z_0}}{m_{H_0}^2}\int {\frac{{{d^4}k}}{{{{\left( {2\pi } \right)}^4}}}
\frac{{-i\left( { - {g_{\mu \alpha }} + \frac{{{k_\mu }{k_\alpha }}}{{{m_{Z_0}^2}}}} \right)}}{{\left( {{k^2} - m_{Z_0}^2} \right)}}\frac{{\left( { - {g_{\nu \b }} + 
\frac{{{{\left( {k + p} \right)}_\nu }{{\left( {k + p} \right)}_\b }}}{{{m_{Z_0}^2}}}} \right)}}{{\left( {{{\left( {k + p} \right)}^2} - m_{Z_0}^2} \right)}} } .
\eea
Expanding the numerator it becomes
 \bea\label{M4_H2}
{\cal M}^{U,ZZ}_H &=& d \frac{\l_0 m^4_{Z_0}}{m_{H_0}^2}\int {\frac{{{d^4}k}}{{{{\left( {2\pi } \right)}^4}}}\frac{-4i}{\left({{k^2} - m_{Z_0}^2}\right){\left( {{{\left( {k + p} \right)}^2} - m_{Z_0}^2} \right)}}} \nonumber\\
&+& 4 \frac{\l_0 m^2_{Z_0}}{m_{H_0}^2}\int {\frac{{{d^4}k}}{{{{\left( {2\pi } \right)}^4}}}\frac{i{{k^2 } } }{\left({{k^2} - m_{Z_0}^2}\right){\left( {{{\left( {k + p} \right)}^2} - m_{Z_0}^2} \right)}}} \nonumber\\
&+& 4 \frac{\l_0 m^2_{Z_0}}{m_{H_0}^2}\int {\frac{{{d^4}k}}{{{{\left( {2\pi } \right)}^4}}}\frac{i{{(k  +  p)^2 } } }{\left({{k^2} - m_{Z_0}^2}\right){\left( {{{\left( {k + p} \right)}^2} - m_{Z_0}^2} \right)}}} \nonumber\\
&- &4 \frac{ \l_0 }{m_{H_0}^2}\int {\frac{{{d^4}k}}{{{{\left( {2\pi } \right)}^4}}}\frac{ik^2 (k + p)^2}{\left({{k^2} - m_{Z_0}^2}\right){\left( {{{\left( {k + p} \right)}^2} - m_{Z_0}^2} \right)}}} \nonumber\\
&+ &4 \frac{ \l_0 }{m_{H_0}^2}\int {\frac{{{d^4}k}}{{{{\left( {2\pi } \right)}^4}}}\frac{ip^2 k^2}{\left({{k^2} - m_{Z_0}^2}\right){\left( {{{\left( {k + p} \right)}^2} - m_{Z_0}^2} \right)}}} \nonumber\\
&- &4 \frac{ \l_0 }{m_{H_0}^2}\int {\frac{{{d^4}k}}{{{{\left( {2\pi } \right)}^4}}}\frac{i(k \cdot p)^2}{\left({{k^2} - m_{Z_0}^2}\right){\left( {{{\left( {k + p} \right)}^2} - m_{Z_0}^2} \right)}}}
 \eea
and in DR
\bea\label{final M4H2}
{\cal M}^{U,ZZ}_H &=& \mu ^{\ve}{}\Biggl \{ 4 d \frac{\l_0 m^4_{Z_0}}{m_{H_0}^2} B_0(p,m_{Z_0},m_{Z_0}) \nonumber\\
&-& 4 \frac{\l_0 m^2_{Z_0}}{m_{H_0}^2}{g_{\mu\nu}}B_{k+p}^{\mu\nu}(p,m_{Z_0},m_{Z_0})
+ 4 \frac{ \l_0 }{m_{H_0}^2}m_{Z_0}^2 A_0(m_{Z_0}) + 4 \frac{ \l_0 }{m_{H_0}^2}m_{Z_0}^4  \nonumber\\
&-&4 \frac{ \l_0 }{m_{H_0}^2} p^2 g_{\m\n}B^{\mu\nu}(p,m_{Z_0},m_{Z_0}) + 4 \frac{ \l_0 }{m_{H_0}^2} p_\m p_\n B^{\mu\nu}(p,m_{Z_0},m_{Z_0})    \Biggr \} \, ,
\eea
where we have defined
\bea
g_{\m\n}B_{k+p}^{\m\n}(p,m_{Z_0},m_{Z_0}) = \int {\frac{{{d^4}k}}{{{{\left( {2\pi } \right)}^4}}}\frac{-i{{(k  +  p)^2 } } }{\left({{k^2} - m_{Z_0}^2}\right){\left( {{{\left( {k + p} \right)}^2} - m_{Z_0}^2} \right)}}}\, .
\eea
Adding up \eq{final M1H2}, \eq{final M2H2}, \eq{final M3H2} and \eq{final M4H2} we obtain
\begin{eqnarray}\label{full MHUn}
{\cal M}^U_H(p)& = &\mu ^{\ve} \Biggl \{ {6}\frac{\l m^2_{Z_0}}{m_{H_0}^2}A_0(m_{Z_0}) + \frac{2\l m^4_{Z_0}}{m_{H_0}^2} + {3\l_0{{}}}{{{}}}A_0(m_{H_0}) \nonumber \\
 &+& 9\l_0 m_{H_0}^2{}B_0(p,m_{H_0},m_{H_0}) + \frac{\l_0 m^2_{Z_0}}{m_{H_0}^2}\Big \{ 4 d m_{Z_0}^2 B_0(p,m_{Z_0},m_{Z_0}) \nonumber\\
&-& 4{ {}}{{}}{g_{\mu\nu}}B_{k+p}^{\mu\nu}(p,m_{Z_0},m_{Z_0}) + 4 A_0(m_{Z_0})  - 4 \frac{p^2}{m_{Z_0}^2}{ {}}{{}}{g_{\mu\nu}}B^{\mu\nu}(p,m_{Z_0},m_{Z_0}) \nonumber\\
&+& 4 m_{Z_0}^2 + 4 \frac{p_\m p_\n}{m_{Z_0}^2} B^{\mu\nu}(p,m_{Z_0},m_{Z_0})    \Big \} \Biggr \}.
\end{eqnarray}


\section{Finite parts}
\label{FiniteParts}

In this Appendix we present the explicit form of the finite diagrams along with the finite parts of the divergent ones, in both $R_\xi$ and Unitary gauges. 
Since the corresponding expressions are quite long, we use for simplicity some shorthand notation. In particular, for the arguments of integrals, we define 
\bea
D4 \equiv (D_1,D_2,D_3,D_4).
\eea
In addition, in agreement with our notation in Appendix \ref{FiniteIntegrals}, we define the following integrals 
\bea
E^{\m,\m\n,\m\n\a,\m\n\a\b} &=& 4! \int_0^1 {\cal D}^5 x \left\langle  \frac{N_{E^{\m,\m\n,\m\n\a,\m\n\a\b}}^{(5)}
(k-\sum_{i=1}^{5} P_ix_{i+1})}{D_1D_2D_3D_4D_5} \right\rangle \nonumber\\
F^{\m,\m\n,\m\n\a,\m\n\a\b,\m\n\a\b\gamma\delta} &=& 5! \int_0^1 {\cal D}^6 x \left\langle  
\frac{N_{F^{\m,\m\n,\m\n\a,\m\n\a\b,\m\n\a\b\gamma\delta}}^{(6)}(k-\sum_{i=1}^{6} P_ix_{i+1})}{D_1D_2D_3D_4D_5D_6} \right\rangle \nonumber\\
G^{\m,\m\n,\m\n\a,\m\n\a\b,\m\n\a\b\gamma\delta,\m\n\a\b\gamma\delta\epsilon} &=& 6! \int_0^1 {\cal D}^7 x \left\langle  
\frac{N_{G^{\m,\m\n,\m\n\a,\m\n\a\b,\m\n\a\b\gamma\delta,\m\n\a\b\gamma\delta\epsilon}}^{(7)}(k-\sum_{i=1}^{7} P_ix_{i+1})}{D_1D_2D_3D_4D_5D_7} \right\rangle \nonumber\\
H^{\m,\m\n,\m\n\a,\m\n\a\b,\m\n\a\b\gamma\delta,\m\n\a\b\gamma\delta\epsilon,\m\n\a\b\gamma\delta\epsilon\theta} 
&=& 7! \int_0^1 {\cal D}^8 x \left\langle  \frac{N_{H^{\m,\m\n,\m\n\a,\m\n\a\b,\m\n\a\b\gamma\delta\epsilon\theta}}^{(8)}(k-\sum_{i=1}^{8} P_ix_{i+1})}{D_1D_2D_3D_4D_5D_8} \right\rangle \nonumber\\
\eea

\subsection{$R_\xi$ gauge}

In the $R_\xi$ gauge we have the following finite parts:

Finite parts of the Triangle diagrams 
\bea
( K^{R_\xi, ZZZ}_{H})_f &=& 16\sqrt{2}\frac{m^6_{Z_0} \l_0^{3/2}}{m^3_{H_0}}   \m^{\ve}\Biggl \{ (4- \varepsilon) C_0(P_1,m_{Z_0},m_{Z_0},m_{Z_0}) \nonumber\\
&-& (1-\xi) \Bigl \{ g_{\m\n} D^{\m\n}(P_1,P_2,m_{Z_0},m_{Z_0},m_{Z_0},m_{\chi_0})  \nonumber\\
&+& g_{\m\n} D^{\m\n}(P_1,P_2,m_{Z_0},m_{Z_0},m_{Z_0},m_{\chi_0})  \nonumber\\
&+& g_{\m\n} D^{\m\n}(P_1,P_2,m_{Z_0},m_{Z_0},m_{Z_0},m_{\chi_0}) + 2 P_1^\m D^{\m}(P_1,P_2,m_{Z_0},m_{Z_0},m_{Z_0},m_{\chi_0})  \nonumber\\
&+&2 P_2^\m D^{\m}(P_1,P_2,m_{Z_0},m_{Z_0},m_{Z_0},m_{\chi_0})  +  P_1^2 D_0(P_1,P_2,m_{Z_0},m_{Z_0},m_{Z_0},m_{\chi_0}) \nonumber\\
&+&P_2^2 D_0 (P_1,P_2,m_{Z_0},m_{Z_0},m_{Z_0},m_{\chi_0})     \Bigr \}\nonumber\\
&+&(1-\xi)^2 \Bigl \{ E_4(D4,D_5(0,m_{\chi_0})) + E_4(D4,D_5(P_1,m_{\chi_0})) \nonumber\\
&+& E_4(D_1,D_2,D_3,D_4(0,m_{\chi_0}),D_5(P_1,m_{\chi_0}))\nonumber\\
&+& 2 P_{1\m} E_3^\m(D4,D_5(0,m_{\chi_0})) + 2 P_{1\m} E_3^\m(D_1,D_2,D_3,D_4(0,m_{\chi_0}),D_5(P_1,m_{\chi_0})) \nonumber\\
&+&2 P_{2\m} E_3^\m(D4,D_5(P_1,m_{\chi_0})) + 2 P_{2\m} E_3^\m(D_1,D_2,D_3,D_4(0,m_{\chi_0}),D_5(P_1,m_{\chi_0})) \nonumber\\
&+&2P_{1\m}P_{2\n} E_2^{\m\n}(D_1,D_2,D_3,D_4(0,m_{\chi_0}),D_5(P_1,m_{\chi_0})) \nonumber\\
&+& P_{2\m}P_{2\n}E_2^{\m\n}(D4,D_5(P_1,m_{\chi_0})) \nonumber\\
&+&P_{2\m}P_{2\n}E_2^{\m\n}(D_1,D_2,D_3,D_4(0,m_{\chi_0}),D_5(P_1,m_{\chi_0})) \nonumber\\
&+& 2P_1 \cdot P_2 E_2(D_1,D_2,D_3,D_4(0,m_{\chi_0}),D_5(P_1,m_{\chi_0})) \nonumber\\
&+& 2P_1 \cdot P_2 (P_1+P_2)_\m E_1^{\m}(D_1,D_2,D_3,D_4(0,m_{\chi_0}),D_5(P_1,m_{\chi_0}))  \nonumber\\
&+&  (P_1 \cdot P_2)^2E_0(D_1,D_2,D_3,D_4(0,m_{\chi_0}),D_5(P_1,m_{\chi_0}))  \Bigr \} \nonumber\\
&-& (1-\xi)^3 \Bigl \{ F_6(D4,D_5(0,m_{\chi_0},D_6(P_1,m_{\chi_0})) \nonumber\\
&+& 2(P_1 + P_2)_\m F_5^\m (D4,D_5(0,m_{\chi_0},D_6(P_1,m_{\chi_0})) \nonumber\\
&+&(P_{1\m}P_{1\n} +3 P_{1\m} P_{2\n} + P_{2\m}P_{2\n}) F_4^{\m\n}(D4,D_5(0,m_{\chi_0},D_6(P_1,m_{\chi_0})) \nonumber\\
&+&(P_{1\m}P_{1\n}P_{2\a} + P_{2\m}P_{2\n}P_{1\a}) F_3^{\m\n\a}(D4,D_5(0,m_{\chi_0},D_6(P_1,m_{\chi_0})) \nonumber\\
&+& P_1\cdot P_2(P_1 + P_2)_\m F_3^\m(D4,D_5(0,m_{\chi_0},D_6(P_1,m_{\chi_0})) \nonumber\\
&+& P_1\cdot P_2 P_{1\m} P_{2\n }F_2^{\m\n}(D4,D_5(0,m_{\chi_0},D_6(P_1,m_{\chi_0})) \nonumber\\
&+&  P_1 \cdot P_2 F_4(D4,D_5(0,m_{\chi_0},D_6(P_1,m_{\chi_0}))   \Bigr \}    \Biggr \}
\eea
Finite parts of the Box diagrams 
\bea
( B^{R_\xi, ZZZZ}_{H})_f  &=& 64  \frac{m_{Z_0}^8 \l_0^2}{m_{H_0}^4} \m^{\ve}\Biggl \{ d D_0(D_1,D_2,D_3,D_4) \nonumber\\
&-& (1-\xi) \Biggl ( E_2(D4,D_5(0,m_{\chi_0})) + E_2(D4,D_5(P_1,m_{\chi_0})) \nonumber\\
&+&E_2(D4,D_5(P_2,m_{\chi_0})) +E_2(D4,D_5(P_3,m_{\chi_0})) \nonumber\\
&+& 2 P_{1,\m}E_1^\m (D4,D_5(P_1,m_{\chi_0})) +2 P_{2,\m}E_1^\m (D4,D_5(P_2,m_{\chi_0}))  \nonumber\\
&+&2 P_{3,\m}E_1^\m (D4,D_5(P_1,m_{\chi_0})) + P_1^2 E_0(D4,D_5(P_1,m_{\chi_0})) \nonumber\\
&+& P_2^2 E_0(D4,D_5(P_2,m_{\chi_0})) + P_2^2 E_0(D4,D_5(P_3,m_{\chi_0}))      \Biggr) \nonumber\\
&+& (1-\xi)^2 \Biggl ( F_4(D4,D_5(0,m_{\chi_0}),D_6(P_1,m_{\chi_0})) +  F_4(D4,D_5(0,m_{\chi_0}),D_6(P_2,m_{\chi_0})) \nonumber\\
&+&  F_4(D4,D_5(P_1,m_{\chi_0}),D_6(P_2,m_{\chi_0})) +  F_4(D4,D_5(P_1,m_{\chi_0}),D_6(P_3,m_{\chi_0})) \nonumber\\
&+& F_4(D4,D_5(0,m_{\chi_0}),D_6(P_3,m_{\chi_0})) + 2P_{1,\m}F_3^\m(D4,D_5(0,m_{\chi_0}),D_6(P_3,m_{\chi_0})) \nonumber\\
&+&2P_{1,\m}F_3^\m(D4,D_5(P_1,m_{\chi_0}),D_6(P_3,m_{\chi_0})) + 2P_{2,\m}F_3^\m(D4,D_5(0,m_{\chi_0}),D_6(P_2,m_{\chi_0})) \nonumber\\
&+&2P_{2,\m}F_3^\m(D4,D_5(P_1,m_{\chi_0}),D_6(P_3,m_{\chi_0})) + 2P_{1,\m}F_3^\m(D4,D_5(P_1,m_{\chi_0}),D_6(P_2,m_{\chi_0})) \nonumber\\
&+&2P_{2,\m}F_3^\m(D4,D_5(P_2,m_{\chi_0}),D_6(P_3,m_{\chi_0})) + 2P_{3,\m}F_3^\m(D4,D_5(0,m_{\chi_0}),D_6(P_3,m_{\chi_0})) \nonumber\\
&+&2P_{3,\m}F_3^\m(D4,D_5(P_1,m_{\chi_0}),D_6(P_3,m_{\chi_0})) + 2P_{3,\m}F_3^\m(D4,D_5(P_2,m_{\chi_0}),D_6(P_3,m_{\chi_0})) \nonumber\\
&+&  P_{1,\m}P_{1,\n} \Bigl \{ F_2^{\m\n}(D4,D_5(0,m_{\chi_0}),D_6(P_1,m_{\chi_0})) + F_2^{\m\n}(D4,D_5(P_1,m_{\chi_0}),D_6(P_2,m_{\chi_0})) \nonumber\\
&+& F_2^{\m\n}(D4,D_5(P_1,m_{\chi_0}),D_6(P_3,m_{\chi_0})) \Bigr \} +  2P_{1,\m}P_{2,\n}  F_2^{\m\n}(D4,D_5(0,m_{\chi_0}),D_6(P_1,m_{\chi_0}))\nonumber\\
&+&  P_{2,\m}P_{2,\n} \Bigl \{ F_2^{\m\n}(D4,D_5(0,m_{\chi_0}),D_6(P_1,m_{\chi_0})) + F_2^{\m\n}(D4,D_5(P_1,m_{\chi_0}),D_6(P_2,m_{\chi_0})) \nonumber\\
&+& F_2^{\m\n}(D4,D_5(P_1,m_{\chi_0}),D_6(P_3,m_{\chi_0})) \Bigr \} \nonumber\\
&+&  P_{3,\m}P_{3,\n} \Bigl \{ F_2^{\m\n}(D4,D_5(0,m_{\chi_0}),D_6(P_1,m_{\chi_0})) + F_2^{\m\n}(D4,D_5(P_1,m_{\chi_0}),D_6(P_2,m_{\chi_0})) \nonumber\\
&+& F_2^{\m\n}(D4,D_5(P_1,m_{\chi_0}),D_6(P_3,m_{\chi_0})) \Bigr \} + 2P_{1,\m}P_{2,\n}  F_2^{\m\n}(D4,D_5(0,m_{\chi_0}),D_6(P_1,m_{\chi_0})) \nonumber\\
&+& 2P_{1,\m}P_{3,\n}  F_2^{\m\n}(D4,D_5(P_1,m_{\chi_0}),D_6(P_3,m_{\chi_0})) \nonumber\\
&+& 2P_{2,\m}P_{3,\n}  F_2^{\m\n}(D4,D_5(P_2,m_{\chi_0}),D_6(P_3,m_{\chi_0})) \nonumber\\
&+&2\Bigl \{ P_1 \cdot P_2 F_2(D4,D_5(P_1,m_{\chi_0}),D_6(P_2,m_{\chi_0})) \nonumber\\
&+& P_1 \cdot P_3 F_2(D4,D_5(P_1,m_{\chi_0}),D_6(P_3,m_{\chi_0})) \nonumber\\
&+& P_2 \cdot P_3 F_2(D4,D_5(P_2,m_{\chi_0}),D_6(P_3,m_{\chi_0})) \Bigr \} \nonumber\\
&+&  2 P_1 \cdot P_2 \Bigl \{ P_{1,\m} + P_{2,\m}  \Bigr \}F_1^\m(D4,D_5(P_1,m_{\chi_0}),D_6(P_2,m_{\chi_0})) \nonumber\\
&+&  2 P_1 \cdot P_3 \Bigl \{ P_{1,\m} + P_{3,\m}  \Bigr \}F_1^\m(D4,D_5(P_1,m_{\chi_0}),D_6(P_3,m_{\chi_0})) \nonumber\\  
&+&  2 P_2 \cdot P_3 \Bigl \{ P_{2,\m} + P_{3,\m}  \Bigr \}F_1^\m(D4,D_5(P_2,m_{\chi_0}),D_6(P_3,m_{\chi_0})) \nonumber\\ 
&+&  (P_1 \cdot P_2)^2  F_0(D4,D_5(P_1,m_{\chi_0}),D_6(P_2,m_{\chi_0})) \nonumber\\
&+&  (P_1 \cdot P_3)^2  F_0(D4,D_5(P_1,m_{\chi_0}),D_6(P_3,m_{\chi_0})) \nonumber\\
&+&   (P_2 \cdot P_3)^2  F_0(D4,D_5(P_2,m_{\chi_0}),D_6(P_3,m_{\chi_0}))     \Biggr) \nonumber\\
&-&(1-\xi)^3 \Biggl(  G_6(D4,D_5(0,m_{\chi_0}),D_6(P_1,m_{\chi_0}),D_7(P_2,m_{\chi_0})) \nonumber\\
&+&  G_6(D4,D_5(0,m_{\chi_0}),D_6(P_1,m_{\chi_0}),D_7(P_3,m_{\chi_0})) \nonumber\\
&+&G_6(D4,D_5(P_1,m_{\chi_0}),D_6(P_2,m_{\chi_0}),D_7(P_3,m_{\chi_0})) \nonumber\\
&+&\sum_{l=1}^3 2 P_{l,\m}G^\m_5(D4,D_5(0,m_{\chi_0}),D_6(P_1,m_{\chi_0}),D_7(P_2,m_{\chi_0})) \nonumber\\
&+&\sum_{l=1}^3 2 P_{l,\m}G^\m_5(D4,D_5(0,m_{\chi_0}),D_6(P_1,m_{\chi_0}),D_7(P_3,m_{\chi_0})) \nonumber\\
&+&\sum_{l=1}^3 2 P_{l,\m}G^\m_5(D4,D_5(P_1,m_{\chi_0}),D_6(P_2,m_{\chi_0}),D_7(P_3,m_{\chi_0}))\nonumber\\
&+&\sum_{l,m=1}^3 2  P_{l,\m} P_{m,\n} G^{\m\n}_4(D4,D_5(0,m_{\chi_0}),D_6(P_1,m_{\chi_0}),D_7(P_2,m_{\chi_0})) \nonumber\\
&+& \sum_{l,m=1}^3 2 P_{l,\m} P_{m,\n} G^{\m\n}_4(D4,D_5(0,m_{\chi_0}),D_6(P_1,m_{\chi_0}),D_7(P_3,m_{\chi_0})) \nonumber\\
&+& \sum_{l,m=1}^3 2  P_{l,\m} P_{m,\n} G^{\m\n}_4(D4,D_5(P_1,m_{\chi_0}),D_6(P_2,m_{\chi_0}),D_7(P_3,m_{\chi_0}))\nonumber\\
&+&\sum_{l,m,n=1}^3  P_{l,\m} P_{m,\n} P_{n,\a} G^{\m\n\a}_3(D4,D_5(0,m_{\chi_0}),D_6(P_1,m_{\chi_0}),D_7(P_2,m_{\chi_0})) \nonumber\\
&+& \sum_{l,m,n=1}^3  P_{l,\m} P_{m,\n} P_{n,\a} G^{\m\n\a}_3(D4,D_5(0,m_{\chi_0}),D_6(P_1,m_{\chi_0}),D_7(P_3,m_{\chi_0})) \nonumber\\
&+&\sum_{l,m,n=1}^3  P_{l,\m} P_{m,\n} P_{n,\a} G^{\m\n\a}_3(D4,D_5(P_1,m_{\chi_0}),D_6(P_2,m_{\chi_0}),D_7(P_3,m_{\chi_0}))\nonumber\\
&+&\sum_{l,m=1, m \ne l }^3  P_l \cdot P_m G_4(D4,D_5(0,m_{\chi_0}),D_6(P_1,m_{\chi_0}),D_7(P_3,m_{\chi_0})) \nonumber\\
&+&\sum_{l,m=1, m \ne l }^3  P_l \cdot P_m G_4(D4,D_5(0,m_{\chi_0}),D_6(P_2,m_{\chi_0}),D_7(P_3,m_{\chi_0})) \nonumber\\
&+&\sum_{l=1}^3  P_{l,\m}G^\m_3(D4,D_5(0,m_{\chi_0}),D_6(P_1,m_{\chi_0}),D_7(P_2,m_{\chi_0})) \nonumber\\
&+&\sum_{l=1}^3  P_{l,\m}G^\m_3(D4,D_5(0,m_{\chi_0}),D_6(P_1,m_{\chi_0}),D_7(P_3,m_{\chi_0})) \nonumber\\
&+&\sum_{l=1}^3  P_{l,\m}G^\m_3(D4,D_5(0,m_{\chi_0}),D_6(P_2,m_{\chi_0}),D_7(P_3,m_{\chi_0})) \nonumber\\
&+&\sum_{l,m=1}^3  P_l \cdot P_m P_{m,\m}G^\m_3(D4,D_5(0,m_{\chi_0}),D_6(P_1,m_{\chi_0}),D_7(P_3,m_{\chi_0})) \nonumber\\
&+&\sum_{l,m=1}^3  P_l \cdot P_m P_{m,\m}G^\m_3(D4,D_5(0,m_{\chi_0}),D_6(P_2,m_{\chi_0}),D_7(P_3,m_{\chi_0})) \nonumber\\
&+& \sum_{l,m=1, m \ne l }^3 P_l \cdot P_m P_{l,\m} P_{m,\n}  G^{\m\n}_2(D4,D_5(0,m_{\chi_0}),D_6(P_1,m_{\chi_0}),D_7(P_3,m_{\chi_0})) \nonumber\\
&+& \sum_{l,m=1, m \ne l }^3 P_l \cdot P_m P_{l,\m} P_{m,\n}  G^{\m\n}_2(D4,D_5(P_1,m_{\chi_0}),D_6(P_2,m_{\chi_0}),D_7(P_3,m_{\chi_0}))\nonumber\\
&+& \sum_{l,m=1, m \ne l }^3 P_l \cdot P_m P_{l,\m} P_{m,\n}  G^{\m\n}_2(D4,D_5(0,m_{\chi_0}),D_6(P_1,m_{\chi_0}),D_7(P_3,m_{\chi_0})) \nonumber\\
&& \Bigl ( P_1 \cdot P_2 P_2 \cdot P_3 + P_2 \cdot P_3 P_1 \cdot P_3   \Bigr)G_2(D4,D_5(0,m_{\chi_0}),D_6(P_1,m_{\chi_0}),D_7(P_3,m_{\chi_0})) \nonumber\\
&+& \sum_{l,m,n=1, n \ne m \ne l }^3 P_l \cdot P_m P_m \cdot P_n ( P_{m} + P_{n})_\m G^{\m}_1(D4,D_5(P_1,m_{\chi_0}),D_6(P_2,m_{\chi_0}),D_7(P_3,m_{\chi_0}))\nonumber\\
&+& \sum_{l,m,n=1, n \ne m \ne l }^3 P_l \cdot P_m P_m \cdot P_n ( P_{m} + P_{n})_\m  G^{\m}_1(D4,D_5(0,m_{\chi_0}),D_6(P_1,m_{\chi_0}),D_7(P_3,m_{\chi_0})) \nonumber\\
&+& \sum_{l,m=1, m \ne l }^3 P_l \cdot P_m P_{l,\m} P_{m,\n}  G^{\m\n}_2(D4,D_5(P_1,m_{\chi_0}),D_6(P_2,m_{\chi_0}),D_7(P_3,m_{\chi_0}))\nonumber\\
&+& P_1 \cdot P_2 P_2 \cdot P_3  P_1 \cdot P_3 G_0(D4,D_5(P_1,m_{\chi_0}),D_6(P_2,m_{\chi_0}),D_7(P_3,m_{\chi_0})) \Biggr) \nonumber\\
&+&(1-\xi)^4 \Biggl (H_8(D4,D_5(0,m_{\chi_0}),D_6(P_1,m_{\chi_0}),D_7(P_2,m_{\chi_0}),D_8(P_3,m_{\chi_0})) \nonumber\\
&+& 2(P_1 + P_2 +P_3 )_\m H_7^\m(D4,D_5(0,m_{\chi_0}),D_6(P_1,m_{\chi_0}),D_7(P_2,m_{\chi_0}),D_8(P_3,m_{\chi_0})) \nonumber\\
&+& \Bigr \{ P_{1\m} P_{1\n} + 3P_{1\m} P_{2\n} + P_{2\m} P_{2\n} \nonumber\\
&+& 4 P_{1\m} P_{3\n} + 3 P_{2\m} P_{3\n} + P_{3\m} P_{3\n}   \Bigl \} \cdot \nonumber\\
&& H_6^{\m\n}(D4,D_5(0,m_{\chi_0}),D_6(P_1,m_{\chi_0}),D_7(P_2,m_{\chi_0}),D_8(P_3,m_{\chi_0})) \nonumber\\
&+& ( P_1 \cdot P_2 + P_2 \cdot P_3) H_6(D4,D_5(0,m_{\chi_0}),D_6(P_1,m_{\chi_0}),D_7(P_2,m_{\chi_0}),D_8(P_3,m_{\chi_0})) \nonumber\\
&+& \Bigr \{ P_{1\m} P_{1\n}P_{2\a} +P_{1\m} P_{2\n}P_{2\a} +2P_{1\m} P_{1\n}P_{3\a} + 4P_{1\m} P_{2\n}P_{3\a} \nonumber\\
&+& P_{2\m} P_{2\n}P_{3\a} + 2P_{1\m} P_{3\n}P_{3\a} + P_{2\m} P_{3\n}P_{3\a} \Bigl \} \cdot \nonumber\\
&& H_5^{\m\n\a}(D4,D_5(0,m_{\chi_0}),D_6(P_1,m_{\chi_0}),D_7(P_2,m_{\chi_0}),D_8(P_3,m_{\chi_0}))\nonumber\\
&+&\Bigr \{  P_1 \cdot P_2 (P_1 + P_2 +2 P_3 )_\m  + P_2 \cdot P_3 (2P_1 + P_2 + P_3 )_\m \Bigl \} \cdot \nonumber\\
&& H_5^\m(D4,D_5(0,m_{\chi_0}),D_6(P_1,m_{\chi_0}),D_7(P_2,m_{\chi_0}),D_8(P_3,m_{\chi_0}))  \nonumber\\
& +& \Bigr \{ P_{1\m} P_{1\n}P_{2\a}P_{3\b} + P_{1\m} P_{2\n}P_{2\a}P_{3\b} + P_{1\m} P_{1\n}P_{3\a}P_{3\b} + P_{1\m} P_{2\n}P_{3\a}P_{3\b} \Bigl \} \cdot \nonumber\\
&& H_4^{\m\n\a\b}(D4,D_5(0,m_{\chi_0}),D_6(P_1,m_{\chi_0}),D_7(P_2,m_{\chi_0}),D_8(P_3,m_{\chi_0})) \nonumber\\
&+& \Bigr \{ P_1 \cdot P_2 [P_{1\m} P_{1\n}  +2 P_{1\m} P_{3\n} +  P_{2\m} P_{3\n} + P_{3\m} P_{3\n} ] \nonumber\\
&+& P_2 \cdot P_3 [P_{1\m} P_{1\n} + P_{1\m} P_{2\n} + 2 P_{1\m} P_{3\n} + P_{2\m} P_{3\n}  ]   \Bigl \} \cdot \nonumber\\
&& H_4^{\m\n}(D4,D_5(0,m_{\chi_0}),D_6(P_1,m_{\chi_0}),D_7(P_2,m_{\chi_0}),D_8(P_3,m_{\chi_0}))\nonumber\\
&+& ( P_1 \cdot P_2 \times P_2 \cdot P_3 ) H_4(D4,D_5(0,m_{\chi_0}),D_6(P_1,m_{\chi_0}),D_7(P_2,m_{\chi_0}),D_8(P_3,m_{\chi_0}))  \Biggl )          \Biggr \}\nonumber\\
\eea
\bea
( B^{R_\xi, \chi ZZZ}_{H})_f &=& 64  \frac{m_{Z_0}^6 \l_0^2}{m_{H_0}^4} \m^{\ve}\Biggl \{ - C_0(D_1,D_2,D_3) - m_{Z_0}^2 D_0(D_1,D_2,D_3,D_4) \nonumber\\
&+& (1-\xi)  \Biggl ( E_4(D4,D_5(0,m_{\chi_0})) +  E_4(D4,D_5(0,m_{\chi_0})) \nonumber\\
&+&  E_4(D4,D_5(P_1,m_{\chi_0})) +  E_4(D4,D_5(P_1,m_{\chi_0})) \nonumber\\
&+& E_4(D4,D_5(0,m_{\chi_0})) + 2P_{1,\m}E_3^\m(D4,D_5(0,m_{\chi_0})) \nonumber\\
&+& 2P_{2,\m}E_3^\m(D4,D_5(0,m_{\chi_0})) + 2P_{3,\m}E_3^\m(D4,D_5(P_1,m_{\chi_0})) \nonumber\\
&+&2P_{2,\m}E_3^\m(D4,D_5(P_2,m_{\chi_0})) +2P_{3,\m}E_3^\m(D4,D_5(P_1,m_{\chi_0})) \nonumber\\
&+&  P_{1,\m}P_{1,\n} \Bigl \{ E_2^{\m\n}(D4,D_5(0,m_{\chi_0})) +   2P_{1,\m}P_{2,\n}  E_2^{\m\n}(D4,D_5(0,m_{\chi_0}))\nonumber\\
&+&  P_{2,\m}P_{2,\n}  E_2^{\m\n}(D4,D_5(0,m_{\chi_0})) + E_2^{\m\n}(D4,D_5(P_1,m_{\chi_0}))  \nonumber\\
&+&  P_{3,\m}P_{3,\n}  E_2^{\m\n}(D4,D_5(0,m_{\chi_0}))  + 2P_{1,\m}P_{2,\n}  E_2^{\m\n}(D4,D_5(0,m_{\chi_0}),D_6(P_1,m_{\chi_0})) \nonumber\\
&+& 2P_{1,\m}P_{3,\n}  E_2^{\m\n}(D4,D_5(P_1,m_{\chi_0})) \nonumber\\
&+&2 P_1 \cdot P_2 E_2(D4,D_5(P_1,m_{\chi_0})) \nonumber\\
&+&  2 P_1 \cdot P_2 \Bigl \{ P_{1,\m} + P_{2,\m}  \Bigr \}E_1^\m(D4,D_5(P_1,m_{\chi_0})) \nonumber\\
&+&  (P_1 \cdot P_2)^2  E_0(D4,D_5(P_1,m_{\chi_0}))     \Biggr) \nonumber\\
&-& (1-\xi)^2 \Biggl ( F_4(D4,D_5(0,m_{\chi_0}),D_6(P_3,m_{\chi_0})) +  F_4(D4,D_5(0,m_{\chi_0}),D_6(P_3,m_{\chi_0})) \nonumber\\
&+&  F_4(D4,D_5(P_1,m_{\chi_0}),D_6(P_3,m_{\chi_0})) +  F_4(D4,D_5(P_1,m_{\chi_0}),D_6(P_3,m_{\chi_0})) \nonumber\\
&+& F_4(D4,D_5(0,m_{\chi_0}),D_6(P_3,m_{\chi_0})) + 2P_{1,\m}F_3^\m(D4,D_5(0,m_{\chi_0}),D_6(P_3,m_{\chi_0})) \nonumber\\
&+& 2P_{2,\m}F_3^\m(D4,D_5(0,m_{\chi_0}),D_6(P_3,m_{\chi_0})) + 2P_{3,\m}F_3^\m(D4,D_5(P_1,m_{\chi_0})) \nonumber\\
&+&2P_{2,\m}F_3^\m(D4,D_5(P_2,m_{\chi_0}),D_6(P_3,m_{\chi_0})) +2P_{3,\m}F_3^\m(D4,D_5(P_1,m_{\chi_0}),D_6(P_3,m_{\chi_0})) \nonumber\\
&+&  P_{1,\m}P_{1,\n} \Bigl \{ F_2^{\m\n}(D4,D_5(0,m_{\chi_0}),D_6(P_3,m_{\chi_0})) +   2P_{1,\m}P_{2,\n}  F_2^{\m\n}(D4,D_5(0,m_{\chi_0}),D_6(P_3,m_{\chi_0}))\nonumber\\
&+&  P_{2,\m}P_{2,\n}  F_2^{\m\n}(D4,D_5(0,m_{\chi_0}),D_6(P_3,m_{\chi_0})) + F_2^{\m\n}(D4,D_5(P_1,m_{\chi_0}),D_6(P_3,m_{\chi_0}))  \nonumber\\
&+&  P_{3,\m}P_{3,\n}  F_2^{\m\n}(D4,D_5(0,m_{\chi_0}),D_6(P_3,m_{\chi_0}))  + 2P_{1,\m}P_{2,\n}  F_2^{\m\n}(D4,D_5(0,m_{\chi_0}),D_6(P_1,m_{\chi_0})) \nonumber\\
&+& 2P_{1,\m}P_{3,\n}  F_2^{\m\n}(D4,D_5(P_1,m_{\chi_0}),D_6(P_3,m_{\chi_0})) \nonumber\\
&+&2 P_1 \cdot P_2 F_2(D4,D_5(P_1,m_{\chi_0}),D_6(P_3,m_{\chi_0})) \nonumber\\
&+&  2 P_1 \cdot P_2 \Bigl \{ P_{1,\m} + P_{2,\m}  \Bigr \}F_1^\m(D4,D_5(P_1,m_{\chi_0}),D_6(P_3,m_{\chi_0})) \nonumber\\
&+&  (P_1 \cdot P_2)^2  F_0(D4,D_5(P_1,m_{\chi_0}),D_6(P_3,m_{\chi_0}))     \Biggr) \nonumber\\
&+&(1-\xi)^3 \Biggl(  G_6(D4,D_5(0,m_{\chi_0}),D_6(P_1,m_{\chi_0}),D_7(P_2,m_{\chi_0})) \nonumber\\
&+&  G_6(D4,D_5(0,m_{\chi_0}),D_6(P_1,m_{\chi_0}),D_7(P_3,m_{\chi_0})) \nonumber\\
&+&G_6(D4,D_5(P_1,m_{\chi_0}),D_6(P_2,m_{\chi_0}),D_7(P_3,m_{\chi_0})) \nonumber\\
&+&\sum_{l=1}^3 2 P_{l,\m}G^\m_5(D4,D_5(0,m_{\chi_0}),D_6(P_1,m_{\chi_0}),D_7(P_2,m_{\chi_0})) \nonumber\\
&+&\sum_{l=1}^3 2 P_{l,\m}G^\m_5(D4,D_5(0,m_{\chi_0}),D_6(P_1,m_{\chi_0}),D_7(P_3,m_{\chi_0})) \nonumber\\
&+&\sum_{l=1}^3 2 P_{l,\m}G^\m_5(D4,D_5(P_1,m_{\chi_0}),D_6(P_2,m_{\chi_0}),D_7(P_3,m_{\chi_0}))\nonumber\\
&+&\sum_{l,m=1}^3 2  P_{l,\m} P_{m,\n} G^{\m\n}_4(D4,D_5(0,m_{\chi_0}),D_6(P_1,m_{\chi_0}),D_7(P_2,m_{\chi_0})) \nonumber\\
&+& \sum_{l,m=1}^3 2 P_{l,\m} P_{m,\n} G^{\m\n}_4(D4,D_5(0,m_{\chi_0}),D_6(P_1,m_{\chi_0}),D_7(P_3,m_{\chi_0})) \nonumber\\
&+& \sum_{l,m=1}^3 2  P_{l,\m} P_{m,\n} G^{\m\n}_4(D4,D_5(P_1,m_{\chi_0}),D_6(P_2,m_{\chi_0}),D_7(P_3,m_{\chi_0}))\nonumber\\
&+&\sum_{l,m,n=1}^3  P_{l,\m} P_{m,\n} P_{n,\a} G^{\m\n\a}_3(D4,D_5(0,m_{\chi_0}),D_6(P_1,m_{\chi_0}),D_7(P_2,m_{\chi_0})) \nonumber\\
&+& \sum_{l,m,n=1}^3  P_{l,\m} P_{m,\n} P_{n,\a} G^{\m\n\a}_3(D4,D_5(0,m_{\chi_0}),D_6(P_1,m_{\chi_0}),D_7(P_3,m_{\chi_0})) \nonumber\\
&+&\sum_{l,m,n=1}^3  P_{l,\m} P_{m,\n} P_{n,\a} G^{\m\n\a}_3(D4,D_5(P_1,m_{\chi_0}),D_6(P_2,m_{\chi_0}),D_7(P_3,m_{\chi_0}))\nonumber\\
&+&\sum_{l,m=1, m \ne l }^3  P_l \cdot P_m G_4(D4,D_5(0,m_{\chi_0}),D_6(P_1,m_{\chi_0}),D_7(P_3,m_{\chi_0})) \nonumber\\
&+&\sum_{l,m=1, m \ne l }^3  P_l \cdot P_m G_4(D4,D_5(0,m_{\chi_0}),D_6(P_2,m_{\chi_0}),D_7(P_3,m_{\chi_0})) \nonumber\\
&+&\sum_{l=1}^3  P_{l,\m}G^\m_3(D4,D_5(0,m_{\chi_0}),D_6(P_1,m_{\chi_0}),D_7(P_2,m_{\chi_0})) \nonumber\\
&+&\sum_{l=1}^3  P_{l,\m}G^\m_3(D4,D_5(0,m_{\chi_0}),D_6(P_1,m_{\chi_0}),D_7(P_3,m_{\chi_0})) \nonumber\\
&+&\sum_{l=1}^3  P_{l,\m}G^\m_3(D4,D_5(0,m_{\chi_0}),D_6(P_2,m_{\chi_0}),D_7(P_3,m_{\chi_0})) \nonumber\\
&+&\sum_{l,m=1}^3  P_l \cdot P_m P_{m,\m}G^\m_3(D4,D_5(0,m_{\chi_0}),D_6(P_1,m_{\chi_0}),D_7(P_3,m_{\chi_0})) \nonumber\\
&+&\sum_{l,m=1}^3  P_l \cdot P_m P_{m,\m}G^\m_3(D4,D_5(0,m_{\chi_0}),D_6(P_2,m_{\chi_0}),D_7(P_3,m_{\chi_0})) \nonumber\\
&+& \sum_{l,m=1, m \ne l }^3 P_l \cdot P_m P_{l,\m} P_{m,\n}  G^{\m\n}_2(D4,D_5(0,m_{\chi_0}),D_6(P_1,m_{\chi_0}),D_7(P_3,m_{\chi_0})) \nonumber\\
&+& \sum_{l,m=1, m \ne l }^3 P_l \cdot P_m P_{l,\m} P_{m,\n}  G^{\m\n}_2(D4,D_5(P_1,m_{\chi_0}),D_6(P_2,m_{\chi_0}),D_7(P_3,m_{\chi_0}))\nonumber\\
&+& \sum_{l,m=1, m \ne l }^3 P_l \cdot P_m P_{l,\m} P_{m,\n}  G^{\m\n}_2(D4,D_5(0,m_{\chi_0}),D_6(P_1,m_{\chi_0}),D_7(P_3,m_{\chi_0})) \nonumber\\
&& \Bigl ( P_1 \cdot P_2 P_2 \cdot P_3 + P_2 \cdot P_3 P_1 \cdot P_3   \Bigr)G_2(D4,D_5(0,m_{\chi_0}),D_6(P_1,m_{\chi_0}),D_7(P_3,m_{\chi_0})) \nonumber\\
&+& \sum_{l,m,n=1, n \ne m \ne l }^3 P_l \cdot P_m P_m \cdot P_n ( P_{m} + P_{n})_\m G^{\m}_1(D4,D_5(P_1,m_{\chi_0}),D_6(P_2,m_{\chi_0}),D_7(P_3,m_{\chi_0}))\nonumber\\
&+& \sum_{l,m,n=1, n \ne m \ne l }^3 P_l \cdot P_m P_m \cdot P_n ( P_{m} + P_{n})_\m  G^{\m}_1(D4,D_5(0,m_{\chi_0}),D_6(P_1,m_{\chi_0}),D_7(P_3,m_{\chi_0})) \nonumber\\
&+& \sum_{l,m=1, m \ne l }^3 P_l \cdot P_m P_{l,\m} P_{m,\n}  G^{\m\n}_2(D4,D_5(P_1,m_{\chi_0}),D_6(P_2,m_{\chi_0}),D_7(P_3,m_{\chi_0}))\nonumber\\
&+& P_1 \cdot P_2 P_2 \cdot P_3  P_1 \cdot P_3 G_0(D4,D_5(P_1,m_{\chi_0}),D_6(P_2,m_{\chi_0}),D_7(P_3,m_{\chi_0})) \Biggr)         \Biggr \}
 \eea
\bea
( B^{R_\xi, ZZ \chi \chi }_{H})_{f,1} = 64  \frac{m_{Z_0}^4 \l_0^2}{m_{H_0}^4} \m^{\ve}\Biggl \{ 2 P_{2,\m} g_{\n\a} D^{\m\n\a}(D1,D2,D3,D4) + P_{2,\m} P_{2,\n}  D^{\m\n}(D1,D2,D3,D4) \Biggr \}\nonumber\\
 \eea
\bea
( B^{R_\xi, ZZ \chi \chi }_{H})_{f,2} &=& 64  \frac{m_{Z_0}^4 \l_0^2}{m_{H_0}^4} \m^{\ve}\Biggl \{ - 2P_{3,\m} E^\m_5(D4,D_5(P_1,m_{\chi_0})) \nonumber\\
&+&  (m^2_{\chi_0} - P_{1}^2)E_6(D4,D_5(P_1,m_{\chi_0})) +m^2_{\chi_0}E_6(D4,D_5(0,m_{\chi_0})) \nonumber\\
&+&  E_4(D4,D_5(P_1,m_{\chi_0})) +E_4(D4,D_5(0,m_{\chi_0}))  \nonumber\\
&+& 2P_{2,\m}E_3^\m(D4,D_5(P_1,m_{\chi_0})) + 2P_{1,\m}E_3^\m(D4,D_5(0,m_{\chi_0})) \nonumber\\
&+&  P_{1,\m}P_{1,\n}  E_2^{\m\n}(D4,D_5(0,m_{\chi_0})) +   2P_{1,\m}P_{2,\n}  E_2^{\m\n}(D4,D_5(0,m_{\chi_0}))\nonumber\\
&+&  P_{2,\m}P_{2,\n}  \Biggl \{  E_2^{\m\n}(D4,D_5(0,m_{\chi_0})) + E_2^{\m\n}(D4,D_5(P_1,m_{\chi_0})) \Biggr \} \nonumber\\
&+&  P_{3,\m}P_{3,\n}  E_2^{\m\n}(D4,D_5(0,m_{\chi_0}))  + 2P_{1,\m}P_{2,\n}  E_2^{\m\n}(D4,D_5(P_1,m_{\chi_0})) \nonumber\\
&+& 2P_{1,\m}P_{3,\n}  E_2^{\m\n}(D4,D_5(P_1,m_{\chi_0})) + 2 P_1 \cdot P_2 E_2(D4,D_5(P_1,m_{\chi_0})) \nonumber\\
&+&  2 P_1 \cdot P_2 \Bigl \{ P_{1,\m} + P_{2,\m}  \Bigr \}E_1^\m(D4,D_5(P_1,m_{\chi_0}))  \Biggr \} 
\eea
\bea
( B^{R_\xi, ZZ \chi \chi }_{H})_{f,3} &=& 64  \frac{m_{Z_0}^4 \l_0^2}{m_{H_0}^4} \m^{\ve}\Biggl \{  F_4(D4,D_5(0,m_{\chi_0}),D_6(0,m_{\chi_0})) \nonumber\\
&+&  F_4(D4,D_5(0,m_{\chi_0}),D_6(P_2,m_{\chi_0})) \nonumber\\
&+&  F_4(D4,D_5(P_1,m_{\chi_0}),D_6(P_2,m_{\chi_0})) +  F_4(D4,D_5(P_1,m_{\chi_0}),D_6(P_2,m_{\chi_0})) \nonumber\\
&+& F_4(D4,D_5(0,m_{\chi_0}),D_6(P_2,m_{\chi_0})) + 2P_{1,\m}F_3^\m(D4,D_5(0,m_{\chi_0}),D_6(P_2,m_{\chi_0})) \nonumber\\
&+& 2P_{2,\m}F_3^\m(D4,D_5(0,m_{\chi_0}),D_6(P_2,m_{\chi_0})) \nonumber\\
&+& 2P_{2,\m}F_3^\m(D4,D_5(0,m_{\chi_0}),D_6(P_2,m_{\chi_0})) \nonumber\\
&+&2P_{2,\m}F_3^\m(D4,D_5(P_2,m_{\chi_0}),D_6(P_2,m_{\chi_0})) \nonumber\\
&+&2P_{2,\m}F_3^\m(D4,D_5(P_1,m_{\chi_0}),D_6(P_2,m_{\chi_0})) \nonumber\\
&+&  P_{1,\m}P_{1,\n}  F_2^{\m\n}(D4,D_5(0,m_{\chi_0}),D_6(P_2,m_{\chi_0})) \nonumber\\
&+&   2P_{1,\m}P_{2,\n}  F_2^{\m\n}(D4,D_5(0,m_{\chi_0}),D_6(P_2,m_{\chi_0}))\nonumber\\
&+&  P_{2,\m}P_{2,\n}  F_2^{\m\n}(D4,D_5(0,m_{\chi_0}),D_6(P_2,m_{\chi_0}))  \nonumber\\
&+&  P_{2,\m}P_{2,\n}  F_2^{\m\n}(D4,D_5(0,m_{\chi_0}),D_6(P_2,m_{\chi_0}))  \nonumber\\
&+& 2P_{1,\m}P_{2,\n}  F_2^{\m\n}(D4,D_5(0,m_{\chi_0}),D_6(P_1,m_{\chi_0})) \nonumber\\
&+& 2P_{1,\m}P_{2,\n}  F_2^{\m\n}(D4,D_5(P_1,m_{\chi_0}),D_6(P_2,m_{\chi_0})) \nonumber\\
&+&2 P_1 \cdot P_2 F_2(D4,D_5(P_1,m_{\chi_0}),D_6(P_2,m_{\chi_0})) \nonumber\\
&+&  2 P_1 \cdot P_2 \Bigl \{ P_{1,\m} + P_{2,\m}  \Bigr \}F_1^\m(D4,D_5(P_1,m_{\chi_0}),D_6(P_2,m_{\chi_0})) \nonumber\\
&+&  (P_1 \cdot P_2)^2  F_0(D4,D_5(P_1,m_{\chi_0}),D_6(P_2,m_{\chi_0})) \Biggr \}
 \eea
\bea
( B^{R_\xi, \chi \chi \chi Z}_{H})_{f,1} &=& 64  \frac{m_{Z_0}^2 \l_0^2}{m_{H_0}^4} \m^{\ve}\Bigl \{ g_{\m\n} D^{\m\n}(P_1,P_2,m_{Z_0},m_{Z_0},m_{Z_0},m_{\chi_0})  \nonumber\\
&+& g_{\m\n} D^{\m\n}(P_1,P_2,m_{Z_0},m_{Z_0},m_{Z_0},m_{\chi_0})  \nonumber\\
&+& g_{\m\n} D^{\m\n}(P_1,P_2,m_{Z_0},m_{Z_0},m_{Z_0},m_{\chi_0}) + 2 P_1^\m D^{\m}(P_1,P_2,m_{Z_0},m_{Z_0},m_{Z_0},m_{\chi_0})  \nonumber\\
&+&2 P_2^\m D^{\m}(P_1,P_2,m_{Z_0},m_{Z_0},m_{Z_0},m_{\chi_0})  +  P_1^2 D_0(P_1,P_2,m_{Z_0},m_{Z_0},m_{Z_0},m_{\chi_0}) \nonumber\\
&+&P_2^2 D_0 (P_1,P_2,m_{Z_0},m_{Z_0},m_{Z_0},m_{\chi_0})     \Bigr \}
 \eea
\bea
( B^{R_\xi, \chi \chi \chi Z}_{H})_{f,2} &=&64  \frac{m_{Z_0}^2 \l_0^2}{m_{H_0}^4} \m^{\ve} \Biggl \{ E_4(D4,D_5(0,m_{\chi_0})) + E_4(D4,D_5(P_1,m_{\chi_0})) \nonumber\\
&+& E_4(D_1,D_2,D_3,D_4(0,m_{\chi_0}),D_5(P_1,m_{\chi_0}))\nonumber\\
&+& 2 P_{1\m} E_3^\m(D4,D_5(0,m_{\chi_0})) + 2 P_{1\m} E_3^\m(D_1,D_2,D_3,D_4(0,m_{\chi_0}),D_5(P_1,m_{\chi_0})) \nonumber\\
&+&2 P_{2\m} E_3^\m(D4,D_5(P_1,m_{\chi_0})) + 2 P_{2\m} E_3^\m(D_1,D_2,D_3,D_4(0,m_{\chi_0}),D_5(P_1,m_{\chi_0})) \nonumber\\
&+&2P_{1\m}P_{2\n} E_2^{\m\n}(D_1,D_2,D_3,D_4(0,m_{\chi_0}),D_5(P_1,m_{\chi_0})) \nonumber\\
&+& P_{2\m}P_{2\n}E_2^{\m\n}(D4,D_5(P_1,m_{\chi_0})) \nonumber\\
&+&P_{2\m}P_{2\n}E_2^{\m\n}(D_1,D_2,D_3,D_4(0,m_{\chi_0}),D_5(P_1,m_{\chi_0})) \nonumber\\
&+& 2P_1 \cdot P_2 E_2(D_1,D_2,D_3,D_4(0,m_{\chi_0}),D_5(P_1,m_{\chi_0})) \nonumber\\
&+& 2P_1 \cdot P_2 (P_1+P_2)_\m E_1^{\m}(D_1,D_2,D_3,D_4(0,m_{\chi_0}),D_5(P_1,m_{\chi_0}))  \nonumber\\
&+&  (P_1 \cdot P_2)^2E_0(D_1,D_2,D_3,D_4(0,m_{\chi_0}),D_5(P_1,m_{\chi_0}))  \Biggr \}
 \eea

\subsection{Unitary gauge}

In the Unitary gauge the finite parts are:

Finite parts of the Box diagrams 
\bea
( B^{U, ZZZZ }_{H})_{f,1} &=& 64  \frac{m_{Z_0}^4 \l_0^2}{m_{H_0}^4} \m^{\ve}\Biggl \{ 6 g_{\n\a} ( P_1 + P_2 + P_3 )_\m D^{\m\n\a}(P_1,P_2,m_{Z_0},m_{Z_0},m_{Z_0},m_{\chi_0}) \nonumber\\
&+&  \Bigl ( 3 P_1^\m P_1^\n  +2 P_1^\m P_2^\n + 3P_2^\m P_2^\n \nonumber\\
&+& 2 P_1^\m P_3^\n + 2P_2^\m P_3^\n + 3  P_3^\m P_3^\n \nonumber\\
&+& 2( P_1 \cdot P_2 + P_1 \cdot P_2 +P_1 \cdot P_2 )g_{\m\n} \Bigr ) D^{\m\n}(P_1,P_2,m_{Z_0},m_{Z_0},m_{Z_0},m_{\chi_0}) \nonumber\\
&+& \Bigl ( 2 P_1\cdot P_2 (P_1^\m + P_2^\m ) +  2 P_1\cdot P_2 (P_1^\m + P_2^\m ) \nonumber\\
&+&  2 P_1\cdot P_2 (P_1^\m + P_2^\m ) \Bigr ) D^{\m\n}(P_1,P_2,m_{Z_0},m_{Z_0},m_{Z_0},m_{\chi_0}) \nonumber\\
&+&  ((P_1\cdot P_2)^2 + (P_1\cdot P_3)^2 +(P_2\cdot P_3)^2 )D_0(P_1,P_2,m_{Z_0},m_{Z_0},m_{Z_0},m_{\chi_0})    \Biggr \}\nonumber\\
\eea
\bea
( B^{U, ZZZZ }_{H})_{f,2} &=& 64  \frac{m_{Z_0}^2 \l_0^2}{m_{H_0}^4} \m^{\ve}\Biggl \{ \Bigl ( 2 \sum_{l,m,n =1}^3 P_{l,\m} P_{m,\n} P_{n,\a}  \nonumber\\
&+& \sum_{l,m,n =1,n \ne m \ne l }^3 2 (P_l \cdot P_m \nonumber\\
&+& P_l \cdot P_n + P_n \cdot P_m) (P_l +P_m + P_n)_\m      \Bigr )D^{\m\n\a}(P_1,P_2,m_{Z_0},m_{Z_0},m_{Z_0},m_{\chi_0})  \nonumber\\
&+& P_1 \cdot P_2 \sum_{l,m =1} P_{l,\m} P_{m,\n} D^{\m\n}(P_1,P_2,m_{Z_0},m_{Z_0},m_{Z_0},m_{\chi_0}) \nonumber\\
&+& P_2 \cdot P_3 \sum_{l,m =1} P_{l,\m} P_{m,\n} D^{\m\n}(P_1,P_2,m_{Z_0},m_{Z_0},m_{Z_0},m_{\chi_0}) \nonumber\\  
&+& \sum_{l,m,n =1,n \ne m \ne l } P_{l} \cdot P_{m} P_{l} \cdot P_{n} g_{\m\n}D^{\m\n}(P_1,P_2,m_{Z_0},m_{Z_0},m_{Z_0},m_{\chi_0}) \nonumber\\  
&+& ( P_1 \cdot P_2 P_2 \cdot P_3 (P_1 +P_3)_\m \nonumber\\
&+&P_1 \cdot P_3 P_2 \cdot P_3 (P_1 +P_3)_\m  )D^{\m}(P_1,P_2,m_{Z_0},m_{Z_0},m_{Z_0},m_{\chi_0}) \nonumber\\  
&+&  P_1\cdot P_2 P_1\cdot P_3 P_2\cdot P_3 D_0(P_1,P_2,m_{Z_0},m_{Z_0},m_{Z_0},m_{\chi_0})   \Biggr \}
\eea
\bea
( B^{U, ZZZZ }_{H})_{f,3} &=& 64  \frac{ \l_0^2}{m_{H_0}^4} \m^{\ve}\Biggl \{ \Bigl ( P_1\cdot P_2 (P_{1,\m} P_{2,\n} P_{3,\a} + P_{1,\m} P_{3,\n} P_{3,\a}  ) + P_2\cdot P_3 (P_{1,\m} P_{1,\n} P_{3,\a}  \nonumber\\
&+& P_{1,\m} P_{2,\n} P_{3,\a}  ) + P_1\cdot P_2 g_{\m\n} (P_1 + P_3 )_\m  \Bigr )D^{\m\n\a}(P_1,P_2,m_{Z_0},m_{Z_0},m_{Z_0},m_{\chi_0})  \nonumber\\
&+&  P_1\cdot P_2 P_2\cdot P_3 P_{1,\m} P_{3,\n} D^{\m\n}(P_1,P_2,m_{Z_0},m_{Z_0},m_{Z_0},m_{\chi_0})    \Biggr \}
\eea

\section{On-shell results}
\label{Onshell}

For completeness, in this Appendix we demonstrate the results for the divergent parts of the two- three- and four-point one-loop functions
calculated on-shell, in both $R_\xi$ and Unitary gauges. 
We start with the case of $R_\xi$ gauge where we will demonstrate the results for the two-, 
three- and four-point functions calculated at on-shell, i.e. at $p^2 = m_Z^2$ for the $Z$-mass and at $p_i^2 = m_H^2$ for everything else. 
Tadpoles are external momentum independent objects.
\begin{itemize} 
\item Vacuum polarization of the $Z$-boson
\begin{center}
\begin{tikzpicture}[scale=0.9]
\draw [thick] [fill=gray] (0,0) circle [radius=0.9];
\draw [photon] (-2.5,0)--(-0.9,0);
\draw [photon] (2.5,0)--(0.9,0);
\node at (7.8,0) {$=\,\,  [{\cal M}_Z^{R_\xi} ]_{\ve} + \left\{ {\cal M}_Z^{R_\xi} \right\}_{\ve} +[ {\cal M}_Z^{R_\xi}]_{f} + \left\{ {\cal M}_Z^{R_\xi} \right\}_{f} $ };
\end{tikzpicture}
\end{center}
with 
\be\label{MZRxionshell}
\ve [{\cal M}_Z^{R_\xi} ]_{\ve}  = \frac{40}{3} \frac{\l_0 m_{Z_0}^4}{m_{H_0}^2}, \hskip 1cm \left\{ {\cal M}_Z^{R_\xi} \right\}_{\ve} =0\, 
\ee
and 
\be\label{dARxionshell}
\ve [\d A^{R_\xi} ]_{\ve}  = \frac{4}{3} \l_0 \frac{m_{Z_0}^2}{m_{H_0}^2},  \hskip 1cm \left\{ \d A^{R_\xi} \right\}_{\ve} = 0\, .
\ee
\item One-loop corrections to the Higgs propagator
\begin{center}
\begin{tikzpicture}[scale=0.9]
\draw [thick] [fill=gray] (0,0) circle [radius=0.9];
\draw [dashed,thick] (-2.5,0)--(-0.9,0);
\draw [dashed,thick] (2.5,0)--(0.9,0);
\node at (7.8,0) {$=\,\,  [{\cal M}^{R_\xi }_H ]_{\ve} + \left\{ {\cal M}^{R_\xi }_H \right\}_{\ve} +[ {\cal M}^{R_\xi }_H ]_{f} + \left\{ {\cal M}^{R_\xi }_H \right\}_{f}  $ };
\end{tikzpicture}
\end{center}
with 
\be\label{MHRxionshell}
\ve [{\cal M}^{R_\xi }_H ]_{\ve} = 26 \l_0 m_{H_0}^2 -12 \l_0 m_{Z_0}^2 +36 \frac{\l_0 m_{Z_0}^4}{m_{H_0}^2}, \hskip 1cm 
\ve \left\{ {\cal M}^{R_\xi }_H \right\}_{\ve} =  12 \frac{\l_0 m_{Z_0}^4}{m_{H_0}^2} \xi^2
\ee
and 
\be\label{dphiRxionshell}
\ve [\d \phi^{R_\xi} ]_{\ve}  = -4\l_0 + 12 \l_0 \frac{m_{Z_0}^2}{m_{H_0}^2}, \hskip 1cm \left\{ \d \phi^{R_\xi} \right\}_{\ve} = 0
\ee
\item One-loop corrections to the Higgs three-point vertex
\vskip .5cm
\begin{center}
\begin{tikzpicture}
\draw [dashed,thick] (0.8,0.5)--(2.3,1.3);
\draw [dashed,thick] (0.8,-0.5)--(2.3,-1.3);
\draw [dashed,thick] (-2.5,0)--(-0.9,0);
\draw [thick] [fill=gray] (0,0) circle [radius=0.9];
\node at (6,0) {$=\,\,   [{\cal K}^{R_\xi}_H ]_{\ve} + \left\{{\cal K}^{R_\xi}_H \right\}_{\ve} +[ {\cal K}^{R_\xi}_H ]_{f} + \left\{ {\cal K}^{R_\xi}_H  \right\}_{f} $ };
\end{tikzpicture}
\end{center}
\vskip .5cm
where 
\bea\label{KHRxionshell}
\ve [{\cal K}^{R_\xi}_H ]_{\varepsilon} &=& \ve [{\cal K}^{R_\xi,{\rm red.}}_{H} + {\cal K}^{R_\xi,{\rm irred.}}_{R_\xi} ]_{\ve}  =  
\frac{m_{H_0}}{\sqrt{2 \l_0}}  \left( 84  \l_0^2 + 144 \frac{ \l_0^2 m_{Z_0}^4}{m_{H_0}^4} \right) \nonumber\\
\left\{ {\cal K}^{R_\xi}_H \right\}_{\ve} &=& \left\{ {\cal K}^{R_\xi,{\rm red.}}_{H} + {\cal K}^{R_\xi,{\rm irred.}}_{H} \right\}_{\ve} = 0.
\eea
\item One-loop corrections to Higgs quartic coupling
\vskip .5cm
\begin{center}
\begin{tikzpicture}
\draw [dashed,thick] (0.8,0.5)--(2.3,1.3);
\draw [dashed,thick] (0.8,-0.5)--(2.3,-1.3);
\draw [dashed,thick] (-0.8,0.5)--(-2.3,1.3);
\draw [dashed,thick] (-0.8,-0.5)--(-2.3,-1.3);
\draw [thick] [fill=gray] (0,0) circle [radius=0.9];
\node at (6,0) {$=\,\,   [  {\cal B}^{R_\xi}_{H}  ]_{\ve} + \left\{  {\cal B}^{R_\xi}_{H}  \right\}_{\ve} + [ {\cal B}^{R_\xi}_{H}  ]_{f} + \left\{  {\cal B}^{R_\xi}_{H} \right\}_{f} $ };
\end{tikzpicture}
\end{center}
\vskip .5cm
with 
\bea\label{BHRxionshell}
\ve [{\cal B}^{R_\xi}_H ]_{\ve} &=& 252 \l_0^2 + 144 \frac{ \l_0^2  m^4_{Z_0}}{m^4_{H_0}}, \hskip 1cm \left\{  {\cal B}^{R_\xi}_H  \right\}_{\ve} = 0.
\eea
\end{itemize}
Next we list on-shell results for the Unitary gauge.
\begin{itemize} 
\item Vacuum polarization of the $Z$-boson
\vskip .5cm
\begin{center}
\begin{tikzpicture}[scale=0.9]
\draw [thick] [fill=gray] (0,0) circle [radius=0.9];
\draw [photon] (-2.5,0)--(-0.9,0);
\draw [photon] (2.5,0)--(0.9,0);
\node at (7.8,0) {$=\,\,  [{\cal M}^U_{Z} ]_{\ve} + \left\{ {\cal M}^U_{Z} \right\}_{\ve} +[ {\cal M}^U_{Z}]_{f} + \left\{ {\cal M}^U_{Z} \right\}_{f} $ };
\end{tikzpicture}
\end{center}
\vskip .5cm
with 
\be\label{MZRxionshell}
\ve [{\cal M}^U_{Z} ]_{\ve} = \frac{40}{3} \frac{\l_0 m_{Z_0}^4}{m_{H_0}^2}, \hskip 1cm \left\{ {\cal M}_U^{Z} \right\}_{\ve} = 0
\ee
and 
\be\label{dARxionshell}
\ve [\d A^{U} ]_{\ve}  = \frac{4}{3} \l_0 \frac{m_{Z_0}^2}{m_{H_0}^2}, \hskip 1cm \left\{ \d A^{U} \right\}_{\ve} = 0
\ee
\item One-loop corrections to the Higgs propagator
\vskip .5cm
\begin{center}
\begin{tikzpicture}[scale=0.9]
\draw [thick] [fill=gray] (0,0) circle [radius=0.9];
\draw [dashed,thick] (-2.5,0)--(-0.9,0);
\draw [dashed,thick] (2.5,0)--(0.9,0);
\node at (7.8,0) {$=\,\,  [{\cal M}^{U}_H ]_{\ve} + \left\{ {\cal M}^{U }_H \right\}_{\ve} +[ {\cal M}^{U }_H ]_{f} + \left\{ {\cal M}^{U }_H \right\}_{f}  $ };
\end{tikzpicture}
\end{center}
\vskip .5cm
with 
\be\label{MHRxionshell}
\ve [{\cal M}^{U }_H ]_{\ve} = 26 \l_0 m_{H_0}^2 -12 \l_0 m_{Z_0}^2 +36 \frac{\l_0 m_{Z_0}^4}{m_{H_0}^2}, \hskip 1cm \left\{ {\cal M}^{U }_H \right\}_{\ve} = 12 \frac{\l_0 m_{Z_0}^4}{m_{H_0}^2} \xi^2
\ee
and 
\be\label{dphiRxionshell}
\ve [\d \phi^{U} ]_{\ve}  = -4\l_0 + 12 \l_0 \frac{m_{Z_0}^2}{m_{H_0}^2}, \hskip 1cm \left\{ \d \phi^{U} \right\}_{\ve} = 0
\ee
\item One-loop corrections to the Higgs three-point vertex
\vskip .5cm
\begin{center}
\begin{tikzpicture}
\draw [dashed,thick] (0.8,0.5)--(2.3,1.3);
\draw [dashed,thick] (0.8,-0.5)--(2.3,-1.3);
\draw [dashed,thick] (-2.5,0)--(-0.9,0);
\draw [thick] [fill=gray] (0,0) circle [radius=0.9];
\node at (6,0) {$=\,\,   [{\cal K}^{U}_H ]_{\ve} + \left\{{\cal K}^{U}_H \right\}_{\ve} +[ {\cal K}^{U}_H ]_{f} + \left\{ {\cal K}^{U}_H  \right\}_{f} $ };
\end{tikzpicture}
\end{center}
\vskip .5cm
where 
\bea\label{KHRxionshell}
\ve [{\cal K}^{U}_H ]_{\varepsilon} &=& \ve [{\cal K}^{U,{\rm red.}}_{H} + {\cal K}^{U,{\rm irred.}}_{H} ]_{\ve}  =  \frac{m_{H_0}}{\sqrt{2 \l_0}}  \left( 84  \l_0^2 + 144 \frac{ \l_0^2 m_{Z_0}^4}{m_{H_0}^4} \right) \nonumber\\
\left\{ {\cal K}^{U}_H \right\}_{\ve} &=& \left\{ {\cal K}^{U,{\rm red.}}_{H} + {\cal K}^{U,{\rm irred.}}_{H, U} \right\}_{\ve} = 0.
\eea

\item One-loop corrections to Higgs quartic coupling
\vskip .5cm
\begin{center}
\begin{tikzpicture}
\draw [dashed,thick] (0.8,0.5)--(2.3,1.3);
\draw [dashed,thick] (0.8,-0.5)--(2.3,-1.3);
\draw [dashed,thick] (-0.8,0.5)--(-2.3,1.3);
\draw [dashed,thick] (-0.8,-0.5)--(-2.3,-1.3);
\draw [thick] [fill=gray] (0,0) circle [radius=0.9];
\node at (6,0) {$=\,\,   [  {\cal B}^U_{H}  ]_{\ve} + \left\{  {\cal B}^U_{H}  \right\}_{\ve} + [ {\cal B}^U_{H}  ]_{f} + \left\{  {\cal B}^U_{H}\right\}_{f} $ };
\end{tikzpicture}
\end{center}
\vskip .5cm
with 
\be\label{BHRxionshell}
\ve [{\cal B}^{U}_H ]_{\ve} = 252 \l_0^2 + 144 \frac{ \l_0^2  m^4_{Z_0}}{m^4_{H_0}}, \hskip 1cm \left\{  {\cal B}^{U}_H  \right\}_{\ve} = 0\, .
\ee
\end{itemize}
We observe that the results of the one-loop diagrams in $R_\xi$ and in the Unitary gauge are the same. 
\end{appendices}

\end{document}